\documentclass{article}
\usepackage{jheppub,esint,shuffle,psfrag}
\usepackage[utf8]{inputenc}

\usepackage{varioref}
\usepackage{amsmath,amsfonts,amsthm,mathrsfs}
\usepackage{enumerate}
\usepackage{fancyvrb}
\usepackage{verbatim}
\usepackage{wrapfig}
\usepackage{appendix}
\usepackage{amstext}
\usepackage{amssymb}
\usepackage{graphicx}
\usepackage{color}
\usepackage{varioref}
\usepackage{multirow,graphics}
\usepackage{epstopdf}
\usepackage{bbm}
\usepackage{slashed}
\usepackage{relsize}

\numberwithin{equation}{section}

\usepackage{tikz}
\usetikzlibrary{plotmarks,calc,decorations, decorations.pathmorphing, patterns}
\usetikzlibrary{arrows}
\tikzstyle dynkin node=[very thick,shape=circle,draw,inner sep=0pt,minimum size=5mm]
\tikzstyle dynkin line=[very thick]
\tikzstyle inverse line=[gray,line width=1.46pt,line cap=round, dash pattern=on 0pt off 2\pgflinewidth]
\tikzstyle red phase=[red,decoration={snake,amplitude=0.1mm,segment length=1.6mm},decorate]
\tikzstyle blue phase=[blue,decoration={snake,amplitude=0.1mm,segment length=0.9mm},decorate]
\tikzstyle green phase=[green,decoration={snake,amplitude=0.1mm,segment length=0.9mm},decorate]
\tikzstyle brown phase=[brown,decoration={snake,amplitude=0.1mm,segment length=0.9mm},decorate]
\newcommand{\boundellipse}[3]
{(#1) ellipse (#2 and #3)
}
\usetikzlibrary{decorations.pathmorphing}
\tikzstyle arrow=[thick,rounded corners=18pt,-latex]
\tikzstyle box=[draw,rounded corners,outer sep=4pt]
\tikzstyle B node=[outer sep=0pt]
\tikzstyle Q node=[inner sep=1pt,outer sep=0pt]
\definecolor{purple_nice}{rgb}{0.4,0.2,0.7}
\definecolor{fuel_blue}{RGB}{42,162,185}

\def\II{\hbox{{1}\kern-.25em\hbox{l}}}

\def\<{\langle}
\def\>{\rangle}

\newcommand{\p}{\partial}
\newcommand{\tr}{{\text{tr}}}

\newcommand{\sig}{\boldsymbol{\sigma}}
\newcommand{\bsig}{\boldsymbol{\bar \sigma}}

\makeatletter
\@ifundefined{usebibtex}{} {}
\makeatother

\def\Tr{\text{Tr}~}

\title{
\Large Exactly solvable single-trace four point correlators in $\chi$CFT$_4$}

\author[a]{Sergey Derkachov,} 
\author[b]{~Enrico Olivucci}
\emailAdd{derkach@pdmi.ras.ru, enrico.olivucci@desy.de}

\affiliation[a]{St. Petersburg Department of the Steklov Mathematical Institute
of Russian Academy of Sciences,
Fontanka 27, 191023 St. Petersburg, Russia}
\affiliation[b]{II. Institut f\"ur Theoretische Physik, Universit\"at Hamburg, Luruper Chaussee 149, 22761
Hamburg, Germany}

\abstract{
  In this paper we study a wide class of planar single-trace four point correlators in the chiral conformal field theory ($\chi$CFT$_4$) arising as a double scaling limit of the $\gamma$-deformed $\mathcal{N}=4$ SYM theory. In the planar (t'Hooft) limit, each of such correlators is described by a single Feynman integral having the bulk topology of a square lattice ``fishnet" and/or of an honeycomb lattice of Yukawa vertices. The computation of this class of Feynmann integrals at any loop is achieved by means of an exactly-solvable spin chain magnet with $SO(1,5)$ symmetry. In this paper we explain in detail the solution of the magnet model as presented in our recent letter and we obtain a general formula for the representation of the Feynman integrals over the spectrum of the separated variables of the magnet, for any number of scalar and fermionic fields in the corresponding correlator. For the particular choice of scalar fields only, our formula reproduces the conjecture of B.~Basso and L.~Dixon for the fishnet integrals.
}

\usepackage{esint}
\usepackage{breqn}
\def \Tr {\mathop{\rm Tr}\nolimits}
\def \tr {\mathop{\rm tr}\nolimits}

\def\numberbysection{\@addtoreset{equation}{section}
                     \def\theequation{\thesection.\arabic{equation}}}

\begin{document}

\hskip9cm\preprint{ZMP-HH/20-15}
\maketitle

\flushbottom

\newpage

\section{Introduction}\label{intro}
\begin{figure}
 \begin{center}
 \includegraphics[scale=0.27]{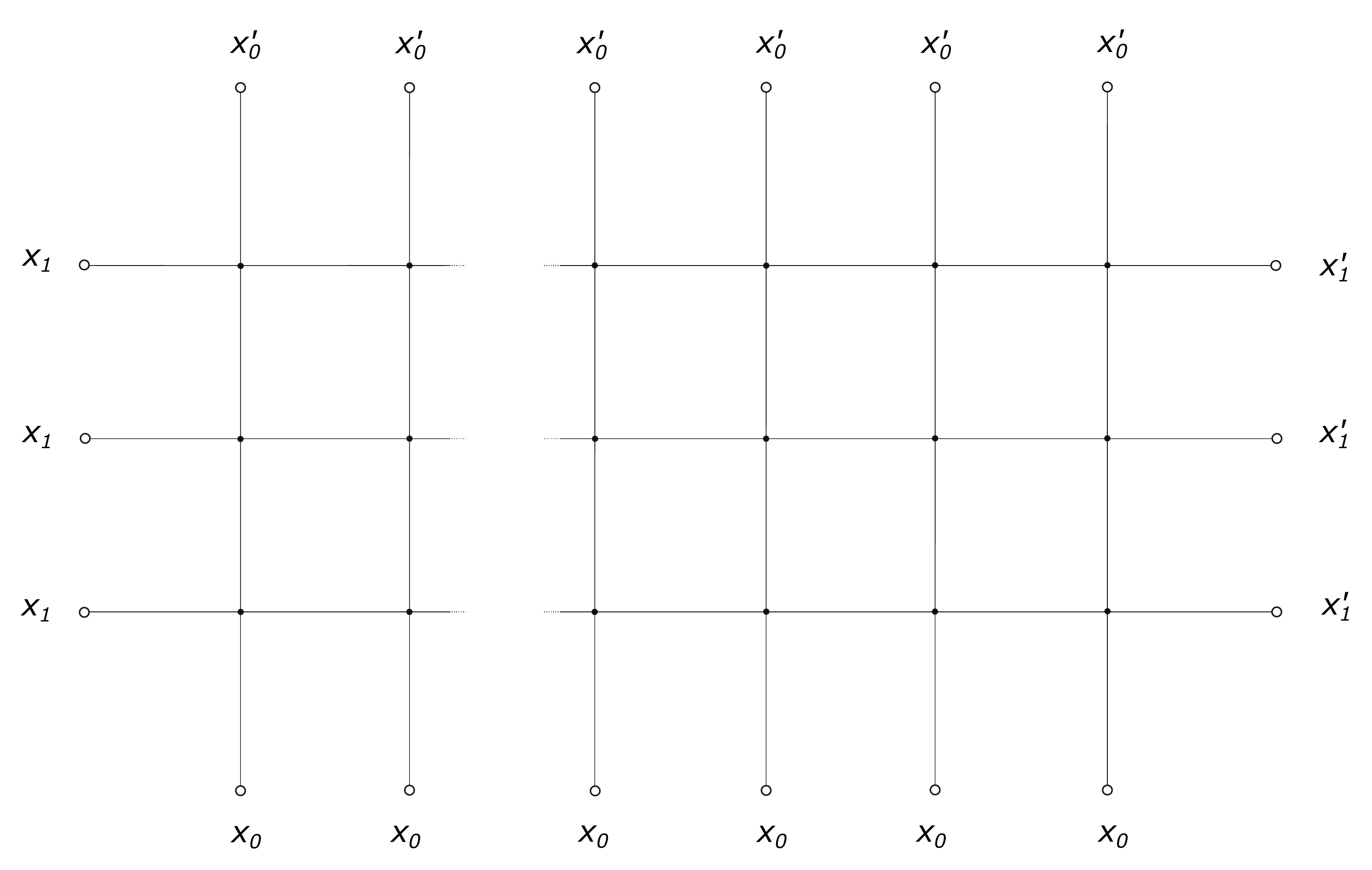}
 \end{center}
 \caption{Graphic representation of the fishnet Feynman integral \eqref{basso-dixon_intro} for a size of the square lattice bulk $N\times L$ with $N=3$ horizontal and $L$ vertical lines. Solid lines are the standard scalar propagators $1/(x-y)^2$ where $x$ and $y$ are the two endpoints of each segment. Intersection points between two lines (black dots) are integrated over. The topology of the bulk is the one of a square-lattice ``fishnet", and the boundary points are identified into four points $(x_0,x_1,x'_0,x'_1)$.}
  \label{basso-dixon_intro}
\end{figure}

The starting point of our research is the elegant explicit expression conjectured by B.~Basso and L.~Dixon for a specific, conformal planar Feynman graph with square-lattice topology (``fishnet")  
 \cite{Basso}, having \(N\) rows and \(L\) columns, and thus \((N+1)(L+1)-4\) loops. This graph is presented on Fig.\ref{basso-dixon_intro}, and its expression - modulo a finite normalization constant - is \begin{align}
 \label{I_BD}
 \begin{aligned}
 &I_{BD}(x_0,x_1,x'_0,x'_1)=\\&=\int \left(\prod_{k=1}^{N}\prod_{h=1}^L \frac{d^4 y_{h,k}}{(y_{h,k}-y_{h+1,k})^2(y_{h+1,k}-y_{h+1,k+1})^2}\right)\left(\prod_{k=1}^{N}\frac{1}{(y_{L+1,k}-x'_1)^2}\right)\left(\prod_{h=1}^{L}\frac{1}{(x_0-y_{h+1,1})^2}\right)\,,
 \end{aligned}
 \end{align}
 where $y_{1,k}\equiv x_1$ and $y_{h,N}\equiv x'_0$.
  It was explained in \cite{Basso} that this Basso-Dixon (BD) formula
takes the form of an \(N\times N\) determinant of explicitly known ``ladder" integrals~\cite{Usyukina:1993ch,Broadhurst:1993ib,Isaev2003}, and it is one of very few examples of explicit results for Feynman graphs with arbitrary many loops. 
 
\begin{figure}
 \includegraphics[scale=0.20]{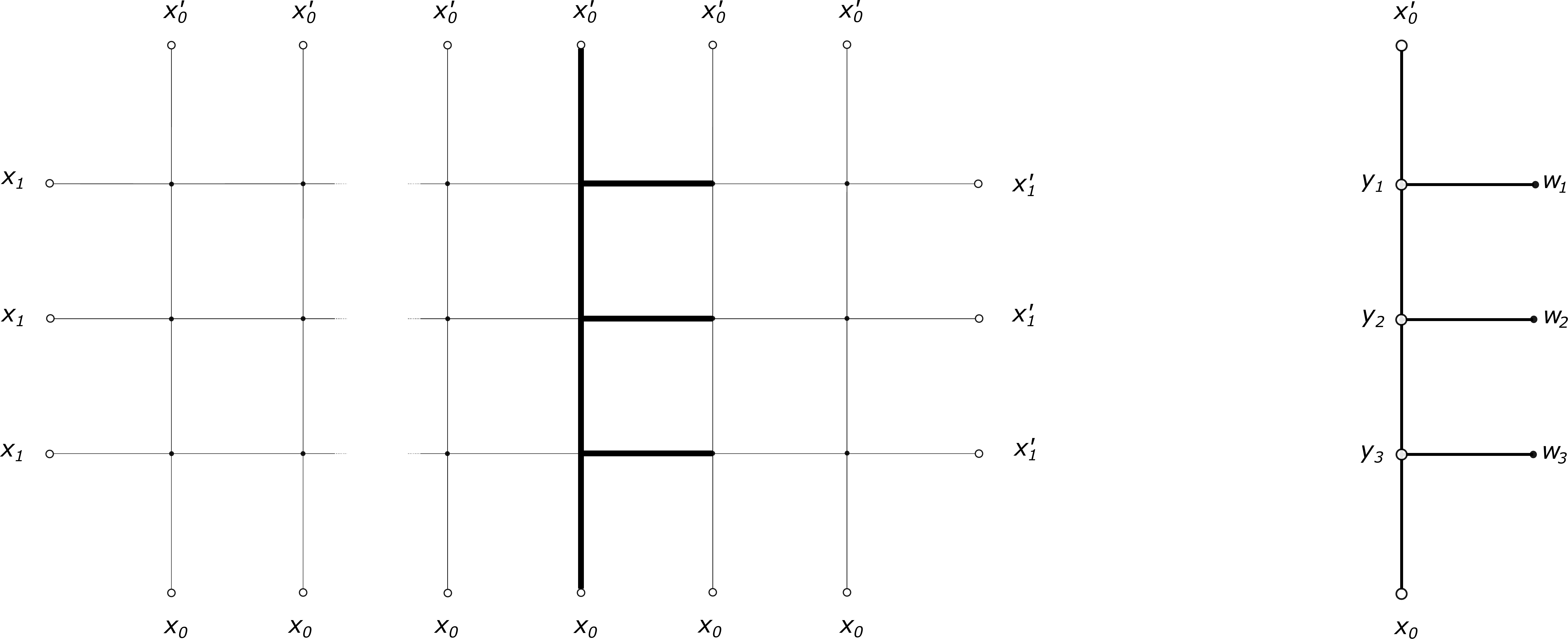}
 \caption{On the left, we highlight a transfer matrix inside the square-lattice of the fishnet integral \eqref{I_BD}. On the right, we represent it separately, as the kernel of an integral operator, where the black dots are the integration point, according to formula \eqref{time_evo_Q}. The transfer matrix makes a function propagate throughout the bulk of the diagram, from right to left.}
  \label{basso-dixon_evo}
 \end{figure}

The Feynman integral \eqref{I_BD} is relevant in the context of the four-dimensional Fishnet conformal field theory \cite{Gurdogan:2015csr}
\begin{equation}
\label{bi-scalarL}
   \mathcal{L}_{bi-scalar} ={}N_c
\,\Tr\Bigl[-\frac{1}{2}\partial_{\mu} \phi_1\partial^{\mu} \phi_1^{\dagger}-\frac{1}{2}\partial_{\mu} \phi_2\partial^{\mu} \phi_2^{\dagger}+ \xi^2\,\phi_1^\dagger \phi_2^\dagger \phi_1\phi_2 \Bigr]\,,
\end{equation}
as it is the only planar and connected integral entering the perturbative expansion in the coupling $\xi^2$ of the four-point correlator
\begin{equation}
\label{G_BD}
G_{BD}(x_0,x_1,x'_0,x'_1)=\bigg \langle \text{Tr}\left[(\phi_1(x_1))^N(\phi_2(x'_0))^L(\phi_1^{\dagger}(x'_1))^N(\phi_2^{\dagger}(x_0))^L\right]\bigg\rangle\,.
\end{equation}
According to the conjecture of Basso and Dixon, which has been proven by direct computation in our last letter \cite{Derkachov_Oliv}, the integral \eqref{I_BD} can be expressed for any $N$ and $L$ as a sum over $N$ separated variables $Y_k \in \mathbb{C}$\begin{equation}
\label{sep_var}
Y_k=\frac{\ell_k}{2}+i\nu_k\,,\,\,\,\,\ell_k\in \mathbb{Z}\,,\,\nu_k\in \mathbb{R}\,,\,\,\,\,k=1,\dots,N\,,
\end{equation}
\begin{align}
\begin{aligned}
\label{BD_formula_SoV}
I_{BD}&(x_0,x_1,x'_0,x'_1)=\\&=\frac{(x_0-x'_0)^{-2(N+L)}}{(x_1-x_0)^{2N}(x'_1-x_0)^{2N}}\left[\frac{(z\bar z)^{\frac{1}{2}}}{z-\bar z}\right]^N\sum_{\boldsymbol{\ell}\in \mathbb{Z}}\int d\boldsymbol{\nu} 
\,\mu(\boldsymbol{\nu},\boldsymbol{\ell})\, \prod_{k=1}^N\,\frac{z^{i\nu_k+\frac{\ell_k+1}{2}}{\bar z}^{i\nu_k-\frac{\ell_k+1}{2}}}{\left(\frac{(\ell_k+1)^2}{4}+\nu_k^2\right)^{N+L}}\,,
\end{aligned}
\end{align}
where the variables $z,\bar z$ are conformal invariants expressed in terms of the cross ratios
\begin{equation}
u=\frac{x_{1'0}^2 x_{10'}^2}{x_{10}^2 x_{1'0'}^2}\,\,\,\,\text{and}\,\,\,\, v=\frac{x_{11'}^2 x_{00'}^2}{x_{10}^2 x_{1'0'}^2}\,,
\end{equation}  as
\begin{equation}
u=z\bar z\,,\,\,\,\,v=(1-z)(1-\bar z)\,.
\end{equation}
We call the expression \eqref{BD_formula_SoV} a separated variables representation in the sense of \cite{Sklyanin1989,Sklyanin1991,Sklyanin1995,Sklyanin1996}, since for a graph with $N$ rows the integrand in the r.h.s. of \eqref{BD_formula_SoV} is factorized into $N$ contributions, each depending on one of the variables $Y_k$, and the non-factorizable part is collected by the Plancherel measure
\begin{equation}
\label{Plancherel_measure}
\mu(\boldsymbol{\nu},\boldsymbol{\ell})\,=\,\prod_{k=1}^N \frac{\ell_k+1}{2\pi^{2N+1}} \prod_{h>k}{\left((\nu_k-\nu_h)^2+\frac{(\ell_k-\ell_h)^2}{4}\right)
\left((\nu_k-\nu_h)^2+\frac{(\ell_k+\ell_h+2)^2}{4}\right)}\,.
\end{equation}
In this paper we provide in full detail the direct derivation of the BD formula \eqref{BD_formula_SoV} first summarized in \cite{Derkachov_Oliv}. To start with, we interpret each column of \eqref{I_BD}, as highlighted in figure \ref{basso-dixon_evo}, as a transfer matrix operator acting on $N$ points in $4D$, and which propagates a wave function $\Phi(x_1,\dots,x_N)$ throughout the bulk of the diagram, from right to left. This propagation is parametrized by a spectral parameter $u$ which plays the role of time interval, and is set to $u=-1$. The Hamiltonian operator which defines this discrete evolution can, as usual \cite{Tarasov:1983cj,Sklyanin1991,Faddeev2016}, be extracted from the evolution operator - the transfer matrix at general $u$ 
\begin{eqnarray}
\label{time_evo_Q}
\left[Q(u)\Phi\right](y_1\,,\ldots,y_N)&& =
\left(\frac{4^u\Gamma(2+u)}{\pi^{2}\Gamma(-u)}\right)^N\,
\prod_{k=1}^{N}
(y_k-y_{k+1})^{-2} \times\,
\\ && \times
\int d^4 w_1\cdots d^4 w_N\,\prod_{k=1}^{N}(w_k-y_k)^{-2(2+u)}(w_k-y_{k+1})^{2(u+1)}
\,\Phi(w_1\,,\ldots\,,w_N)\,,\nonumber
\end{eqnarray}
by taking a logarithmic derivative in the parameter $u$. The resulting Hamiltonian turns out to be equivalent to the $4D$ version of the open conformal spin chain  introduced in $2D$ by L.~Lipatov \cite{Lipatov:2009nt,Bartels:2011nz} for the study of the scattering amplitudes of high-energy gluons in the multi-Regge-kinematics (MRK) in $\mathcal{N}=4$ SYM theory\footnote{First the relation between Regge asymptotics and integrable model was noticed in quantum chromodynamics by \cite{Lipatov:1993qn,Lipatov:1993yb,Faddeev:1994zg}.}. Therefore, as we explain in section \ref{sec:magnet} the hamiltonian is that of a chain of nearest-neighbor interacting states in the quantum spaces $\mathbb{V}_k = L^2(x_k,d^4x_k)$ for each site $x^{\mu}_k$, all of whom carry the same irreducible representation of the conformal group $SO(1,5)$ defined by a scaling dimension $\Delta=1$ and spins $\ell=\dot{\ell}=0$
\begin{align}
\label{ham_intro}
\begin{aligned}
\mathbb H = 2\ln\left(x^2_{12}x^2_{23}\cdots x^2_{N0}\right)+
\frac{1}{x_{N0}^{2}}
\ln\left(p_N^2\right) x_{N0}^{2}+
\ln\left(p_1^2\right)+ \sum_{k=1}^{N-1}
\frac{1}{\left(x_{k\,k+1}^2\right)}
\ln\left(p_k^2\,p_{k+1}^2\right) x_{k\,k+1}^{2}
\,.
\end{aligned}
\end{align}
In this context, formula \eqref{BD_formula_SoV} is nothing but the representation of \eqref{I_BD} over a basis of eigenfunctions of the spin-chain Hamiltonian \eqref{ham_intro}, where
\begin{equation}
\prod_{k=1}^N\,{\left(\frac{(\ell_k+1)^2}{4}+\nu_k^2\right)^{-1}}\,,
\end{equation}
 is the eigenvalue of the transfer matrix \eqref{time_evo_Q} at $u=-1$, and $Y_k=(\nu_k,\ell_k)$ are the quantum numbers of the eigenfunction for the model of length $N$. The eigenfunctions of the model with $N$ sites can be regarded as bound states of the spin chain of $N$ scalar particles of scaling dimension $\Delta=1$; their construction and properties are presented in detail in section \ref{sec:eigenf}. The representation \eqref{BD_formula_SoV} is obtained injecting a complete basis of eigenfunctions at the point $x'_1$ of the diagram and letting it evolve from $x'_1$ to $x_1$ by the action of transfer matrices \eqref{time_evo_Q}, where the Plancherel measure \eqref{Plancherel_measure} is the overlap of two eigenfunctions, and its computation is presented step-by-step in section \ref{sec:scalprod}.
 \\
 
 The need to achieve deep understanding of formula \eqref{BD_formula_SoV} has two compelling reasons. The first is that the spin chain magnet with four-dimensional conformal symmetry \eqref{ham_intro} is an integrable model, as discussed in section \ref{sec:magnet}, which is a rare example of exactly solvable integrable model formulated in space-time dimension $d>2$. In particular, the solution of the open spin chain points towards the definition of Baxter operators and subsequent separation of variables for the - harder - closed spin chain model, in the spirit of the $2D$ technique of \cite{Derkachov2001}. In this context, the crucial formulae underling our results in \cite{Derkachov_Oliv}, that is the generalization of star-triangle integral identity \cite{DEramo:1971hnd} to propagators of non-zero spin fields\footnote{Other suggestive star-triangle identities for integrable lattice models in relation to conformal field theories were studied in \cite{Au_Yang_1999}.} in $4D$, are presented in a handful graphical notation in section \ref{sec:startri} and any detail of their derivation can be found in Appendix \ref{app:chainstar}. 

The second reason lies in the fact that the Fishnet CFT can be derived as a strong deformation limit \cite{Gurdogan:2015csr,Caetano:2016ydc} of $\mathcal{N}=4$ SYM theory. In the paper \cite{Basso} the formula \eqref{BD_formula_SoV} was conjectured via the AdS/CFT correspondance, where the separated variables $(\nu_k,\ell_k)$ are interpreted as rapidities $\nu_k$ and bound state indices $\ell_k$ labeling the mirror excited states of the dual string theory. In this perspective our computations provide one of many checks of the correspondence in the Fishnet CFT limit of $\mathcal{N}=4$ SYM (see \cite{Gurdogan:2015csr,Gromov:2017cja,Grabner:2017pgm,Gromov_2019} for other examples) and may be developed to provide similar checks and new results in the realm of the recently developed techniques for the exact computation of planar $n$-point conformal correlators by decomposition in polygonal building blocks \cite{Basso:2013vsa,Basso:2015zoa,Eden:2016xvg,Fleury:2016ykk,Fleury:2017eph,Coronado:2018cxj,Kostov:2019stn,Fleury:2020ykw,Belitsky:2019fan,Belitsky:2020qrm} - and its application to the Fishnet theory \cite{Basso2018}.

It is worth to mention here that the integrability of $\mathcal{N}=4$ SYM theory is based on the conjectured holography, and it is realized by very sophisticated techniques of integrability (see \cite{Beisert:2010jr} and references therein) which are still partially obscure as they lack a rigorous derivation. Therefore, the Fishnet CFT - where the integrability can be realized at the same time by the deformation of the Quantum Spectral Curve of $\mathcal{N}=4$ SYM (see \cite{Gromov:2017cja,Kazakov:2018ugh,Levkovich-Maslyuk:2020rlp} and references therein)  and by direct, clear, spin chain methods - is a perfect playground to start the unveiling of the integrability of $\mathcal{N}=4$ SYM theory. A recent example in this direction is the formulation of the Thermodynamic Bethe Ansatz (TBA) \cite{Basso:2019xay} for the Fishnet CFT defined in arbitrary spacetime dimension \cite{Kazakov2018}. The form of the S matrix of Fishnet CFT  - equal to Zamolodchikov's R matrix \cite{Zamolodchikov:1977nu,Zamolodchikov:1978xm} modulo a phase factor - is conjectured starting from the form of the eigenfunctions of the spin magnet \eqref{ham_intro} at $N=2$. In sections \ref{zamol} and \ref{sec:eigenf} we provide, for the $4D$ model, a rigorous derivation of the R matrix by direct, systematic computations via star-triangle identities.

The large amount of results obtained in the last few years in Fishnet CFT, demands the exploration to go on and take a first step towards the superconformal theory, i.e. to study the integrability features of the general double scaling limit of $\gamma$-deformed $\mathcal N=4$ SYM - dubbed as $\chi$CFT$_4$ -rather than its bi-scalar reduction \eqref{bi-scalarL}. In this spirit, this paper continues the program of \cite{Kazakov2019}, exploring another class of exactly-solvable four-point functions which generalizes \eqref{G_BD}. The integraction Lagrangian of the $\chi$CFT$_4$ is
\begin{equation}
\label{chiCFT4_intro}
\begin{aligned}
    \mathcal{L}_{\rm int} ={}N_c
\,\Tr\Bigl[\xi_1^2\,\phi_2^\dagger \phi_3^\dagger &
\phi_2\phi_3\!+\!\xi_2^2\,\phi_3^\dagger \phi_1^\dagger 
\phi_3\phi_1\!+\!\xi_3^2\,\phi_1^\dagger \phi_2^\dagger \phi_1\phi_2\!+\!i\sqrt{\xi_2\xi_3}(\psi_3 \phi_1 \psi_{ 2}+ \bar\psi_{ 3} \phi^\dagger_1 \bar\psi_2 )\\
& +i\sqrt{\xi_1\xi_3}(\psi_1 \phi_2 \psi_{ 3}+ \bar\psi_{ 1} \phi^\dagger_2 \bar\psi_3 )
 +i\sqrt{\xi_1\xi_2}(\psi_2 \phi_3 \psi_{ 1}+ \bar\psi_{ 2} \phi^\dagger_3 \bar\psi_1 )\,\Bigr]\,,
\end{aligned}
\end{equation}
where the summation over $j=1,2,3$ is assumed. This class of planar exactly-solvable correlators of $\chi$CFT$_4$ is obtained by admitting also fermionic fields at the points $x_2$ and $x_4$
    \begin{equation}
  \label{Full_correlator_intro}
  G_{\chi}(x_0,x_1,x'_0,x'_1)=\bigg \langle \text{Tr}\left[(\phi_1(x_1))^N  \,\mathcal{O}_{L_1,L_2,M_1,M_2}(x'_0) \,(\phi_1^{\dagger}(x'_1))^N \,\mathcal{O}^{\dagger}_{L_1,L_2,M_1,M_2}(x_0)\,\right]\bigg \rangle\,,
  \end{equation}
  where 
  \begin{equation}
  \label{O_operators_intro}
  \mathcal{O}_{L_1,L_2,M_1,M_2}(x) = (\phi_2(x))^{L_1}(\phi_3^{\dagger}(x))^{L_2}({\psi}_2(x))^{M_1}(\bar{\psi}_3(x))^{M_2}\,.
  \end{equation}
    \begin{figure}
 \begin{center}
 \includegraphics[scale=0.27]{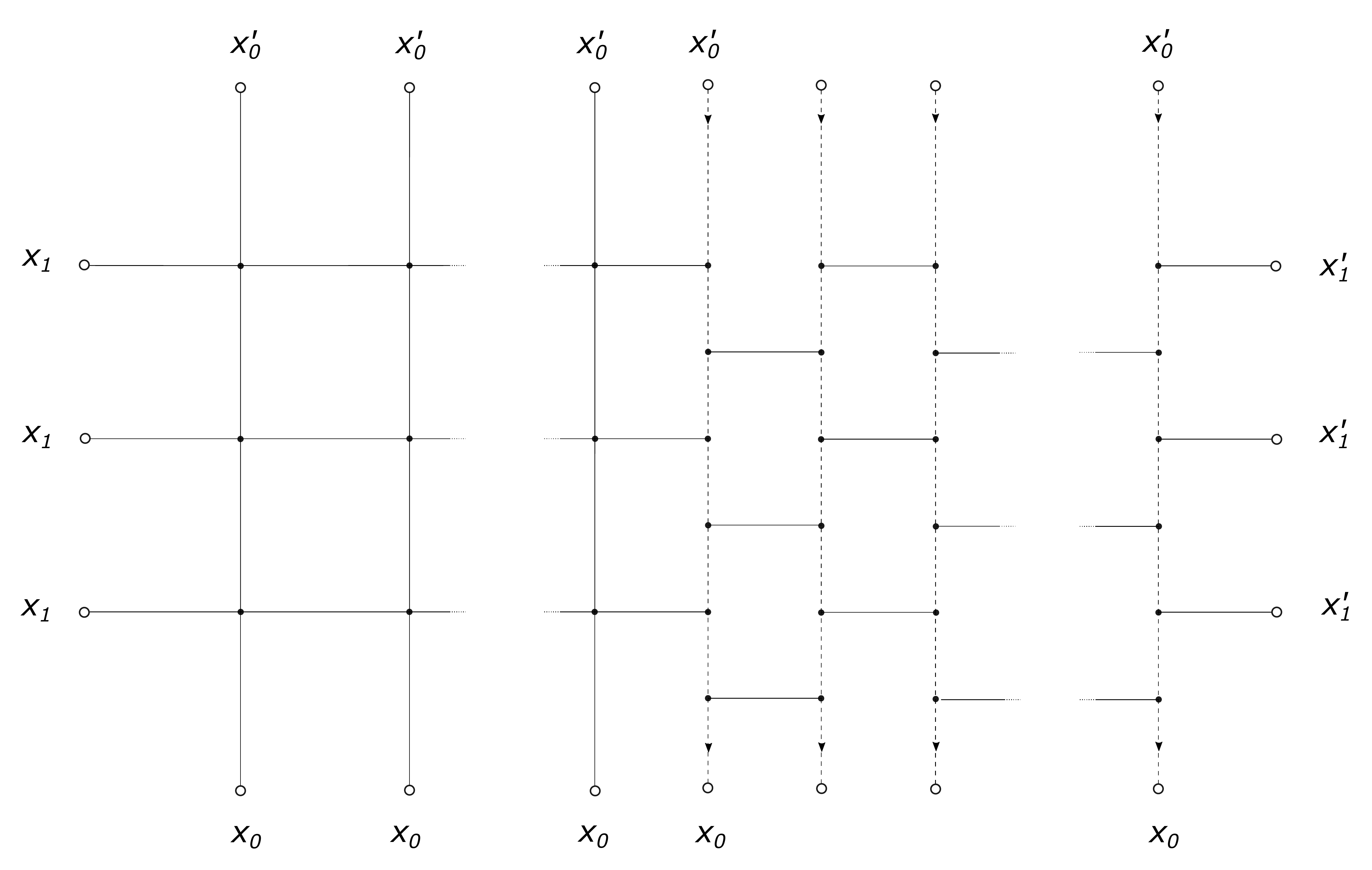}
 \end{center}
 \caption{Graphic representation of the Feynmann integral $I_{\chi}(x_0,x_1,x'_0,x'_1)$ for a size of the square lattice bulk $N\times (L_1+L_2)$ with $N=3$ horizontal and $L_1+L_2$ vertical solid lines, and a Yukawa lattice of size $N\times (M_1+M_2)$, where $M_1+M_2$ is the number of fermionic (dashed) vertical lines appearing on the right part of the graph and corresponding to the $M_1$ insertions of $\psi_2(x_0')\bar \psi_2(x_0)$ and the $M_2$ insertions of  $\bar \psi_3(x_0')\bar \psi_3(x_0)$. Every dashed segment along a line is alternatively given by $\sig^{\mu}x_{\mu}/(x^2)^2$ or $\bsig^{\mu}x_{\mu}/(x^2)^2$, and the arrows define the order of matrices $\sig^{\mu}$ and $\bsig^{\nu}$ along a line. At the boundary the points are identified into four points $(x_0,x_1,x'_0,x'_1)$.}
 \label{chiBD_fig}
 \end{figure}
  As for the scalar correlators \eqref{I_BD} in the bi-scalar theory \eqref{bi-scalarL}, also the correlator \eqref{Full_correlator_intro} receive a single contribution in the weak couplings expansion, given - modulo a finite normalization constant - by the Feynman integrals $I_{\chi}(x_0,x_1,x'_0,x'_1)$
 which are depicted in figure \ref{chiBD_fig} and whose simple topology is a mixture of a square-lattice of scalar vertices and an hexagonal lattice of Yukawa vertices. Extending the logic of the bi-scalar case, in section \ref{sect:Qspin} we explain how the computation of such Feynman integrals is mapped to the diagonalization of the same spin chain magnet \eqref{ham_intro}.
  \begin{figure}
 \begin{center}
 \includegraphics[scale=0.20]{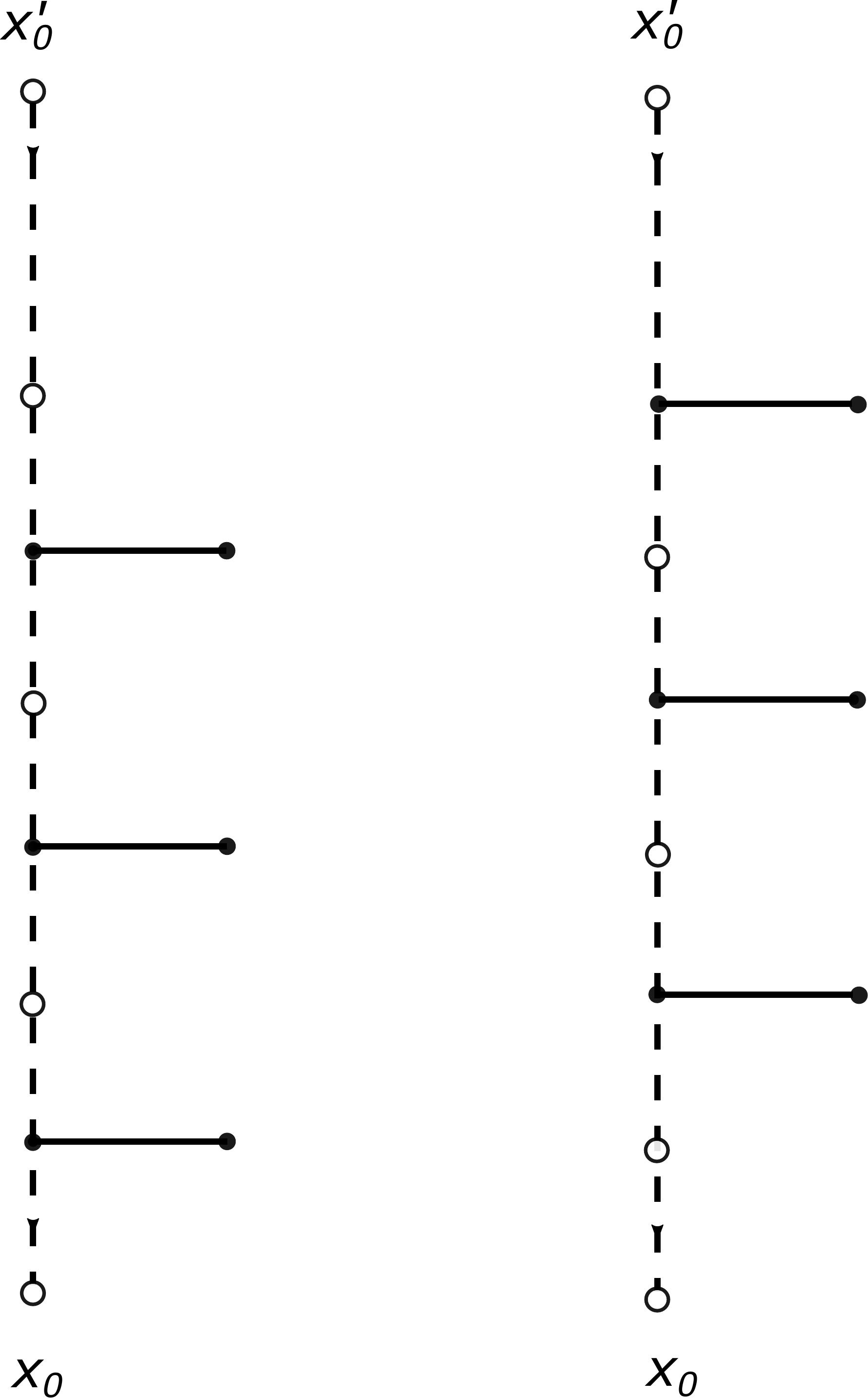}
 \end{center}
 \caption{Graphic representation of the integral kernel of fermionic graph-building operators appearing in the general Feynman integral of figure \ref{chiBD_fig}. These two operators commute with each other as explained in section \ref{sect:Qspin} and with the scalar transfer matrix depicted in figure \ref{basso-dixon_evo}. In particular, their expansion over the basis of eigenfunctions of the model \eqref{ham_intro} has a relatively simple form worked out in section \ref{sec:specQ}.}
  \label{gen_evo}
 \end{figure}
The main result of our paper is given by the  representation over separated variables for the correlator of $\chi$CFT$_4$ theory defined in \eqref{Full_correlator_intro} for any choice of positive integers $N$ and $L_1,L_2,M_1,M_2$:
\begin{align}
\label{Final_G_graph_intro}
\begin{aligned}
 I_{\chi}(&x_0,x_1,x'_0,x_1')=\frac{1}{(x_1-x_0)^{2N}(x_1'-x_0)^{2N}(x_0-x'_0)^{2(N+L_1+L_2-3/2(M_1+M_2))}}\times\\  &\times\sum_{\boldsymbol{\ell}}\,\tilde W_{\boldsymbol \ell}\left(x_0,x_1,x'_0,x_1'\right)\int d\boldsymbol{\nu} \,\mu(\boldsymbol{\nu},\boldsymbol{\ell}) \prod_{k=1}^{N} (z\bar z)^{i\nu_k}
\tau_+(0,Y_k)^{N+L_1+L_2}\,\tau_+(1,Y_k)^{M_1} \,\tau_-(1,Y_k)^{M_2}\,.
\end{aligned}
\end{align}
The quantities
\begin{equation}
\tau_{+}(0,Y)=\frac{4 \pi^2 }{4\nu^2+(\ell+1)^2}\,,\,\,\,\,\,\tau_{\pm}(1,Y)=\frac{8\pi^2}{\left(1\pm 2i\nu +{\ell}\right)\left(1\mp 2i\nu-{\ell}\right)\left(3 \mp 2i\nu +\ell\right)} \,,
\end{equation}
are the eigenvalues of the scalar and fermionic transfer matrices at $N=1$ and are computed in section \ref{sec:specQ}. The functions $\tilde{W}_{\boldsymbol{\ell}}$ are polynomials carrying the spinor indices of fermionic fields in the correlator \eqref{Full_correlator_intro}. In order to deliver a compact expression for $\tilde{W}_{\boldsymbol{\ell}}$, we shall already introduce the notations
\begin{equation}
(\mathbf{x-y})^{\dot a}_{a} = \frac{(\sigma^{\mu})^{\dot a}_{a}(x-y)_{\mu}}{|x-y|}\,,\,\,\,\, (\overline{\mathbf{x-y}})^{a}_{\dot a} = \frac{(\bar{\sigma}^{\mu})^{a}_{\dot a}(x-y)_{\mu}}{|x-y|}\,,
\end{equation}  
together with the harmonic polynomials $[(\mathbf{x-y})]^{\ell}$, $[(\mathbf{\overline{x-y}})]^{\ell}$ defined as 
\begin{equation}
\label{garmonic}
[(\mathbf{x-y})]^{{\dot a_1\dots \dot a_{\ell}}}_{{a_1\dots a_{\ell}}} = \frac{1}{N!}\sum_{\sigma\in\mathbb{S}_\ell}(\mathbf{x-y})^{\dot a_{\sigma(1)}}_{a_1}\cdots (\mathbf{x-y})^{\dot a_{\sigma(\ell)}}_{a_{\ell}} = \frac{1}{N!}\sum_{\sigma\in\mathbb{S}_\ell}(\mathbf{x-y})^{\dot a_{1}}_{a_{\sigma(1)}}\cdots (\mathbf{x-y})^{\dot a_{\ell}}_{a_{\sigma(\ell)}}\,,
\end{equation}
and two solutions of Yang-Baxter equation acting on the symmetric spinors of degree $\ell$ and $1$ (see Appendix \ref{app:R} and sections \ref{sec:startri} and \ref{sec:final})
\begin{equation}
\label{Rmat_intro}
{\left(\mathbf{R}^+_{h,k}\right)}^{\dot c_{h} \,(\dot r_1\dots \dot r_{\ell_k})}_{\dot a_{h}\, (\dot s_1\dots \dot  s_{\ell_k})}\,,\,\,\,\,\,\,{\left(\mathbf{R}^-_{n,k}\right)}^{c_{n} \, (r_1\dots  r_{\ell_k})}_{ a_{n}\,(s_1\dots  s_{\ell_k})}\,.
\end{equation}
It follows that the polynomials $\tilde{W}_{\boldsymbol{\ell}}$ are given by a trace over the space of symmetric spinors of degrees $\ell_1,\dots,\ell_N$, of a combination of harmonic polynomials \eqref{garmonic} and R matrices \eqref{Rmat_intro}
\begin{align}
\begin{aligned}
 &(\tilde{W}_{\boldsymbol \ell})_{a_1,\dots,a_{M_1},\dot a_1,\dots,\dot a_{M_2}}^{\dot b_1,\dots,\dot b_{M_1}, b_1,\dots, b_{M_2}}\left(x_0,x_1,x'_0,x'_1\right)=\\&=\text{Tr}_{\ell_1,\dots,\ell_N}\Bigg [\left(\overleftarrow{\prod_{k=1}^{N}}[(\mathbf{{x_{0'1}}\overline{x_{10}}})]^{\ell_k}\overleftarrow{\prod_{h=1}^{M_2}}([\mathbf{\overline{x_{00'}}}]\mathbf{R}^{+}_{h,k})_{\dot a_h}^{b_h}\right)   \left(\mathbf{\overrightarrow{\prod_{k=1}^{N}}[(x_{1'0}\overline{x_{10'}}})]^{\ell_k} \overrightarrow{\prod_{n=1}^{M_1}}(\mathbf{R}^-_{n,k} [\mathbf{{x_{00'}}}])_{a_n}^{\dot b_n}\right)\Bigg]\,.
\end{aligned}
\end{align}
The latter formula simplifies in the scalar case $M_1=M_2=0$, when the integral $I_{\chi}$ reduces to the Basso-Dixon integral of size $N\times (L_1+L_2)$ and \eqref{Final_G_graph_intro} reduces - a part a finite normalization constant - to formula \eqref{BD_formula_SoV}.\\

The paper is organized as follows: in section \ref{sec:magnet} we introduce in detail the spin-chain hamiltonian \eqref{ham_intro}, its relation with the fishnet graph \eqref{I_BD} and the quantum integrability of the model. In section \ref{ladder} we find the spectrum and eigenfunctions of the spin chain in the simplest case $N=1$, and compute the integrals \eqref{I_BD} for $N=1$ and any $L$ - that is the ladder diagrams of length $L$ \cite{Usyukina:1993ch,Broadhurst:1993ib} - bringing them into the form \eqref{BD_formula_SoV}. In section \ref{sec:startri} we derive the chain-rule and star-triangle identities in $4D$ which generalizes the well-known scalar relation \cite{DEramo:1971hnd,Isaev2003,Vasilev2004} to massless propagators of fields with non-zero spin. In particular, in the subsection \ref{zamol} we show how the fused Yang R matrix which mixes spinor indices in the star-triangle identity is related to the Zamolodchikov's R matrix for the $O(4)$ model. Section \ref{sect:Qspin} deals with the introduction of the generalization of \eqref{time_evo_Q} to any spin, including the graph-building operators depicted in figure \ref{gen_evo} for spin $1/2$. In section \ref{sec:eigenf} we present the construction for the eigenfunction of the spin chain model for any size $N$, the symmetry properties of the eigenfunctions respect to permutation of their quantum labels (separated variables) (section \ref{sec:symm_eig}) and we compute the overlapping of two functions obtaining the Plancherel measure \eqref{Plancherel_measure} in section \ref{sec:scalprod}.
In the last two sections \ref{sec:specQ} and \ref{sec:final} we find the expansion of the graph-building operators introduced in section \ref{sect:Qspin} over the eigenfunctions of the spin chain. We apply these representations to the computation of the Feynman integral contributing to the single trace correlators \eqref{G_BD}, delivering our final results.\\ Our aim in this paper is to present every computation in a rather explicit and easy-to-check way, both via analytic computations or via graphical techniques developed throughout the paper starting from the star-triangle identities. Nevertheless, we left several cumbersome calculations  to the appendices \ref{app:spinors}-\ref{app:N3scalar}, to which we refer along the main text.

\section{Integrable hamiltonian and ladder diagrams}
\label{sec:magnet}
\begin{figure}
\begin{center}
\includegraphics[scale=0.45]{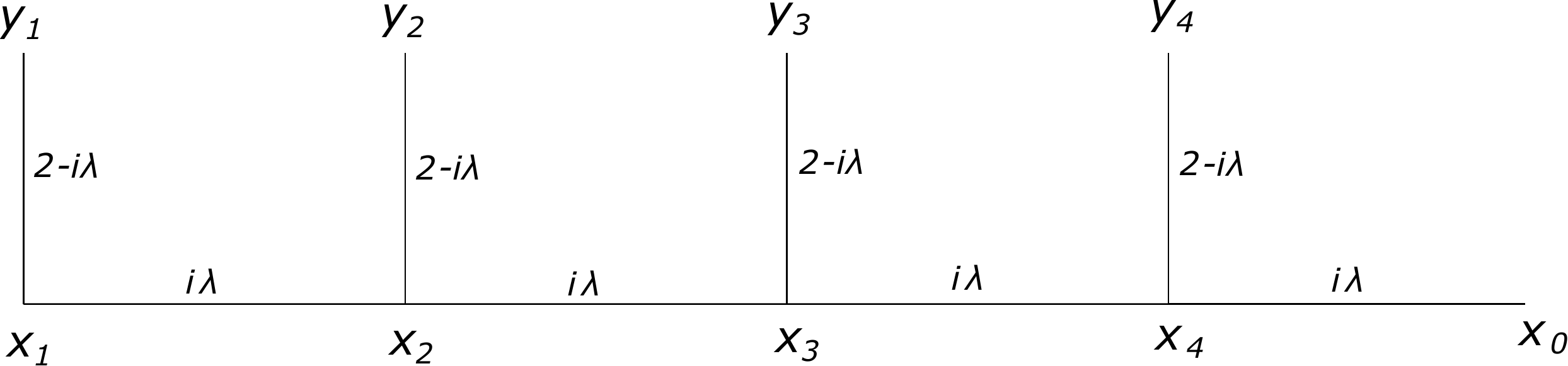}
\end{center}
\caption{Graphical representation of the kernel $Q(x_1,\dots,x_N,x_0|y_1,\dots,y_N; -i\lambda)$ of the operator $Q(-i\lambda)$ at the size $N=4$. The lines are functions $1/x^2$ to the power written aside; integration points $y_k$ are marked by black dots and external points $x_k$ are marked by grey dots. For $\lambda=-i$ the kernel is a proportional to a portion of scalar Feynman integral in $4d$.}
\label{figQbuild}
\end{figure}

In this section we introduce the spin chain model which underlies the computation of the four-point functions under study in this paper. Namely, we start from the definition of the Hamiltonian operator acting on a collection $x_1,\dots,x_N$ of nearest-neighbor interacting sites, each of them carring the representation of zero spins $\ell=\dot{\ell}=0$ and scaling dimension $\Delta=2-i\lambda$ of the group $SO(1,5)$ of Euclidean conformal transformation in $4d$. The Hamiltonian operator acts on the tensor product of Hilbert spaces $\mathbb{V}_k = L^2(x_k,d^4x_k)$ and has the following expression in terms of coordinates $x_k$ and momenta $p_k = -i\partial_k$
\begin{align}\label{4Dlocal}
\begin{aligned}
\mathbb H = 2\ln\left(x^2_{12}x^2_{23}\cdots x^2_{N0}\right)+
\frac{1}{x_{N0}^{2i\lambda}}
\ln\left(p_N^2\right) x_{N0}^{2i\lambda}+
\ln\left(p_1^2\right)+ \sum_{k=1}^{N-1}
\frac{1}{\left(x_{k\,k+1}^2\right)^{i\lambda}}
\ln\left(p_k^2\,p_{k+1}^2\right) x_{k\,k+1}^{2i\lambda}
\,,
\end{aligned}
\end{align}
where $x_{k k+1}=x_k-x_{k+1}$, $p^2_{k} = -\partial_k \cdot\partial_{k}$ and
$x_{N+1}=x_0$. All $x_k$ and $p_k$ are vectors in
the 4-dimensional Euclidean space and $x^{2} = x^{\mu}x_{\mu}$.
The point $x_0$ is effectively a parameter for the model,
and we will always omit it from the set of coordinates.
The spin chain \eqref{4Dlocal} is the four-dimensional version of the open $SL(2,\mathbb{C})$ Heisenberg magnet which describes the scattering amplitudes of high energy gluons in the Regge limit of $\mathcal{N}=4$ SYM theory \cite{Lipa:1993pmr,Lipatov2004}, and for periodic boundary conditions it was studied in \cite{Chicherin2013a}. The quantum integrability of \eqref{4Dlocal} is realized by the commutative family  of operators $Q(u)$ labeled by the spectral parameter $u\in \mathbb{C}$
\begin{align}
Q(u)Q(v) = Q(v)Q(u) \,,\ \qquad \mathbb H\, Q(u) = Q(u)\,\mathbb H\,,
\end{align}
where
\begin{align}
\label{TN}
& Q(u)= Q_{12}(u)\,Q_{23}(u)\cdots Q_{N0}(u)\,,\\
&Q_{ij}(u)= (x_{ij}^2)^{-i\lambda}(p_i^2)^{u}(x_{ij}^2)^{u+i\lambda} = (p_{i}^2)^{u+i\lambda}(x_{ij}^2)^{u}(p_{i}^2)^{-i\lambda}\,.
\end{align}
The equivalence of the two representations for the
operator $Q_{ij}(u)$ follows from the star-triangle relation~\cite{Isaev2003}
\begin{align}\label{STR}
(x^2)^{a}(p^2)^{a+b}(x^2)^{b} = (p^2)^{b}(x^2)^{a+b}(p^2)^{a}\,.
\end{align}
The explicit expression for the Q-operator
\begin{align}\label{Q1}
Q(u) = \left(x^2_{12}x^2_{23}\cdots x^2_{N0}\right)^{-i\lambda}
\left(p^2_1\right)^{u}\left(x^2_{12}\right)^{u+i\lambda}
\left(p^2_2\right)^{u}\left(x^2_{23}\right)^{u+i\lambda}
\cdots
\left(p^2_{N}\right)^{u}\left(x^2_{N0}\right)^{u+i\lambda}
\end{align}
clearly shows that there are two special values
of spectral parameter $u=0$ and $u = -i\lambda$ for which the operator simplifies.
At the point $u=0$ we have the reduction to the
identity operator, while at the point $u = -i\lambda$
we obtain the graph-building operator represented in figure \ref{figQbuild}
\begin{align}
Q(-i\lambda) = \left(x^2_{12}x^2_{23}\cdots x^2_{N0}\right)^{-i\lambda}
\left(p^2_1p^2_2\cdots p_N^2\right)^{-i\lambda}\,,
\end{align}
which is an integral operator whose kernel is a portion of square lattice ``fishnet" Feynman integral
\begin{figure}
\begin{center}
\includegraphics[scale=0.5]{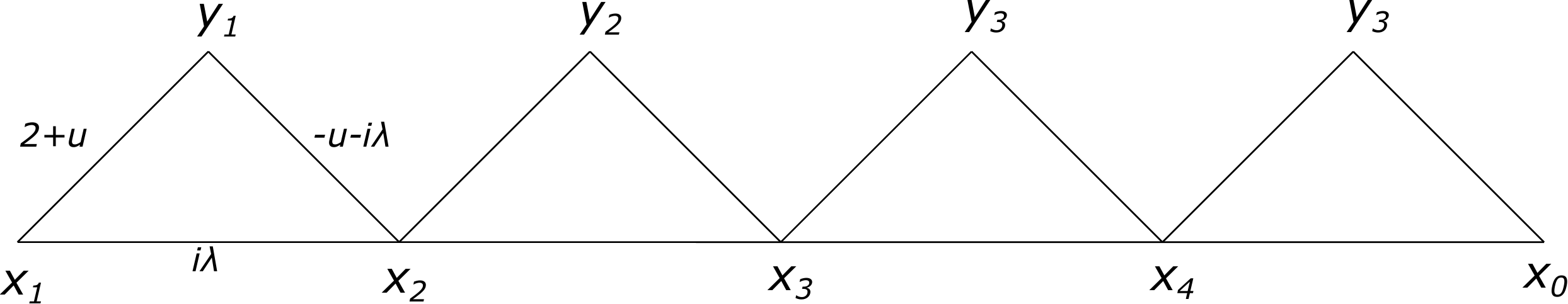}
\end{center}
\caption{Graphical representation of the kernel $Q(x_1,\dots,x_N,x_0|y_1,\dots,y_N; u)$ of the operator $Q(u)$ for the size $N=4$. The lines are functions $1/x^2$ to the power written aside; integration points $y_k$ are marked by black dots and external points $x_k$ are marked by grey dots.}\label{Qgen}
\end{figure}
The Hamiltonian can be obtained as the sum of the logarithmic derivatives of the operator $Q(u)$ at these special points \cite{Tarasov:1983cj,Sklyanin1991,Faddeev2016}
\begin{align}
\mathbb H = Q(0)^{-1}Q^{\prime}(0) + Q(-i\lambda)^{-1}Q^{\prime}(-i\lambda)\,.
\end{align}
It is simple to check that the expansion around $u=0$ in~(\ref{Q1}) gives
\begin{align}
\begin{aligned}
&
Q(\varepsilon) = \II + \varepsilon \mathbb H_1 +o(\varepsilon)\,,\\
&
\mathbb H_1 = \ln\left(x^2_{12}x^2_{23}\cdots x^2_{N0}\right)+ \frac{1}{\left(x_{12}^2\right)^{i\lambda}}
\ln\left(p_1^2\right)\, \left(x_{12}^2\right)^{i\lambda}
+\ldots+\frac{1}{\left(x_{N0}^2\right)^{i\lambda}}
\ln\left(p_N^2\right)\, \left(x_{N0}^2\right)^{i\lambda}\,.
\end{aligned}
\end{align}
Similarly, the expansion around the point $u=-i\lambda$
can be performed after some transformations
based on star-triangle relation~(\ref{STR})
\begin{align*}
\begin{aligned}
Q^{-1}(-i\lambda)\,Q(u) &=
\left(p^2_1p^2_2\cdots p_N^2\right)^{i\lambda}\left(p^2_1\right)^{u}\left(x^2_{12}\right)^{u+i\lambda}
\left(p^2_2\right)^{u}\left(x^2_{23}\right)^{u+i\lambda}
\cdots
\left(p^2_{N}\right)^{u}\left(x^2_{N0}\right)^{u+i\lambda}   \\
&=\left(p^2_1\right)^{u+i\lambda}\,
\left(x^2_{12}\right)^{u}\left(p^2_2\right)^{u+i\lambda}
\left(x^2_{12}\right)^{i\lambda}
\cdots
\left(x^2_{N-1 N}\right)^{u}
\left(p^2_{N}\right)^{u+i\lambda}
\left(x^2_{N-1 N}\right)^{i\lambda}\,
\left(x^2_{N0}\right)^{u+i\lambda}\,,
\end{aligned}
\end{align*}
and therefore it reads
\begin{align}
\begin{aligned}
&
Q^{-1}(-i\lambda)Q(-i\lambda+\varepsilon) =
\II + \varepsilon \mathbb  H_2 +o(\varepsilon)\,,\\
&
\mathbb H_2 = \ln\left(x^2_{12}x^2_{23}\cdots x^2_{N0}\right)+ \frac{1}{x_{N-1\,N}^{2i\lambda}}
\ln\left(p_N^2\right)\, x_{N-1\,N}^{2i\lambda}
+\ldots+\frac{1}{x_{12}^{2i\lambda}}
\ln\left(p_2^2\right)\, x_{12}^{2i\lambda}+
\ln\left(p_1^2\right)\,.
\end{aligned}
\end{align}
Finally the sum $\mathbb H = \mathbb H_1+\mathbb H_2$ coincides with~(\ref{4Dlocal}).

In explicit form, the action of the operator $Q(u)$ on a function
$\Phi(x_1\,,\ldots,x_N)$ can be expressed as an integral operator
\begin{eqnarray}\label{Q2}
\left[Q(u)\Phi\right](x_1\,,\ldots,x_N)&& =
\left(\frac{4^u\Gamma(2+u)}{\pi^{2}\Gamma(-u)}\right)^N\,
\prod_{k=1}^{N}
(x_k-x_{k+1})^{-2i\lambda} \times\,
\\ && \times
\int d^4 w_1\cdots d^4 w_N\,\prod_{k=1}^{N}(w_k-x_k)^{-2(2+u)}(w_k-x_{k+1})^{2(u+i\lambda)}
\,\Phi(w_1\,,\ldots\,,w_N)\,,\nonumber
\end{eqnarray}
where by definition $x_{N+1} = x_0$ and
$(x)^{2\alpha} = \left(x^{\mu}x_{\mu}\right)^{\alpha}$ and we consider 4-dimensional vector $x$ in Euclidean signature.
Note that the operator $Q(u)$ maps the function of
$N$ variables to the function of $N$ variables and the
$x_0$ plays some special role of external variable.

We shall use the standard Feynman diagram notations in the coordinate space
from the quantum field theory.
It is very useful because some nontrivial transformations of the
integrals can be visualized graphically in this way.
The diagrammatic representation for the kernel
of the integral operator $Q(u)$ is shown
schematically on the Fig.\ref{Qgen}.

Relations between
operators are equivalent to the corresponding relations between operator's kernels. The most convenient way to check such
relations is to prove the equivalence of the corresponding diagrams (kernels). It can be done diagrammatically with the help
of several simple identities~--~integration rules.
Below we give some of these rules (see also ref.~\cite{Derkachov2001,Chicherin2013a,Vasilev2004,Fradkin:1978pp}) which are also implemented in a Wolfram Mathematica package \cite{Preti:2018vog,Preti:2019rcq}.
All internal vertices contain the integration with respect to the variable
attached to this vertex.

\begin{itemize}
\item The function $(x-y)^{-2\alpha} = \left((x-y)^{\mu}(x-y)_{\mu}\right)^{-\alpha}$ is represented by
the line with index $\alpha$ connecting points $x$ and $y$:
\vspace*{0.2cm}
\begin{center}
{\includegraphics[scale=0.5]{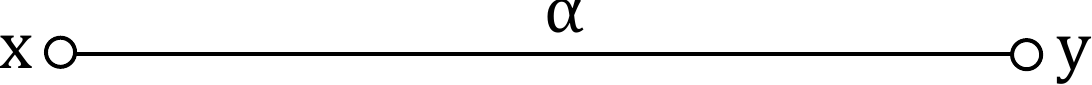}}
\end{center}
%
\item Chain rule
\begin{align}\label{chain}
\int d^4 z \frac{1}{(x-z)^{2\alpha}(z-y)^{2\beta}} =
\pi^2 a(\alpha,\beta,4-\alpha-\beta)\,\frac{1}
{(x-y)^{2(\alpha+\beta-2)}}\,,
\end{align}
where $a(\alpha) = \frac{\Gamma(2-\alpha)}{\Gamma(\alpha)}$ and
$a(\alpha,\beta,\cdots,\gamma) = a(\alpha)\,a(\beta)\cdots a(\gamma)$.
Its diagrammatic form is
\vspace*{0.2cm}
\begin{center}{\includegraphics[scale=0.5]{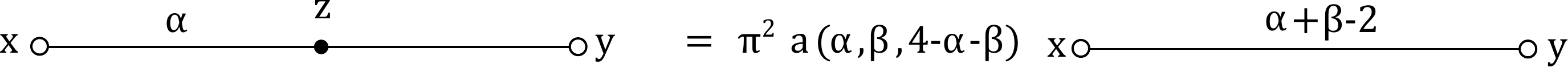}}
\end{center}

\noindent For the special case $\beta\to 4-\alpha$ one gets

\begin{align}\label{delta}
\int d^4 z \frac{1}{(x-z)^{2\alpha}(z-y)^{2(4-\alpha)}} =
\pi^4 a(\alpha,4-\alpha)\,\delta^{(4)}(x-y)
\end{align}

\item Star-triangle relation $\alpha+\beta+\gamma = 4$

\begin{align}\label{str0}
\int d^4 w \frac{1}{(x-w)^{2\alpha}(y-w)^{2\beta}(z-w)^{2\gamma}} =
\frac{\pi^2 a(\alpha,\beta,\gamma)}
{(y-z)^{2(2-\alpha)}
(z-x)^{2(2-\beta)}
(x-y)^{2(2-\gamma)}}
\end{align}
\vspace*{0.2cm}
\begin{center}
{\includegraphics[scale=0.5]{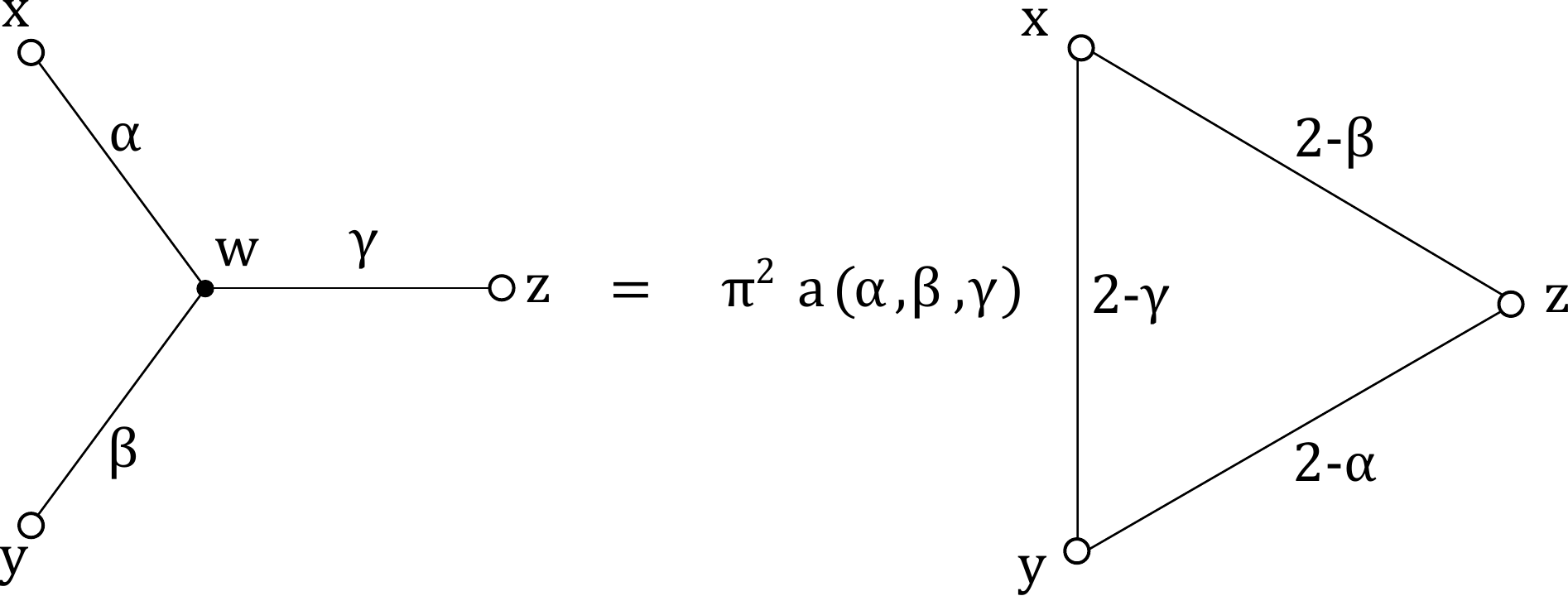}}
\end{center}
\item Cross relation
\begin{center}
{\includegraphics[scale=0.5]{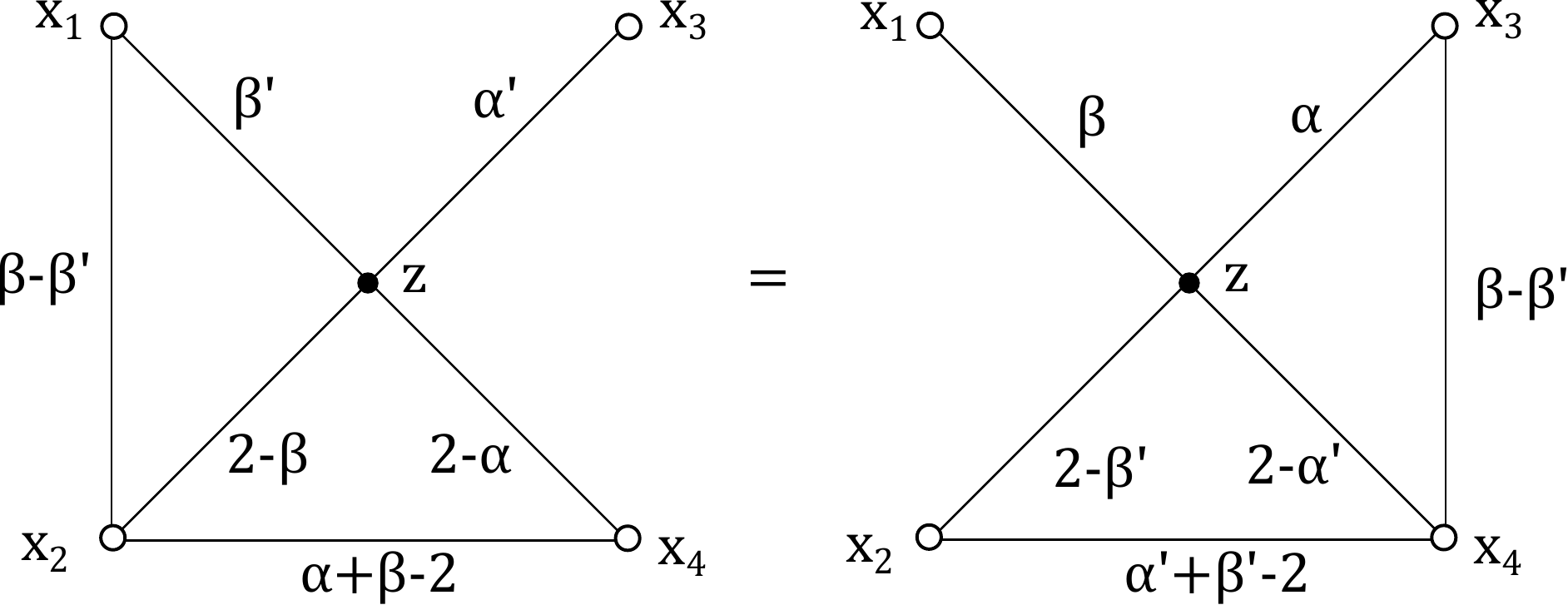}}
\end{center}


\noindent
where $\alpha+\beta=\alpha'+\beta'$.

\end{itemize}

The proof of the commutativity $Q(u)Q(v) = Q(v)Q(u)$
is equivalent to the proof of the corresponding relation for
the kernels which is demonstrated, in more general form, in section \ref{sect:Qspin}. The proof is presented there diagrammatically,
with the help of cross relation.

The transformation from representation~(\ref{Q1})
to the representation~(\ref{Q2}) is based on the fact the
operator $\left(p^2\right)^{u}$ can be expressed as an integral operator in coordinate representation
\begin{align*}
\left[\left(p^2\right)^{u}\Phi\right](x) = \frac{4^u\Gamma(2+u)}{\pi^{2}\Gamma(-u)}\,
\int d^4 y \frac{1}{(x-y)^{2(2+u)}}\,\Phi(y)\,,
\end{align*}
and this formula is derived from the standard Fourier transformation
\begin{align*}
\int d^4 p \frac{e^{ipx}}{p^{2\alpha}} =
\frac{\pi^2\,\Gamma(2-\alpha)}{4^{\alpha-2}\Gamma(\alpha)}\,
\frac{1}{x^{2(2-\alpha)}}\,.
\end{align*}
The kernel of the arbitrary power of the graph-building operator
$Q^L(-i\lambda)$ is shown schematically on the Fig.\ref{figQbuild} and coincides essentially with the diagram of the BD type.

Due to integrability of the model, that is the existence of the commuting family $Q(u)$, the diagonalization of the Hamiltonian \eqref{4Dlocal} and the calculation of the diagram
of BD type are reduced to the problem of the
diagonalization of the operators $Q(u)$.

\section{Conformal ladder integrals}
\label{ladder}

Now we are going to study the spectral problem for the operator  $Q(u)$ in  the simplest case of length $N=1$, namely
\begin{align}\label{Q2n1}
\left[Q(u)\Phi\right](x) =
\frac{4^u\Gamma(2+u)}{\pi^{2}\Gamma(-u)}\,
(x-x_{0})^{-2i\lambda}\,\int d^4 w\,
(w-x)^{-2(2+u)}(w-x_{0})^{2(u+i\lambda)}
\,\Phi(w)\,.
\end{align}
For simplicity in this section we put $x_0=0$.
Due to evident translation invariance of $Q(u)$
this choice does not lead to the loss
of the generality and $x_0$ can be restored in
any final formula.

\subsection{Eigenfunctions in tensor notations}

In this section we shall use some standard formulae from the
so-called Gegenbauer polynomial technique for the evaluation of
Feynman diagrams~\cite{Kotikov:1995cw,Chetyrkin:1980pr}.
Let us introduce a class of symmetric traceless tensor
$x^{\mu_1\cdots\mu_n}$ connected with the usual product
$x^{\mu_1}x^{\mu_2}\cdots x^{\mu_n}$ by the following relation
\begin{align}\label{x}
x^{\mu_1\cdots\mu_n} = \,\Pi^{\mu_1,\dots,\mu_n}_{\nu_1,\dots,\nu_n}\, x^{\nu_1} \cdots x^{\nu_n} \,\equiv\,\hat{S}\, \sum_{p\geq 0}
\frac{(-1)^p(n-p)!}{4^p p!(n-2p)!}\,
\delta^{\mu_1\mu_2}\cdots\delta^{\mu_{2p-1}\mu_{2p}}
\,x^{2p}\,x^{\mu_{2p+1}}\cdots x^{\mu_n}\,,
\end{align}
where $\hat{S}$ is the symmetrization over all indices
\begin{align}
\hat{S}\,\Phi^{\mu_1\cdots\mu_n} =
\frac{1}{n!} \sum_{p\in S_n} \Phi^{\mu_{p(1)}\cdots\mu_{p(n)}}\,.
\end{align}
By construction this tensor is symmetric and traceless
$\delta_{\mu_i\mu_k}\,x^{\mu_1\cdots\mu_i\cdots\mu_k\cdots\mu_n}=0$.
The basic formula~\cite{Kotikov:1995cw,Chetyrkin:1980pr} for the convolution of the tensor propagator
\begin{align}
\label{eig1}
\Psi^{\mu_1\cdots\mu_n}(x) = \frac{x^{\mu_1\cdots\mu_n}}
{x^{2(1+\frac{i\lambda}{2}+i\nu+\frac{n}{2})}} =
\frac{\hat{x}^{\mu_1\cdots\mu_n}}
{x^{2(1+\frac{i\lambda}{2}+i\nu)}}\ \ ;\ \ \hat{x}^{\mu} = \frac{x^{\mu}}{\sqrt{x^2}}
\end{align}
with a scalar propagator $1/x^{2\beta}$ is given by
\begin{align}\label{chainn}
\int d^4 x \frac{x^{\mu_1\cdots\mu_n}}{x^{2\alpha}(z-x)^{2\beta}} =
\pi^2 A^{n,0}(\alpha,\beta)\,\frac{z^{\mu_1\cdots\mu_n}}
{z^{2(\alpha+\beta-2)}}\,,
\end{align}
where
\begin{align}
A^{n,m}(\alpha,\beta) = \frac{a_n(\alpha)\,a_m(\beta)}
{a_{n+m}(\alpha+\beta-2)} \ \ ;\ \
a_n(\alpha) = \frac{\Gamma(2-\alpha+n)}{\Gamma(\alpha)}\,.
\end{align}
Using~(\ref{chainn}) it is easy to check that \eqref{eig1} is eigenfunction of the operator
$Q(u) = x^{-2i\lambda}\,p^{2u}\,x^{2(u+i\lambda)}$
and to calculate the corresponding eigenvalue.
We have
\begin{align}\label{N=1}
\left[Q(u)\Psi^{\mu_1\cdots\mu_n}\right](x) =
\frac{4^u}{\pi^{2}\,a_0(2+u)}\,x^{-2i\lambda}\,
\int d^4 y \frac{1}{(x-y)^{2(2+u)}y^{2(-u-i\lambda)}}
\frac{y^{\mu_1\cdots\mu_n}}
{y^{2(1+\frac{i\lambda}{2}+i\nu+\frac{n}{2})}} = \\
=4^u\,\frac{a_n\left(1-u-\frac{i\lambda}{2}+i\nu+\frac{n}{2}\right)}
{a_n\left(1-\frac{i\lambda}{2}+i\nu+\frac{n}{2}\right)}
\,\frac{x^{\mu_1\cdots\mu_n}}
{x^{2(1+\frac{i\lambda}{2}+i\nu+\frac{n}{2})}} =
\tau(u,\nu,n)\,\frac{x^{\mu_1\cdots\mu_n}}
{x^{2(1+\frac{i\lambda}{2}+i\nu+\frac{n}{2})}}
\end{align}

{We see that eigenvalue $\tau(u,\nu,n)$ depends on $\nu$
and $n$ so that the corresponding eigenspace is $(n+1)^2$-dimensional,
where $(n+1)^2$ is dimension of the space of symmetric and traceless
tensors $x^{\mu_1\cdots\mu_n}$ of the rank $n$.
This degeneracy of the spectrum is dictated by the $SO(4)$-symmetry and
each eigenspace coincides to the $(n+1)^2$-dimensional irreducible representation of the group $SO(4)$.}

The system of functions $\Psi_{\mu_1\cdots\mu_n}(x)$
is orthogonal and complete:
the orthogonality relation has the following form
\begin{align}\label{ort1}
\int \,d^4 x \,
\frac{x^{\mu_1\cdots\mu_n}}
{x^{2(1+\frac{i\lambda}{2}+i\nu+\frac{n}{2})}}
\frac{x_{\nu_1\cdots\nu_{n'}}}
{x^{2(1-\frac{i\lambda}{2}-i\nu^{\prime}+\frac{n^{\prime}}{2})}} = c_n\,
\delta_{n n'}\,\Pi^{\mu_1,\dots,\mu_n}_{\nu_1,\dots,\nu_n}\,,
\end{align}
and the completeness relation is
\begin{align}\label{compl1}
\sum_{n\geq 0}\, \frac{1}{c_n}\,\int\limits_{-\infty}^{+\infty} d\nu\,
\frac{x^{\mu_1\cdots\mu_n}}
{x^{2(1+\frac{i\lambda}{2}+i\nu+\frac{n}{2})}}
\frac{y^{\mu_1\cdots\mu_{n}}}
{y^{2(1-\frac{i\lambda}{2}-i\nu+\frac{n}{2})}} =
\delta^{4}\left(x - y\right)\ ;\ \ c_n = \frac{2^{1-n}\pi^3}{n+1}\,.
\end{align}
{Relations~(\ref{ort1}) and~(\ref{compl1}) can be proved in
a straightforward way but we should note that in fact they are
consequences of Peter-Weyl theorem for the group $SU(2)$ (see Appendix \ref{app:SU(2)}). Moreover, the connection to the representation
theory explains in a natural way the appearance of the
factor $n+1$ in $c_n$: it is dimension of the irreducible representation
of the group $SU(2)$.}

{Finally, we obtain two equivalent representation for the kernel
$Q_{u}(x,y)$ of the integral operator $Q(u)$
\begin{align}\nonumber
&Q_{u}(x,y) =
\frac{4^u}{\pi^{2}\,a_0(2+u)}\,
\frac{1}{x^{2i\lambda}}\,
\frac{1}{(x-y)^{2(2+u)}}\,
\frac{1}{y^{2(-u-i\lambda)}} = \\
\label{Qspec}
&\frac{1}{2\pi^3}\,\sum_{n\geq 0}\, 2^n(n+1)\,
\int\limits_{-\infty}^{+\infty} d\nu\,
\tau(u,\nu,n)\,\frac{x^{\mu_1\cdots\mu_n}}
{x^{2(1+\frac{i\lambda}{2}+i\nu+\frac{n}{2})}}
\frac{y^{\mu_1\cdots\mu_{n}}}
{y^{2(1-\frac{i\lambda}{2}-i\nu+\frac{n}{2})}}
\end{align}
The first expression for the kernel is by
definition of the operator $Q(u)$ and the second expression is the spectral decomposition obtained by inserting of the resolution of the unity.}

\subsection{Diagrams computation}

{Now it is possible to use the spectral representation
of the operator $Q(u)$ and reduce the generic ladder
diagram to the expression containing only one integration over $\nu$
and summation over $n$.}

The kernel $Q_{u_1\cdots u_{L+1}}(x,y)$ of the integral
operator $Q(u_1)\cdots Q(u_{L+1})$ is given by the convolutions of the kernels
of the integral operators $Q(u_k)$
\begin{align*}
Q_{u_1\cdots u_{L+1}}(x,y) = \frac{1}{x^{2i\lambda}}\,
\prod_{k=1}^{L}\frac{4^{u_k}}{\pi^{2}\,a_0(2+u_k)}\,\int d^4 x_k
\frac{1}{x_{k-1 k}^{2(2+u_k)}}\,
\frac{1}{x_k^{-2u_k}}\,
\frac{1}{(x_{L}-y)^{2(2+u_{L+1})}}\,
\frac{1}{y^{2(-u_{L+1}-i\lambda)}}
\end{align*}
where $x_{ik} = x_i-x_k$ and we put $x_0=x$.
The functions \eqref{eig1} are common eigenfunctions
of all commuting operators $Q(u_k)$
\begin{align}\nonumber
\left[Q(u_k)\Psi^{\mu_1\cdots\mu_n}\right](x) =
\tau(u_k,\nu,n)\,\Psi^{\mu_1\cdots\mu_n}(x)\,,
\end{align}
so that the conformal ladder integral
\begin{align*}
&\frac{1}{x^{2i\lambda}}\,
\prod_{k=1}^{L}\int d^4 x_k
\frac{1}{x_{k-1 k}^{2(2+u_k)}}\,
\frac{1}{x_k^{-2u_k}}\,
\frac{1}{(x_{L}-y)^{2(2+u_{L+1})}}\,
\frac{1}{y^{2(-u_{L+1}-i\lambda)}}
\end{align*}
can be calculated by inserting of the resolution of identity
\begin{align*}
\frac{1}{2\pi^3}\,\sum_{n\geq 0}\, 2^n(n+1)\,
\int\limits_{-\infty}^{+\infty} d\nu\,
\frac{x^{\mu_1\cdots\mu_n}}
{x^{2(1+\frac{i\lambda}{2}+i\nu+\frac{n}{2})}}
\frac{y^{\mu_1\cdots\mu_{n}}}
{y^{2(1-\frac{i\lambda}{2}-i\nu+\frac{n}{2})}} =
\delta^{4}\left(x - y\right)\,,
\end{align*}
and after transition to the unit vectors $x^{\mu}= |x|\,\hat{x}^{\mu}$ and
$y^{\mu}= |y|\,\hat{y}^{\mu}$ we obtain the following expression
\begin{align*}
\frac{\pi^{2L-1}}{2}\, \sum_{n\geq 0}\, 2^n(n+1)\,
\int\limits_{-\infty}^{+\infty} d\nu\,
\frac{\hat{x}^{\mu_1\cdots\mu_n}}
{x^{2(1+\frac{i\lambda}{2}+i\nu)}}
\frac{\hat{y}^{\mu_1\cdots\mu_{n}}}
{y^{2(1-\frac{i\lambda}{2}-i\nu)}}
\prod_{k=1}^{L+1} 4^{-u_k}\,a_0(u_k+2)\,\tau(u_k,\nu,n)\,.
\end{align*}
The convolution $\hat{x}^{\mu_1\cdots\mu_n} \hat{y}^{\mu_1\cdots\mu_n}$
can be calculated explicitly.
We use evident formula
$$
\hat{x}^{\mu_1\cdots\mu_n} \hat{y}^{\mu_1\cdots\mu_n} = \hat{x}^{\mu_1}\cdots\hat{x}^{\mu_n} \hat{y}^{\mu_1\cdots\mu_n} = \hat{x}^{\mu_1\cdots\mu_n} \hat{y}^{\mu_1}\cdots\hat{y}^{\mu_n}\,,
$$
and the explicit expression~(\ref{x}) to get
\begin{align}
\label{Gegenbauer}
\hat{x}^{\mu_1\cdots\mu_n} \hat{y}^{\mu_1\cdots\mu_n} =
\sum_{p\geq 0}
\frac{(-1)^p(n-p)!}{4^p p!(n-2p)!}\,
\left(\hat{x}\,,\hat{y}\right)^{n-2p} = 2^{-n}\,
C^{1}_n\left(\cos\theta\right) = 2^{-n}\,\frac{\sin\left((n+1)\theta\right)}{\sin\theta}\,,
\end{align}
where $C^{1}_n\left(x\right)$ is Gegenbauer polynomial and
it is useful to express everything in terms of angle $\theta$ between
two unit vectors $\hat{x}$ and $\hat{y}$:
$\left(\hat{x}\,,\hat{y}\right) = \cos\theta$.

Collecting everything together we obtain the following generalization of the relation~(\ref{Qspec})
\begin{align}\label{QspecL}
&\frac{1}{x^{2i\lambda}}\,
\prod_{k=1}^{L}\int d^4 x_k
\frac{1}{x_{k-1 k}^{2(2+u_k)}}\,
\frac{1}{x_k^{-2u_k}}\,
\frac{1}{(x_{L}-y)^{2(2+u_{L+1})}}\,
\frac{1}{y^{2(-u_{L+1}-i\lambda)}} = \\
\nonumber
& \frac{\pi^{2L-1}}{2}\, \sum_{n\geq 0}\,(n+1)\,
\int\limits_{-\infty}^{+\infty} d\nu\,
\frac{\hat{x}^{\mu_1\cdots\mu_n}}
{x^{2(1+\frac{i\lambda}{2}+i\nu)}}
\frac{\hat{y}^{\mu_1\cdots\mu_{n}}}
{y^{2(1-\frac{i\lambda}{2}-i\nu)}}
\prod_{k=1}^{L+1} 4^{-u_k}\,a_0(u_k+2)\,\tau(u_k,\nu,n)\,
\frac{\sin\left((n+1)\theta\right)}{\sin\theta}
\end{align}
A particular case of the general ladder integral is recovered for parameters $\lambda =0\,,u_k = -1$, when the propagators become that of a bare massless scalar field. In this case the ladder of lenght $L$ coincides with the BD diagram of size $1 \times L$, that is 
\vspace{0.25cm}
\begin{center}
\includegraphics[scale=0.6]{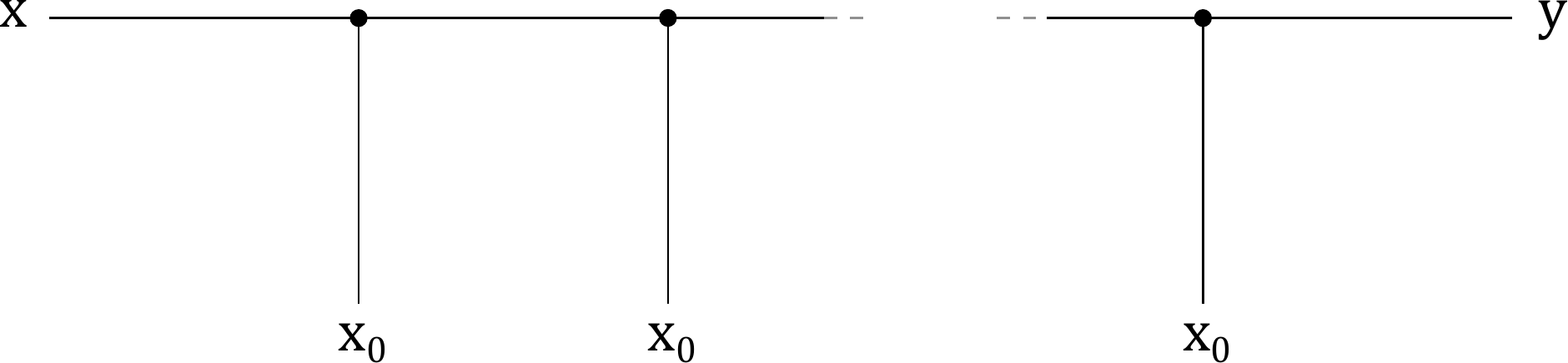}
\end{center}
In this case the eigenvalue becomes
$$
4^{-u}\,a_0(u+2)\,\tau(u,\nu,n) \xrightarrow{\lambda =0\,,\, u = -1}
\frac{1}{\left(\frac{n}{2}-i\nu\right)
\left(\frac{n}{2}+i\nu+1\right)}\,,
$$
and the ladder integral is reduced to the form
\begin{align*}
&\prod_{k=1}^{L}
\int d^4 x_k
\frac{1}{x_{k-1 k}^{2}}\,
\frac{1}{x_k^{2}}\,
\frac{1}{(x_{L}-y)^{2}} = \\
&\frac{\pi^{2L-1}}
{2x^2}\,
\sum_{n\geq 0}\, (n+1)\,\int\limits_{-\infty}^{+\infty} d\nu\,
\left(\frac{y^2}
{x^{2}}\right)^{i\nu}
\frac{1}{\left(\frac{n}{2}-i\nu\right)^{L+1}
\left(\frac{n}{2}+i\nu+1\right)^{L+1}}\,
\frac{\sin\left((n+1)\theta\right)}{\sin\theta} = \\
&\frac{\pi^{2L-1}}
{2x^2}\,
\sum_{n > 0}\, n\,\int\limits_{-\infty}^{+\infty} d\nu\,
\left(\frac{y^2}
{x^{2}}\right)^{i\nu}
\frac{1}{\left(\frac{n}{2}-\frac{1}{2}-i\nu\right)^{L+1}
\left(\frac{n}{2}+\frac{1}{2}+i\nu\right)^{L+1}}\,
\frac{\sin\left(n\theta\right)}{\sin\theta}=\\
&\frac{\pi^{2L-1}}
{4 x^2}\,
\sum_{n\in Z}\, n\,\int\limits_{-\infty}^{+\infty} d\nu\,
\left(\frac{y^2}
{x^{2}}\right)^{i\nu}
\frac{1}{\left(\left(\nu-\frac{i}{2}\right)^2+\frac{n^2}{4}\right)^{L+1}}\,
\frac{\sin\left(n\theta\right)}{\sin\theta}\,,
\end{align*}
{where we used the shift $n \to n-1$ and then
symmetry with respect to $n \to -n$.}

After the last transformation: change of variable $\nu \to \nu +\frac{i}{2}$ and deformation of the integration contour to the initial position we obtain
\begin{align}
\prod_{k=1}^{L}
\int d^4 x_k
\frac{1}{x_{k-1 k}^{2}}\,
\frac{1}{x_k^{2}}\,
\frac{1}{(x_{L}-y)^{2}} =
\frac{\pi^{2L-1}}
{4 \left(x^2y^2\right)^{\frac{1}{2}}}\,
\sum_{n\in Z}\, n\,\int\limits_{-\infty}^{+\infty} d\nu\,
\left(\frac{y^2}
{x^{2}}\right)^{i\nu}
\frac{1}{\left(\nu^2+\frac{n^2}{4}\right)^{L+1}}\,
\frac{\sin\left(n\theta\right)}{\sin\theta}\,.
\end{align}
The initial Basso-Dixon integral is produced by the shift $x\to x-x_0$
and $y\to y-x_0$ but the angle remains the same. It is important to note that all previous formulae are not specific for the four dimensional case but can be generalized to $d$ dimensions, as the formula \eqref{chainn} can be immediately generalized.

The peculiarity of the four-dimensional case lies in the fact that it is possible to convert tensors
to spinors and back. As we shall show in the rest of
the paper the expression for eigenfunctions in a spinor
form admits generalization to any number of sites $N$ in
model with Hamiltonian \eqref{4Dlocal}.

\subsection{Eigenfunctions in spinor notations}
\label{eispin}

{In the rest of the paper we shall use the spinor representation for the eigenfunctions of the Q-operator. Now we are going to illustrate the main notations and tricks using the simplest example $N=1$.}

Let us convert the tensor $x^{\mu_1\cdots\mu_n}$ to the spinor
\begin{align}
\psi_{a_1\ldots a_n}^{\dot{a}_1\ldots \dot{a}_n} =
\left(\boldsymbol{\sigma}_{\mu_1}\right)_{a_1}^{\dot{a}_1}\ldots
\left(\boldsymbol{\sigma}_{\mu_n}\right)_{a_n}^{\dot{a}_n}\,
x^{\mu_1\cdots\mu_n}\,.
\end{align}
This spinor is symmetric with respect to $a_1\ldots a_n$ and
$\dot{a}_1\ldots \dot{a}_n$ independently. Note that our metric is Euclidean and we shall use notation from Appendix \ref{app:spinors}.

It is useful to introduce a generating function using the convolution with two
external spinors
\begin{align}
\psi_{a_1\ldots a_n}^{\dot{a}_1\ldots \dot{a}_n} \,\alpha^{a_1}\cdots\alpha^{a_n}
\,\beta_{\dot{a}_1}\cdots \beta_{\dot{a}_n} =
|x|^{n}\langle\alpha|\boldsymbol{x}|\beta\rangle^n\,,
\end{align}
where we introduce the notation
\begin{equation}
\boldsymbol{x}=\sig^{\mu} \hat{x}_{\mu}\, ,\,\,\,  \bar{\boldsymbol{x}}=\bsig^{\mu}\hat{x}_{\mu}\, ,\,\,\, x^{\mu} = |x|\,\hat{x}_{\mu}
\end{equation}
$\boldsymbol{x}$ is
the two-dimensional matrix and
$\langle\alpha|\boldsymbol{x}|\beta\rangle =
\alpha^{a}\boldsymbol{x}_{a}^{\dot{a}}\beta_{\dot{a}}$.
Note that this quantity can be represented in the equivalent form
$\langle\alpha|\boldsymbol{x}|\beta\rangle = x^{\mu}c_{\mu}$,
where the vector $c_{\mu} = \langle\alpha|\boldsymbol{\sigma}_{\mu}|\beta\rangle$
is automatically a null-vector $\left(c\,, c\right)=0$ due to the Fierz identity
$\boldsymbol{\sigma}_{\mu}\otimes\boldsymbol{\sigma}^{\mu} = 2(\II-\mathbb{P})$.
The spinor $\psi_{a_1\ldots a_n}^{\dot{a}_1\ldots \dot{a}_n}$ can be reconstructed from the generating
function via differentiation
\begin{align}
\psi_{a_1\ldots a_n}^{\dot{a}_1\ldots \dot{a}_n} = \frac{1}{(n!)^2}\, \partial_{\alpha^{a_1}}\cdots\partial_{\alpha^{a_n}}\,
\partial_{\beta_{\dot{a}_1}}\cdots\partial_{\beta_{\dot{a}_n}}\,
\langle\alpha|\boldsymbol{x}|\beta\rangle^n
\end{align}
and we shall consider the components of these
auxiliary spinors as generic independent variables.
Note that initial spinors have lower indices $\alpha_{a}$ and $\beta_{\dot{a}}$ and the index raising operation is defined as complex conjugation $\alpha^{a} =(\alpha_{a})^*=\bar{\alpha}_{a}$ and
$\beta^{\dot{a}} =(\beta_{\dot{a}})^*=\bar{\beta}_{\dot{a}}$ so that the rules of hermitian conjugation are
\begin{align}
\boldsymbol{\sigma}^{\dagger}_{\mu} = \overline{\boldsymbol{\sigma}}_{\mu}
\ \ ;\ \ \alpha^a = \bar{\alpha}_{a} \ \ ;\ \
\beta^{\dot{a}} = \bar{\beta}_{\dot{a}} \ \ ;\ \
\langle\alpha|\boldsymbol{\sigma}_{\mu}|\beta\rangle^{\dagger} = \langle\beta|\overline{\boldsymbol{\sigma}}_{\mu}|\alpha\rangle
\end{align}
in agreement with usual transition $\langle \alpha| \rightleftarrows |\alpha\rangle$ which means transposition and complex conjugation.

The orthogonality relation~(\ref{ort1})
\begin{align*}
\int \,d^4 x \,
\frac{x^{\mu_1\cdots\mu_n}}
{x^{2(1+\frac{i\lambda}{2}+i\nu+\frac{n}{2})}}
\frac{x^{\nu_1\cdots\nu_{n'}}}
{x^{2(1-\frac{i\lambda}{2}-i\nu^{\prime}+\frac{n^{\prime}}{2})}} = \frac{2^{1-n}\pi^3}{n+1}\,
\delta_{nn'}\,\delta_{\mu_1\cdots\mu_n ,\nu_1\cdots\nu_{n}}\delta(\nu-\nu')\,,
\end{align*}
has the following spinor counterpart
\begin{align}\label{ort1sp}
\int \,d^4 x \,
\frac{\langle\alpha|\boldsymbol{x}|\beta\rangle^n}
{x^{2(1+\frac{i\lambda}{2}+i\nu)}}
\frac{\langle\beta'|\overline{\boldsymbol{x}}|\alpha'\rangle^{n^{\prime}}}
{x^{2(1-\frac{i\lambda}{2}-i\nu^{\prime})}} =
\frac{2\pi^3}{n+1}\,\langle\alpha|\alpha'\rangle^n\,
\langle\beta'|\beta\rangle^n\,
\delta_{n n^{\prime}}\,\delta(\nu-\nu^{\prime})\,.
\end{align}
The previous l.h.s. is an integral of the type
\begin{align*}
\int \,d^4 x \,
\frac{\left(A\,, x\right)^n}
{x^{2(1+\frac{i\lambda}{2}+i\nu+\frac{n}{2})}}
\frac{\left(B\,, x\right)^{n^{\prime}}}
{x^{2(1-\frac{i\lambda}{2}-i\nu^{\prime}+\frac{n^{\prime}}{2})}} =
c_n\,\delta(\nu-\nu^{\prime})\,\delta_{n n^{\prime}} \left(A\,, B\right)^n \,,
\end{align*}
where in our particular case
$A_{\mu} = \langle\alpha|\boldsymbol{\sigma}_{\mu}|\beta\rangle\,, B_{\mu} = \langle\beta'|\overline{\boldsymbol{\sigma}}_{\mu}|\alpha'\rangle$ so that
$A^{\mu} A_{\mu} = B^{\mu} B_{\mu} = 0$ and $A^{\mu} B_{\mu} = \left(A\,, B\right) = 2\langle\alpha|\alpha'\rangle
\langle\beta'|\beta\rangle$. Indeed, in components we have
$$
\left(A\,, B\right) = \alpha^{a}
\left(\boldsymbol{\sigma}_{\mu}\right)_{a}^{\dot{a}} \beta_{\dot{a}}
\,{\beta'}^{\dot{b}}
\left(\overline{\boldsymbol{\sigma}}_{\mu}\right)_{\dot{b}}^{\ b}
\alpha'_{b} = 2\,\left(\alpha^{a}\,{\alpha'}_{a}\right)\,
\left({\beta'}^{\dot{a}}\,\beta_{\dot{a}}\right) = 2\langle\alpha|\alpha'\rangle
\langle\beta'|\beta\rangle\,,
$$
due to Fierz identity \begin{align}
\boldsymbol{\sigma}_{\mu}\otimes\overline{\boldsymbol{\sigma}}_{\mu} = 2\,\mathbb{P} \,\ \,\,\Longleftrightarrow\,\,
\left(\boldsymbol{\sigma}_{\mu}\right)_{a}^{\dot{a}}
\left(\overline{\boldsymbol{\sigma}}_{\mu}\right)_{\dot{b}}^{b} =
2\delta_{a}^{b}\,\delta_{\dot{b}}^{\dot{a}}\,,
\end{align}
and the pairing between spinors is the standard scalar product in $\mathbb{C}^2$
\begin{align}
\langle\alpha|\alpha'\rangle =
\alpha^{a} \alpha'_{a} = \bar{\alpha}_{a} \alpha'_{a}\, ,\, \,\,\,
\langle\beta'|\beta\rangle = {\beta'}^{\dot{a}}\beta_{\dot{a}} =
\bar{\beta}'_{\dot{a}}\beta_{\dot{a}}\,.
\end{align}

{As explained in Appendix \ref{app:spinors} the dotted and un-dotted indices distinguish spinors that are transformed according to representations
of two different copies of $SU(2)$ in our Euclidean case.}

{In many situations it will be useful to
adopt more condensed notations.
Let $A: \mathbb{C}^2\to \mathbb{C}^2$ is some two by two matrix. We denote $A\otimes\cdots\otimes A$ by
$[A]^n$ -- it is the operator acting in the space $Sym\, \mathbb{C}^2\otimes\cdots\otimes\mathbb{C}^2 = Sym\, \mathbb{C}^{\otimes n}$ according to the notation
\begin{align*}
&[A]^n = A\otimes\cdots\otimes A\,,\\
&[A]^{n_1}\,[B]^{n_2} = A\otimes\cdots\otimes A\otimes B\otimes\cdots\otimes B\,,
\end{align*}
and clearly the matrices with
enclosed in different brackets $[\dots]$ are acting on the symmetric spinors of different spaces.
Using these notations we can rewrite relation~(\ref{ort1sp}) in the following way
\begin{align}\label{ort1sp1}
\int \,d^4 x \,
\frac{[\boldsymbol{x}]^n}
{x^{2(1+\frac{i\lambda}{2}+i\nu)}}
\frac{[\overline{\boldsymbol{x}}]^{n^{\prime}}}
{x^{2(1-\frac{i\lambda}{2}-i\nu^{\prime})}} =
2^n\,c_n\,
\delta_{n n^{\prime}}\,\delta(\nu-\nu^{\prime})\,\mathbb{P}\,,
\end{align}
where $\mathbb{P}$ is operator of permutation:
$\mathbb{P} \,|\beta\rangle^{\otimes n}\otimes |\alpha\rangle^{\otimes n} = |\alpha\rangle^{\otimes n}\otimes |\beta\rangle^{\otimes n}$.
}
For full clarity, we should write here for once the explicit spinor indices:
\begin{align}
\nonumber
([\boldsymbol{x}]^n)^{\dot{\mathbf{a}}}_{\mathbf{a}}=([\boldsymbol{x}]^n)^{(\dot{a}_1\dots \dot{a}_{n})}_{(a_1\dots a_{n})}
\qquad  ; \qquad ([\overline{\boldsymbol{x}}]^{n^{\prime}})^{{\mathbf{b}}}_{\dot{\mathbf{b}}}=
([\overline{\boldsymbol{x}}]^{n^{\prime}})^{(b_1\dots b_{n^{\prime}})}_{(\dot{b}_1\dots \dot{b}_{n^{\prime}})}\,,\\
\label{spin_index}
\int \,d^4 x \,
\frac{([\boldsymbol{x}]^n)^{\dot{\mathbf{a}}}_{\mathbf{a}}}
{x^{2(1+\frac{i\lambda}{2}+i\nu)}}
\frac{([\overline{\boldsymbol{x}}]^{n^{\prime}})^{{\mathbf{b}}}_{\dot{\mathbf{b}}}}
{x^{2(1-\frac{i\lambda}{2}-i\nu^{\prime})}} =
2^n\,c_n\,
\delta_{n n^{\prime}}\,\delta(\nu-\nu^{\prime})\,\delta^{\dot{\mathbf{a}}}_{\dot{\mathbf{b}}}\,
\delta^{\mathbf{b}}_{\mathbf{a}}\,.
\end{align}
In order to transform the completeness relation in tensorial form~(\ref{compl1}) to the spinorial form, we exploit the property
$\left(\boldsymbol{\sigma}_{\mu}\right)_{a}^{\dot{a}}
\left(\overline{\boldsymbol{\sigma}}_{\nu}\right)_{\dot{a}}^{a} = 2\delta_{\mu\nu}\,,$
on the first stage, then transform the convolution with respect to spatial indices to
the convolution with respect to spinor indices and on the last step to use Gaussian integration for
substitution of convolution with respect to spinor indices
\begin{align*}
&x^{\mu_1\cdots\mu_n}\,y^{\mu_1\cdots\mu_n} = \frac{1}{2^n}\,
\left(\boldsymbol{\sigma}_{\mu_1}\right)_{a_1}^{\dot{a}_1}\cdots
\left(\boldsymbol{\sigma}_{\mu_n}\right)_{a_n}^{\dot{a}_n}\,
x^{\mu_1\cdots\mu_n}\,
\left(\overline{\boldsymbol{\sigma}}_{\nu_1}\right)_{\dot{a}_1}^{a_1}
\cdots
\left(\overline{\boldsymbol{\sigma}}_{\nu_n}\right)_{\dot{a}_n}^{a_n}\,
y^{\nu_1\cdots\nu_n} = \\
&\frac{|x|^n\,|y|^n}{2^n\,(n!)^2}\, \int D\alpha D\beta\, \langle\alpha|\boldsymbol{x}|\beta\rangle^n\,
\langle\beta|\overline{\boldsymbol{y}}|\alpha\rangle^{n}\,.
\end{align*}
This substitution is based on the standard Gaussian integration \cite{Perelomov:1980tt,Faddeev:1980be,Isaev:2018xcg} over $\alpha$ and $\beta$ 
\begin{align}
\label{spinor_measure}
\int D\alpha = \frac{1}{\pi^{2}}\,\int d^2 \alpha_1 d^2 \alpha_2\, e^{-|\alpha_1|^2-|\alpha_2|^2}\,.
\end{align}
We have the following Gaussian integrals
\begin{align}\label{Gauss}
\int D\alpha\, \alpha^{a_1}\cdots\alpha^{a_n}\,
\alpha_{c_1}\cdots\alpha_{c_n} =
\int D\alpha\, \bar{\alpha}_{a_1}\cdots\bar{\alpha}_{a_n}\,
\alpha_{c_1}\cdots\alpha_{c_n} =
n!\,\hat{S}\,
\delta^{a_1}_{c_1}\,\delta^{a_2}_{c_2}\cdots
\delta^{a_n}_{c_n}\,,\\
\int D \beta\, \beta^{\dot{a}_1}\cdots\beta^{\dot{a}_n}\,
\beta_{\dot{c}_1}\cdots\beta_{\dot{c}_n} =
\int D \beta\, \bar{\beta}_{\dot{a}_1}\cdots\bar{\beta}_{\dot{a}_n}\,
\beta_{\dot{c}_1}\cdots\beta_{\dot{c}_n} =
n!\,\hat{S}\,
\delta^{\dot{a}_1}_{\dot{c}_1}\,\delta^{\dot{a}_2}_{\dot{c}_2}\cdots
\delta^{\dot{a}_n}_{\dot{c}_n}\,,
\end{align}
where $\hat{S}$ is the symmetrization over all spinor indices
\begin{align}
\hat{S}\,\Phi^{a_1\cdots a_n} =
\frac{1}{n!} \sum_{p\in S_n} \Phi^{a_{p(1)}\cdots a_{p(n)}}\,.
\end{align}
Finally, after all transformations one obtains the
completeness relation in spinor form
\begin{align}
\sum_{n\geq 0}\, \frac{1}{c_n}\,\frac{1}{2^n\,(n!)^2}\,
\int\limits_{-\infty}^{+\infty} d\nu\,
\int D\alpha D\beta\,
\frac{\langle\alpha|\boldsymbol{x}|\beta\rangle^n}
{x^{2(1+\frac{i\lambda}{2}+i\nu)}}
\frac{\langle\beta|\overline{\boldsymbol{y}}|\alpha\rangle^{n}}
{y^{2(1-\frac{i\lambda}{2}-i\nu)}} =
\delta^{4}\left(x - y\right)\,.
\end{align}

\section{Star-triangle identity in four dimensions}
\label{sec:startri}
The computation of the spectrum of \eqref{4Dlocal}, or equivalently of \eqref{Q1}, can be done exactly  constructing first a basis of eigenfunctions and then obtaining the eigenvalues by direct application of the operator to the eigenfunctions. The recipe needed to do all this procedure follows closely the one elaborated for the $2d$ analogue of the model under study in \cite{Derkachov2001,Derkachov2014}, which is ultimately based on the chain-rule identity (or its equivalent star-triangle form)
\begin{align}
\label{chain2d}
\begin{aligned}
&\int \frac{d^2 w}{[z_1-w]^{a}[z_2-w]^{b}}=\pi\frac{\Gamma\left(1-\bar a\right)\Gamma\left(1-\bar b\right)\Gamma\left(\bar a +\bar b -1\right)}{\Gamma\left(a\right)\Gamma\left(b\right)\Gamma\left(2-a-b\right)}
\frac{1}{[z_{1}-z_{2}]^{(a+b-1)}}
\,,
\end{aligned}
\end{align} 
where $a,b$, $\bar a,\bar b$ are complex numbers subject to the constraint $a-\bar a\,,\, b-\bar b \in \mathbb{Z}$, and we used the notation for two-dimensional conformal propagator
\begin{equation}
\label{propag2d}
\frac{1}{[z]^a} = \frac{1}{(z)^a(z^*)^{\bar a}}\,,
\end{equation}
where $\Delta=(a+\bar{a})/2$ is the scaling dimension and $\ell=a-\bar a$ is the spin. 
The analogue of $\eqref{propag2d}$ in $4d$ is given by 
\begin{align}
\label{spin_prog4d}
\frac{(n^{\mu} x_{\mu})^{\ell}}{(x^2)^{\Delta}}\,,\,\,\, n^{\mu} n_{\mu} =0\,,
\end{align}
and the null vector $n$ can be constructed by means of auxiliary spinors, as explained in Appendix \ref{app:spinors}
\begin{equation}
n_{\mu}= \langle \alpha|\sig_{\mu}|\beta \rangle\,,\,\,\text{or}\,\,\,\,
n_{\mu}= \langle \beta|\bsig_{\mu}|\alpha \rangle\,.
\end{equation}
Now, for the case of $\ell=0$ the formula \eqref{chain2d} can be immediately generalized to any space-time dimension $d$ as
\begin{align}
\label{scalar_chain_d}
\begin{aligned}
&\int \frac{d^d y}{(x_1-y)^{2a}(x_2-y)^{2b}}=\pi^{\frac{d}{2}} \frac{\Gamma\left(\frac{d}{2}-a\right)\Gamma\left(\frac{d}{2}- b\right)\Gamma\left(a + b -\frac{d}{2} \right)}{\Gamma\left(a\right)\Gamma\left(b\right)\Gamma\left({d}-a-b\right)}\frac{1}{x_{12}^{2(a+b-\frac{d}{2})}}
\,.
\end{aligned}
\end{align} 
Nevertheless, it is crucial for us to deal with the general case \eqref{spin_prog4d}, as made evident by the discussion about the spin chain model \eqref{4Dlocal} at $N=1$ of section \ref{ladder}. For this reason the rest of the section is devoted to the computation of \eqref{scalar_chain_d} for $d=4$ generalized to the case of any spins $\ell$ and $\ell'$
\begin{align}
\begin{aligned}
&\int d^4 y\frac{[\langle \alpha_1|\sig_{\mu}|\beta_1 \rangle(x_1-y)^{\mu}]^{\ell}[\langle \beta_2|\bsig_{\nu}|\alpha_2 \rangle (x_2-y)^{\nu}]^{\ell'}}{(x_1-y)^{2(a+\ell/2)}(x_2-y)^{2(b+\ell'/2)}}\,.
\end{aligned}
\end{align}
In order to formulate a generalization of the chain-rule identity and star-triangle relation \eqref{str0} which includes spin degrees of freedom, we should take into account that two-components spinors can undergo linear transformations by means of matrices. In particular, we will see in the next section that the spinorial degrees of freedom in the generalized star-triangle relation are mixed by an R-operator.
\subsection{Spinors mixing}
The R-operator is a solution of the Yang-Baxter equation (YBE) \cite{Baxter1982,Takhtajan:1979iv}
\begin{equation}
\label{YB} R_{12}(u) R_{13}(u+v)
R_{23}(v) = R_{23}(v) R_{13}(u+v) R_{12}(u)\,,
\end{equation}
and it is function of a spectral parameter $u$. We recall that the YBE is defined on the tensor product of three vector spaces $V_1\otimes V_2\otimes V_3$, and the indices show that $R_{12}(u)$ acts nontrivially
in the space $V_1\otimes V_2$ and is the identity operator on $V_3$. In order to fix the ideas let us set $V_1=V_2=V_3$ to be the the space of two-spinors $V=\mathbb{C}^2$. Then the solution of the YBE is the so-called Yang R-matrix
\begin{equation} \label{RYang}
R^{c\, d}_{a\, b}(u) = \frac{1}{u+1}\,\left(u\, \delta_{a}^{c}\,\delta_{b}^{d}+
\delta_{a}^{d}\,\delta_{b}^{c}\right)\,,
\end{equation}
where $a,b,c,d =1,2$.
In the following sections we will need to pick $V_k$  to be the space of the symmetric spinors $\Psi_{(a_1\ldots a_{n_k})}$ that is the space of the $(n_k+1)$-dimensional representation of the group $\mathrm{SU}(2)$, corresponding to spin $\frac{n_k}{2}$ and $(\ldots)$ is the standard notation for the symmetric structure
\begin{equation}
(a_1\ldots a_i \ldots a_j \ldots a_{n_k}) =(a_1\ldots a_j \ldots a_i \ldots a_{n_k})\, .
\end{equation}
Thus, the general operator $R_{12}(u)$ acts in a tensor product of two representations with spins $\frac{n_1}{2}$ and $\frac{n_2}{2}$, namely the space of spinors $\Psi_{(a_1\ldots a_{n_1})(b_1\ldots b_{n_2})}$. In the matrix notations we have
\begin{align}
\left[R_{12}(u)\Psi\right]_{(a_1\ldots a_{n_1})(b_1\ldots b_{n_2})} =
R^{(c_1\ldots c_{n_1})(d_1\ldots d_{n_2})}_{(a_1\ldots a_{n_1})(b_1\ldots b_{n_2})}(u)\,
\Psi_{(c_1\ldots c_{n_1})(d_1\ldots d_{n_2})}\,,
\end{align}
where the summation over repeated indices is assumed.
For simplicity we skip indices $12$ in the matrix of operator $R_{12}(u)$.
The standard procedure for constructing finite-dimensional higher-spin $\mathrm{R}$-operators out of
the Yang R-matrix is the {\it fusion procedure}.
Following the recipe of~\cite{Kulish1981,Kulish82} we form the product of the Yang R-matrices
\begin{equation} \label{n_1}
\mathbf{R}_{(a_1\ldots a_{n_1})\,b}^{(c_1 \ldots c_{n_1})\,d}
\textstyle\left(u+\frac{n_1-1}{2}\right) =
\mathrm{Sym}\,
R^{c_1\, d_1}_{a_1\, b}(u)\,R^{c_2\, d_2}_{a_2\, d_1}(u+1)
\,\cdots\,R^{c_{n_1}\, d}_{a_{n_1}\, d_n}(u+n_1-1),
\end{equation}
where $\mathrm{Sym}$ implies symmetrization with respect to groups of indices $c_1 \ldots c_{n_1}$.
The symmetry with respect to groups of indices $a_1 \ldots a_{n_1}$ then follows from 
the Yang-Baxter relations for the Yang R-matrix.
In such a way one obtains an operator acting on the space of symmetric rank $n_1$
spinors, i.e. on the space of spin $\frac{n_1}{2}$ representation, and on the
two-dimensional space of spin $\frac{1}{2}$ representation.
Next the R-operator~(\ref{n_1}) is used as building block and
repetition of the same procedure increases spin of the representation in the second space from $\frac{1}{2}$ to $\frac{n_2}{2}$,
\begin{equation}
\label{n_2}
\mathbf{R}_{(a_1\ldots a_{n_1})\,(b_1\ldots b_{n_2})}^{(c_1 \ldots c_{n_1})\,(d_1 \ldots d_{n_2})}
\textstyle\left(u+\frac{n_1-1}{2}\right) =
\mathrm{Sym}\,
\mathbf{R}^{(c'_1 \ldots c'_{n_1})\, d_1}_{(a_1\ldots a_{n_1})\, b_1}(u)\,\mathbf{R}^{(c''_1 \ldots c''_{n_1})\, d_2}_{(c'_1 \ldots c'_{n_1})\, b_2}(u+1)
\cdots\mathbf{R}^{(c_1 \ldots c_{n_1})\, d_{n_2}}_{(a_1\ldots a_{n_1})\, b_{n_2}}(u+n_2-1).
\end{equation}
In a more compact notation we will refer to the fused $\mathbf R$-matrix acting on symmetric spinors of rank $n_1$ and $n_2$ as $\mathbf{R}_{n_1,n_2}(u)$ or 
$\mathbf{R}_{(a_1\ldots a_{n_1})\,(b_1\ldots b_{n_2})}^{(c_1 \ldots c_{n_1})\,(d_1 \ldots d_{n_2})}\to \mathbf{R}_{\boldsymbol{a}\boldsymbol{b}}^{\boldsymbol{c}\boldsymbol{d}}$ in the case when we will need explicit indieces.
In Appendix \ref{app:R}  we derive explicit compact expression for the genearal R-matrix 
$\mathbf{R}_{n_1,n_2}(u)$ in a form which is especially adapted to the calculation 
of the Feynman diagrams.

Note that all formulae in this section are written for the spinors with un-dotted indices.
Of course there are analogous formulae for the spinors with dotted indices -- in all formulae one should change un-dotted indices by the dotted ones. 
We do not write all that explicitly just to avoid the non informative doubling. 
In the next section we will work with all spinors with dotted and un-dotted indices.

\subsection{Zamolodchikov's $R$-matrix}
\label{zamol}
The Yang $R$-matrix \eqref{RYang} can be used to define solutions of the Yang-Baxter equation acting on a couple of space-time indices $\mu,\nu=0,\dots,d-1$. Indeed (as explained in detail in the appendix \eqref{ts}-\eqref{st}) an arbitrary tensor of rank two $t^{\mu\nu}$ can be converted into a spinor of rank four according to the isomorphism defined by
\begin{align}\label{ts2}
\psi_{ab}^{\dot{a}\dot{b}} =
\left(\boldsymbol{\sigma}_{\mu}\right)_{a}^{\dot{a}}
\left(\boldsymbol{\sigma}_{\nu}\right)_{b}^{\dot{b}}\, t^{\mu\nu}\,,
\end{align}
whose inverse formula, transforming a rank-four spinor to the corresponding rank-two tensor, reads
\begin{align}\label{st2}
t^{\mu \nu} = \frac{1}{4}\,
\left(\overline{\boldsymbol{\sigma}}^{\mu}\right)_{\dot{a}}^{a}
\left(\overline{\boldsymbol{\sigma}}^{\nu}\right)_{\dot{b}}^{b}\,
\psi_{ab}^{\dot{a}\dot{b}}\,.
\end{align}
It follows from $\eqref{st2}$ that the action of the Yang $R$-matrix on the space of rank-two spinors
\begin{align}
\label{actr2}
\phi_{ab} \xrightarrow{R(u)} \phi^{\prime}_{ab} =
R_{a b}^{cd}(u)\, \phi_{cd}\,,
\end{align}
can be used to induce an action on the space of rank-two tensors $t^{\mu\nu}$. 
Indeed, let the action on rank-four spinors be the direct product of \eqref{actr2}
\begin{align}
\phi_{ab}^{\dot{a}\dot{b}} \rightarrow
{\phi^{\prime}}_{ab}^{\dot{a}\dot{b}} =
R_{ab}^{cd}(u)\,\phi_{cd}^{\dot{c}\dot{d}}\,
R^{\dot{a}\dot{b}}_{\dot{c}\dot{d}}(u)
\end{align}
then we can define $\mathcal{R}_{\mu \nu}^{\alpha\beta}$ such that
\begin{align}
t_{\mu\nu} \rightarrow t^{\prime}_{\mu\nu} = 
\mathcal{R}_{\mu\nu}^{\alpha\beta}(u)\, t_{\alpha\beta}\,,
\end{align}
imposing through relations \eqref{ts2} and \eqref{st2} that
\begin{equation}
t'_{\mu\nu}=\mathcal{R}_{\mu\nu}^{\alpha\beta}(u)t_{\alpha\beta}=\frac{1}{4}\,
\left(\overline{\boldsymbol{\sigma}}_{\mu}\right)_{\dot{a}}^{a}
\left(\overline{\boldsymbol{\sigma}}_{\nu}\right)_{\dot{b}}^{b}\,
{\phi'}_{ab}^{\dot{a}\dot{b}}\,,\,\,\,\,\,\text{and}\,\,\,\,\,\,\,
\phi_{cd}^{\dot{c}\dot{d}}=\left(\boldsymbol{\sigma}^{\alpha}\right)_{c}^{\dot{c}}
\left(\boldsymbol{\sigma}^{\beta}\right)_{d}^{\dot{d}}\,t_{\alpha\beta}\,.
\end{equation}
Finally we obtain that
\begin{align}
\label{ZamoR}
\mathcal{R}_{\mu\nu}^{\alpha\beta}(u) = \frac{1}{4}\,
\left(\overline{\boldsymbol{\sigma}}_{\mu}\right)_{\dot{c}}^{c}
\left(\overline{\boldsymbol{\sigma}}_{\nu}\right)_{\dot{d}}^{d}\,
R_{cd}^{ab}(u)\,
\left(\boldsymbol{\sigma}^{\alpha}\right)_{a}^{\dot{a}}
\left(\boldsymbol{\sigma}^{\beta}\right)_{b}^{\dot{b}}\,
R^{\dot{c}\dot{d}}_{\dot{a}\dot{b}}(u)
\end{align}
which is a solution of the Yang-Baxter equation on the rank-two tensors. The expression on the r.h.s. of \eqref{ZamoR} can be computed explicitly \begin{align*}
&\mathcal R^{\alpha\beta}_{\mu\nu}(u) = \frac{1}{4(u+1)^2}\,
\left(\overline{\boldsymbol{\sigma}}_{\alpha}\right)_{\dot{c}}^{ c}
\left(\overline{\boldsymbol{\sigma}}_{\beta}\right)_{\dot{d}}^{d}\,
\left[u\,\delta_{c}^{a}\delta_{d}^{b} +
\delta_{c}^{b}\delta_{d}^{a}\right]\,
\left(\boldsymbol{\sigma}_{\mu}\right)_{a}^{\dot{a}}
\left(\boldsymbol{\sigma}_{\nu}\right)_{b}^{\dot{b}}\,
\left[u\,\delta^{\dot{c}}_{\dot{a}}\delta^{\dot{d}}_{\dot{b}} +
\delta^{\dot{c}}_{\dot{b}}\delta^{\dot{d}}_{\dot{a}}\right]= \\
&=\frac{1}{(u+1)^2}\left(\frac{u^2}{4}
\tr\left(\overline{\boldsymbol{\sigma}}_{\alpha}\boldsymbol{\sigma}_{\mu}\right)
\tr\left(\overline{\boldsymbol{\sigma}}_{\beta}\boldsymbol{\sigma}_{\nu}\right)
+
\frac{1}{4}
\tr\left(\overline{\boldsymbol{\sigma}}_{\alpha}\boldsymbol{\sigma}_{\nu}\right)
\tr\left(\overline{\boldsymbol{\sigma}}_{\beta}\boldsymbol{\sigma}_{\mu}\right)
+\frac{u}{4}
\tr\left(\boldsymbol{\sigma}_{\mu}{\overline{\boldsymbol{\sigma}}_{\alpha}
\boldsymbol{\sigma}_{\nu}}\overline{\boldsymbol{\sigma}}_{\beta}
\right)+\frac{u}{4}
\tr\left(\boldsymbol{\sigma}_{\mu}{\overline{\boldsymbol{\sigma}}_{\beta}
\boldsymbol{\sigma}_{\nu}}\overline{\boldsymbol{\sigma}}_{\alpha}
\right)\right) = \\ 
&=\frac{1}{(u+1)}\left(u\,\delta_{\alpha \mu}\delta_{\beta \nu} +
\delta_{\alpha \nu}\delta_{\beta \mu} -\frac{u}{u+1}\,
\delta_{\alpha \beta}\delta_{\mu \nu}\right)\,.
\end{align*}
Note that in our Euclidean situation it is possible not distinguish lower 
and upper tensor indices $\mu\,,\nu\,,\alpha\,,\beta\,\ldots$
The last formula can be rewritten in the standard form in terms of identity $\mathbbm{1}$, permutation $\mathbb{P}$ and contraction $\mathbb{K}$ operators
\begin{equation}
\label{ZamoR1}
\mathcal{R}(u) =\frac{1}{(u+1)}\left(u\,\II +\mathbb{P} -\frac{u}{u+1}\,\mathbb{K}\right)\,,
\end{equation}
and we recognize that it coincides with the $R$-matrix of the $O(d)$ model by A. Zamolodchikov and Al. Zamolodchikov for the space-time dimension $d=4$ \cite{Zamolodchikov:1977nu,Zamolodchikov:1978xm} (see also, in relation to fishnet graphs: \cite{Zamolodchikov:1980mb}).

The same fusion procedure used in order to define an $R$-matrix over the spaces of $\ell$ and $\ell'$ symmetric spinors starting from \eqref{RYang}, can be applied in order to generalize the Zamolodchikov's $R$-matrix  \eqref{ZamoR} to rank $\ell$ and $\ell'$ symmetric tensors. The fused matrix $\mathcal{R}_{\ell,\ell'}(u)$ is given through the fused matrix $\mathbf{R}_{\ell,\ell'}(u)$ over spinors \eqref{n_2}, according to the isomorphism \eqref{ts}-\eqref{st}, as 
\begin{align}
\begin{aligned}
&\mathcal{R}_{(\mu_1,\dots,\mu_{\ell})(\nu_1,\dots,\nu_{\ell'})}^{(\alpha_1,\dots,\alpha_{\ell})(\beta_1,\dots,\beta_{\ell'})}(u) =\frac{1}{2^{\ell+\ell'}}\,
\left(\overline{\boldsymbol{\sigma}}_{\mu_1}\right)_{\dot{c}_1}^{c_1}\cdots\left(\overline{\boldsymbol{\sigma}}_{\mu_{\ell}}\right)_{\dot{c}_{\ell}}^{c_{\ell}}
\left(\overline{\boldsymbol{\sigma}}_{\nu_1}\right)_{\dot{d}_1}^{d_1}\cdots \left(\overline{\boldsymbol{\sigma}}_{\nu_{\ell'}}\right)_{\dot{d}_{\ell'}}^{d_{\ell'}}\cdot\\&\cdot \mathbf{R}_{(c_1,\dots,c_{\ell})(d_1,\dots,d_{\ell'})}^{(a_1,\dots,a_{\ell})(b_1,\dots,b_{\ell'})}(u)\,
 \left(\boldsymbol{\sigma}_{\alpha_1}\right)_{a_1}^{\dot{a}_1}\cdots\left(\boldsymbol{\sigma}_{\alpha_{\ell}}\right)_{a_{\ell}}^{\dot{a}_{\ell}}
\left(\boldsymbol{\sigma}_{\beta_1}\right)_{b_1}^{\dot{b}_1}\cdots \left(\boldsymbol{\sigma}_{\beta_{\ell'}}\right)_{b_{\ell'}}^{\dot{b}_{\ell'}}\,\mathbf{R}^{(\dot c_1,\dots,\dot c_{\ell})(\dot d_1,\dots,\dot d_{\ell'})}_{(\dot a_1,\dots,\dot a_{\ell})(\dot b_1,\dots,\dot b_{\ell'})}(u)\,,
\end{aligned}
\end{align}
which in more compact notation we will denote by
\begin{equation}
\mathcal{R}_{\boldsymbol{\mu}\boldsymbol{\nu}}^{\boldsymbol{\alpha}\boldsymbol{\beta}}(u) =\frac{1}{2^{\ell+\ell'}}\ \text{Tr}_{\ell\ell'}\left(
[\overline{\boldsymbol{\sigma}}_{\boldsymbol{\mu}}]^{\ell}\,
[\overline{\boldsymbol{\sigma}}_{\boldsymbol{\nu}}]^{\ell'}\,
\mathbf{R}_{\ell,\ell'}(u) 
[{\boldsymbol{\sigma}}_{\boldsymbol{\alpha}}]^{\ell}\,
[{\boldsymbol{\sigma}}_{\boldsymbol{\beta}}]^{\ell'}\,
\overline{\mathbf{R}}_{\ell,\ell'}(u)\right)\,,
\end{equation}
and for $\ell=\ell'=1$ coincides with \eqref{ZamoR1}. 

To avoid misunderstanding we use in the last equation notation $\mathbf{R}_{\ell,\ell'}(u)$ for R-matrix acting on the un-dotted spinors and $\overline{\mathbf{R}}_{\ell,\ell'}(u)$ for R-matrix acting on the dotted spinors.  
Note that below for the sake of simplicity we shall use  universal 
notation $\mathbf{R}_{\ell,\ell'}(u)$ for both R-matrices because explicit indices 
can be restored unambiquously when needed.

\subsection{Integral identities} 
In this section we prove the star-triangle identities which generalize the well-known scalar relation \eqref{str0} for a vertex of three scalar bare propagators, to the case of propagators in the irreducible tensors representation.
We shall use notation
\begin{equation}
\mathbf{x}=\sig^{\mu}\frac{x_{\mu}}{|x|}\, ,\,\,\,  \bar{\mathbf{x}}=\bsig^{\mu}\frac{x_{\mu}}{|x|}\,,
\end{equation}
so hat the propagators of two massless fields with bare dimension $\Delta_1$ and $\Delta_2$ representation of spins $(\ell,\dot{\ell})$= $(0,n_1/2), (n_2/2,0)$, apart from inessential constants are
\begin{equation}
\label{propags}
\frac{[\mathbf{x-z}]^{n_1} }{(x-z)^{2\Delta_1}}\,,\,\,\,\,\, \frac{\left[\,\mathbf{\overline{z-y}}\,\right]^{n_2}}{ (y-z)^{2\Delta_2}}\,.
\end{equation}
Such objects already appeared in the section \ref{eispin} as eigenfunctions of the $Q$-operator for $N=1$. 
In order to introduce the eigenfunctions for any length $N$ we follow the scheme used in two-dimensions, thus we need the convolution of the two propagators \eqref{propags}, going under the name of chain-rule
\begin{align}
\label{chain_rule}
\begin{aligned}
&\int d^4z\, \frac{[\mathbf{x-z}]^{n_1} \left[\,\mathbf{\overline{z-y}}\,\right]^{n_2}}{(x-z)^{2\Delta_1} (y-z)^{2\Delta_2}}=
C^{n_1,n_2}(\Delta_1,\Delta_2)\frac{[\mathbf{{y-x}}]^{n_1}\mathbf{R}_{n_1,n_2}\left(2-\Delta_1-\Delta_2\right)[\,\mathbf{\overline{y-x}}\,]^{n_2}}{(x-y)^{2\left(\Delta_1+\Delta_2-2\right)}}\,.
\end{aligned}
\end{align}
The explicit expression for the coefficient $C^{n_1,n_2}(\Delta_1,\Delta_2)$ is given below~(\ref{coeff}). 
We should note that there is some freedom in the representation of the right hand 
side due to invarianve of R-matrix 
\begin{align}
[\mathbf{{x}}]^{n_1}\,[\mathbf{{x}}]^{n_2}\,
\mathbf{R}_{n_1,n_2}(u) = \mathbf{R}_{n_1,n_2}(u)\, [\mathbf{{x}}]^{n_1}\,[\mathbf{{x}}]^{n_2}
\end{align}
and evident relation $\mathbf{x}\,\bar{\mathbf{x}} = \II$. Using these relations it is possible to transform expression $[\mathbf{{x}}]^{n_1}\,
\mathbf{R}_{n_1,n_2}\,[\mathbf{{\overline{x}}}]^{n_2}$ as follows 
\begin{align*}
[\mathbf{{x}}]^{n_1}\,
\mathbf{R}_{n_1,n_2}\,[\mathbf{{\overline{x}}}]^{n_2} = 
[\mathbf{{\overline{x}}}]^{n_2}\,[\mathbf{{x}}]^{n_1} [\mathbf{{x}}]^{n_2}\,
\mathbf{R}_{n_1,n_2}\,[\mathbf{{\overline{x}}}]^{n_2}  = 
[\mathbf{{\overline{x}}}]^{n_2}\,\mathbf{R}_{n_1,n_2}\,
[\mathbf{{x}}]^{n_1} [\mathbf{{x}}]^{n_2}\,[\mathbf{{\overline{x}}}]^{n_2}  = 
[\mathbf{{\overline{x}}}]^{n_2}\,\mathbf{R}_{n_1,n_2}\,
[\mathbf{{x}}]^{n_1}
\end{align*}
It is instructive to represent~(\ref{chain_rule}) in a more explicit form using spinor indices in full analogy with~(\ref{spin_index})
\begin{align}
\label{chain_rule_ind}
\int d^4z\, \frac{\left([\mathbf{x-z}]^{n_1}\right)^{\dot{\mathbf{a}}}_{\mathbf{a}} 
\left(\left[\,\mathbf{\overline{z-y}}\,\right]^{n_2}\right)^{{\mathbf{b}}}_{\dot{\mathbf{b}}}}
{(x-z)^{2\Delta_1} (y-z)^{2\Delta_2}}=C^{n_1,n_2}(\Delta_1,\Delta_2)
\frac{\left([\mathbf{y-x}]^{n_1}\right)^{\dot{\mathbf{c}}}_{\mathbf{a}} 
\mathbf{R}_{\dot{\mathbf{c}}\dot{\mathbf{b}}}^{\dot{\mathbf{a}}\dot{\mathbf{d}}}
\left(2-\Delta_1-\Delta_2\right)
\left(\left[\,\mathbf{\overline{y-x}}\,\right]^{n_2}\right)^{{\mathbf{b}}}_{\dot{\mathbf{d}}}}
{(x-y)^{2\left(\Delta_1+\Delta_2-2\right)}}
\end{align}
This formula is the fundamental relation of this paper and appeared for the first time in the work \cite{Derkachov_Oliv}. Its detailed derivation is provided in the Appendix \ref{app:chainstar}. 

\begin{figure}
\begin{center}
\includegraphics[scale=0.45]{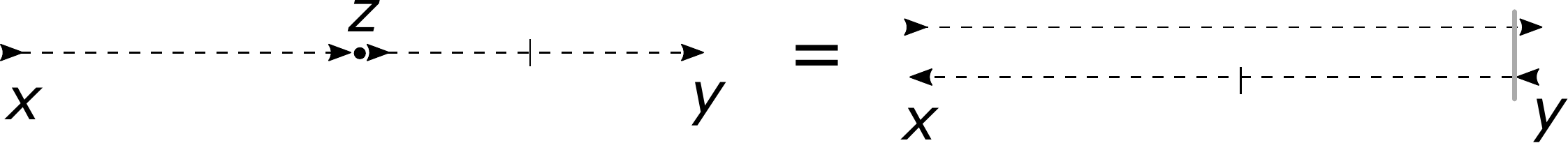}
\end{center}
\caption{Chain rule for the convolution of two conformal propagators. The matrices $\bsig$ are denoted by the barred line, and the arrows indicate the flow of matrices in the two spaces. The action of the $\mathbf R$-matrix is represented along the spinorial structure by a grey line.}
\label{figchain}
\end{figure}

It is useful to recast the chain-rule in the form of a star-triangle identity by means of a conformal inversion respect to the point $x_0^{\mu}=0$ followed by a translation of vector $-t^{\mu}$. After a few manipulations (see Appendix \ref{app:chainstar}), the star-triangle relation reads
\begin{align}
\notag &\int d^4z\, \frac{[ (\mathbf{\overline{x- z}})\mathbf{(z-t)}]^{n_1} [(\mathbf{\overline{t- z}})(\mathbf{z-y})]^{n_2}}{(x-z)^{2\Delta_1} (y-z)^{2 \Delta_2}(z-t)^{2(4-\Delta_1-\Delta_2)}} = 
C^{n_1,n_2}(\Delta_1,\Delta_2)\times\\
& \qquad\qquad\qquad\qquad\qquad\qquad\times\frac{[(\mathbf{\overline{x-y}})(\mathbf{y-t})]^{n_1}\mathbf{R}_{n_1,n_2}\left(2-\Delta_1-\Delta_2\right)[(\mathbf{\overline{t-x}})(\mathbf{x-y})]^{n_2}}{(x-y)^{2(\Delta_1+\Delta_2-2)}(x-t)^{2(2-\Delta_2)}(y-t)^{2(2-\Delta_1)}}\,.\notag
\end{align}
Star-triangle relation holds in the general form, under the uniqueness constraint $a+b+c=4$
\begin{align}
\label{STRopp}
\begin{aligned}
&\int d^4z\, \frac{[ (\mathbf{\overline{x-z}})\mathbf{(z-t)}]^{n_1} [(\mathbf{\overline{t-z}})(\mathbf{z-y})]^{n_2}}{(x-z)^{2a} (y-z)^{2b}(z-t)^{2c}}= 
C^{n_1,n_2}(a,b)\times\\& \qquad\qquad\qquad\qquad\qquad\qquad\times\frac{[(\mathbf{\overline{x-y}})(\mathbf{y-t})]^{n_1}\mathbf{R}_{n_1,n_2}\left(c-2\right)[(\mathbf{\overline{t-x}})(\mathbf{ x-y})]^{n_2}}{(x-y)^{2(2-c)}(x-t)^{2(2-b)}(y-t)^{2(2-a)}}\,,
\end{aligned}
\end{align}
where the explicit expression for the coefficient $C^{n_1,n_2}(a,b)$ has the following form
\begin{align}\label{coeff}
\begin{aligned}
&C^{n_1,n_2}(a,b) =  \pi^2\frac{\Gamma\left(2-a+\frac{n_1}{2}\right)\Gamma\left(2-b+\frac{n_2}{2}\right)\Gamma\left(2-c+\frac{n_2-n_1}{2}\right)}{\Gamma\left(a+\frac{n_1}{2}\right)\Gamma\left(b+\frac{n_2}{2}\right)\Gamma\left(c-1+\frac{n_2-n_1}{2}\right)}\frac{(-1)^{n_1}}{\left(c-1+\frac{n_1+n_2}{2}\right)}\,.
\end{aligned}
\end{align}
\begin{figure}
\includegraphics[scale=0.40]{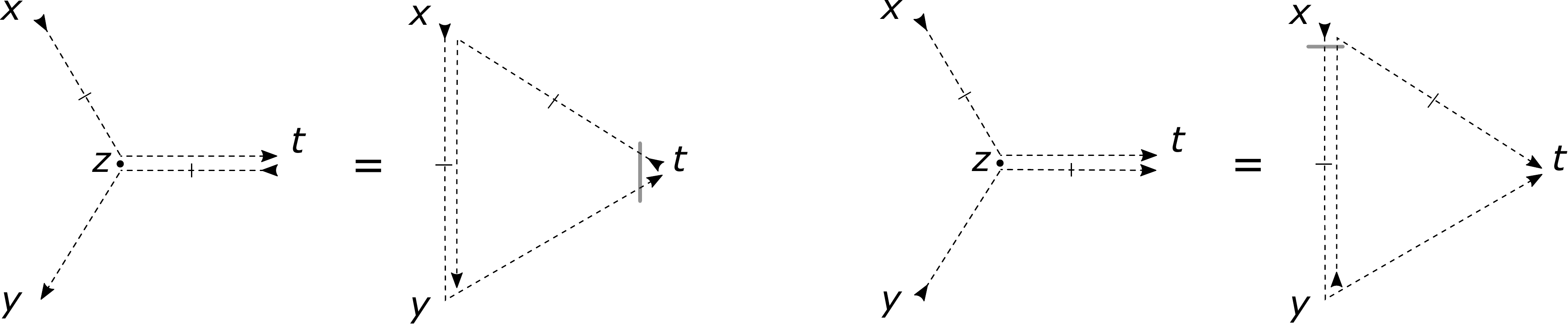}\caption{Star-triangle relations for "opposite flow" (left) and "same flow" (right), where the order of matrices $\sig$, $\bsig$ in the product is given by the arrows and the position of $\bsig$ is denoted by the barred lines. The action of the $\mathbf R$-matrix is represented along the spinorial structure by a grey line.}
\label{figstr}
\end{figure}
We point out that in the case $n_2=0$ (and similarly for $n_1=0$), the identity \eqref{STRopp} boils down to
\begin{align}
\label{STR_0}
\begin{aligned}
&\int d^4z\, \frac{[ (\mathbf{\overline{x-z}})\mathbf{(z-t)}]^{n}}{(x-z)^{2a} (y-z)^{2b}(z-t)^{2c}}&= C^{n,0}(a,b)\frac{[(\mathbf{\overline{x-y}})(\mathbf{y-t})]^{n}}
{(x-y)^{2(2-c)}(x-t)^{2(2-b)}(y-t)^{2(2-a)}}\,,
\end{aligned}
\end{align}
with coefficient 
\begin{align}
\begin{aligned}
&C^{n,0}(a,b) = \pi^2\frac{\Gamma\left(2-a+\frac{n}{2}\right)\Gamma\left(2-b\right)\Gamma\left(2-c+\frac{n}{2}\right)}{\Gamma\left(a+\frac{n}{2}\right)\Gamma\left(b\right)\Gamma\left(c+\frac{n}{2}\right)}
\end{aligned}
\end{align}
and was first worked out in \cite{Chicherin2013a}.
\begin{figure}
\begin{center}
\includegraphics[scale=0.45]{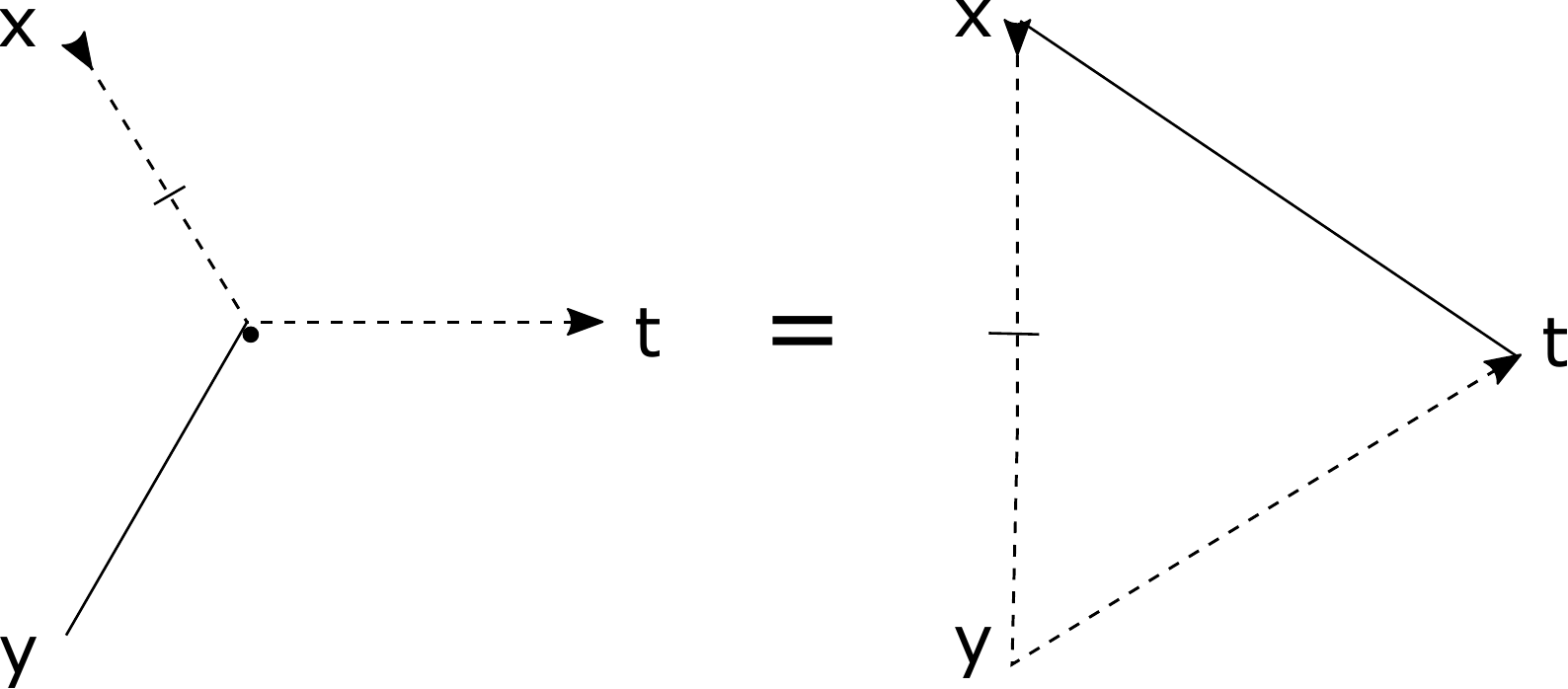}
\end{center}
\caption{Star-triangle relation \eqref{STR_0}. Differently from Fig.\ref{figstr}, there is no action of the $\mathbf R$-matrix, as there are no spinors to mix ($n_2=0$).}
\end{figure}
In the formula \eqref{STRopp} the matrices $\sig, \bsig$ are ordered in the product from $x$ to $t$ in the $n_1$-space and from $t$ to $y$ in the $n_2$-space, and we refer to this relation as star-triangle with ``opposite flow" of sigma matrices. It is possible to obtain another star-triangle relation starting from the chain-rule, for which the the sigma matrices have the ``same flow" as the matrix products are ordered from $x$ to $t$ in the $n_1$-space and from $y$ to $t$ in the $n_2$-space (see Fig.\ref{figstr}).
\begin{align}
\begin{aligned}
\label{STRsame}
 &\int d^4z\, \frac{[(\overline{\mathbf{x-z}})(\mathbf{z-t})]^{n_1} [(\mathbf{y-z}) (\overline{\mathbf{z-t}})]^{n_2}}{(x-z)^{2a} (y-z)^{2b}(z-t)^{2c}}=
 C^{n_1,n_2}(a,b)(-1)^{n_1}\times\\& \qquad\qquad\qquad\qquad\qquad\qquad\times\frac{[\overline{\mathbf{ y- x}}]^{n_1} \mathbf{R}_{n_1,n_2}\left(c-2\right)[\mathbf{y-t}]^{n_1}[(\mathbf{y-x})(\overline{\mathbf{x-t}})]^{n_2}}{(x-y)^{2(2-c)}(x-t)^{2(2-b)}(y-t)^{2(2-a)}}\,.
\end{aligned}
\end{align}
Following the lines of \cite{Derkachov2001}, we list a remarkable consequence of the star-triangle formula \eqref{STRsame} which we refer to as exchange-relation. It will be frequently used in the study of the eigenfunctions of the operators \eqref{Q1}. Under the constraint $a'+b'=a+b$, the exchange relation reads:
\begin{align}
\label{exch_same}
\begin{aligned}
 &\int d^4z\, \frac{[(\mathbf{x_0- z})(\mathbf{\overline{ z-x}})(\mathbf{x-x_0'})]^{\ell} [(\mathbf{x_0-y}) (\mathbf{\overline{ y- z}})(\mathbf{z-x_0'})]^{\ell'}}{(x_0-y)^{2(2-b')}(y-z)^{2(2-a')}(x_0-z)^{2a}(x-z)^{2b}(z-x_0')^{2(2-b')}(x-x_0')^{2(2-b)}}=\\&= C \int d^4z\, \frac{\mathbf{R} [(\mathbf{x_0-y})(\overline{\mathbf{y-z}})(\mathbf{z-x_0'})]^{\ell} [\mathbf{(x_0-z)}(\overline{\mathbf{z-x}})(\mathbf{ x- x_0'})]^{\ell'}\mathbf{R}^{-1}}{(x_0-y)^{2(2-b)}(y-z)^{2(2-a)}(x_0-z)^{2a'}(x-z)^{2b'}(z-x_0')^{2(2-b)}(x-x_0')^{2(2-b')}}\,,
\end{aligned}
\end{align}
where $\mathbf{R}=\mathbf{R}_{\ell,\ell'}(b-b')$, and the coefficient $C$ is given by
\begin{equation}
\label{C_coeff}
C=\frac{\Gamma\left(2-b+\frac{\ell}{2}\right)\Gamma\left(2-a+\frac{\ell}{2}\right)\Gamma\left(b'+\frac{\ell'}{2}\right)\Gamma\left(a'+\frac{\ell'}{2}\right)}{\Gamma\left(a+\frac{\ell}{2}\right)\Gamma\left(b+\frac{\ell}{2}\right)\Gamma\left(2-a'+\frac{\ell'}{2}\right)\Gamma\left(2-b'+\frac{\ell'}{2}\right)}\,.
\end{equation}
The identity \eqref{exch_same} shows that the exchange of parameters $(a,b,\ell)\,,\, (a',b',\ell')$ between the l.h.s. and the r.h.s. of the equation produces a coefficient and the mixing of external spinors by the fused-$\mathbf{R}$ matrix. The graphical representation of this identity in a Feynman diagram formalism is
\begin{center}
\includegraphics[scale=0.5]{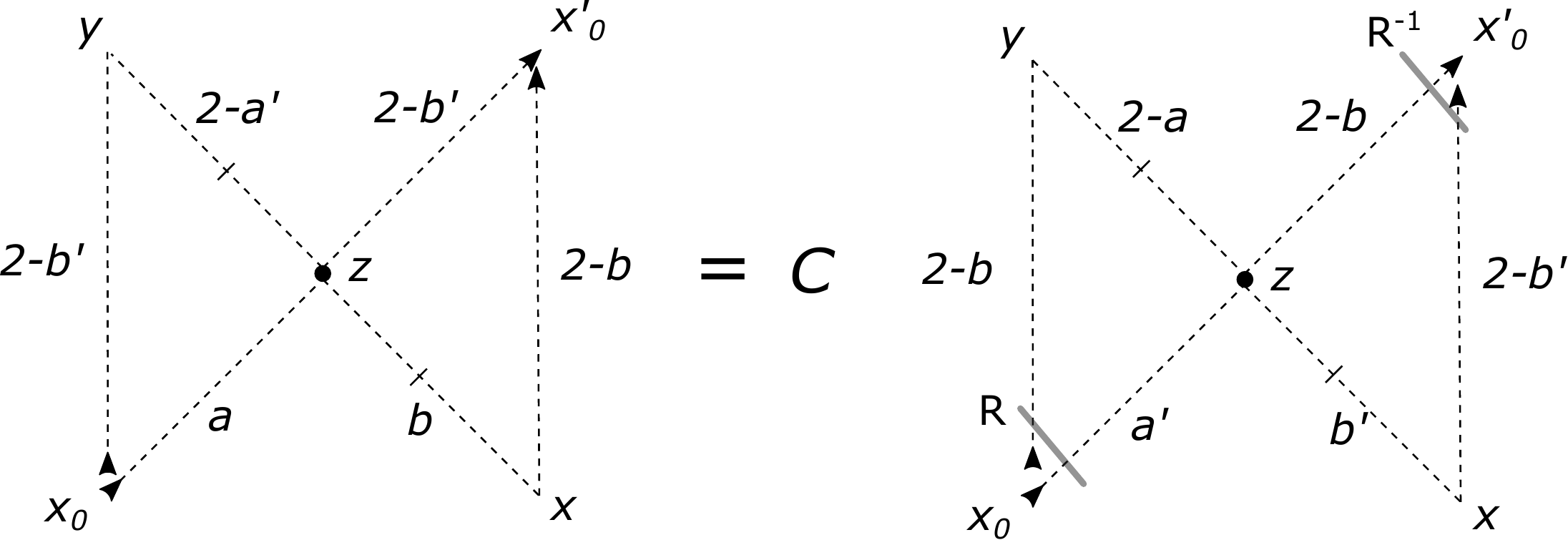}
\end{center}
The proof of \eqref{exch_same} is a simple consequence of the star-triangle relation \eqref{STRsame}, and it is explained in detail in the Appendix \ref{app:exchange_I}.
For completeness we write explicitly the reduction of \eqref{exch_same} to the case $\ell'=0$, as it is ubiquitous in the computation of the eigenvalues (see section \ref{sec:eigenf})
\begin{align}
\label{exch_0}
\begin{aligned}
 &\int d^4z\, \frac{[(\mathbf{x_0- z})(\overline{\mathbf{z-x}})(\mathbf{x-x_0'})]^{\ell} }{(x_0-y)^{2(2-b')}(y-z)^{2(2-a')}(x_0-z)^{2a}(x-z)^{2b}(z-x_0')^{2(2-b')}(x-x_0')^{2(2-b)}}=\\&= C \int d^4z\, \frac{ [(\mathbf{x_0-y})(\overline{\mathbf{y-z}})(\mathbf{z-x_0'})]^{\ell}}{(x_0-y)^{2(2-b)}(y-z)^{2(2-a)}(x_0-z)^{2a'}(x-z)^{2b'}(z-x_0')^{2(2-b)}(x-x_0')^{2(2-b')}}\,,
\end{aligned}
\end{align}
and can be represented as
\begin{center}
\includegraphics[scale=0.5]{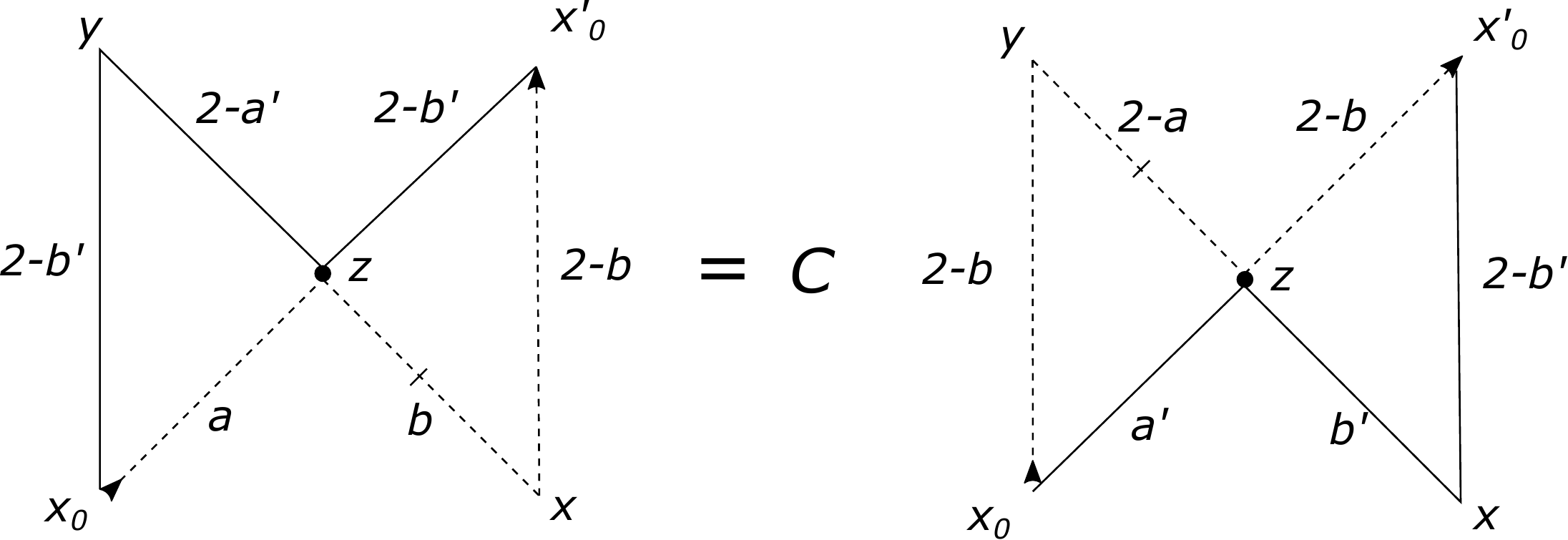}
\end{center}
It can be useful to state \eqref{exch_same} in a different form, such that there is no coefficient $C$ between the l.h.s. and r.h.s. and is the direct generalization to $4d$ of the two-dimensional identities of \cite{Derkachov2001}:
\begin{align}
\label{exch_same_1}
\begin{aligned}
&\frac{1}{(x-x_0)^{2(2-a-b)}}\int d^4z\, \frac{[(\mathbf{x_0-y})\mathbf{R}^{-1} (\mathbf{\overline{y-z}})(\mathbf{z-x_0'})]^{\ell'}[(\mathbf{\overline{y-x_0}})(\mathbf{x_0- z})(\mathbf{\overline{z-x}})]^{\ell} }{(x_0-y)^{2(a'-a)}(y-z)^{2(2-a')}(x_0-z)^{2a}(x-z)^{2b}(z-x_0')^{2(2-b')}}=\\&= \frac{1}{(y-x_0')^{2(2-a-b)}}\int d^4z\, \frac{[\mathbf{(x_0-z)}(\overline{\mathbf{z-x}})(\mathbf{ x- x_0'})]^{\ell'}[(\overline{\mathbf{y-z}})(\mathbf{z-x_0'})\mathbf{R}^{-1}(\overline{\mathbf{x_0'-x}})]^{\ell} }{(y-z)^{2 b}(x_0-z)^{2(2-b')}(x-z)^{2(2-a')}(z-x_0')^{2 a}(x-x_0')^{2(b-b')}}\,,
\end{aligned}
\end{align}
\begin{center}
\includegraphics[scale=0.5]{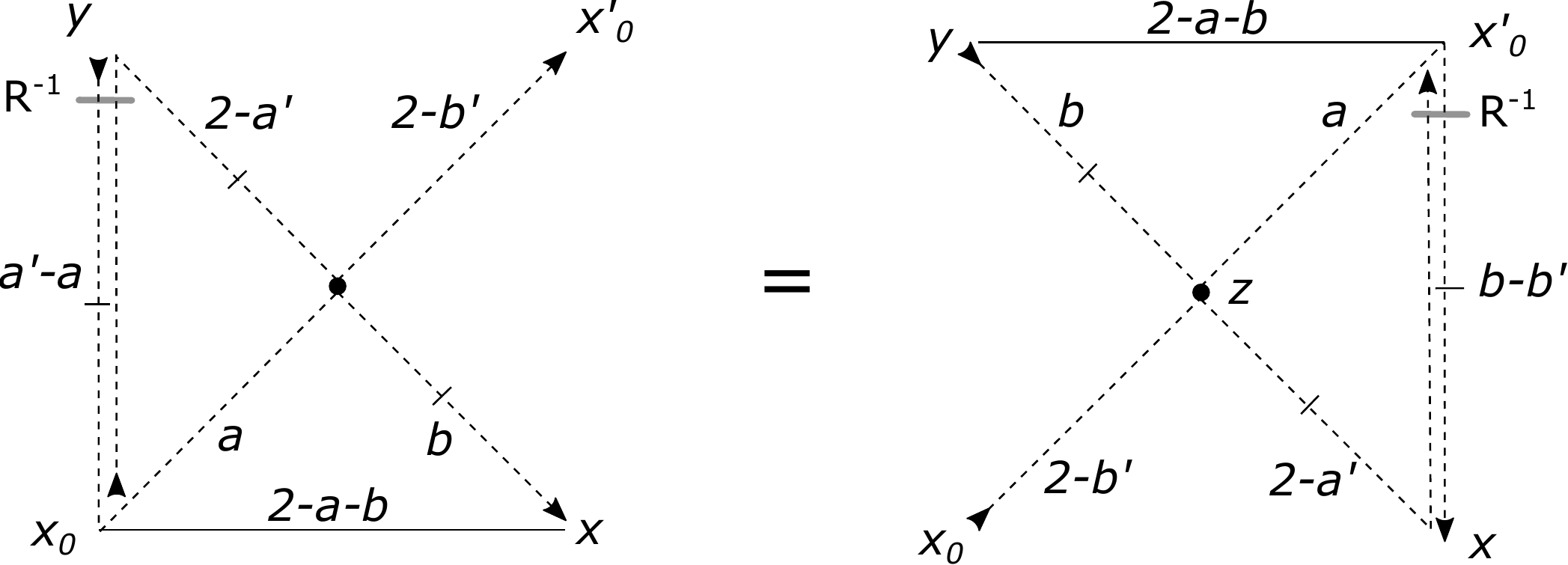}
\end{center}
As it was for the star-triangle identity, also for the exchange relation it is possible to state it for two different choices of the relative flow of $\sig$ and $\bsig$ matrices. Therefore, while in \eqref{exch_same_1} both spinor structure go from the left $(x_0,y)$ to the right $(x,x_0')$ of the picture, there exist another exchange relation with opposite flows.
This additional relation, valid under the constraint $a+a'=b+b'$, reads
\begin{align}
\label{exchange_opposite}
\begin{aligned}
&
\frac{1}{(y-x_0)^{2(a'-a)}}\int d^4 z  \, \frac{[\mathbf{(x_0-y)\overline{(y-z)}
(z-x_0')}]^{\ell'}\,
[\mathbf{\overline{(x-z)}(z-x_0)\overline{(x_0-y)}}]^{\ell}}
{(y-z)^{2\left(2-a'\right)}
(z-x_0')^{2\left(2-b'\right)}
(z-x)^{2\left(b\right)}(z-x_0)^{2\left(a\right)}}\, = \\
&
\frac{C}{(x-x_0')^{2\left(b-b'\right)}}
\int d^4 z \frac{[\mathbf{(x_0-z)\overline{(z-x)}(x-x_0')}]^{\ell'}\,
[\mathbf{\overline{\mathbf{(x-x_0')}} 
(x_0'-z)\overline{(z-y)}}]^{\ell}}
{\left(y-z\right)^{2(2-a)}(z-x_0')^{2\left(2-b\right)}
(z-x)^{2\left(b'\right)}(z-x_0)^{2\left(a'\right)}}\,,
\end{aligned}
\end{align}
where the coefficient $C$ is the same as \eqref{C_coeff}.
\begin{center}
\includegraphics[scale=0.5]{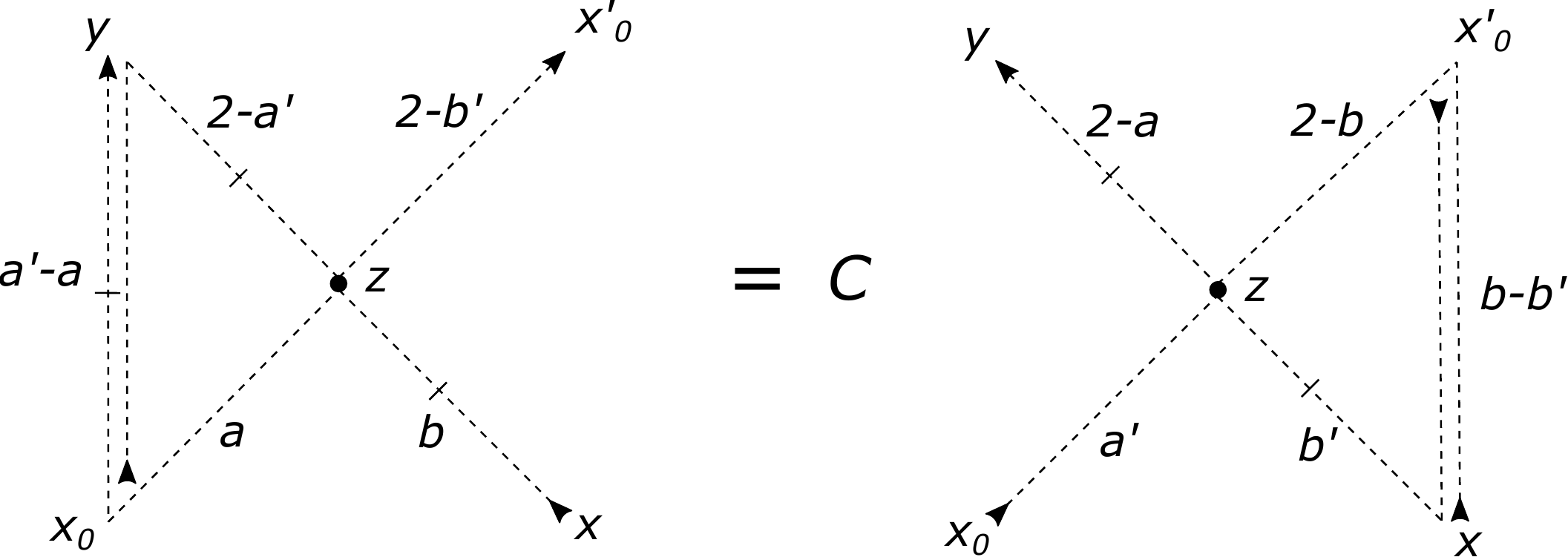}
\end{center}
The proof of relation \eqref{exchange_opposite} is not a consequence of the star-triangle identity already introduced, but it follows from a new, different identity of star-triangle type, as we explain in detail in Appendix \ref{app:exchange_II}.
It is possible to re-cast the equation \eqref{exchange_opposite} in an equivalent way such that the factor $C$ defined in \eqref{C_coeff} disappears:
\begin{align}
\label{exch_opp_1}
\begin{aligned}
&\int d^4z\, \frac{[(\mathbf{x_0-y})(\mathbf{\overline{y-z}})(\mathbf{z-x_0'})]^{\ell'}[(\mathbf{\overline{z-x}})(\mathbf{z-x_0})(\mathbf{\overline{x_0-y}})]^{\ell} }{(x_0-x)^{2(2-a-b)}(x_0-y)^{2(a'-a)}(y-z)^{2(2-a')}(x_0-z)^{2a}(x-z)^{2b}(z-x_0')^{2(2-b')}}=\\&= \int d^4z\, \frac{[\mathbf{(x_0-z)}(\overline{\mathbf{z-x}})(\mathbf{ x- x_0'})]^{\ell'}[(\overline{\mathbf{x-x_0'}})(\mathbf{x_0'-z})(\overline{\mathbf{z-y}})]^{\ell} }{(y-x_0')^{2(2-a-b)}(y-z)^{2(2-a)}(x_0-z)^{2a'}(x-z)^{2b'}(z-x_0')^{2(2-b)}(x-x_0')^{2(b-b')}}\,,
\end{aligned}
\end{align}
which we can represent graphically as
\begin{center}
\includegraphics[scale=0.5]{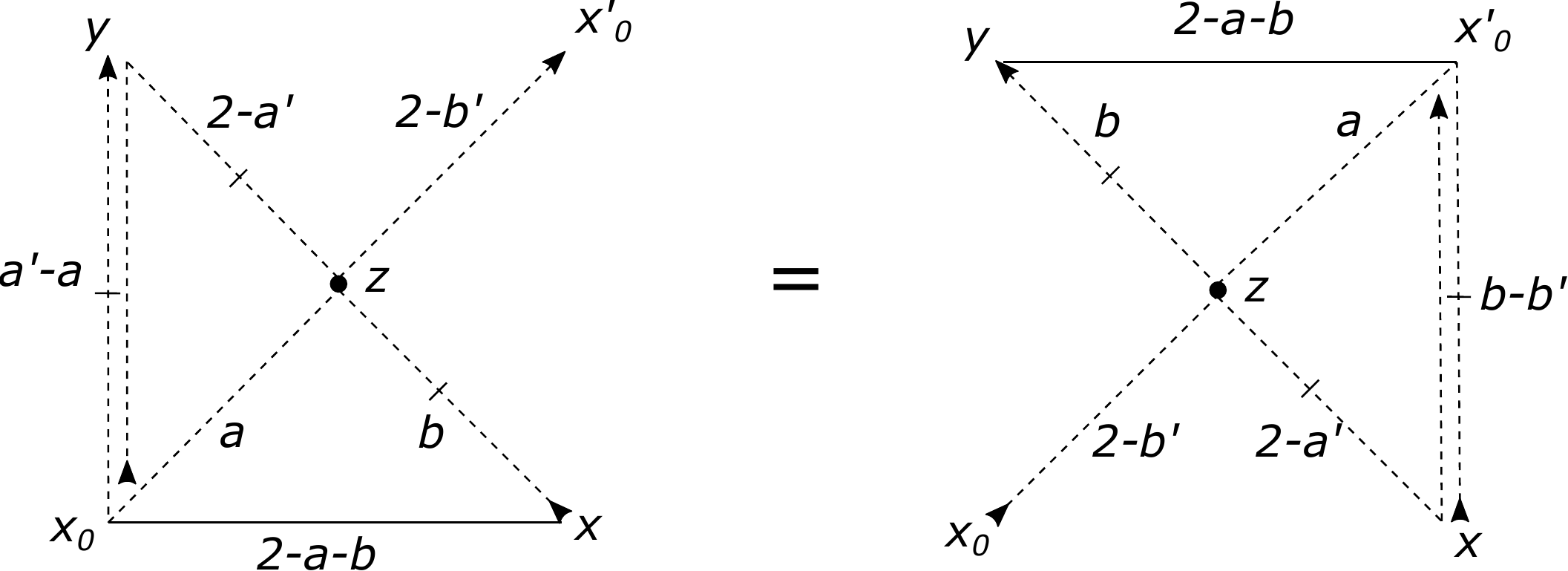}
\end{center}
and which will be the basic block for the computation of the scalar product between eigenfunctions.
The identity \eqref{exch_opp_1} is obtained starting from the multiplication by $(x-x_0)^{2(a+b-2)}$ of both sides in equation \eqref{exchange_opposite}. Then we move such line in the r.h.s. from $(x-x_0)$ to $(y-x_0')$, by means of triangle-star transformation in $(x_0,x,z)$ followed by a star-triangle involving the star of vertices $x_0'$ and $y$. The coefficient produced by such transformation is exactly $C^{-1}$, and cancels with $C$ in the r.h.s. of \eqref{exchange_opposite}.

\section{$Q$-operators with spin}
\label{sect:Qspin}
\begin{figure}
\begin{center}
\includegraphics[scale=0.4]{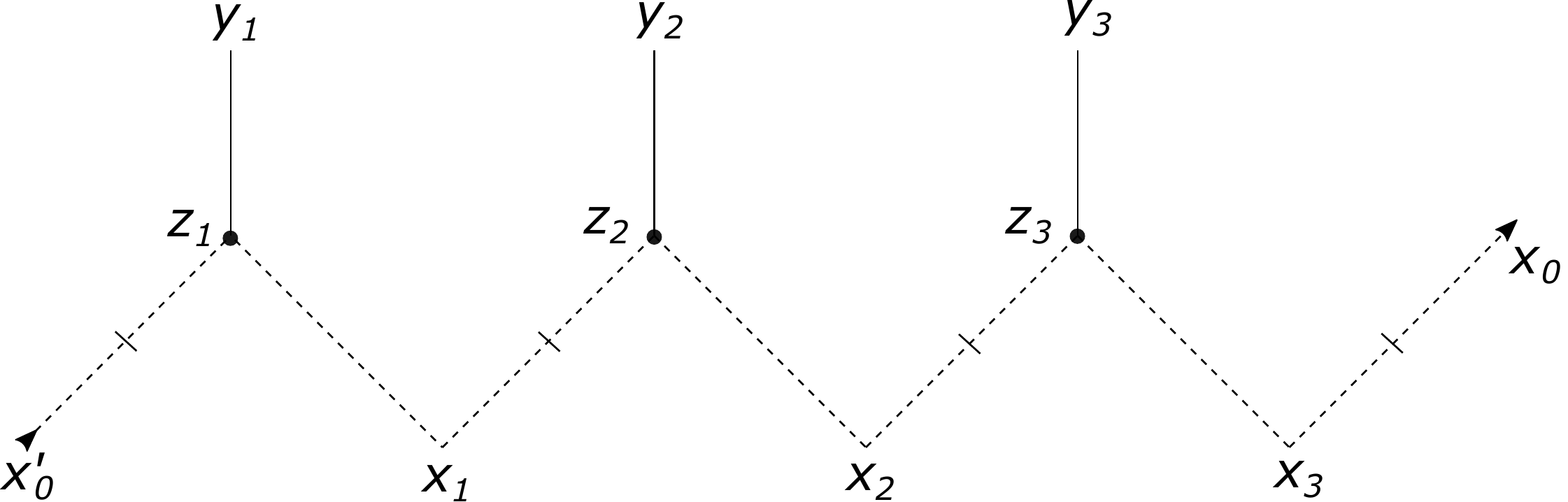}
\end{center}
\caption{Graphic representation of the integral kernel ${\hat Q}^{(3)}_{0,1}(\mathbf{x}|\mathbf y)$ for $\lambda=-i$. The solid lines stand for scalar propagators $1/(z_k-y_k)^2$, the dashed lines stand for $(\sig^{\mu}(z-x)_{\mu})/((z-x)^2)^2$ while the dashed lines with a bar stand for $(\bsig^{\mu}(x-z)_{\mu})/((x-z)^2)^2$. The arrows denote the order of the matrices $\sig,\bsig$ in the product.}
\label{Yuk_1}
\end{figure}

\begin{figure}
\begin{center}
\includegraphics[scale=0.4]{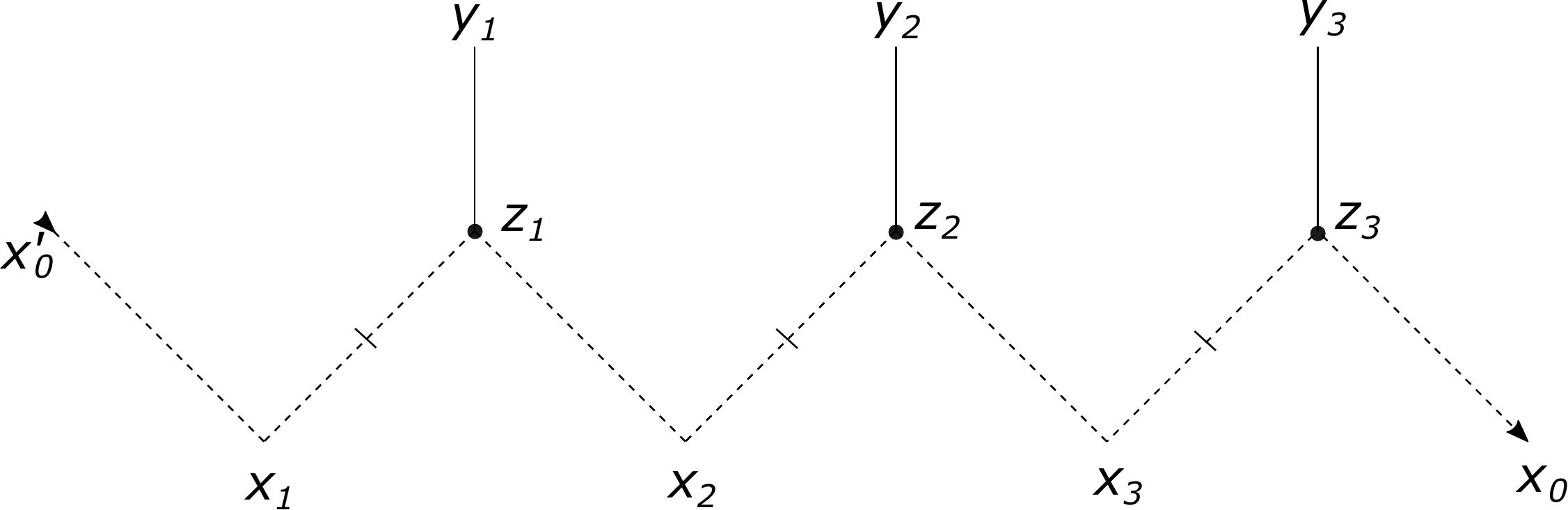}
\end{center}
\caption{Graphic representation of the integral kernel ${\check Q}^{(3)}_{0,1}(\mathbf{x}|\mathbf y)$ for $\lambda=-i$. The solid lines stand for scalar propagators $1/(z_k-y_k)^2$, the dashed lines stand for $(\sig^{\mu}(x-y)_{\mu})/((x-y)^2)^2$ while the dashed lines with a bar stand for $(\bsig^{\mu}(x-y)_{\mu})/((x-y)^2)^2$. The arrows denote the order of the matrices $\sig,\bsig$ in the product.}
\label{Yukk}
\end{figure}

In the section~\ref{ladder} we illustrated how to translate the computation of massless ladder Feynman integrals to the diagonalization of the family of operators $Q(u)$ \eqref{Q1}, which is the generating function for the commuting operators including the Hamiltonian \eqref{4Dlocal}, and realize the integrability of the model. 
The same technique can be applied to the computation of massless Feynman integrals with the topology of a square lattice, which indeed can be reduced to the diagonalization of the operators \eqref{Q1} at special value for the parameter $u=-i\lambda$. Moreover, this logic can be extended to the more general situation including fermions or other fields with higher spin in the theory, provided that the topology of massless Feynmann integrals is still the one of a regular lattice with conformal symmetry (dimensionless vertices). In order to do so we must consider a more general class of integral operators for any spin, which in the scalar case reduce to the operators introduced in \eqref{Q1}.
Therefore we define two families of operators
\begin{align}
\label{Qhatdef}
\begin{aligned}
(\hat{Q}_{\nu,\ell})_{\dot{\mathbf{a}}}^{\mathbf{a}} :\, &L^2(d^4 x_1 \cdots d^4 x_N)\,\longrightarrow  L^2(d^4 x_1 \cdots d^4 x_N)\,,\\ &\phi(x_1,\dots,x_N)\,\mapsto\, \int d^4y_1\cdots d^4y_N\, (\hat{Q}_{\nu,\ell})_{\dot{\mathbf{a}}}^{\mathbf{a}}(x_0',x_1,\dots,x_N,x_0|y_1,\dots,y_N) \cdot \phi(y_1,\dots,y_N) \,,
\end{aligned}
\end{align}
and
\begin{align}
\label{Qcheckdef}
\begin{aligned}
(\check{Q}_{\nu,\ell})_{{\mathbf{a}}}^{\dot{\mathbf{a}}}  :\,&L^2(d^4 x_1 \cdots d^4 x_N)\,\longrightarrow  L^2(d^4 x_1 \cdots d^4 x_N)\,,\\ &\phi(x_1,\dots,x_N)\,\mapsto\, \int d^4y_1\cdots d^4y_N\,(\check{Q}_{\nu,\ell})_{{\mathbf{a}}}^{\dot{\mathbf{a}}} (x_0',x_1,\dots,x_N,x_0|y_1,\dots,y_N) \cdot \phi(y_1,\dots,y_N) \,.
\end{aligned}\end{align}
\begin{figure}
\begin{center}
\includegraphics[scale=0.4]{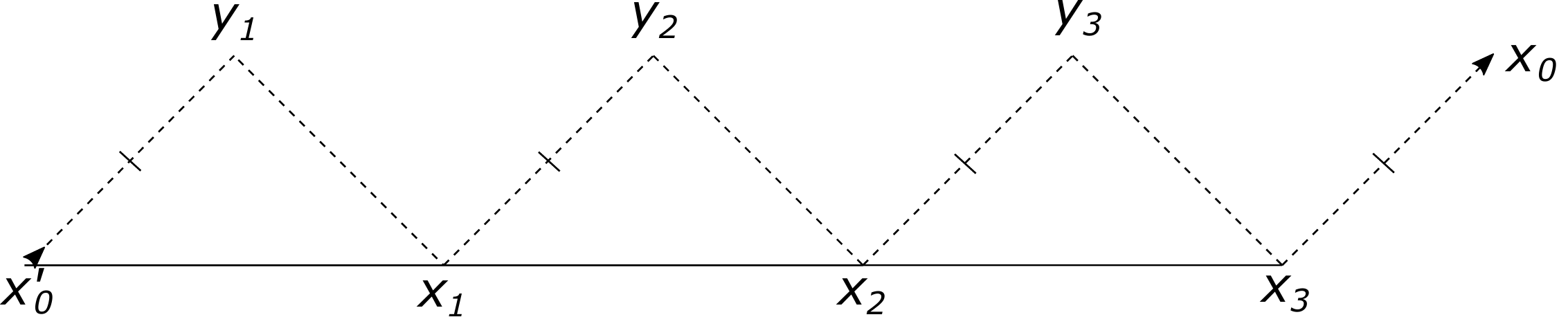}\vspace*{5mm}
\includegraphics[scale=0.4]{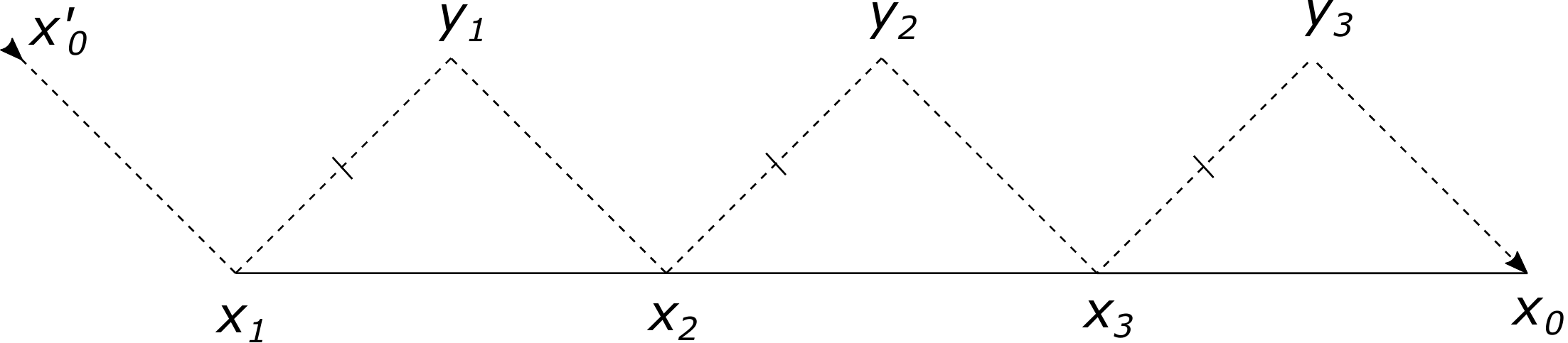}
\end{center}
\caption{Graphic representation of the integral kernels ${\hat Q}_{\nu,\ell}(\mathbf{x}|\mathbf y)$ (up) and ${\check Q}_{\nu,\ell}(\mathbf{x}|\mathbf y)$ (down) for the length $N=3$. The solid lines stand for scalar propagators $((x_k-x_h)^2)^{-2i\lambda}$, the dashed lines denotes the numerators $[\mathbf{{x-y}}]^{\ell}/((x-y)^2)^{1-i\lambda-i\nu}$ while the dashed lines with a bar stand for $[\mathbf{\overline{x-y}}]^{\ell}/((x-y)^2)^{1-i\lambda+i\nu}$. The arrows denote the order of the matrices $\sig,\bsig$ in the product.}
\end{figure}
For full clarity, we should write here for once the explicit spinor indices of the operators under study:
\begin{equation}
\label{Q_spin_index}
(\hat{Q}_{\nu,\ell})_{\dot{\mathbf{a}}}^{\mathbf{a}}=(\hat{Q}_{\nu,\ell})_{(\dot{a}_1\dots \dot{a}_{\ell})}^{(a_1\dots a_{\ell})}\qquad\text{and}\qquad (\check{Q}_{\nu,\ell})_{{\mathbf{a}}}^{\dot{\mathbf{a}}}=(\check{Q}_{\nu,\ell})_{(a_1\dots a_{\ell})}^{(\dot{a}_1\dots \dot{a}_{\ell})}\,,
\end{equation}
where the $(\dots)$ notation means the symmetry respec to any exchange $a_i \leftrightarrow a_j$ or $\dot{a}_i \leftrightarrow \dot{a}_j$. 
In the following we will always omit the spinor indices since we will the state relations valid for any choice of $\mathbf{a}$ and $\dot{\mathbf{a}}$, i.e. for the full matrix. Moreover we will add the superscript $N$ as in $\hat{Q}_{\nu,\ell}^{(N)}$ or $\check{Q}_{\nu,\ell}^{(N)}$ only in the relations which involve at the same time the $Q$-operators for the model at different sizes.
The kernel functions in the r.h.s. of \eqref{Qhatdef} and \eqref{Qcheckdef} are respectively given, in matrix form, by
\begin{align}
\label{Qhat_ker}
\begin{aligned}
&{\hat Q}_{\nu,\ell}(x_0',x_1,\dots,x_N,x_0|y_1,\dots,y_N) = \prod_{k=0}^{N-1} \frac{1}{(x_k-x_{k+1})^{2(i\lambda)}}\times\\&\times\frac{[(\mathbf{\overline {x_0'- y_1}})(\mathbf{y_1- x_1})(\mathbf{\overline {x_1- y_2}})\cdots (\mathbf {y_N-x_N})(\overline{\mathbf{x_N- x_0}})]^{\ell}}{(x_0'-y_1)^{2(1-i\frac{\lambda}{2}+i\nu)}(y_1-x_1)^{2(1-i\frac{\lambda}{2}-i\nu)}\cdots (y_N-x_N)^{2(1-i\frac{\lambda}{2}-i\nu)}(x_N-x_0)^{2(1+i\frac{\lambda}{2}+i\nu)}}\,,
\end{aligned}
\end{align}
and
\begin{align}
\label{Qcheck_ker}
\begin{aligned}
&{\check Q}_{\nu,\ell}(x_0',x_1,\dots,x_N,x_0|y_1,\dots,y_N) = \prod_{k=1}^{N} \frac{1}{(x_k-x_{k+1})^{2(i\lambda)}}\times\\&\times\frac{[(\mathbf{{x_0'- x_1}})(\overline{\mathbf{x_1- y_1}})(\mathbf{{y_1- x_2}})\cdots (\mathbf{\overline {x_{N}-y_N}})({\mathbf{y_N- x_0}})]^{ \ell}}{(x_0'-x_1)^{2(1+i\frac{\lambda}{2}-i\nu)}(x_1-y_1)^{2(1-i\frac{\lambda}{2}+i\nu)}\cdots (x_N-y_N)^{2(1-i\frac{\lambda}{2}+i\nu)}(y_N-x_0)^{2(1-i\frac{\lambda}{2}-i\nu)}}\,,
\end{aligned}
\end{align}
where $x_{N+1}\equiv x_0$. The spinor indices described in \eqref{Q_spin_index} are evident in formulas \eqref{Qhat_ker} and \eqref{Qcheck_ker} due to the matrices $\sig_{a}^{\dot{a}}$ 
and $\bsig_{\dot{a}}^{a}$, and the symmetry in the exchange of spinor indices is encoded in the notation $[\cdots]^{\ell}$ introduced in section \ref{eispin}.
\begin{figure}
\begin{center}
\includegraphics[scale=0.4]{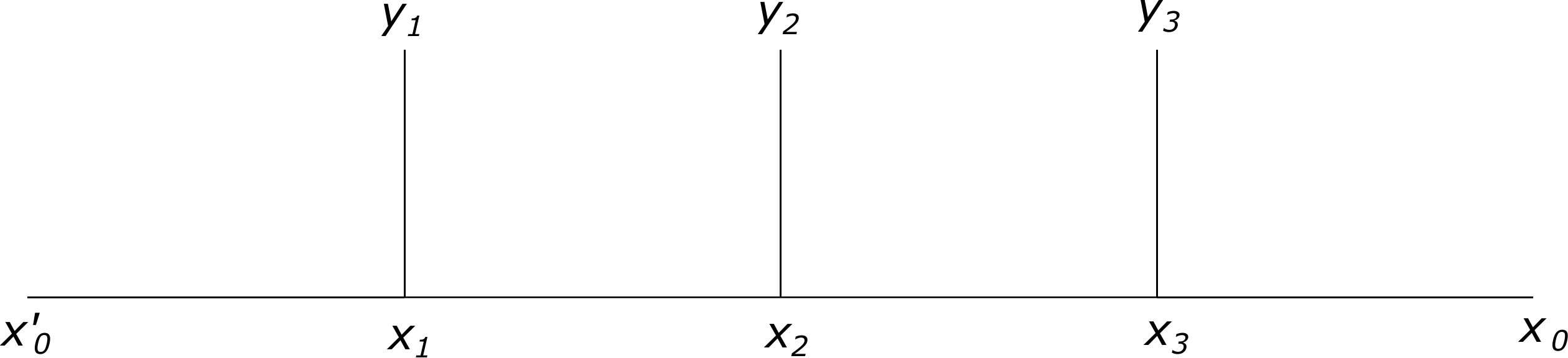}
\end{center}
\caption{Graphic representation of the integral kernel ${\hat Q}^{(3)}_{i/2,0}(\mathbf{x}|\mathbf y)={\check Q}^{(3)}_{-i/2,0}(\mathbf{x}|\mathbf{y})$ at $\lambda=-i$. The solid lines stand for the standard scalar propagators $(x_k-x_h)^{-2}$ and $(x_k-y_k)^{-2}$.}
\label{lambda_red}
\end{figure}
Such integral operators scale respectively with dimension $\Delta=1+i\frac{\lambda}{2}+i\nu$ and $\Delta=1+i\frac{\lambda}{2}-i\nu$. It is possible to recover from \eqref{Qhat_ker} or \eqref{Qcheck_ker} more familiar objects in the realm of Feynmann integrals fixing $\lambda=-i$. Indeed, setting $(\nu,\ell)=(i/2,0)$ in \eqref{Qhat_ker} or, respectively, $(-i/2,0)$ in \eqref{Qcheck_ker} those two formulas simplify to a portion of square-lattice graph with external fixed points $x_0$ and $x_0'$:
\begin{align}
\label{squarelat}
&{\hat Q}_{i/2,0}(\mathbf{x}|\mathbf y)={\check Q}_{-i/2,0}(\mathbf{x}|\mathbf{y}) = \frac{1}{(x_0'-x_1)^2}\prod_{k=1}^{N} \frac{1}{(x_k-x_{k+1})^{2}(x_k-y_{k})^{2}}\,.
\end{align}
Similarly, in order to obtain spin-$\frac{1}{2}$ fermionic propagators, we fix $(\nu,\ell)=(0,1)$ bot in \eqref{Qhat_ker} and \eqref{Qcheck_ker}, so that they reduce to a row of planar Yukawa (hexagonal) lattice (see fig.\ref{Yuk_1} and fig.\ref{Yukk})
\begin{align}
\label{Yukhat}
&\notag {\hat Q}_{0,1}(\mathbf{x}|\mathbf{y}) = \frac{(\mathbf{\overline {x_0'-y_1}})(\mathbf{y_1- x_1})\cdots (\mathbf {y_N-x_N})(\overline{\mathbf{x_N- x_0}})}{(x_0'-y_1)^{2(\frac{1}{2})}(y_1-x_1)^{2(\frac{1}{2})}\cdots (y_N-x_N)^{2(\frac{1}{2})}(x_N-x_0)^{2(\frac{3}{2})}}\prod_{k=0}^{N-1} \frac{1}{(x_{k}-x_{k+1})^{2}}\,=\\&=\frac{\pi^{N}}{\Gamma\left(\frac{3}{2}\right)^{-2N}} \int d^4 z_1\cdots d^4 z_N\,\frac{(\mathbf{\overline {x_0'- z_1}})(\mathbf{z_1-x_1})\cdots (\mathbf {z_N-x_N})(\overline{\mathbf{x_N- x_0}})}{(x_0'-z_1)^{2(\frac{3}{2})}(z_1-x_1)^{2(\frac{3}{2})}\cdots (z_N-x_N)^{2(\frac{3}{2})}(x_N-x_0)^{2(\frac{3}{2})}}\prod_{k=1}^{N} \frac{1}{(z_k-y_{k})^{2}}\,,
\end{align}
\begin{align}
\label{Yukcheck}
&\notag {\check Q}_{0,1}(\mathbf{x}|\mathbf{y}) = \frac{(\mathbf{\overline {x_0'- x_1}})(\mathbf{x_1- y_1})\cdots (\mathbf {x_N-y_N})(\overline{\mathbf{y_N- x_0}})}{(x_0'-x_1)^{2(\frac{3}{2})}(x_1-y_1)^{2(\frac{1}{2})}\cdots (x_N-y_N)^{2(\frac{1}{2})}(y_N-x_0)^{2(\frac{1}{2})}}\prod_{k=1}^{N} \frac{1}{(x_{k}-x_{k+1})^{2}}\,=\\&=\frac{\pi^{N}}{\Gamma\left(\frac{3}{2}\right)^{-2N}} \int d^4 z_1\cdots d^4 z_N\,\frac{(\mathbf{\overline {x_0'- x_1}})(\mathbf{x_1-z_1})\cdots (\mathbf {x_N-z_N})(\overline{\mathbf{z_N- x_0}})}{(x_0'-z_1)^{2(\frac{3}{2})}(z_1-x_1)^{2(\frac{3}{2})}\cdots (z_N-x_N)^{2(\frac{3}{2})}(x_N-x_0)^{2(\frac{3}{2})}}\prod_{k=1}^{N} \frac{1}{(z_k-y_{k})^{2}}\,.
\end{align}

We can use the formula \eqref{exch_same} to compute the commutator of two $\hat Q$-operators at the same value of the parameters $\lambda$ and $N$. Starting from the composed operator $(\hat{Q}_{\nu,\ell})_{\mathbf{\dot{a}}}^{\mathbf a} (\hat{Q}_{\nu',\ell'})_{\mathbf{\dot{b}}}^{\mathbf b}$ represented in the left picture, we want to commute the two operators by means of the integral identity \eqref{exch_same}. The first step is to open the triangles containing a vertex $y_k$ into star integrals by means of \eqref{STR_0} (right picture)
\begin{center}
\includegraphics[scale=0.3]{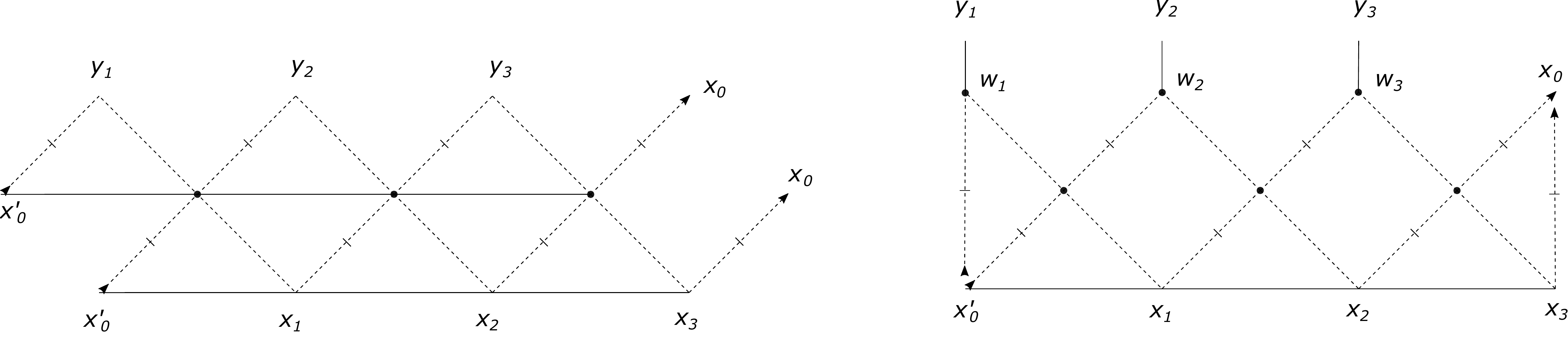}
\end{center}
Then we insert a couple of lines for each $k<N$ odd $[(\mathbf{\overline{w_k-x_k}})(\mathbf{{w_k-x_k}})]^{\ell}=1$  and for each $k<N$ even $[(\mathbf{\overline{w_k-x_k}})(\mathbf{{w_k-x_k}})]^{\ell'}=1$ (left picture). Then it is possible to apply the exchange identity \eqref{exch_same} to each square of vertices $(x_k,x_{k+1},w_{k+1},w_{k+2})$, so that the labels $(\nu,\ell)$ and $(\nu',\ell')$ are interchanged and there appear the matrix $\mathbf{R}\equiv \mathbf{R}_{\ell,\ell'}(i\nu-i\nu')$ (right picture).
\begin{center}
\includegraphics[scale=0.3]{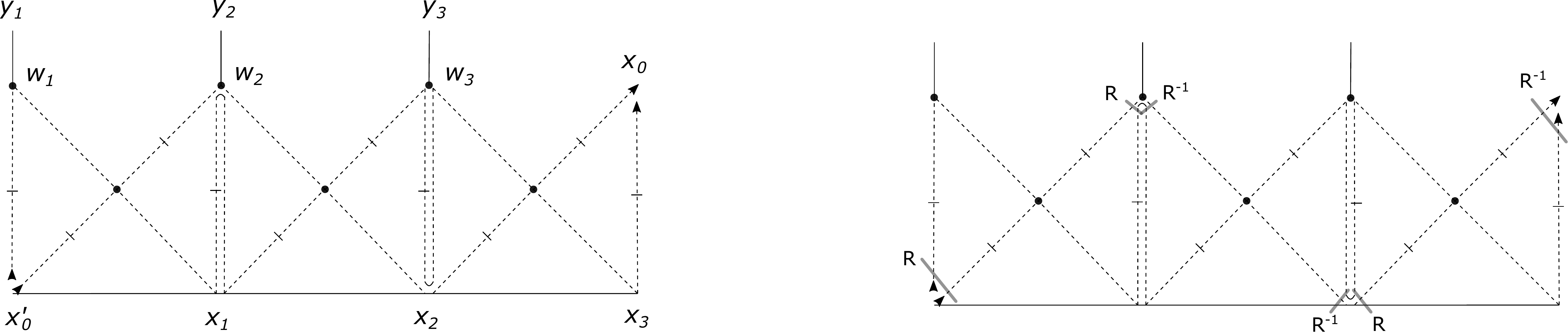}
\end{center}
Finally we notice that all the $\mathbf{R}$-matrices at vertices $w_k$ or $x_k$ cancel, $\mathbf{R}\mathbf{R}^{-1}=\mathbbm{1}$, and we remove the extra lines $[(\mathbf{\overline{w_k-x_k}})(\mathbf{{w_k-x_k}})]=1$ (left picture). As a last step we integrate the points $w_k$ by means of the star-triangle identity (right picture).
\begin{center}
\includegraphics[scale=0.3]{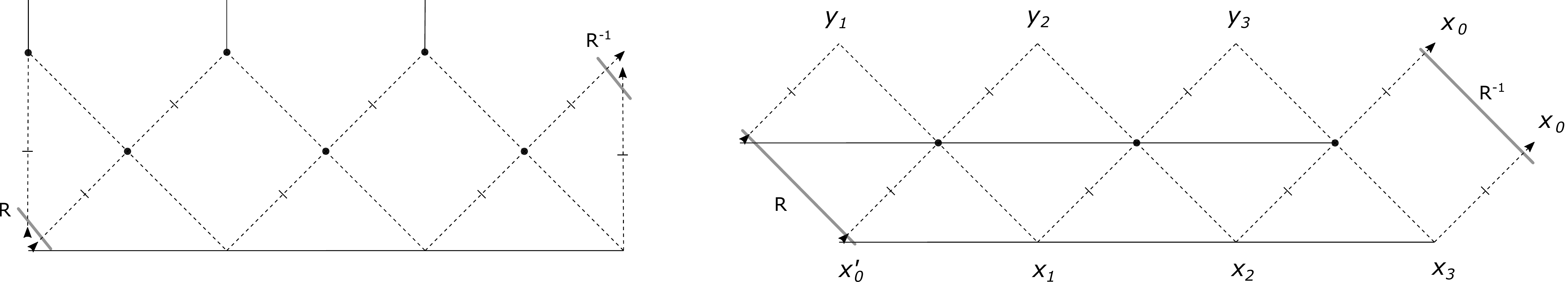}
\end{center}
The result of the procedure is that we obtained the algebra of $\hat{Q}$-operators
\begin{equation}
\label{hatcomm_ind}
(\hat{Q}_{\nu,\ell})_{\mathbf{\dot{a}}}^{\mathbf a} (\hat{Q}_{\nu',\ell'})_{\mathbf{\dot{b}}}^{\mathbf b}=\mathbf{R}^{\mathbf{\dot c \dot d}}_{\mathbf{\dot a \dot b}}(i\nu'-i\nu)\,(\hat{Q}_{\nu',\ell'})_{\mathbf{\dot{d}}}^{\mathbf d} (\hat{Q}_{\nu,\ell})_{\mathbf{\dot{c}}}^{\mathbf c}\,\mathbf{R}^{\mathbf{ab}}_{\mathbf{cd}}(i\nu-i\nu')\,,
\end{equation}
or in more compact notation
\begin{equation}
\label{hatcomm}
\hat{Q}_{\nu,\ell} \otimes \hat{Q}_{\nu',\ell'}=\mathbf{R}^{-1}\,\hat{Q}_{\nu',\ell'}\otimes \hat{Q}_{\nu,\ell}\,\mathbf{R}\,.
\end{equation}
Note that sign of the tensor product indicates the tensor product with respect to spinor indices and we use the important property of R-matrix: $\mathbf{R}^{-1}(u) = \mathbf{R}(-u)$ 
(see Appendix \ref{app:R}).

The same relation holds for $\check{Q}$-operators, and the proof follows the very same lines of the previous one
\begin{align}
\label{checkcomm}
(\check{Q}_{\nu,\ell})^{\mathbf{\dot{a}}}_{\mathbf a}\, 
(\check{Q}_{\nu',\ell'})^{\mathbf{\dot{b}}}_{\mathbf b}=
\mathbf{R}_{\mathbf{ab}}^{\mathbf{cd}}(i\nu'-i\nu)\,
(\check{Q}_{\nu',\ell'})^{\mathbf{\dot{d}}}_{\mathbf d}\, 
(\check{Q}_{\nu,\ell})^{\mathbf{\dot{c}}}_{\mathbf c}\,
\mathbf{R}_{\mathbf{\dot c \dot d}}^{\mathbf{\dot a \dot b}}(i\nu-i\nu')\\
\check{Q}_{\nu,\ell} \otimes \check{Q}_{\nu',\ell'}=\mathbf{R}^{-1}\,\check{Q}_{\nu',\ell'}\otimes \check{Q}_{\nu,\ell}\,\mathbf{R}\,.
\end{align}
A straightforward consequence is that for $\ell=0$ or $\ell'=0$ the equations \eqref{checkcomm} and \eqref{hatcomm} boil down to the commutation relations
\begin{equation}
\label{l0comm}
\hat{Q}_{\nu,0} \,\hat{Q}_{\nu',\ell'}=\hat{Q}_{\nu',\ell'}\, \hat{Q}_{\nu,0}\,,\,\qquad\qquad \check{Q}_{\nu,0} \,\check{Q}_{\nu',\ell'}=\check{Q}_{\nu',\ell'}\, \check{Q}_{\nu,0}\,,
\end{equation} and in particular the scalar fishnet kernel \eqref{squarelat} commutes with both Yukawa kernels \eqref{Yukhat} and \eqref{Yukcheck}.
Finally, in order to complete the algebra of $Q$-operators we can state the commutation 
\begin{align}
\label{hatcheckcomm}
(\hat{Q}_{\nu,\ell})_{\mathbf{\dot{a}}}^{\mathbf a} \,
(\check{Q}_{\nu',\ell'})^{\mathbf{\dot{b}}}_{\mathbf b} = 
(\check{Q}_{\nu',\ell'})^{\mathbf{\dot{b}}}_{\mathbf b}\,
(\hat{Q}_{\nu,\ell})_{\mathbf{\dot{a}}}^{\mathbf a}\\ 
\hat{Q}_{\nu,\ell} \otimes \check{Q}_{\nu',\ell'}=\check{Q}_{\nu',\ell'} \otimes\hat{Q}_{\nu,\ell}\,.
\end{align}
Let us start from the graphical representation of $\hat{Q}_{\nu,\ell} \otimes \check{Q}_{\nu',\ell'}$ on the left picture. As a first step we open the triangles containing a vertex $y_k$ into star integrals, according to \eqref{STR_0}.
\begin{center}
\includegraphics[scale=0.3]{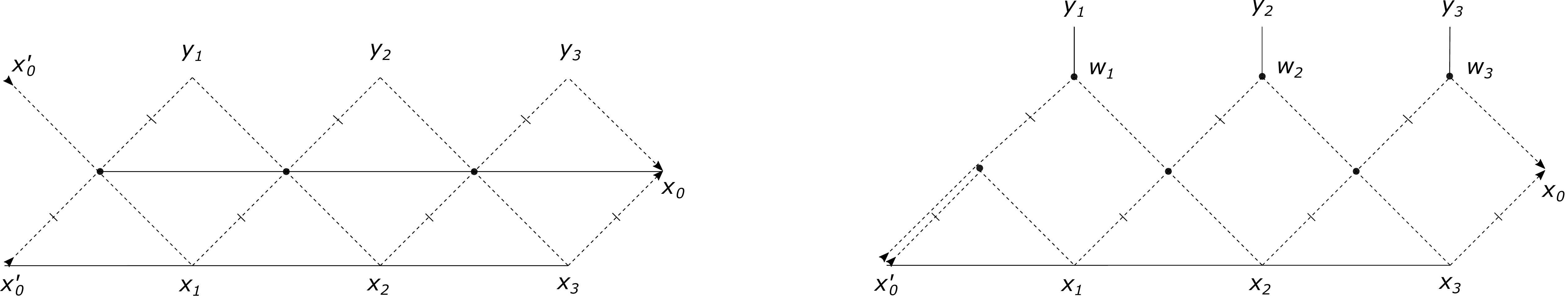}
\end{center}
Then we integrate the point $z_1$ by means of star-triangle identity obtaining a pair of vertical lines $[(\mathbf{w_1-x_1})]^{\ell}\otimes [(\overline{\mathbf{w_1-x_1}})]^{\ell'}$ (left picture). We can move vertical lines from $k=1$ to $k=2$ by means of the exchange relation \eqref{exch_same} (middle figure). The procedure can be iterated further, moving the lines to $k=N$ (right figure).
\begin{center}
\includegraphics[scale=0.22]{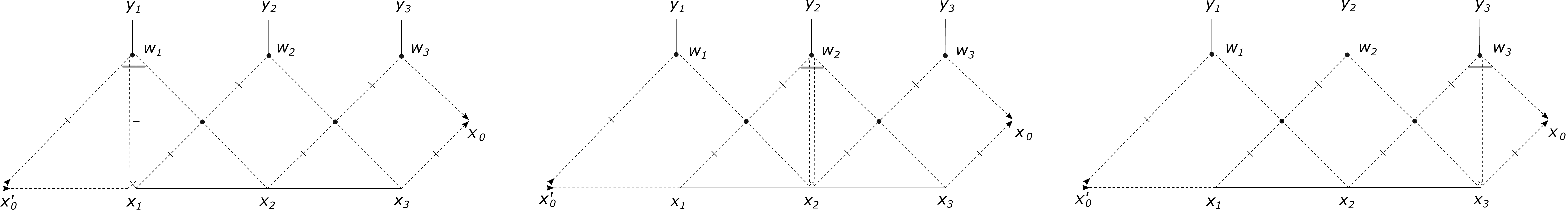}
\end{center}
\begin{center}
\includegraphics[scale=0.28]{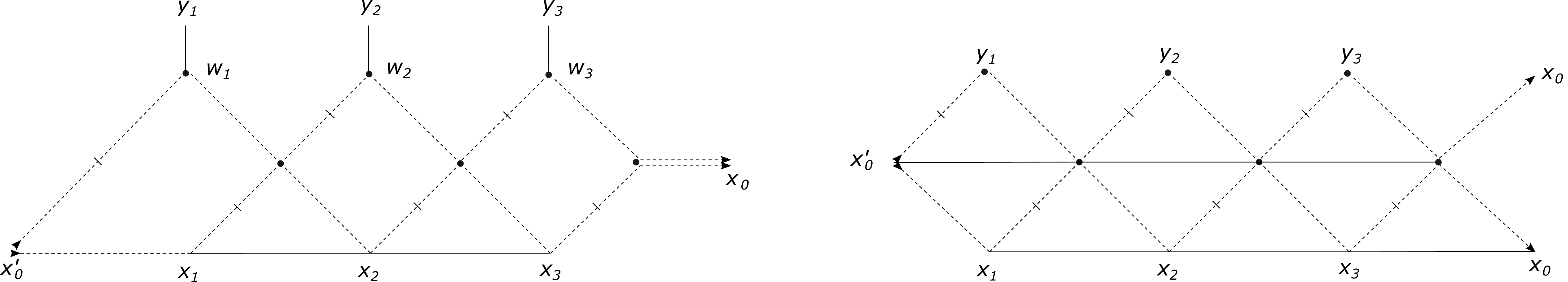}
\end{center}
The result of the procedure is that we obtained the algebra of $\hat{Q}$ and $\check{Q}$-operators.

We can also dress the operators \eqref{Qhat_ker} and  \eqref{Qcheck_ker} with symmetric spinors $\alpha^{\otimes \ell}$ and $\beta^{\otimes \ell}$
\begin{align}
\label{dressed}
\begin{aligned}
&\langle \beta|\hat{Q}_{\nu,\ell}|\alpha \rangle\equiv \,\beta^{\dot{c}_1}\cdots\beta^{\dot{c}_{\ell} }(\hat{Q}_{\nu,\ell})_{(\dot{c}_1\dots \dot{c}_{\ell})}^{({c}_1\dots {c}_{\ell})}\, \alpha_{c_1}\cdots \alpha_{c_{\ell}}\,,\\
&\langle \alpha|\check{Q}_{\nu,\ell}|\beta \rangle\equiv \,\alpha^{{c}_1}\cdots\alpha^{{c}_{\ell} }(\check{Q}_{\nu,\ell})_{({c}_1\dots {c}_{\ell})}^{(\dot{c}_1\dots \dot{c}_{\ell})}\, \beta_{\dot c_1}\cdots \beta_{\dot c_{\ell}}\,.
\end{aligned}
\end{align}
The kernel of the operators \eqref{dressed} are invariant under $SO(4)$ rotations and covariant respect to scaling of coordinates. The numerator of \eqref{Qhat_ker} after dressing becomes 
\begin{equation}
\label{harmpol}
\langle\beta|(\overline{\mathbf{x_0'-y_1}})(\mathbf{ y_1-x_1})(\overline{\mathbf{x_1- y_2}})\cdots (\mathbf{y_N- x_N})(\overline{\mathbf{x_N- x_0}})|\alpha\rangle^{\ell}\,.
\end{equation}
This expression depends on all unit vectors $a_i^{\mu} = (y_i-x_i)^{\mu}/|y_i-x_i|$ and $b_i^{\mu} = (y_i-x_{i-1})^{\mu}/|y_i-x_{i-1}|$ and is an harmonic polynomial of degree $\ell$ in each of them. For example, let us extract some reference vector $b_i^{\mu}$ and then it is easy to see 
that the general structure of our expression has the following form   
\begin{align}\label{harm}
b_i^{\mu_1}\cdots b_i^{\mu_{\ell}}\, C^{\mu_1}\cdots C^{\mu_{\ell}} = b_i^{\mu_1}\cdots b_i^{\mu_{\ell}}\, C^{\mu_1\ldots\mu_{\ell}} = 
b_i^{\mu_1\ldots\mu_{\ell}}\,C^{\mu_1\ldots\mu_{\ell}} = 
b_i^{\mu_1\ldots\mu_{\ell}}\,\, C^{\mu_1}\cdots C^{\mu_{\ell}}
\end{align}
where we use notation \eqref{x} for the traceless symmetric tensor and 
introduce vector $C^{\mu}$ which has the general form 
$C^{\mu} =\langle\beta|A\,\bsig^{\mu}\,B|\alpha\rangle$ with some  
matrices $A$ and $B$ absorbing all not important for us at the moment 
products of matrices from initial expression.
All transformations in \eqref{harm} are evident and based on the property $C^{\mu} C_{\mu}  = 0$
which is a consequence of the Fierz identity 
$\bsig^{\mu} \otimes\bsig^{\mu} = 2(\mathbbm{1}-\mathbb{P})$ 
\begin{align*}
\langle\beta|A\,\bsig^{\mu}\,B|\alpha\rangle\langle\beta|A\,\bsig_{\mu} \,B|\alpha\rangle = 2 \langle\beta|A\otimes \langle\beta| A\, (\mathbbm{1}-\mathbb{P})\, 
B|\alpha\rangle\otimes B|\alpha\rangle=0\,.
\end{align*} 
Relation \eqref{harm} shows that all dependence on the vector $b_i^{\mu}$ accumulated 
in an symmetric and traceless tensor $b_i^{\mu_1\ldots\mu_{\ell}}$ that gives 
harmonic polynomial of degree $\ell$ after convolution with $C^{\mu_1}\cdots C^{\mu_{\ell}}$.

Relations \eqref{hatcomm},\eqref{checkcomm},\eqref{hatcheckcomm} can be rewritten by inserting spinors as
\begin{align*}
&\langle \beta| \hat{Q}_{\nu,\ell}|\alpha \rangle \,\langle\beta'|\hat{Q}_{\nu',\ell'}|\alpha'\rangle\,=\,\langle \beta',\beta|\mathbf R^{-1}\, \hat{Q}_{\nu',\ell'}  \otimes \hat{Q}_{\nu,\ell}\,\mathbf R|\alpha',\alpha\rangle\,,\\
&\langle \alpha| \check{Q}_{\nu,\ell}|\beta \rangle \, 
\langle\alpha'|\check{Q}_{\nu',\ell'}|\beta'\rangle\,=\,
\langle \alpha',\alpha|\mathbf R^{-1}\, \check{Q}_{\nu',\ell'}  \otimes \check{Q}_{\nu,\ell}\,\mathbf |\beta',\beta\rangle\,,\\
&\langle \beta| \hat{Q}_{\nu,\ell}|\alpha \rangle \, \langle\alpha'|\check{Q}_{\nu',\ell'}|\beta'\rangle\,=\,\langle \alpha'| \check{Q}_{\nu',\ell'}|\beta'\rangle\,\langle\beta|\hat{Q}_{\nu,\ell}|\alpha\rangle.
\end{align*}
In the choice $\nu=\nu'$ and $\ell=\ell'$ we check that $\mathbb{R}_{\ell,\ell}(0)=\mathbb{P}_{\ell,\ell'}$ and the first two formulas simplify as
\begin{align*}
&\langle \beta| \hat{Q}_{\nu,\ell}|\alpha \rangle \,\langle\beta'|\hat{Q}_{\nu,\ell}|\alpha'\rangle\,=\,\langle \beta',\beta|\mathbb{P}\, \hat{Q}_{\nu,\ell}  \otimes \hat{Q}_{\nu,\ell}\,\mathbb{P}|\alpha',\alpha\rangle\,=\,\langle \beta|\hat{Q}_{\nu,\ell} |\alpha\rangle \, \langle \beta'|\hat{Q}_{\nu,\ell}\,|\alpha'\rangle\,,\\
&\langle \alpha| \check{Q}_{\nu,\ell}|\beta \rangle \, \langle\alpha'|\check{Q}_{\nu,\ell}|\beta'\rangle\,=\,\langle \alpha',\alpha|\mathbb{P}\, \check{Q}_{\nu,\ell}  \otimes \check{Q}_{\nu,\ell}\,\mathbb{P}|\beta',\beta\rangle\,=\,\langle \alpha|\check{Q}_{\nu,\ell} |\beta\rangle \, 
\langle \alpha'|\check{Q}_{\nu,\ell}\,|\beta'\rangle\,,
\end{align*}
which stand as a consistency check for \eqref{hatcomm} and \eqref{checkcomm}.

\section{Eigenfunctions}
\label{sec:eigenf}
The construction of the eigenfunctions for the spin chain of length $N$ follows the same iterative technique explained in \cite{Derkachov2014} for the two-dimensional case.
First of all we introduce the integral kernels
\begin{align}
\label{lambda_kernel}
\Lambda_{\nu,\ell}^{(N)}(\mathbf{x}|\mathbf{y}) = r(\nu,\ell)^{N-1}\,\hat{Q}^{(N-1)}_{\nu,\ell}(x_1,x_2,\dots,x_N,x_0|y_1,\dots,y_{N-1})\,,\\
\notag
r(\nu,\ell)=
\frac{\Gamma\left(1+i\frac{\lambda}{2}+i\nu+\frac{\ell}{2}\right)
\Gamma\left(1+i\frac{\lambda}{2}-i\nu+\frac{\ell}{2}\right)}
{\Gamma\left(1-i\frac{\lambda}{2}-i\nu+\frac{\ell}{2}\right)
\Gamma\left(1-i\frac{\lambda}{2}+i\nu+\frac{\ell}{2}\right)}\,,
\end{align}
which carry symmetric spinor indices
$(\Lambda_{\nu,\ell})^{(a_1\dots a_{\ell})}_{(\dot a_1\dots \dot a_{\ell})} =
(\Lambda_{\nu,\ell})_{\mathbf{\dot{a}}}^{\mathbf a}$ and $\hat{Q}$-kernel is defined in \eqref{Qhat_ker}.


The corresponding integral operator is
defined as follows
\begin{align*}
\left[\Lambda_{\nu,\ell}^{(N)}\phi\right](x_1,\dots,x_N) = \int d^4y_1\cdots d^4y_{N-1}\, {\hat Q}^{(N-1)}_{\nu,\ell}(x_1,x_2,\dots,x_N,x_0|y_1,\dots,y_{N-1}) \,\phi(y_1,\dots,y_{N-1}) \,,
\end{align*}
The integral operator and its kernel carry symmetric spinor indices, therefore for each choice of indices $\mathbf{a}$ and $\dot{\mathbf{a}}$ the operator $(\Lambda_{\nu,\ell})_{\mathbf{\dot{a}}}^{\mathbf a}$ maps
a function of $N-1$ variables $x_1,\cdots,x_{N-1}$ to a function of $N-1$ variables $x_1,\cdots,x_{N}$ which carry additional
symmetric spinor indices $\mathbf{a}$ and $\mathbf{\dot{a}}$.
In analogy with the particle creation operator in quantum field theory, the operator $(\Lambda_{\nu,\ell})_{\mathbf{\dot{a}}}^{\mathbf a}$ creates the
dependence on a new variable $x_N$ and symmetric spinor indices $\mathbf{\dot{a}}$ and $\mathbf a$.
The operator $\Lambda_{\nu,\ell}^{(N)}$ is a close relative of the
$\hat{Q}$-operator.
The integral kernel ${\hat Q}^{(N)}_{\nu,\ell}(x_0',x_1,\dots,x_N,x_0|y_1,\dots,y_N)$ is defined in \eqref{Qhat_ker} and after corresponding changes it is easy to obtain explicit form of the integral kernel of the operator $\Lambda_{\nu,\ell}^{(N)}$. For simplicity we present it here
\begin{align}
\nonumber
&{\hat Q}^{(N-1)}_{\nu,\ell}(x_1,x_2,\dots,x_N,x_0|y_1,\dots,y_{N-1}) = \prod_{k=0}^{N-1} \frac{1}{(x_k-x_{k+1})^{2(i\lambda)}}\times\\
\label{Lambda}
&\times\frac{[(\mathbf{\overline {x_1- y_1}})(\mathbf{y_1- x_2})(\mathbf{\overline {x_2- y_2}})\cdots (\mathbf {y_{N-1}-x_N})(\overline{\mathbf{x_N- x_0}})]^{\ell}}{(x_1-y_1)^{2(1-i\frac{\lambda}{2}+i\nu)}(y_1-x_2)^{2(1-i\frac{\lambda}{2}-i\nu)}\cdots (y_{N-1}-x_N)^{2(1-i\frac{\lambda}{2}-i\nu)}(x_N-x_0)^{2(1+i\frac{\lambda}{2}+i\nu)}}\,,
\end{align}

The fundamental observation in order to construct iteratively the eigefunctions of the model \eqref{ham_intro} is that the generators of commuting charges $Q(u)$ at lengths $N$ and $N-1$ are intertwined by the operator $\Lambda_{\nu,\ell}$ of length $N$
\begin{equation}
\label{QLambda}
 Q^{(N)}(u) \,\Lambda_{\nu,\ell}^{(N)}= \tau(u,\nu,\ell)\,\Lambda_{\nu,\ell}^{(N)}\, Q^{(N-1)}(u)\,,
\end{equation}
where
\begin{equation}
\tau(u,\nu,\ell) = 4^u \frac{\Gamma\left(1+\frac{\ell}{2}+i\nu-\frac{i }{2}\lambda  \right)\Gamma \left(1+\frac{\ell}{2}-i\nu+u+\frac{i }{2}\lambda \right)}{\Gamma \left(1+\frac{\ell}{2}-i\nu+\frac{i}{2}\lambda \right) \Gamma \left(1+\frac{\ell}{2}+i\nu-u-\frac{i }{2}\lambda \right)}\,.
\end{equation}
The proof of this relation can be given graphically, and makes essentially use of the star-triangle relation \eqref{STRsame}.
Starting from the l.h.s.of equation \eqref{QLambda} (left picture), we  open the triangles of vertex $y_k$ into star integrals (right picture) by means of \eqref{STR_0}
\begin{center}
\includegraphics[scale=0.30]{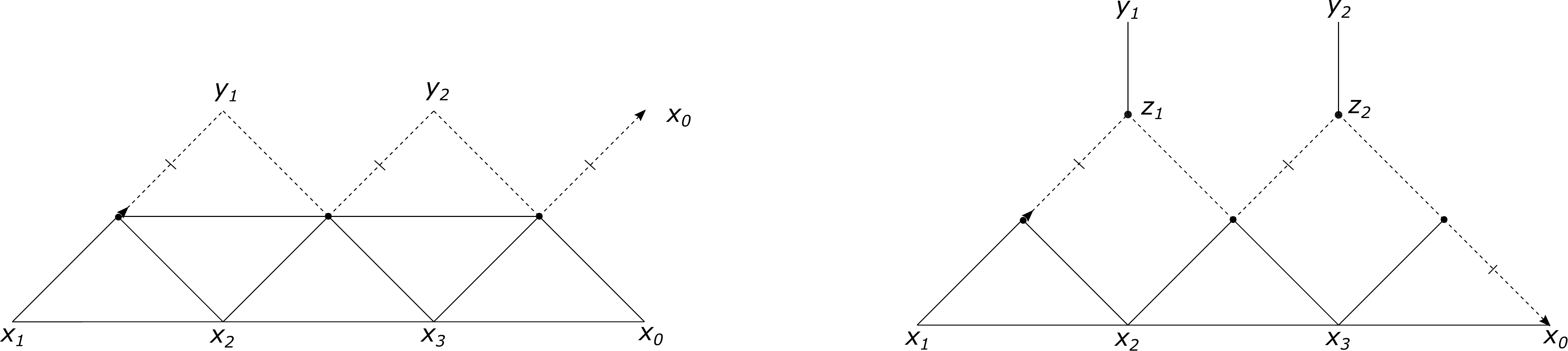}
\end{center}
The next step is to integrate the star-integral of vertices $(z_2,x_0,x_3)$ in order to obtain a vertical line $(z_{2}-x_3)$ (left picture), which can be moved leftwards to $(x_{2}-z_{1})$ by means of the exchange relation \eqref{exch_same_1} (right picture)
\begin{center}
\includegraphics[scale=0.30]{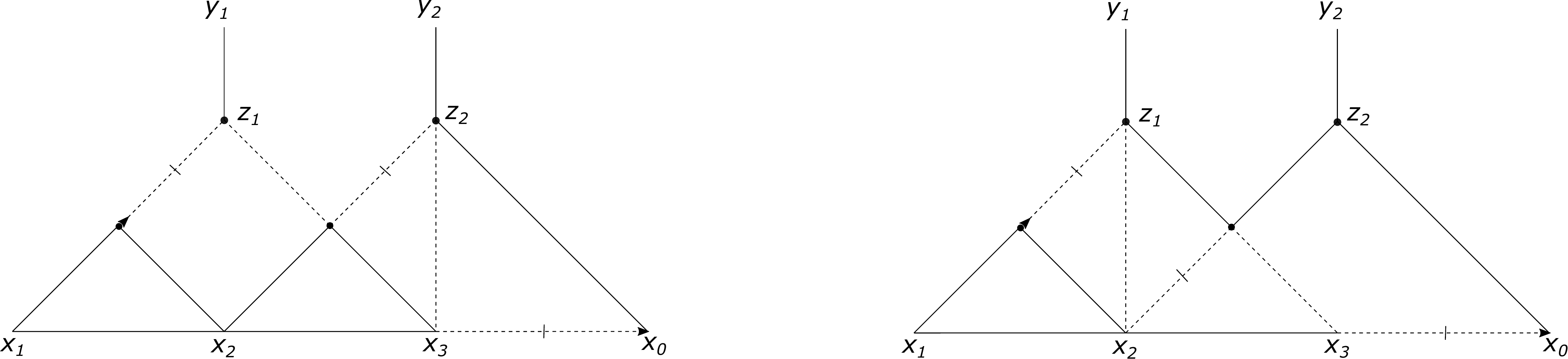}
\end{center}
Finally, the triangle of basis $z_1-x_2$ in the previous picture is opened into a star integral (left picture) and chain-rule integration in $z$ and star-integrals in $z_k$ are performed, obtaining the r.h.s. of \eqref{QLambda} represented in the right picture.
\begin{center}
\includegraphics[scale=0.30]{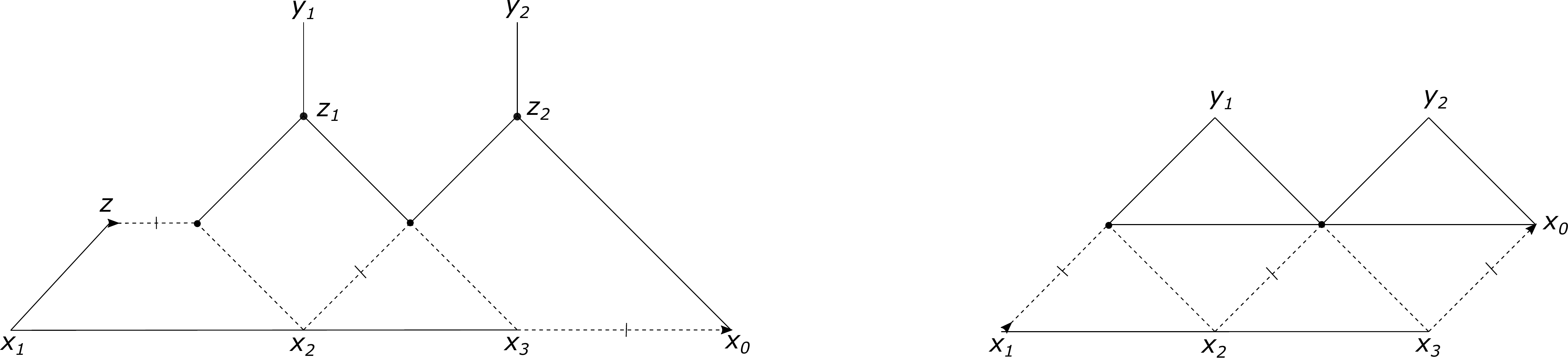}
\end{center}
The coefficient $\tau(u,\nu,\ell)$ results from the collection of the star-triangle coefficients \eqref{coeff} produced step by step along the proof. The condition \eqref{QLambda}, connecting the operator $Q$ at lengths $N$ and $N-1$, allows to have a recursive relation which reduces the problem to the diagonalization of the operator $Q$ of length one. The condition $N=1$ of the recursion
\eqref{QLambda} consists indeed in the eigenvalue equation
\begin{equation}
\label{init}
Q^{(1)}(u)\, \Lambda^{(1)}_{\nu,\ell}(x_1,x_0) = \tau(u,\nu,\ell)\, \Lambda^{(1)}_{\nu,\ell}(x_1,x_0)\,,
\end{equation}
where we notice that the kernel $\Lambda_{\nu,\ell}$ at length one reduces to a function of $x$ only, according to \eqref{lambda_kernel}
\begin{align}
\label{N1}
\Lambda^{(1)}_{\nu,\ell}(x_1,x_0) =
\frac{[(\overline{\mathbf{x_1- x_0}})]^{\ell}}{(x_1-x_0)^{2(1+i\frac{\lambda}{2}+i\nu)}}\,,
\end{align}
Note that it is the same eigenfunction \eqref{eig1}
from the Section \ref{ladder} in spinor form and restored
dependence on $x_0$.

The proof of \eqref{init} is simple and follows from the
chain rule \eqref{chain_rule} at $\ell'=0$.

\begin{center}
\includegraphics[scale=0.4]{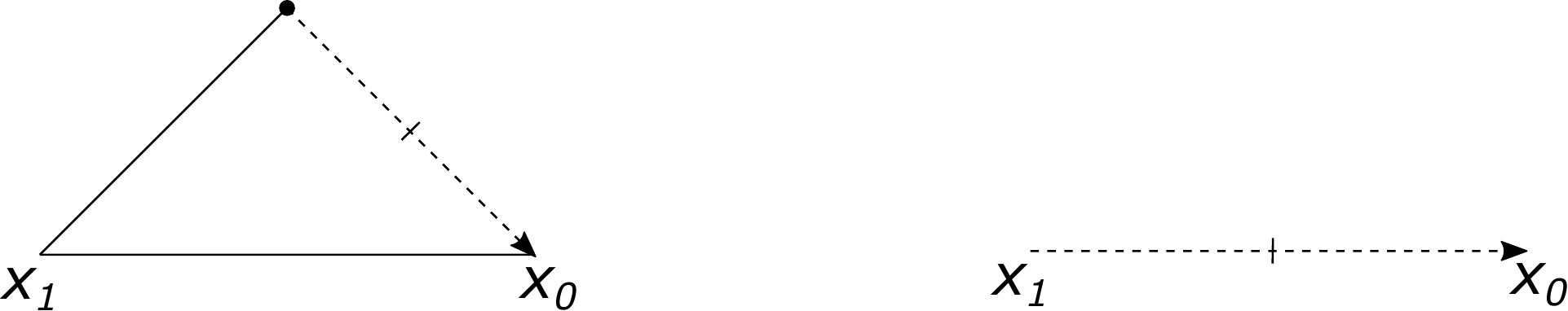}
\end{center}
Of course it is the same explicit calculation  \eqref{N=1} of the eigenvalue $\tau(u,\nu,\ell)$ in the Section \ref{ladder}.

As a consequence of \eqref{QLambda} and \eqref{init}, the function
\begin{align}
\label{def_eigenf}
\Psi_{\nu_1,\dots,\nu_N,\ell_1,\dots,\ell_N}(x_1,\dots,x_N,x_0) = \Lambda^{(N)}_{\nu_N,\ell_N}\Lambda^{(N-1)}_{\nu_{N-1},\ell_{N-1}}\cdots \Lambda^{(2)}_{\nu_2,\ell_2}\Lambda^{(1)}_{\nu_1,\ell_1}\,,
\end{align}
is an eigenfunction of the operator $Q(u)$ at length $N$,
the eigenvalue being $\prod_{k=1}^N \tau(u,\nu_k,\ell_k)$
\begin{equation}
\label{eigenvfact}
Q^{(N)}(u)\,\Lambda^{(N)}_{\nu_N,\ell_N}\Lambda^{(N-1)}_{\nu_{N-1},\ell_{N-1}}\cdots \Lambda^{(2)}_{\nu_2,\ell_2}\Lambda^{(1)}_{\nu_1,\ell_1} =
\prod_{k=1}^N \tau(u,\nu_k,\ell_k)\,
\Lambda^{(N)}_{\nu_N,\ell_N}\Lambda^{(N-1)}_{\nu_{N-1},\ell_{N-1}}\cdots \Lambda^{(2)}_{\nu_2,\ell_2}\Lambda^{(1)}_{\nu_1,\ell_1}\,.
\end{equation}
The formula \eqref{eigenvfact} shows that the spectrum of the spin chain \eqref{ham_intro} at length $N$ is factorized into $N$ identical contributions, each depending on a couple of quantum numbers $(\ell_k,\nu_k)$ or, alternatively, on a complex variable $Y_k$ with the quantization condition
\begin{equation}
\label{SOV}
 Y_k + Y_k^* \in \mathbb{Z}^+\,\Longrightarrow Y_k = \frac{\ell_k}{2}+i \nu_k\,.
\end{equation}
The equivalence stated by \eqref{eigenvfact} between a length-$N$ model and $N$ length-$1$ models realized a separation of variables at a quantum level, being $\{Y_k\}$ the eigenvalues of the quantum separated variables. According to the parametrization \eqref{SOV} we will use the compact notation for the eigenfunctions
\begin{equation}
\label{eigenf_Y}
\Psi^{(N)}_{\mathbf{Y}}(\mathbf{x})\equiv \Psi_{\nu_1,\dots,\nu_N,\ell_1,\dots,\ell_N}(x_1,\dots,x_N,x_0)\,,
\end{equation}
where $\mathbf{Y}=(Y_1,\dots,Y_N)$. It is important to notice that the functions $\Psi^{(N)}$ are not scalar objects but have spinorial indices, inherited from the ones of each layer $\Lambda^{(k)}_{Y_k}$, namely
\begin{equation}
\left(\Psi^{(N)}_{\mathbf{Y}}\right)^{(a_{11},\dots,a_{1\ell_1})\dots(a_{N1},\dots,a_{N\ell_N})}
_{(\dot a_{11},\dots,\dot a_{1\ell_1})\dots(\dot a_{N1},\dots,\dot a_{N\ell_N})} =
\left(\Psi^{(N)}_{\mathbf{Y}}\right)^{\mathbf a_{1}\dots\mathbf a_{N}}_{\dot{\mathbf a}_{1}\dots\dot{\mathbf a}_{N}} =
\left(\Lambda^{(N)}_{Y_N}\right)^{\mathbf a_{N}}_{\dot{\mathbf a}_{N}}
\cdots\left(\Lambda^{(2)}_{Y_2}\right)^{\mathbf a_{2}}_{\dot{\mathbf a}_{2}}
\left(\Lambda^{(1)}_{Y_1}\right)^{\mathbf a_{1}}_{\dot{\mathbf a}_{1}}
\end{equation}
Despite that, the eigenvalue \eqref{eigenvfact} does not depend in any way on the spinor indices, due to the invariance of operators $Q^{(N)}(u)$ under the $SO(4)$ rotations.
As it was done in \eqref{dressed} for the operators $\hat{Q}$ and $\check{Q}$, we can dress the eigenfunction \eqref{def_eigenf} with a couple of symmetric spinors $(\alpha_k,\beta_k)$ of degree $\ell_k$ for each layer $k=1,\dots,N$:
\begin{equation}
\langle \boldsymbol{\beta}|\Psi^{(N)}_{\mathbf{Y}}| \boldsymbol{\alpha}\rangle \,= \langle \beta_N|\Lambda^{(N)}_{Y_N}|\alpha_N\rangle\langle\beta_{N-1}|\Lambda^{(N-1)}_{Y_{N-1}}|\alpha_{N-1}\rangle\cdots \langle\beta_1|\Lambda^{(1)}_{Y_1}|\alpha_1\rangle\,,
\end{equation}
and
\begin{equation*}
 \langle \beta_k|\Lambda^{(k)}_{Y_k}|\alpha_k\rangle=
 (\beta_k)^{\dot{c}_1}\cdots (\beta_k)^{\dot{c}_{\ell_k}}\,
 (\Lambda^{(k)}_{Y_k})_{(\dot{c}_1 \dots \dot{c}_{\ell_k})}^{(c_1\dots c_{\ell_k})}\,
 (\alpha_k)_{c_1} \cdots (\alpha_k)_{c_{\ell_k}}\,.
\end{equation*}

\subsection{Symmetry of the eigenfunctions}
\label{sec:symm_eig}
The eigenvalue \eqref{eigenvfact} does not depend on the order in which the variables $Y_k$ appear in the eigenfunctions, namely any permutation $s \in \mathbb{S}_N$
\begin{equation}
\label{permY}
s\left(Y_1,\dots,Y_N\right) \,=\, \left(Y_{s(1)},\dots,Y_{s(N)}\right)\,,
\end{equation}
in the definition \eqref{eigenf_Y} leads to the same eigenvalue. At the level of the eigenfunctions this reflects in the fact that any such permutation is equivalent to a mixing of the spinorial indices only, and the operators $Q(u)$ are insensitive to the spinor structure.
Any permutation $s$ can be decomposed into a product of elementary exchanges
\begin{equation}
 s_{k}\left(Y_1,\dots,Y_k,Y_{k+1},\dots,Y_N\right)\,=\,
 \left(Y_1,\dots,Y_{k+1},Y_k,\dots,Y_N\right)\,,
\end{equation}
which is interchange of two adjacent layers for the eigenfunction.
This interchange can be compensated by the corresponding transformations of the spinors as stated by the following identity
\begin{equation}
\label{layer_comm0}
\left(\Lambda^{(k+1)}_{Y_{k+1}}\right)_{\mathbf{\dot{a}}}^{\mathbf a}
\left(\Lambda^{(k)}_{Y_{k}}\right)_{\mathbf{\dot{b}}}^{\mathbf b}=
\mathbf{R}^{\mathbf{\dot c \dot d}}_{\mathbf{\dot a \dot b}}(Y_k,Y_{k+1})\,
\mathbf{R}^{\mathbf{ab}}_{\mathbf{cd}}(Y_{k+1},Y_k)
\left(\Lambda^{(k+1)}_{Y_{k}}\right)_{\mathbf{\dot{d}}}^{\mathbf d}
\left(\Lambda^{(k)}_{Y_{k+1}}\right)_{\mathbf{\dot{c}}}^{\mathbf c}\,,
\end{equation}
or in more compact form
\begin{equation}
\label{layer_comm}
\Lambda^{(k+1)}_{Y_{k+1}}\,\Lambda^{(k)}_{Y_{k}} = \mathbf{\bar{R}}_{k+1,k}(Y_{k},Y_{k+1})\,\Lambda^{(k+1)}_{Y_{k}}\,\Lambda^{(k)}_{Y_{k+1}}
\,\mathbf{R}_{k+1,k}(Y_{k+1},Y_{k})\,,
\end{equation}
where for simplicity we denote $\mathbf{R}(Y_n,Y_m) \equiv \mathbf{R}(i\nu_{n}-i\nu_{m})$ and indices $k+1$ and $k$ show that R-matrix act nontrivially in the spaces of dotted symmetric spinors of the ranks $\ell_{k+1}$ and $\ell_k$ for the $\mathbf{\bar{R}}$-matrix and in the spaces of un-dotted symmetric spinors of the ranks $\ell_{k+1}$ and $\ell_k$ for the $\bar{R}$-matrix.
Note that the spaces of the spinors are ordered rigidly from
the right to the left $\ell_N,\ldots,\ell_2,\ell_1$ in correspondence with the ordering of the layers in eigenfunction.

We provide a graphical proof which follows closely the steps of the proof given in sect \ref{sect:Qspin} for \eqref{hatcomm}.  Starting from the l.h.s. of \eqref{layer_comm} (depicted for $k=3$ in the left picture) we open the triangles of vertices $y_k$ into star integrals and insert identities in the spinor space as vertical lines $[\mathbf{(x_h-z_{h-1})(\overline{x_h-z_{h-1}})}]^{\ell_k}$ (right picture).
\begin{center}
\includegraphics[scale=0.35]{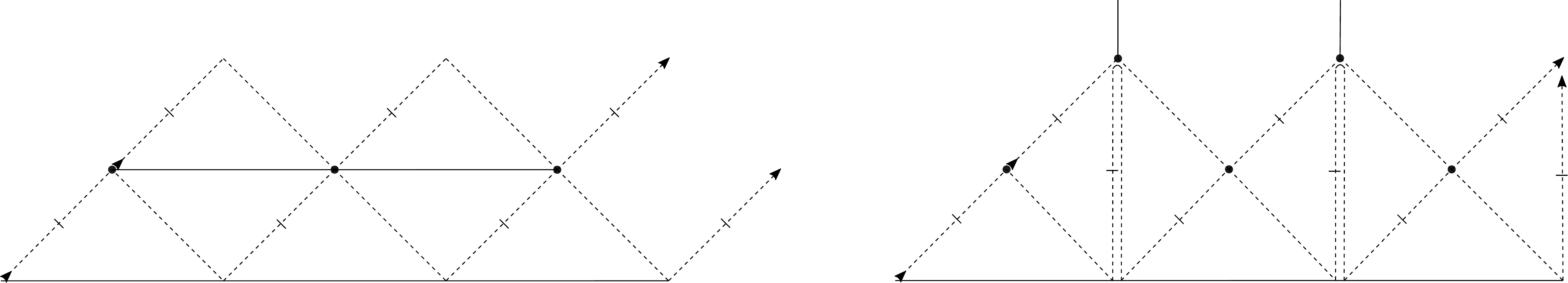}
\end{center}
Then, we apply the exchange relation \eqref{exch_same} to the squares of vertices $(z_{k-1},x_{k},z_{k},x_{k+1})$, exchanging the weights $(\nu_k,\ell_k)\leftrightarrow(\nu_{k+1},\ell_{k+1})$ and producing the $\mathbf{R}$-matrices (left picture). Furthermore, we open the triangle of basis $(z_1-x_2)$ into a star integral according to \eqref{STR}.
\begin{center}
\includegraphics[scale=0.35]{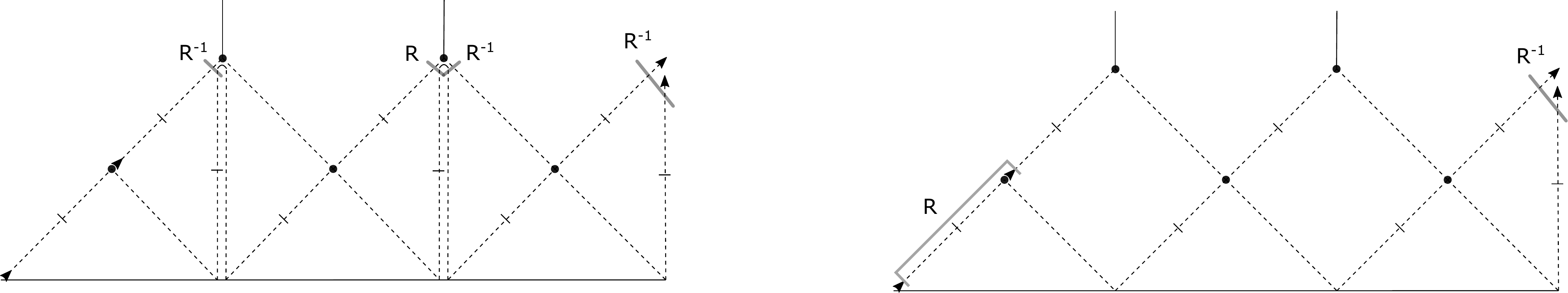}
\end{center}
Finally we take the star integrals in the points $z$ and $z_k$, ending up with the r.h.s. of \eqref{layer_comm}
\begin{center}
\includegraphics[scale=0.45]{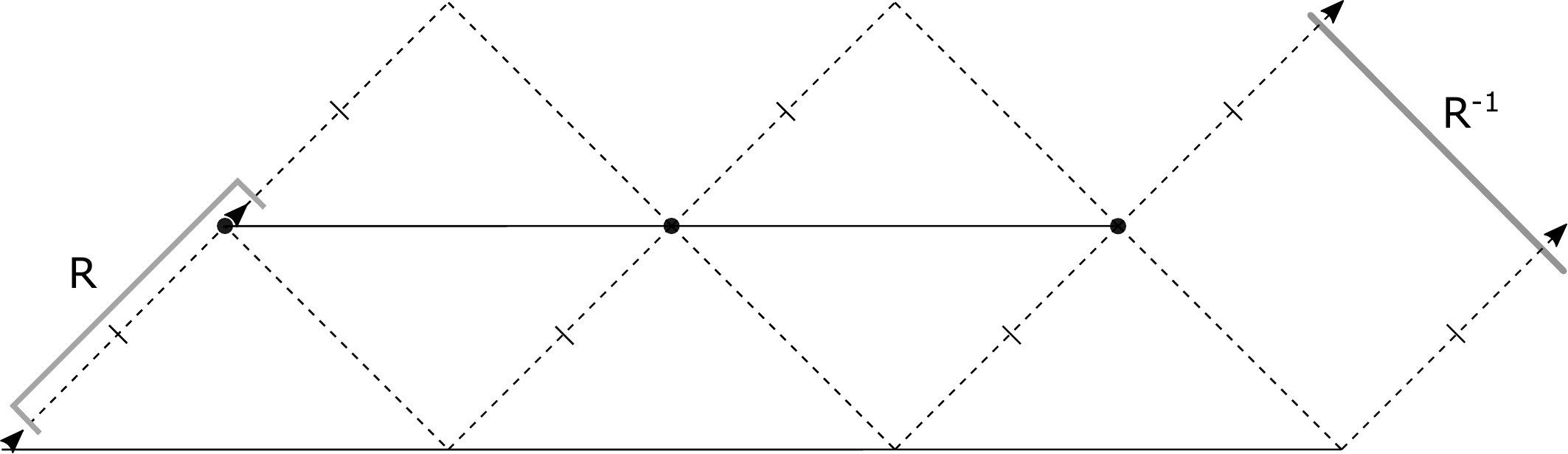}
\end{center}

Formula \eqref{layer_comm} is the statement of the invariance
of the eigenfunction with respect to interchange of layers and corresponding transformation of the spinors. Of course there should be some self-consistency conditions like the Coxeter relations for the generators of the permutation group $\mathbb{S}_N$

Let us consider the simplest nontrivial examples $N=2$ and $N=3$ to show the
appearance of these relations.
In the simplest case $N=2$ we have the following representation for the eigenfunction
\begin{align*}
\left(\Psi_{Y_1,Y_2}\right)^{\mathbf a_{1}\mathbf a_{2}}_{\dot{\mathbf a}_{1}\dot{\mathbf a}_{2}}  =
\left(\Lambda^{(2)}_{Y_2}\right)^{\mathbf a_{2}}_{\dot{\mathbf a}_{2}}
\left(\Lambda^{(1)}_{Y_1}\right)^{\mathbf a_{1}}_{\dot{\mathbf a}_{1}}
\end{align*}
and there exists only one permutation of the layers
\begin{equation}
\left(\Lambda^{(2)}_{Y_2}\right)_{\mathbf{\dot{a}}}^{\mathbf a}
\left(\Lambda^{(1)}_{Y_1}\right)_{\mathbf{\dot{b}}}^{\mathbf b}=
\mathbf{R}^{\mathbf{\dot c \dot d}}_{\mathbf{\dot a \dot b}}(Y_1,Y_2)\,
\mathbf{R}^{\mathbf{ab}}_{\mathbf{cd}}(Y_2,Y_1)
\left(\Lambda^{(2)}_{Y_1}\right)_{\mathbf{\dot{d}}}^{\mathbf d}
\left(\Lambda^{(1)}_{Y_2}\right)_{\mathbf{\dot{c}}}^{\mathbf c}\,.
\end{equation}
The second application of this permutation returns everything
back to the initial eigenfunction
\begin{align*}
\left(\Lambda^{(2)}_{Y_2}\right)_{\mathbf{\dot{a}}}^{\mathbf a}
\left(\Lambda^{(1)}_{Y_{1}}\right)_{\mathbf{\dot{b}}}^{\mathbf b}=
\mathbf{R}^{\mathbf{\dot c \dot d}}_{\mathbf{\dot a \dot b}}(Y_1,Y_{2})\,
\mathbf{R}^{\mathbf{ab}}_{\mathbf{cd}}(Y_{2},Y_1)
\left(\Lambda^{(2)}_{Y_{1}}\right)_{\mathbf{\dot{d}}}^{\mathbf d}
\left(\Lambda^{(1)}_{Y_2}\right)_{\mathbf{\dot{c}}}^{\mathbf c} = \\
\mathbf{R}^{\mathbf{\dot c \dot d}}_{\mathbf{\dot a \dot b}}(Y_1,Y_{2})\,
\mathbf{R}^{\mathbf{ab}}_{\mathbf{cd}}(Y_{2},Y_1) \,\,
\mathbf{R}^{\mathbf{\dot e \dot f}}_{\mathbf{\dot c \dot d}}(Y_{2},Y_{1})\,
\mathbf{R}^{\mathbf{cd}}_{\mathbf{ef}}(Y_{1},Y_{2})
\left(\Lambda^{(2)}_{Y_{2}}\right)_{\mathbf{\dot{e}}}^{\mathbf e}
\left(\Lambda^{(1)}_{Y_{1}}\right)_{\mathbf{\dot{f}}}^{\mathbf f} =
\left(\Lambda^{(2)}_{Y_2}\right)_{\mathbf{\dot{a}}}^{\mathbf a}
\left(\Lambda^{(1)}_{Y_{1}}\right)_{\mathbf{\dot{b}}}^{\mathbf b}
\end{align*}
due to relations for the R-matrices (see Appendix )
\begin{align*}
\mathbf{R}^{\mathbf{\dot c \dot d}}_{\mathbf{\dot a \dot b}}(Y_1,Y_{2})\,
\mathbf{R}^{\mathbf{\dot e \dot f}}_{\mathbf{\dot c \dot d}}(Y_{2},Y_{1}) = \delta^{\mathbf{\dot e }}_{\mathbf{\dot a}}\,
\delta^{\mathbf{\dot f}}_{\mathbf{\dot b}} \\
\mathbf{R}^{\mathbf{cd}}_{\mathbf{ef}}(Y_{1},Y_{2})\,
\mathbf{R}^{\mathbf{ab}}_{\mathbf{cd}}(Y_{2},Y_1) =
\delta_{\mathbf{e }}^{\mathbf{a}}\,
\delta_{\mathbf{f}}^{\mathbf{b}}
\end{align*}
For simplicity in the following we shall use more compact notations and
as an example we give the same formulae in compact notations
\begin{align*}
&\Lambda^{(2)}_{Y_2}\Lambda^{(1)}_{Y_{1}}=
\mathbf{\bar{R}}_{21}(Y_1,Y_2)\,
\Lambda^{(2)}_{Y_{1}}\Lambda^{(1)}_{Y_2}\,
\mathbf{R}_{21}(Y_2,Y_1) = \\
&\mathbf{\bar{R}}_{21}(Y_1,Y_2)\,
\mathbf{\bar{R}}_{21}(Y_2,Y_1)\,
\Lambda^{(2)}_{Y_{2}}\Lambda^{(1)}_{Y_1}\,
\mathbf{R}_{21}(Y_1,Y_2)
\mathbf{R}_{21}(Y_2,Y_1)  = \Lambda^{(2)}_{Y_2}\Lambda^{(1)}_{Y_{1}}
\end{align*}
due to relations for the R-matrices
\begin{align}
\label{R_unitarity}
\mathbf{\bar{R}}_{21}(Y_1,Y_2)\,
\mathbf{\bar{R}}_{21}(Y_2,Y_1) = \II \ ;\ \
\mathbf{R}_{21}(Y_1,Y_2)\,
\mathbf{R}_{21}(Y_2,Y_1) = \II
\end{align}
In the case of $N=3$ the eigenfunction is represented by the
product of three layers $\Lambda^{(3)}_{Y_3}\Lambda^{(2)}_{Y_2}\Lambda^{(1)}_{Y_{1}}$
and there are six permutations. The permutation $\Lambda^{(3)}_{Y_3}\Lambda^{(2)}_{Y_2}\Lambda^{(1)}_{Y_{1}} \to \Lambda^{(3)}_{Y_1}\Lambda^{(2)}_{Y_2}\Lambda^{(1)}_{Y_{3}}$ can
be performed in two ways
\begin{align*}
&\Lambda^{(3)}_{Y_3}\Lambda^{(2)}_{Y_2}\Lambda^{(1)}_{Y_{1}} =
\mathbf{\bar{R}}_{21}(Y_1,Y_2)\,
\Lambda^{(3)}_{Y_3}\Lambda^{(2)}_{Y_{1}}\Lambda^{(1)}_{Y_2}\,
\mathbf{R}_{21}(Y_2,Y_1) = \\
&\mathbf{\bar{R}}_{21}(Y_1,Y_2)\,
\mathbf{\bar{R}}_{31}(Y_1,Y_3)\,
\Lambda^{(3)}_{Y_1}\Lambda^{(2)}_{Y_{3}}\Lambda^{(1)}_{Y_2}\,
\mathbf{R}_{31}(Y_3,Y_1)
\mathbf{R}_{21}(Y_2,Y_1) =\\
&
\mathbf{\bar{R}}_{21}(Y_1,Y_2)\,
\mathbf{\bar{R}}_{31}(Y_1,Y_3)\,\mathbf{\bar{R}}_{32}(Y_2,Y_3)
\Lambda^{(3)}_{Y_1}\Lambda^{(2)}_{Y_{2}}\Lambda^{(1)}_{Y_3}\,
\mathbf{R}_{32}(Y_3,Y_2)\mathbf{R}_{31}(Y_3,Y_1)
\mathbf{R}_{21}(Y_2,Y_1)
\end{align*}
and
\begin{align*}
&\Lambda^{(3)}_{Y_3}\Lambda^{(2)}_{Y_2}\Lambda^{(1)}_{Y_{1}} =
\mathbf{\bar{R}}_{32}(Y_2,Y_3)\,
\Lambda^{(3)}_{Y_2}\Lambda^{(2)}_{Y_{3}}\Lambda^{(1)}_{Y_1}\,
\mathbf{R}_{32}(Y_3,Y_2) = \\
&
\mathbf{\bar{R}}_{32}(Y_2,Y_3)\,
\mathbf{\bar{R}}_{31}(Y_1,Y_3)\,
\Lambda^{(3)}_{Y_2}\Lambda^{(2)}_{Y_{1}}\Lambda^{(1)}_{Y_3}\,
\mathbf{R}_{31}(Y_3,Y_1)
\mathbf{R}_{32}(Y_3,Y_2) = \\
&
\mathbf{\bar{R}}_{32}(Y_2,Y_3)\,
\mathbf{\bar{R}}_{31}(Y_1,Y_3)\,\mathbf{\bar{R}}_{21}(Y_2,Y_1)
\Lambda^{(3)}_{Y_1}\Lambda^{(2)}_{Y_{2}}\Lambda^{(1)}_{Y_3}\,
\mathbf{R}_{21}(Y_2,Y_1)\mathbf{R}_{31}(Y_3,Y_1)
\mathbf{R}_{32}(Y_3,Y_2)
\end{align*}
These two ways lead to the same result due to the validity of the
Yang-Baxter relations for the R-matrices
\begin{align*}
\mathbf{\bar{R}}_{21}(Y_1,Y_2)\,
\mathbf{\bar{R}}_{31}(Y_1,Y_3)\,\mathbf{\bar{R}}_{32}(Y_2,Y_3) =
\mathbf{\bar{R}}_{32}(Y_2,Y_3)\,
\mathbf{\bar{R}}_{31}(Y_1,Y_3)\,\mathbf{\bar{R}}_{21}(Y_1,Y_2)\,,\\
\mathbf{R}_{32}(Y_3,Y_2)\,\mathbf{R}_{31}(Y_3,Y_1)\,
\mathbf{R}_{21}(Y_2,Y_1) = \mathbf{R}_{21}(Y_2,Y_1)\,
\mathbf{R}_{31}(Y_3,Y_1)\,
\mathbf{R}_{32}(Y_3,Y_2)\,.
\end{align*}
Now we define the natural representation of the symmetric group $\mathbb{S}_N$ on the eigenfunctions.
First of all we define in explicit form the action of the generators
\begin{align*}
\hat{s}_k\, \Lambda^{(N)}_{Y_{N}}\cdots\Lambda^{(k+1)}_{Y_{k+1}}\,\Lambda^{(k)}_{Y_{k}} \cdots\Lambda^{(1)}_{Y_{1}}= \mathbf{\bar{R}}_{k+1,k}(Y_{k},Y_{k+1})\,
\Lambda^{(N)}_{Y_{N}}\cdots\Lambda^{(k+1)}_{Y_{k}}\,\Lambda^{(k)}_{Y_{k+1}}
\cdots\Lambda^{(1)}_{Y_{1}}
\,\mathbf{R}_{k+1,k}(Y_{k+1},Y_{k})\,,
\end{align*}
or in explicit index notations
\begin{align}
&\hat{s}_k\, \left(\Lambda^{(N)}_{Y_{N}}\right)^{\mathbf a_{N}}_{\dot{\mathbf a}_{N}}\cdots
\left(\Lambda^{(k+1)}_{Y_{k+1}}\right)^{\mathbf a_{k+1}}_{\dot{\mathbf a}_{k+1}}
\left(\Lambda^{(k)}_{Y_{k}}\right)^{\mathbf a_{k}}_{\dot{\mathbf a}_{k}}
\cdots\left(\Lambda^{(1)}_{Y_{1}}\right)^{\mathbf a_{1}}_{\dot{\mathbf a}_{1}}
=\\
\nonumber
&\mathbf{R}^{\dot{\mathbf{c}}_{k+1}\dot{\mathbf{c}}_k}_{\dot{\mathbf{a}}_{k+1}
\dot{\mathbf{a}}_k} \left(Y_{k},Y_{k+1}\right)\,
\mathbf{R}^{\mathbf{a}_{k+1}\mathbf{a}_k}_{\mathbf{c}_{k+1}\mathbf{c}_k}
\left(Y_{k+1},Y_{k}\right)\,
\left(\Lambda^{(N)}_{Y_{N}}\right)^{\mathbf a_{N}}_{\dot{\mathbf a}_{N}}
\cdots\left(\Lambda^{(k+1)}_{Y_{k}}\right)^{\mathbf c_{k}}_{\dot{\mathbf c}_{k}}
\left(\Lambda^{(k)}_{Y_{k+1}}\right)^{\mathbf c_{k+1}}_{\dot{\mathbf c}_{k+1}}
\cdots\left(\Lambda^{(1)}_{Y_{1}}\right)^{\mathbf a_{1}}_{\dot{\mathbf a}_{1}}
\end{align}
In the most compact form we have defined the action of
generators of the symmetric group as follows
\begin{equation}
\hat{s}_k \Psi_{\mathbf{Y}}=\mathbf{\bar{S}}^{-1}_k(\mathbf{Y})\,
\Psi_{s_k(\mathbf{Y})}\,\mathbf{S}_k(\mathbf{Y})\,,
\end{equation}
where $\mathbf{\bar{S}}_{k}(\mathbf{Y}) = \mathbf{\bar{R}}_{k+1,k}(Y_{k+1},Y_{k})$ and similarly $\mathbf{S}_{k}(\mathbf{Y}) = \mathbf{R}_{k+1,k}(Y_{k+1},Y_{k})$
and
\begin{align*}
s_k\left(\mathbf{Y}\right) = s_k (Y_1\ldots Y_k,Y_{k+1}\ldots Y_N)  = (Y_1\ldots Y_{k+1},Y_k\ldots Y_N)
\end{align*}
More generally, for any element $s\in\mathbb{S}_N$ we have
\begin{equation}
\hat{s}\, \Psi_{\mathbf{Y}}=\mathbf{\bar{S}}^{-1}(\mathbf{Y})\,\Psi_{s(\mathbf{Y})}\,
\mathbf{S}(\mathbf{Y})\,.
\end{equation}
The composition of the transformations is defined in a natural way.
For $s=s_1s_2$ we have
\begin{align*}
&\hat{s}\, \Psi_{\mathbf{Y}} = \hat{s}_1\,\hat{s}_2\, \Psi_{\mathbf{Y}} = \hat{s}_1\,\left[\mathbf{\bar{S}}^{-1}_2(\mathbf{Y})\,\Psi_{s_2(\mathbf{Y})}\,
\mathbf{S}_2(\mathbf{Y})\right] = \mathbf{\bar{S}}^{-1}_2(\mathbf{Y})\,\left[\hat{s}_1\,\Psi_{s_2(\mathbf{Y})}\right]\,
\mathbf{S}_2(\mathbf{Y}) = \\
&
\mathbf{\bar{S}}^{-1}_2(\mathbf{Y})\,
\mathbf{\bar{S}}^{-1}_1(s_2\mathbf{Y})\,\Psi_{s_1(s_2(\mathbf{Y}))}\,
\mathbf{S}_1(s_2\mathbf{Y})\,\mathbf{S}_2(\mathbf{Y}) =  \mathbf{\bar{S}}^{-1}(\mathbf{Y})\,\Psi_{s(\mathbf{Y})}\,\mathbf{S}(\mathbf{Y})\,.
\end{align*}
so that
$$
s = s_1s_2 \to \mathbf{\bar{S}}(\mathbf{Y}) = \mathbf{\bar{S}}_1(s_2\mathbf{Y})\,\mathbf{\bar{S}}_2(\mathbf{Y}) \ \ ; \ \
\mathbf{S}(\mathbf{Y}) = \mathbf{S}_1(s_2\mathbf{Y})\,\mathbf{S}_2(\mathbf{Y})
$$
In a general case a permutation
\begin{equation}
s = s_{h}s_{k}\cdots s_{\ell}s_{r}\,,
\end{equation}
is represented by
\begin{equation}
\mathbf{S}(\mathbf{Y}) = \mathbf S_{h}(s_{k}\cdots s_{\ell}s_{r}\mathbf{Y})
\cdots \mathbf S_{\ell}(s_{r}\mathbf{Y})\,\mathbf S_{r}(\mathbf{Y})\,.
\end{equation}

As we have demonstrated on the simple examples $N=2$ and $N=3$ the
needed Coxeter relation are fulfilled due to the special
properties of R-matrix. In the general case the unitarity of the $\mathbf{R}$-matrix \eqref{R_unitarity}
ensures the involutivity  $\hat{s}_k^2=\II$
\begin{equation}
\mathbf S_{k}(s_k(\mathbf{Y})) \mathbf S_{k}(\mathbf{Y}) =
\mathbf{R}_{k+1,k}(Y_{k},Y_{k+1})
\mathbf{R}_{k+1,k}(Y_{k+1},Y_{k})=\II\,,
\end{equation}
and the Yang-Baxter equation \eqref{YB} ensures the satisfaction of the most complicated qubic Coxeter relations
$\hat{s}_k\hat{s}_{k+1}\hat{s}_k = \hat{s}_{k+1}\hat{s}_k\hat{s}_{k+1}$ which have the following explicit form
\begin{equation}
\mathbf S_{k}(s_{k+1}s_k\mathbf{Y}) \,\mathbf S_{k+1}(s_k\mathbf{Y}) \,\mathbf S_{k}(\mathbf{Y})= \mathbf S_{k+1}(s_k s_{k+1}\mathbf{Y}) \,\mathbf S_{k}(s_{k+1}\mathbf{Y}) \,\mathbf S_{k+1}(\mathbf{Y})\,.
\end{equation}
or in terms of R-matrices
\begin{align*}
\mathbf{R}_{k+2,k+1}(Y_{k+2},Y_{k+1})
\mathbf{R}_{k+2,k}(Y_{k+2},Y_{k})
\mathbf{R}_{k+1,k}(Y_{k+1},Y_{k}) = \\
\mathbf{R}_{k+1,k}(Y_{k+1},Y_{k})
\mathbf{R}_{k+2,k}(Y_{k+2},Y_{k})
\mathbf{R}_{k+2,k+1}(Y_{k+2},Y_{k+1})
\end{align*}
Of course all similar formulae are valid for $\mathbf{\bar{R}}$-matrices also.

In the most general form the formula \eqref{layer_comm} states that our eigenfunction is invariant with respect to the action of the symmetric group $\mathbb{S}_N$: for any element $s\in\mathbb{S}_N$ we have
\begin{equation}
\hat{s}\, \Psi_{\mathbf{Y}}=\mathbf{\bar{S}}^{-1}(\mathbf{Y})\,\Psi_{s(\mathbf{Y})}\,
\mathbf{S}(\mathbf{Y}) = \Psi_{\mathbf{Y}}\,
\end{equation}

\subsection{Scalar product}
\label{sec:scalprod}
In order to complete the study of eigenfunctions we need to compute their scalar product, therefore the spectral measure over the variables $\mathbf{Y}$. First of all we recall that the Hilbert space under consideration is $L^2(d^4 x_1,\dots,d^4 x_N)$ and therefore for each two functions $f,g$ the scalar product is defined by
\begin{equation}
\label{scal_prod}
\langle f,g\rangle\,=\,\int d^4 x_1 \cdots d^4 x_N (f(x_1,\dots,x_N))^*g(x_1,\dots,x_N)\,.
\end{equation}
In this context we must consider the functions $\langle \boldsymbol{\beta}|\Psi_{\mathbf{Y}}|\boldsymbol{\alpha}\rangle$, where the dressing \eqref{dressed} hides the spinor indices, and the auxiliary spinors $\alpha_k$ and $\beta_k$ can be regarded as additional continuous quantum numbers describing the degeneracy of the spectrum \eqref{eigenvfact}. According to \eqref{scal_prod} the conjugate eigenfunction is
\begin{equation}
\langle\boldsymbol{\beta}|\Psi_{\mathbf{Y}}|\boldsymbol{\alpha}\rangle^{*} =\langle\boldsymbol{\alpha}|{\Psi}_{\mathbf{Y}}^{\dagger}|\boldsymbol{\beta}\rangle\,,
\end{equation}
where ${\Psi}^{\dagger}_{\mathbf{Y}}$ is obtained from ${\Psi}_{\mathbf{Y}}$ by transposition of spinor indices and complex conjugation.

As is it for the eigenfunctions in \eqref{eigenf_Y}, also the scalar product of two eigenfunctions can be written in an operatorial form as
\begin{equation}
\label{scalar_1}
\langle \Psi^{(N)}_{\mathbf{Y'}}|\Psi^{(N)}_{\mathbf{Y}}\rangle=\bar{\Lambda}^{(1)}_{Y'_1} \bar {\Lambda}^{(2)}_{Y'_2} \cdots \bar {\Lambda}^{(N)}_{Y'_N} \, \Lambda^{(N)}_{Y_N}\cdots \Lambda^{(2)}_{Y_2} \Lambda^{(1)}_{Y_1}\,,
\end{equation}
where each layer $(\bar{\Lambda}_{Y})_{(a_1,\dots,a_{\ell})}^{(\dot a_1,\dots,\dot a_{\ell})} = (\bar{\Lambda}_{Y})_{\mathbf{a}}^{\dot{\mathbf{a}}}$ is defined as an integral operator
\begin{align}
\begin{aligned}
\left[\bar{\Lambda}_{Y}^{(N)}\phi\right](x_1,\dots,x_{N-1}) =
\int d^4y_1\cdots d^4y_{N}\, {\bar{\Lambda}}^{(N)}_{Y}(x_1,\dots,x_{N-1},x_0|y_1,\dots,y_{N}) \cdot \phi(y_1,\dots,y_{N}) \,,
\end{aligned}
\end{align}
and its kernel being defined as
\begin{align}
\label{Lambda_dag}
\begin{aligned}
&\bar{\Lambda}^{(N)}_{Y}(x_1,\dots,x_{N-1},x_0|y_1,\dots,y_N) = \prod_{k=1}^{N} \frac{1}{(y_k-y_{k+1})^{2(-i\lambda)}}\times\\&\times\frac{[({\mathbf{y_N- x_0}})(\overline{\mathbf {y_N-x_{N-1}}})\cdots(\mathbf{\overline{x_1- y_2}})(\mathbf{ {y_1- x_1}}) ]^{\ell}}{(y_1-x_1)^{2(1+i\frac{\lambda}{2}-i\nu)}(x_1-y_2)^{2(1+i\frac{\lambda}{2}+i\nu)}\cdots (x_{N-1}-y_N)^{2(1+i\frac{\lambda}{2}+i\nu)}(y_N-x_0)^{2(1-i\frac{\lambda}{2}-i\nu)}}\,.
\end{aligned}
\end{align}
Integral operator and its kernel carry symmetric spinor indices but these
indices play passive role and operator $(\bar{\Lambda}_{Y})_{\mathbf{a}}^{\dot{\mathbf{a}}}$ maps
function of $N$ variables $x_1,\cdots,x_{N}$ to the function of $N-1$ variables $x_1,\cdots,x_{N-1}$ which carry additional
symmetric spinor indices $\mathbf{\dot{a}}$ and $\mathbf a$.
There is analogy with annihilation operator in quantum field theory:
operator $(\bar{\Lambda}_{Y})_{\mathbf{a}}^{\dot{\mathbf{a}}}$ annihilates dependence on variable $x_N$.

The scalar product of two eigenfunctions in the simplest case $N=1$ was calculated in the Section \ref{ladder}
\begin{align}
\label{N1scalar}
\langle \Psi^{(1)}_{\mathbf{Y'}}|\Psi^{(1)}_{\mathbf{Y}}\rangle =
\left(\bar{\Lambda}^{(1)}_{Y'} \right)_{{\mathbf a}}^{\dot{\mathbf a}}\left(\Lambda^{(1)}_{Y}\right)^{\mathbf b}_{\dot{\mathbf b}}
= \frac{2 \pi^3}{\ell+1}\,\delta(Y,Y')\,
\delta^{\mathbf b}_{{\mathbf a}}\,\delta_{\dot{\mathbf b}}^{\dot{\mathbf a}}
\end{align}
where $\delta(Y,Y') = \delta_{\ell,\ell'} \delta(\nu-\nu')$.

The computation of a complicated integral \eqref{exchorto1} can be carried out exactly thanks to the following exchange relation, valid for any $k=2,\dots,N$:
\begin{align}
\label{exchorto1}
\left(\bar{\Lambda}^{(k)}_{Y'} \right)_{\mathbf a}^{\dot{\mathbf a}}\,
\left(\Lambda^{(k)}_{Y}\right)^{\mathbf b}_{\dot{\mathbf b}} =
\lambda(Y,Y')\,
\mathbf{R}^{\dot{\mathbf{a}}\dot{\mathbf{s}}}_{\dot{\mathbf{c}}
\dot{\mathbf{b}}} (Y',Y)\,\mathbf{R}^{\mathbf{c}\mathbf{b}}_{\mathbf{a}\mathbf{s}} (Y,Y')\,
\left(\Lambda^{(k-1)}_{Y}\right)^{\mathbf s}_{\dot{\mathbf s}}\,
\left(\bar{\Lambda}^{(k-1)}_{Y'} \right)_{{\mathbf c}}^{\dot{\mathbf c}}\,,
\end{align}
where the coefficient $\lambda(Y,Y')$ has the following explicit form
\begin{align}\label{munu}
\lambda(Y,Y') =\frac{\pi^4 }{\left((\nu-\nu')^2+
\frac{(\ell-\ell')^2}{4}\right)
\left((\nu-\nu')^2+\frac{(\ell+\ell'+2)^2}{4}\right)}\,,
\end{align}
Note the evident symmetry $\lambda(Y,Y') = \lambda(Y',Y)$.

The proof of this identity contains simple graphical steps based on the star-triangle relation \eqref{STR_0} and on the exchange-relation \eqref{exch_opp_1}.
Let us start from the picture of the kernel $\bar{\Lambda}^{(N)}_{Y'}$ (here for $N=3$):
\begin{center}
\includegraphics[scale=0.5]{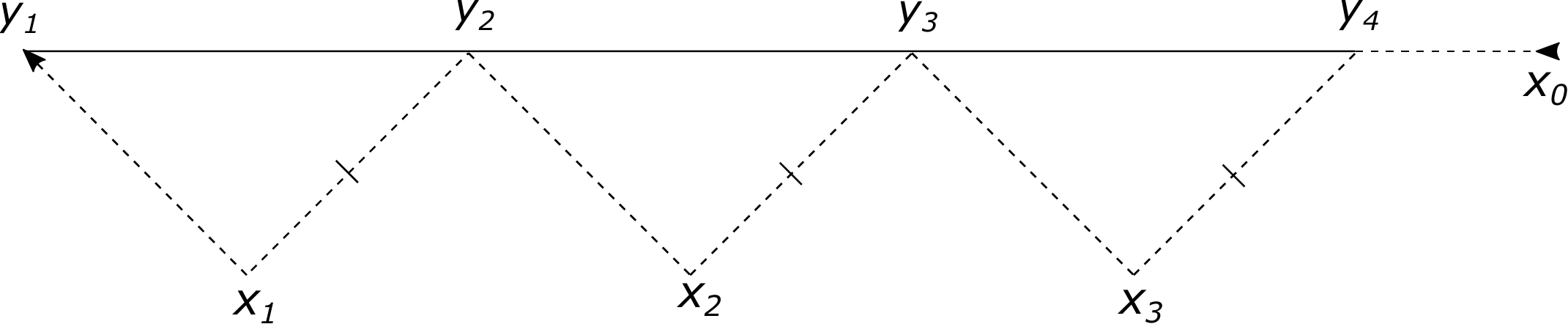}
\end{center}
 In order to prove the identity \eqref{exchorto1} we begin with the picture of its l.h.s. (on the left). We apply the star-triangle identity in order to integrate the rightmost point, producing the first $\mathbf{R}_{\ell,\ell'}$-matrix and a pair of vertical dashed lines $[(\mathbf{\overline{x_3-z_3}})]^{\ell'}[(\mathbf{x_3-z_3})]^{\ell}$. Moreover, we insert on the right picture a scalar line of weight $0=2i\lambda-2i\lambda$ (right picture).
\begin{center}
\includegraphics[scale=0.345]{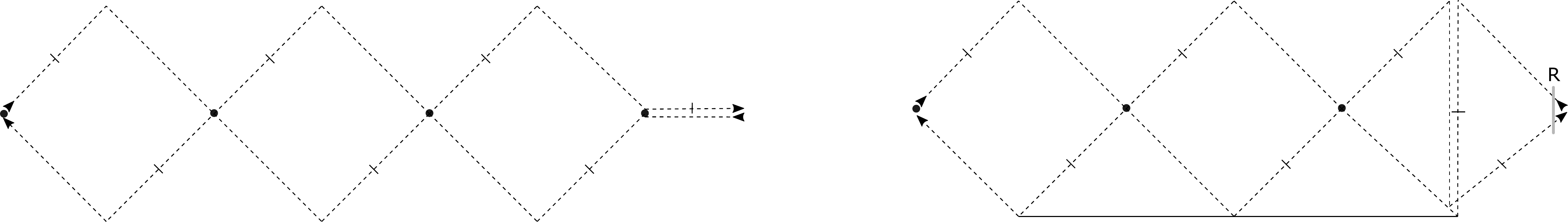}
\end{center}
At this point we apply the exchange relation \eqref{exch_opp_1} as depicted, in order to move leftwards the two dashed vertical lines,  moving at the same time the scalar horizontal line of weight  $-2i\lambda$ upwards (left picture). Finally, we compute the chain-rule integration at the leftmost point, which produces
the matrix $\mathbf{R}^{-1}_{\ell',\ell}$ and cancels the two vertical lines.
\begin{center}
\includegraphics[scale=0.35]{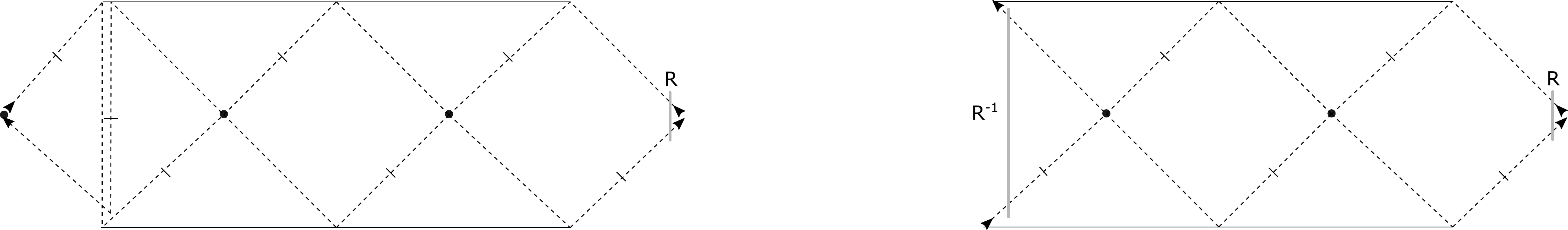}
\end{center}
As the exchange relation \eqref{exch_opp_1} has no extra coefficient, the overall coefficient produced along the proof is given only by the star-triangle on the rightmost point and the chain-rule on the leftmost point, matching with the coefficient in \eqref{munu}.
We should to note that during the derivation we tacitly assume that $Y \neq Y'$.

The exchange relation \eqref{exchorto1} allows to reduce the scalar product of two eigenfunction for general $N$ to the scalar products
of simplest $N=1$ eigenfunctions.
The reduction procedure consist the exchange of the layer $\Lambda^{(N-k)}_{Y_{N-k}}$, for $k=0,1,\dots,N-2$ with the product of layers
$\bar{\Lambda}^{(1)}_{Y'_{k+1}}\cdots\bar{\Lambda}^{(N-k+1)}_{Y'_N}$ by means of \eqref{exchorto1}
\begin{align}\nonumber
\bar{\Lambda}^{(1)}_{Y'_1} \cdot \bar {\Lambda}^{(2)}_{Y'_2} \cdots \bar {\Lambda}^{(N)}_{Y'_N}\Lambda^{(N)}_{Y_N}\Lambda^{(N-1)}_{Y_{N-1}}\cdots \Lambda^{(2)}_{Y_2}\Lambda^{(1)}_{Y_1} \longrightarrow \\
\nonumber
\underline{\bar{\Lambda}^{(1)}_{Y'_1} \Lambda^{(1)}_{Y_N}}\,
\bar {\Lambda}^{(1)}_{Y'_2} \cdots \bar {\Lambda}^{(N-1)}_{Y'_N} \Lambda^{(N-1)}_{Y_{N-1}}\cdots \Lambda^{(2)}_{Y_2}\Lambda^{(1)}_{Y_1} \longrightarrow \\
\nonumber
\underline{\bar{\Lambda}^{(1)}_{Y'_1} \Lambda^{(1)}_{Y_N}}\,
\underline{\bar {\Lambda}^{(1)}_{Y'_2} \Lambda^{(1)}_{Y_{N-1}}}\,
\bar {\Lambda}^{(1)}_{Y'_3} \cdots \bar {\Lambda}^{(N-2)}_{Y'_N} \Lambda^{(N-2)}_{Y_{N-2}}\cdots \Lambda^{(2)}_{Y_2}\Lambda^{(1)}_{Y_1} \longrightarrow \\
\label{strategy}
\underline{\bar{\Lambda}^{(1)}_{Y'_1} \Lambda^{(1)}_{Y_N}}\,
\underline{\bar {\Lambda}^{(1)}_{Y'_2} \Lambda^{(1)}_{Y_{N-1}}}\,
\underline{\bar{\Lambda}^{(1)}_{Y'_3} \Lambda^{(1)}_{Y_{N-2}}}\cdots
\underline{\bar {\Lambda}^{(1)}_{Y'_2} \Lambda^{(1)}_{Y_{N-1}}}\,\,
\underline{\bar{\Lambda}^{(1)}_{Y'_{N-1}} \Lambda^{(1)}_{Y_2}}\,\,
\underline{\bar {\Lambda}^{(1)}_{Y'_{N}} \Lambda^{(1)}_{Y_{1}}}\,
\end{align}

All this procedure is very similar to the Wick theorem in free field theory which allows to reduce calculation of $N$-point Green function to the
product of $N=2$ Green functions.

Let us consider the calculation of the scalar product in the
simplest nontrivial example $N=2$. We have
\begin{equation}
\label{scalar_2}
\langle \Psi^{(2)}_{Y'_1,Y'_2}|\Psi^{(2)}_{Y_1,Y_2}\rangle=
\bar{\Lambda}^{(1)}_{Y'_1}\bar {\Lambda}^{(2)}_{Y'_2}\,
\Lambda^{(2)}_{Y_2}\Lambda^{(1)}_{Y_1}
\end{equation}
Using relation \eqref{exchorto1} it is possible to reduce calculation to the case $N=1$. Schematically we have
$$
\bar{\Lambda}^{(1)}_{Y'_1}\bar {\Lambda}^{(2)}_{Y'_2}\,
\Lambda^{(2)}_{Y_2}\Lambda^{(1)}_{Y_1} \rightarrow
\underline{\bar{\Lambda}^{(1)}_{Y'_1}\Lambda^{(1)}_{Y_2}}\,
\underline{\bar {\Lambda}^{(1)}_{Y'_2}\,\Lambda^{(1)}_{Y_1}}
$$
or explicitly in index form
\begin{align*}
&\left(\bar{\Lambda}^{(1)}_{Y'_1} \right)_{\mathbf a_{1}}^{\dot{\mathbf a}_{1}}\left(\bar{\Lambda}^{(2)}_{Y'_2} \right)_{\mathbf a_{2}}^{\dot{\mathbf a}_{2}}\,
\left(\Lambda^{(2)}_{Y_2}\right)^{\mathbf b_2}_{\dot{\mathbf b}_2}\left(\Lambda^{(1)}_{Y_1}\right)^{\mathbf b_1}_{\dot{\mathbf b}_1} =\\
&\lambda(Y_2,Y'_2)\,
\mathbf{R}^{\dot{\mathbf{a}}_2\dot{\mathbf{s}}}_{\dot{\mathbf{c}_2}
\dot{\mathbf{b}}_2} (Y'_2,Y_2)\,
\mathbf{R}^{\mathbf{c}_2\mathbf{b}_2}_{\mathbf{a}_2\mathbf{s}} (Y_2,Y'_2)\,
\underline{\left(\bar{\Lambda}^{(1)}_{Y'_1} \right)_{\mathbf a_{1}}^{\dot{\mathbf a}_{1}}\left(\Lambda^{(1)}_{Y_2}\right)^{\mathbf s}_{\dot{\mathbf s}}}
\,\underline{\left(\bar{\Lambda}^{(1)}_{Y'_2} \right)_{{\mathbf c}_2}^{\dot{\mathbf c}_2}\left(\Lambda^{(1)}_{Y_1}\right)^{\mathbf b_1}_{\dot{\mathbf b}_1}} = \\
&
\frac{1}{\mu(Y_1,Y_2)}\,
\delta(Y_1,Y'_2)\,\delta(Y_2,Y'_1)\,
\mathbf{R}^{\dot{\mathbf{a}}_2\dot{\mathbf{a}}_1}_{\dot{\mathbf{b}_1}
\dot{\mathbf{b}}_2} (Y_1,Y_2)\,\mathbf{R}^{\mathbf{b}_1\mathbf{b}_2}_{\mathbf{a}_2\mathbf{a}_1} (Y_2,Y_1)
\,,
\end{align*}
where on the last step we use formula \eqref{N1scalar}
for the scalar product in the case $N=1$ and denote
\begin{align*}
\frac{1}{\mu(Y_1,Y_2)} = \frac{2 \pi^3}
{\ell_1+1}\,\frac{2 \pi^3}
{\ell_2+1}\,\lambda(Y_1,Y_2)\,.
\end{align*}
This calculation of the scalar product is based on the
exchange property \eqref{exchorto1} which is derived
under condition $Y\neq Y'$, i.e. $Y_2\neq Y'_2$ in the present situation.
Note that this expression for the scalar product in $N=2$ case cannot
be the complete formula because it is valid for $Y_2 \neq Y'_2$ only and generally is not compatible with the symmetry properties of the eigenfunction.
Indeed due to this symmetry we have
\begin{align*}
&\left(\bar{\Lambda}^{(1)}_{Y'_1} \right)_{\mathbf a_{1}}^{\dot{\mathbf a}_{1}}\left(\bar{\Lambda}^{(2)}_{Y'_2} \right)_{\mathbf a_{2}}^{\dot{\mathbf a}_{2}}\,
\left(\Lambda^{(2)}_{Y_2}\right)^{\mathbf b_2}_{\dot{\mathbf b}_2}\left(\Lambda^{(1)}_{Y_1}\right)^{\mathbf b_1}_{\dot{\mathbf b}_1} = \\
&\mathbf{R}^{\mathbf{\dot c}_2\mathbf{\dot c}_1}_{\mathbf{\dot{b}}_2 \mathbf{\dot{b}}_1}(Y_1,Y_2)\,
\mathbf{R}^{\mathbf{b}_2 \mathbf{b}_1}_{\mathbf{c}_2\mathbf{c}_1}(Y_2,Y_1)
\left(\bar{\Lambda}^{(1)}_{Y'_1} \right)_{\mathbf a_{1}}^{\dot{\mathbf a}_{1}}\left(\bar{\Lambda}^{(2)}_{Y'_2} \right)_{\mathbf a_{2}}^{\dot{\mathbf a}_{2}}\,
\left(\Lambda^{(2)}_{Y_1}\right)_{\mathbf{\dot{c}}_1}^{\mathbf{c}_1}
\left(\Lambda^{(1)}_{Y_2}\right)_{\mathbf{\dot{c}}_2}^{\mathbf{c}_2} =\\
&\frac{1}{\mu(Y_1,Y_2)}\,\delta(Y_1,Y'_1)\,\delta(Y_2,Y'_{2})\
\mathbf{R}^{\mathbf{\dot c}_2\mathbf{\dot c}_1}_{\mathbf{\dot{b}}_2 \mathbf{\dot{b}}_1}(Y_1,Y_2)\,
\mathbf{R}^{\mathbf{b}_2 \mathbf{b}_1}_{\mathbf{c}_2\mathbf{c}_1}(Y_2,Y_1)
\mathbf{R}^{\dot{\mathbf{a}}_2\dot{\mathbf{a}}_1}_{\dot{\mathbf{c}_2}
\dot{\mathbf{c}}_1} (Y_2,Y_1)\,\mathbf{R}^{\mathbf{c}_2\mathbf{c}_1}_{\mathbf{a}_2\mathbf{a}_1} (Y_1,Y_2) = \\
&
\frac{1}{\mu(Y_1,Y_2)}\,
\delta(Y_1,Y'_1)\,\delta(Y_2,Y'_{2})\,
\delta^{\dot{\mathbf{a}}_2}_{\dot{\mathbf{b}}_2}\,
\delta^{\dot{\mathbf{a}}_1}_{\dot{\mathbf{b}}_1}\,
\delta^{\mathbf{b}_2}_{\mathbf{a}_2}\,
\delta^{\mathbf{b}_1}_{\mathbf{a}_1}\,,
\end{align*}
where on the last step we used unitarity of R-matrix
$\mathbf{R}(Y_1,Y_2)\,\mathbf{R}(Y_2,Y_1) = \II $.
This formula is valid for $Y_1 \neq Y'_2$.

The complete symmetric expression should contain
both terms and can be restored by the symmetry in a unique way
\begin{align}
\label{scal_prod2}
&\left(\bar{\Lambda}^{(1)}_{Y'_1} \right)_{\mathbf a_{1}}^{\dot{\mathbf a}_{1}}\left(\bar{\Lambda}^{(2)}_{Y'_2} \right)_{\mathbf a_{2}}^{\dot{\mathbf a}_{2}}\,
\left(\Lambda^{(2)}_{Y_2}\right)^{\mathbf b_2}_{\dot{\mathbf b}_2}\left(\Lambda^{(1)}_{Y_1}\right)^{\mathbf b_1}_{\dot{\mathbf b}_1} = \\
\nonumber
&
\frac{1}{\mu(Y_1,Y_2)}\,
\left[\delta(Y_1,Y'_1)\,\delta(Y_2,Y'_{2})\,
\delta^{\dot{\mathbf{a}}_2}_{\dot{\mathbf{b}}_2}\,
\delta^{\dot{\mathbf{a}}_1}_{\dot{\mathbf{b}}_1}\,
\delta^{\mathbf{b}_2}_{\mathbf{a}_2}\,
\delta^{\mathbf{b}_1}_{\mathbf{a}_1}+\delta(Y_1,Y'_2)\,\delta(Y_2,Y'_1)\,
\mathbf{R}^{\dot{\mathbf{a}}_2\dot{\mathbf{a}}_1}_{\dot{\mathbf{b}_1}
\dot{\mathbf{b}}_2} (Y_1,Y_2)\,\mathbf{R}^{\mathbf{b}_1\mathbf{b}_2}_{\mathbf{a}_2\mathbf{a}_1} (Y_2,Y_1)\right]\,.
\end{align}
In order to achieve the formula for the scalar product at length $N$, one can proceed by induction and compute the $N=3$ case explicitly, as it is done in Appendix \ref{app:N3scalar}.
The generalization is natural and formula for the scalar product for general $N$ have the following form
\begin{align}
\label{ScalProd}
\begin{aligned}
&
\langle \Psi^{(N)}_{\mathbf{Y'}}|\Psi^{(N)}_{\mathbf{Y}}\rangle = \frac{1}{\mu(\mathbf Y)}\sum_{s \in \mathbb{S}_N}\,\delta(\mathbf{Y}'-s(\mathbf{Y}))\, 
\mathbf{\bar{S}}^{-1}(\mathbf Y)\, \mathbf{S}(\mathbf Y)\,,
\end{aligned}
\end{align}
where $\mu(\mathbf Y)$ is symmetric $\mu(\mathbf Y) = \mu(s\mathbf Y)$ and
\begin{equation}
\delta(\mathbf{Y}'-s(\mathbf{Y}))  =
\prod_{k=1}^N \delta_{\ell'_k,\ell_{s(k)}} \delta(\nu'_k-\nu_{s(k)})\,.
\end{equation}
Let us check the symmetry for any $s_1 \in \mathbb{S}_N$
\begin{align*}
&
\langle \Psi^{(N)}_{\mathbf{Y'}}|\hat{s}_1\,\Psi^{(N)}_{\mathbf{Y}}\rangle =
\langle \Psi^{(N)}_{\mathbf{Y'}}|\mathbf{\bar{S}}^{-1}_1(\mathbf{Y})\,
\Psi^{(N)}_{s_1(\mathbf{Y})}\,
\mathbf{S}_1(\mathbf{Y})\rangle =
\mathbf{\bar{S}}^{-1}_1(\mathbf{Y})\,\langle \Psi^{(N)}_{\mathbf{Y'}}|\Psi^{(N)}_{s_1(\mathbf{Y})}\rangle\,
\mathbf{S}_1(\mathbf{Y}) = \\
& \frac{1}{\mu(\mathbf Y)}\sum_{s \in \mathbb{S}_N}\,
\delta(\mathbf{Y}'-s(s_1\mathbf{Y}))\,
\mathbf{\bar{S}}^{-1}_1(\mathbf{Y})\, \mathbf{\bar{S}}^{-1}(s_1(\mathbf Y))\,
\mathbf{S}(s_1(\mathbf Y))
\mathbf{S}_1(\mathbf{Y}) = \\ 
&\frac{1}{\mu(\mathbf Y)}\sum_{s_2 \in \mathbb{S}_N}\,\delta(\mathbf{Y}'-s_2(\mathbf{Y}))\,
\mathbf{\bar{S}}_2^{-1}(\mathbf Y)\, \mathbf{S}_2(\mathbf Y) = 
\langle \Psi^{(N)}_{\mathbf{Y'}}|\Psi^{(N)}_{\mathbf{Y}}\rangle
\end{align*}
Effectively one obtains summation over elements $s_2 = s s_1 \in \mathbb{S}_N$
because 
$$
s_2 = s s_1 \to \mathbf{\bar{S}}_2(\mathbf{Y}) =\mathbf{\bar{S}}(s_1\mathbf{Y})\,\mathbf{\bar{S}}_1(\mathbf{Y})\ \,,\ 
\mathbf{S}_2(\mathbf{Y}) = \mathbf{S}(s_1\mathbf{Y})\,\mathbf{S}_1(\mathbf{Y})
$$
To prove the general formula \eqref{ScalProd} it is enough to calculate one contribution and then the whole answer is restored uniquely by the symmetry.
We shall use the strategy outlined in \eqref{strategy}.

First of all we have the following delta-function
\begin{equation}
\delta(\mathbf{Y}-s(\mathbf{Y}'))  = \delta(Y_1-Y'_N)\,\delta(Y_2-Y'_{N-1})\cdots\delta(Y_N-Y'_1) =
\prod_{k=1}^N \delta_{\ell_k,\ell'_{s(k)}} \delta(\nu_k-\nu'_{s(k)})\,,
\end{equation}
where $\delta(Y-Y') =
\delta_{\ell,\ell'} \delta(\nu-\nu')$ and $s$ is the special permutation which reverse
the order of quantum numbers
\begin{align*}
s\left(\mathbf{Y}\right) = s(Y_1\ldots Y_k Y_{k+1}\ldots Y_N)  = (Y_N\ldots Y_{k+1} Y_{k}\ldots Y_1)
\end{align*}
Secondly, there appears the overall normalization factor
\begin{align*}
\mu^{-1}(\boldsymbol{Y}) = \left(2\pi^3\right)^{N}\frac{1}{(\ell_1+1)\cdots(\ell_N+1)}
\prod_{i<k}\lambda(Y_i,Y_k)
\end{align*}
which shows that the measure over the quantum
numbers $(\nu,\ell)$ is not trivial.
In explicit form we have
\begin{equation}
\mu(\boldsymbol{Y})\,=\,\prod_{k=1}^N \frac{\ell_k+1}{2\pi^{2N+1}} \prod_{h>k}{\left((\nu_k-\nu_h)^2+\frac{(\ell_k-\ell_h)^2}{4}\right)
\left((\nu_k-\nu_h)^2+\frac{(\ell_k+\ell_h+2)^2}{4}\right)}\,.
\end{equation}
The last ingredient is the nontrivial
product of R-matrices. This operator can be
constructed iteratively and we expect from the formula \eqref{ScalProd} 
that it has natural interpretation as $\mathbf{\bar{S}}^{-1}(\mathbf Y)\, \mathbf{S}(\mathbf Y)$ where $s$ is the special permutation which reverse
the order of quantum numbers.

Finally, we have the following induction
\begin{align*}
&\left(\bar{\Lambda}^{(1)}_{Y'_1}\right)_{\mathbf a_{1}}^{\dot{\mathbf a}_{1}}
\cdots \left(\bar{\Lambda}^{(k)}_{Y'_k} \right)_{\mathbf a_{k}}^{\dot{\mathbf a}_{k}}\,
\left(\Lambda^{(k)}_{Y_k}\right)^{\mathbf b_k}_{\dot{\mathbf b}_k}
\cdots
\left(\Lambda^{(1)}_{Y_1}\right)^{\mathbf b_1}_{\dot{\mathbf b}_1} = \frac{1}{\mu(\boldsymbol{Y}_k)}\,\delta(\mathbf{Y}_k-s(\mathbf{Y}'_k))\,
\mathbf{R}^{\dot{\mathbf{a}}_k\cdots\, \dot{\mathbf{a}}_1}_{\dot{\mathbf{b}_1}
\cdots\, \dot{\mathbf{b}}_k} (\mathbf{Y}'_k)\,
\mathbf{R}^{\mathbf{b}_1\cdots\, \mathbf{b}_k}_{\mathbf{a}_k\cdots\, \mathbf{a}_1}
(\mathbf{Y}'_k)\,,
\end{align*}
where $\boldsymbol{Y}_k = (Y_1\ldots Y_k)$ and
$\boldsymbol{Y}'_k = (Y'_1\ldots Y'_k)$.

It is easy to check that iterative application of the relation \eqref{exchorto1} gives
\begin{align*}
\left(\bar{\Lambda}^{(1)}_{Y'_1}\right)_{\mathbf a_{1}}^{\dot{\mathbf a}_{1}}
\left(\bar{\Lambda}^{(2)}_{Y'_2}\right)_{\mathbf a_{2}}^{\dot{\mathbf a}_{2}}
\cdots
\left(\bar{\Lambda}^{(k)}_{Y'_k} \right)_{\mathbf a_{k}}^{\dot{\mathbf a}_{k}}\,
\left(\Lambda^{(k)}_{Y_k}\right)^{\mathbf b}_{\dot{\mathbf b}} =
\mu(Y_k,Y'_k)\cdots\mu(Y_k,Y'_2)\,\\
\mathbf{R}^{\dot{\mathbf{a}}_k\dot{\mathbf{s}}_{k-1}}_{\dot{\mathbf{c}}_k
\dot{\mathbf{b}}} (Y'_k,Y_k)\,
\mathbf{R}^{\dot{\mathbf{a}}_{k-1}\dot{\mathbf{s}}_{k-2}}_{\dot{\mathbf{c}}_{k-1}
\dot{\mathbf{s}}_{k-1}} (Y'_{k-1},Y_k)
\cdots
\mathbf{R}^{\dot{\mathbf{a}}_3\dot{\mathbf{s}}_2}_{\dot{\mathbf{c}_3}
\dot{\mathbf{s}}_3} (Y'_3,Y_k)\,
\mathbf{R}^{\dot{\mathbf{a}}_2\dot{\mathbf{s}}_1}_{\dot{\mathbf{c}_2}
\dot{\mathbf{s}}_2} (Y'_2,Y_k)\,
\underline{\left(\bar{\Lambda}^{(1)}_{Y'_1}\right)_{\mathbf a_{1}}^{\dot{\mathbf a}_{1}}
\left(\Lambda^{(1)}_{Y_k}\right)^{{\mathbf s}_1}_{\dot{\mathbf s}_1}}\,\\
\mathbf{R}^{\mathbf{c}_2\mathbf{s_2}}_{\mathbf{a}_2\mathbf{s}_1} (Y_k,Y'_2)\,
\mathbf{R}^{\mathbf{c}_3\mathbf{s}_3}_{\mathbf{a}_3\mathbf{s}_2} (Y_k,Y'_3)\,
\cdots
\mathbf{R}^{\mathbf{c}_{k-1}\mathbf{s}_{k-1}}_{\mathbf{a}_{k-1}\mathbf{s}_{k-2}} (Y_k,Y'_{k-1})
\mathbf{R}^{\mathbf{c}_k\mathbf{b}}_{\mathbf{a}_k\mathbf{s}_{k-1}} (Y_k,Y'_k)\,
\left(\bar{\Lambda}^{(1)}_{Y'_2}\right)_{\mathbf c_{2}}^{\dot{\mathbf c}_{2}}
\left(\bar{\Lambda}^{(2)}_{Y'_3}\right)_{\mathbf c_{3}}^{\dot{\mathbf c}_{3}}
\cdots
\left(\bar{\Lambda}^{(k-1)}_{Y'_k} \right)_{\mathbf c_{k}}^{\dot{\mathbf c}_{k}}
\end{align*}
so that using $N=1$ formula for the scalar product and supposing the previous $(k-1)$ step of induction we obtain the following recurrence relations
\begin{align*}
\delta(\mathbf{Y}_k-s(\mathbf{Y}'_k)) = \delta(Y'_2-Y_{k-1})\,\delta(Y'_3-Y_{k-2})\cdots\delta(Y'_{k}-Y_1)\,
\delta(Y'_1-Y_k)\,,\\
\frac{1}{\mu(\boldsymbol{Y}_k)} =
\frac{1}{\mu(Y_1,\cdots,Y_k)} = \frac{1}{\mu(Y_1,\cdots,Y_{k-1})}
\frac{2 \pi^3}{\ell_k+1}\mu(Y_k,Y_1)\cdots\mu(Y_k,Y_{k-1})
\end{align*}
which are evidently reproduce the needed expressions for
the $k$ step of induction and the last recurrence relation
which defines the iterative construction of the appearing
R-operators
\begin{align*}
&\mathbf{R}^{\dot{\mathbf{a}}_k\cdots\, \dot{\mathbf{a}}_1}_{\dot{\mathbf{b}_1}
\cdots\, \dot{\mathbf{b}}_k}(Y'_1,\cdots,Y'_k) = \\
&\mathbf{R}^{\dot{\mathbf{a}}_k\dot{\mathbf{s}}_{k-1}}_{\dot{\mathbf{c}}_k
\dot{\mathbf{b}}_k} (Y'_{k},Y'_{1})\,
\mathbf{R}^{\dot{\mathbf{a}}_{k-1}\dot{\mathbf{s}}_{k-2}}_{\dot{\mathbf{c}}_{k-1}
\dot{\mathbf{s}}_{k-1}} (Y'_{k-1},Y'_{1})
\cdots
\mathbf{R}^{\dot{\mathbf{a}}_3\dot{\mathbf{s}}_2}_{\dot{\mathbf{c}_3}
\dot{\mathbf{s}}_3} (Y'_{3},Y'_{1})\,
\mathbf{R}^{\dot{\mathbf{a}}_2\dot{\mathbf{a}}_1}_{\dot{\mathbf{c}_2}
\dot{\mathbf{s}}_2} (Y'_{2},Y'_{1})\,\mathbf{R}^{\dot{\mathbf{c}}_k\cdots\, \dot{\mathbf{c}}_2}_{\dot{\mathbf{b}_1}
\cdots\, \dot{\mathbf{b}}_{k-1}} (Y'_2,\cdots,Y'_k)\\
&\mathbf{R}^{\mathbf{b}_1\cdots\, \mathbf{b}_k}_{\mathbf{a}_k\cdots\, \mathbf{a}_1}
(Y'_1,\cdots,Y'_k) = \\ &\mathbf{R}^{\mathbf{c}_2\mathbf{s_2}}_{\mathbf{a}_2\mathbf{a}_1} (Y'_{1},Y'_{2})\,
\mathbf{R}^{\mathbf{c}_3\mathbf{s}_3}_{\mathbf{a}_3\mathbf{s}_2} (Y'_{1},Y'_{3})\,
\cdots
\mathbf{R}^{\mathbf{c}_{k-1}\mathbf{s}_{k-1}}_{\mathbf{a}_{k-1}\mathbf{s}_{k-2}} (Y'_{1},Y'_{k-1})
\mathbf{R}^{\mathbf{c}_k\mathbf{b}_k}_{\mathbf{a}_k\mathbf{s}_{k-1}} (Y'_{1},Y'_{k})\,
\mathbf{R}^{\mathbf{b}_1\cdots\, \mathbf{b}_{k-1}}_{\mathbf{c}_k\cdots\, \mathbf{c}_2}
(Y'_2,\cdots,Y'_k)
\end{align*}
Of course the operators $\mathbf{R}^{\dot{\mathbf{a}}_k\cdots\, \dot{\mathbf{a}}_1}_{\dot{\mathbf{b}_1}
\cdots\, \dot{\mathbf{b}}_k}(Y'_1,\cdots,Y'_k)$ and $\mathbf{R}^{\mathbf{b}_1\cdots\, \mathbf{b}_k}_{\mathbf{a}_k\cdots\, \mathbf{a}_1}
(Y'_1,\cdots,Y'_k)$ should be connected to the special permutation
$s(Y_1\ldots Y_k)  = (Y_k\ldots Y_1)$ which reverse the order of quantum numbers
\begin{align*}
\Lambda^{(k)}_{Y_k}\cdots\Lambda^{(2)}_{Y_2}\Lambda^{(1)}_{Y_{1}} =  \mathbf{\bar{R}}_{k\cdots21}(Y_1,\cdots,Y_k)\,
\Lambda^{(k)}_{Y_1}\Lambda^{(k-1)}_{Y_2}\cdots\Lambda^{(1)}_{Y_{k}}\,
\mathbf{R}_{k\cdots21}(Y_1,\cdots,Y_k)
\end{align*}
In order to obtain this connection we consider the special permutation in more details. In the simplest case $N=2$ we have
\begin{align*}
\Lambda^{(2)}_{Y_2}\Lambda^{(1)}_{Y_{1}} =
\mathbf{\bar{R}}_{21}(Y_1,Y_2)\,
\Lambda^{(2)}_{Y_{1}}\Lambda^{(1)}_{Y_2}\,
\mathbf{R}_{21}(Y_2,Y_1)
\end{align*}
The case $N=3$ was considered in the previous section
\begin{align*}
&\Lambda^{(3)}_{Y_3}\Lambda^{(2)}_{Y_{2}}\Lambda^{(1)}_{Y_{1}}
= \mathbf{\bar{R}}_{21}(Y_1,Y_2)\,
\mathbf{\bar{R}}_{31}(Y_1,Y_3)\,\mathbf{\bar{R}}_{32}(Y_2,Y_3)
\Lambda^{(3)}_{Y_1}\Lambda^{(2)}_{Y_{2}}\Lambda^{(1)}_{Y_3}\,
\mathbf{R}_{32}(Y_3,Y_2)\mathbf{R}_{31}(Y_3,Y_1)
\mathbf{R}_{21}(Y_2,Y_1)
\end{align*}
In the full analogy with the $N=3$ case
the permutation $\Lambda^{(k)}_{Y_k}\cdots\Lambda^{(1)}_{Y_{1}} \to \Lambda^{(k)}_{Y_1}\cdots\Lambda^{(1)}_{Y_{k}}$ can
be performed moving $Y_1$ from the right to the left at the first stage
\begin{align*}
&\Lambda^{(k)}_{Y_k}\Lambda^{(k-1)}_{Y_{k-1}}\cdots\Lambda^{(1)}_{Y_{1}}
= \\
&\mathbf{\bar{R}}_{21}(Y_{1},Y_{2})\,
\mathbf{\bar{R}}_{31}(Y_{1},Y_{3})\cdots
\mathbf{\bar{R}}_{k1}(Y_{1},Y_{k})\,
\Lambda^{(k)}_{Y_1}\Lambda^{(k-1)}_{Y_{k}}\cdots\Lambda^{(2)}_{Y_2}\,
\mathbf{R}_{k1}(Y_{k},Y_{1})
\cdots
\mathbf{R}_{31}(Y_{3},Y_{1})
\mathbf{R}_{21}(Y_{2},Y_{1})
\end{align*}
which results in the following recurrent formulae
\begin{align*}
&\mathbf{\bar{R}}_{k\cdots21}(Y_1,\cdots,Y_k) =
\mathbf{\bar{R}}_{21}(Y_{1},Y_{2})\,
\mathbf{\bar{R}}_{31}(Y_{1},Y_{3})\cdots
\mathbf{\bar{R}}_{k1}(Y_{1},Y_{k})\,
\mathbf{\bar{R}}_{k\cdots2}(Y_2,\cdots,Y_k)\,,\\
&\mathbf{R}_{k\cdots21}(Y_1,\cdots,Y_k) =
\mathbf{R}_{k\cdots2}(Y_2,\cdots,Y_k)\,
\mathbf{R}_{k1}(Y_{k},Y_{1})
\cdots
\mathbf{R}_{31}(Y_{3},Y_{1})
\mathbf{R}_{21}(Y_{2},Y_{1})\,.
\end{align*}
The simple comparison of two recurrent formulae and 
exact expressions in the case $N=2$ leads to 
the needed identification of the product of R-matrices 
and special element of the permutation group which 
reverse the order of quantum numbers.

Let us now consider the scalar product between the functions dressed with auxiliary spinors (here we omit the length $N$, obvious from the context)
\begin{align}
\begin{aligned}
\langle\boldsymbol{\beta}|\Psi_{\mathbf{Y}}|\boldsymbol{\alpha}\rangle = \langle \beta_N|\Lambda^{(N)}_{Y_N}|\alpha_N\rangle\cdots \langle \beta_2|\Lambda^{(2)}_{Y_2}|\alpha_2\rangle \langle \beta_1|\Lambda^{(1)}_{Y_1}|\alpha_1\rangle \,,
\end{aligned}
\end{align}
and
\begin{align}
\begin{aligned}
\langle\boldsymbol{\alpha}'|\Psi_{\mathbf{Y'}}^{\dagger}|\boldsymbol{\beta}'\rangle =  \langle \alpha'_1|\bar{\Lambda}^{(1)}_{Y'_1}|\beta'_1\rangle \langle \alpha'_2|\bar{\Lambda}^{(2)}_{Y'_2}|\beta'_2\rangle\cdots \langle \alpha'_N|\bar{\Lambda}^{(N)}_{Y'_N}|\beta'_N\rangle \,.
\end{aligned}
\end{align}
The expression \eqref{ScalProd} gets dressed accordingly, and reads
\begin{equation}
\frac{1}{\mu(\mathbf Y)}\sum_{s \in \mathbb{S}_N}\,\delta(\mathbf{Y}'-s(\mathbf{Y})) 
\langle \beta_1,\dots,\beta_N | \mathbf{\bar{S}}(\mathbf Y)^{\dagger}| \beta'_1,\dots,\beta'_N \rangle \langle \alpha'_1,\dots,\alpha'_N  |\mathbf{S}(\mathbf Y)  | \alpha_1,\dots,\alpha_N \rangle\,,
\end{equation}
or in a compact form 
\begin{align}
\begin{aligned}
&\frac{1}{\mu(\mathbf Y)}\sum_{s \in \mathbb{S}_N}\,\delta(\mathbf{Y}-s(\mathbf{Y}')) \langle \boldsymbol{\beta}|\mathbf{\bar{S}}(\mathbf Y)^{\dagger}|\boldsymbol{\beta}' \rangle \langle \boldsymbol{\alpha}'|\mathbf{S}(\mathbf Y)|\boldsymbol{\alpha}\rangle\,.
\end{aligned}
\end{align}
We can conjecture the completeness of the eigenfunctions \eqref{eigenf_Y}, based on the observation that at $N=1$ they are indeed complete and that they are the four-dimensional analogue of the complete basis for the $2d$ problem studied in \cite{Derkachov2001,Derkachov2014}. Therefore, we can write an invertible transform  from the space of coordinates to the space of quantum numbers $(\nu,\ell)$ and symmetric spinors $\alpha,\beta$, so that a generic function $\phi(\mathbf{x})= \phi(x_1,\dots,x_N)$ is mapped into its Fourier coefficient
\begin{equation}
\label{tilphi}
\widetilde{\phi}(\mathbf{Y},\boldsymbol{\alpha},\boldsymbol{\beta})  =\int d^4 x_1\cdots d^4 x_N\,\langle\boldsymbol{\alpha}|{\Psi}_{\mathbf{Y}}^{\dagger}(\mathbf{x})|\boldsymbol{\beta}\rangle \,\phi(\mathbf{x})\,.
\end{equation}
The inverse transform of \eqref{tilphi} provides the expansion of $\phi(\mathbf{x})$ over the basis of eigenfunctions
\begin{equation}
\label{Ytrans}
\medmuskip=1mu
\thinmuskip=0.7mu
\thickmuskip=0.7mu
\phi(\mathbf{x})= \frac{1}{N!}\sum_{\ell_1,\dots,\ell_N} \int d\nu_1\cdots d\nu_N \,\,\,\,\,\,\,\,\mu(\boldsymbol{\nu},\boldsymbol{\ell}) \int D\alpha_1\cdots D\alpha_N D\beta_1\cdots D\beta_N \,\,\,\,\,\,\,\,\,\,\,\,\,\langle\boldsymbol{\beta}|{\Psi}_{\mathbf{Y}}(\mathbf{x})|\boldsymbol{\alpha}\rangle\,\,\,\,\, \widetilde{\phi}(\mathbf{Y},\boldsymbol{\alpha},\boldsymbol{\beta})\,\,\,\,,
\end{equation}
where the sums in $\ell_k$ run over $\mathbb{N}$, the integrations $d\nu_k$ are defined on the real line and the integration in the space of spinors $D\alpha_k$ is defined in \eqref{spinor_measure}.

As a consistency check of the conjectured completeness, we can set $\phi(\mathbf{x})=\langle\boldsymbol{\beta}|{\Psi}_{\mathbf{Y}}(\mathbf{x})|\boldsymbol{\alpha}\rangle$ in \eqref{tilphi} so that \begin{equation}
\widetilde{\phi}(\mathbf{Y}',\boldsymbol{\alpha}',\boldsymbol{\beta}') = \int d\mathbf{x} \,\langle\boldsymbol{\beta'}|{\Psi}_{\mathbf{Y'}}(\mathbf{x})|\boldsymbol{\alpha'}\rangle^* \langle\boldsymbol{\beta}|{\Psi}_{\mathbf{Y}}(\mathbf{x})|\boldsymbol{\alpha}\rangle\,,
\end{equation}
and according to \eqref{Ytrans} its inverse transform becomes
\begin{align}
\begin{aligned}
&\frac{1}{N!}\sum_{\boldsymbol{\ell}'} \int d\boldsymbol{\nu}'\,\int D\boldsymbol{\alpha}'D\boldsymbol{\beta}' \,\,\langle\boldsymbol{\beta}'|{\Psi}_{\mathbf{Y}'}(\mathbf{x})|\boldsymbol{\alpha}'\rangle\, \sum_{s \in \mathbb{S}_N}\prod_{k=1}^N \left[\delta_{\ell_k,\ell'_{s(k)}}\delta(\nu_k-\nu'_{s(k)})\langle \boldsymbol{\alpha}'|\mathbf{S}(\mathbf{Y})|\boldsymbol{\alpha} \rangle \langle \boldsymbol{\beta}|\mathbf{S}(\mathbf{Y})^{\dagger}|\boldsymbol{\beta}'\rangle\right]\,=\\&=\frac{1}{N!} \sum_{s \in \mathbb{S}_N} \langle\boldsymbol{\beta}|\mathbf{S}(\mathbf{Y})^{\dagger}{\Psi}_{s(\mathbf{Y})}(\mathbf{x})\mathbf{S}(\mathbf{Y})|\boldsymbol{\alpha}\rangle=\frac{1}{N!} \sum_{s \in \mathbb{S}_N}  \langle \boldsymbol{\alpha}| \Psi_{\mathbf{Y}}(\mathbf{x})|\boldsymbol{\beta}\rangle =  \langle \boldsymbol{\alpha}| \Psi_{\mathbf{Y}}(\mathbf{x})|\boldsymbol{\beta}\rangle\,.
\end{aligned}
\end{align}

\section{Spectral representation of $\hat{Q}_{\nu,\ell}$ and $\check{Q}_{\nu,\ell}$}
\label{sec:specQ}
In this section we will employ the integral transforms \eqref{tilphi} and \eqref{Ytrans} to provide a representation of the integral operators $\hat{Q}$ and $\check{Q}$ over the separated variables $(\mathbf{Y},\boldsymbol{\alpha},\boldsymbol{\beta})$. As it was shown in \eqref{squarelat} and \eqref{Yukhat},\eqref{Yukcheck}, for $\lambda=-i$ and specific values of $\nu$, these operators become the graph-building kernel of a scalar square lattice for $\ell=0$ or a Yukawa hexagonal lattice, for $\ell=1$. Therefore it is possible to apply the integral transform \eqref{Ytrans} in order to provide a representation of regular, planar Feynmann diagrams on the spectrum of separated variables.

First of all we point out the functions of the basis \eqref{eigenf_Y} were obtained in order to diagonalize the $Q$-operators with one reference point $x_0$, while $\hat{Q}$ and $\check{Q}$ have two reference points $x_0$ and $x_0'$. The operators $Q(u)$ can actually be recoverd as
\begin{equation}
Q\left(-1+i\frac{\lambda}{2}+i\nu\right)\propto \lim_{|x_0'|\to \infty}\check{Q}_{\nu,0}\,,
\end{equation}
More rigorously, the connection between $\hat{Q}_{\nu,0}$ of $\check{Q}_{\lambda,\nu,0}$ and $Q_{\nu}$ is given by a conformal transformation. By means of a translation $x^{\mu} \to x^{\mu}-{x_0'}^{\mu}$ followed by an inversion $x^{\mu} \to x^{\mu}/x^2$, the operator $\check{Q}_{\nu,0}$ gets transformed as
\begin{align}
\begin{aligned}
\label{x0'_transf}
\check{Q}_{\nu,0}  \,\longrightarrow\,&\left(\frac{\pi^{2}\Gamma\left(1-i\frac{\lambda}{2}-i\nu\right)}{4^{-1+i\frac{\lambda}{2}+i\nu}\Gamma(1+i\frac{\lambda}{2}+i\nu)}\right)^N \, (x_0^2)^{1+i\frac{\lambda}{2}+i\nu}\times \\&\qquad\qquad\times \left(\prod_{k=1}^N x_{k}^{2(2+i\lambda)}\right) Q\left(-1+i\frac{\lambda}{2}+i\nu \right) \left(\prod_{k=1}^N x_{k}^{-2(2+i\lambda)}\right)\,,
\end{aligned}
\end{align}
that is, a part from the external parameter $x_0$ and the constant pre-factor, the result is a unitary conjugation of $Q(u)$ depending only on $\lambda$. Having at our disposal a basis of eigenfunctions for $Q(u)$, we can transform all $\hat{Q}$ and $\check{Q}$ operators according to \eqref{x0'_transf} to study their action on the eigenfunctions, and eventually transform back the result. 

\begin{figure}
\begin{center}
\includegraphics[scale=0.35]{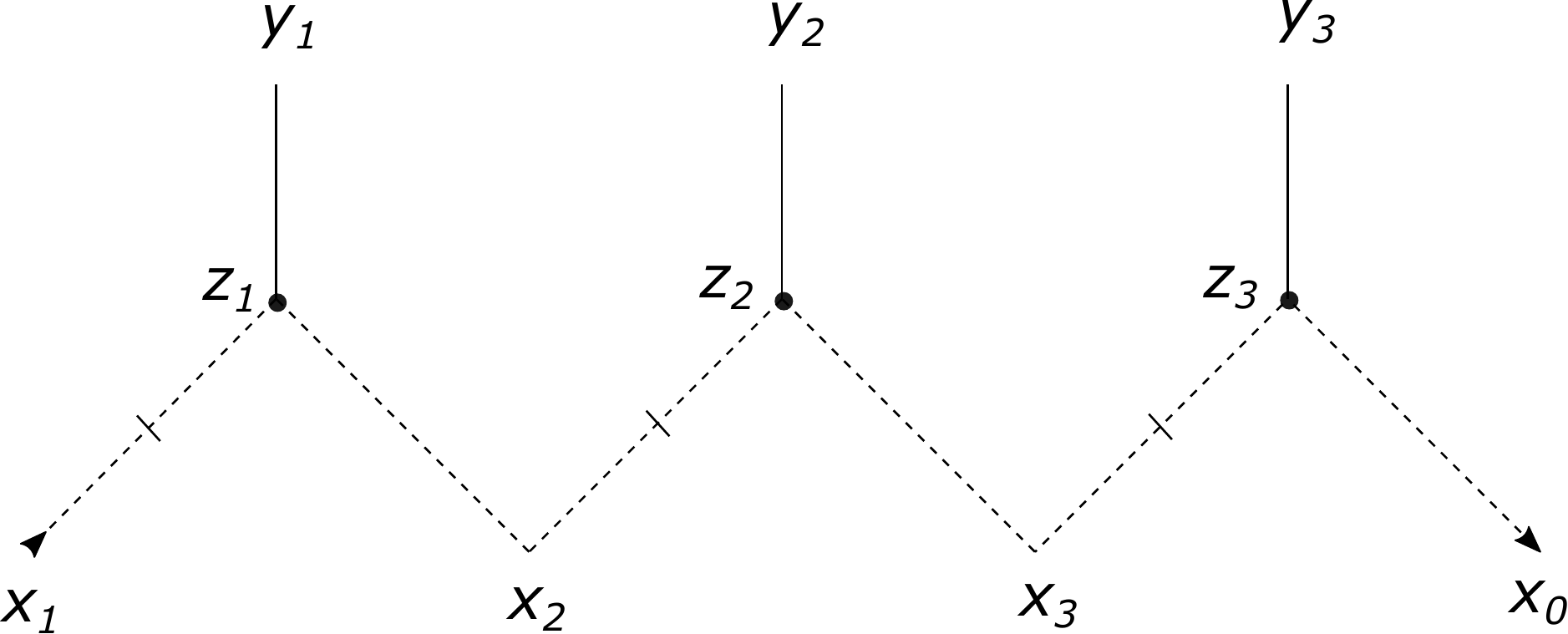}\hspace{10mm}\includegraphics[scale=0.35]{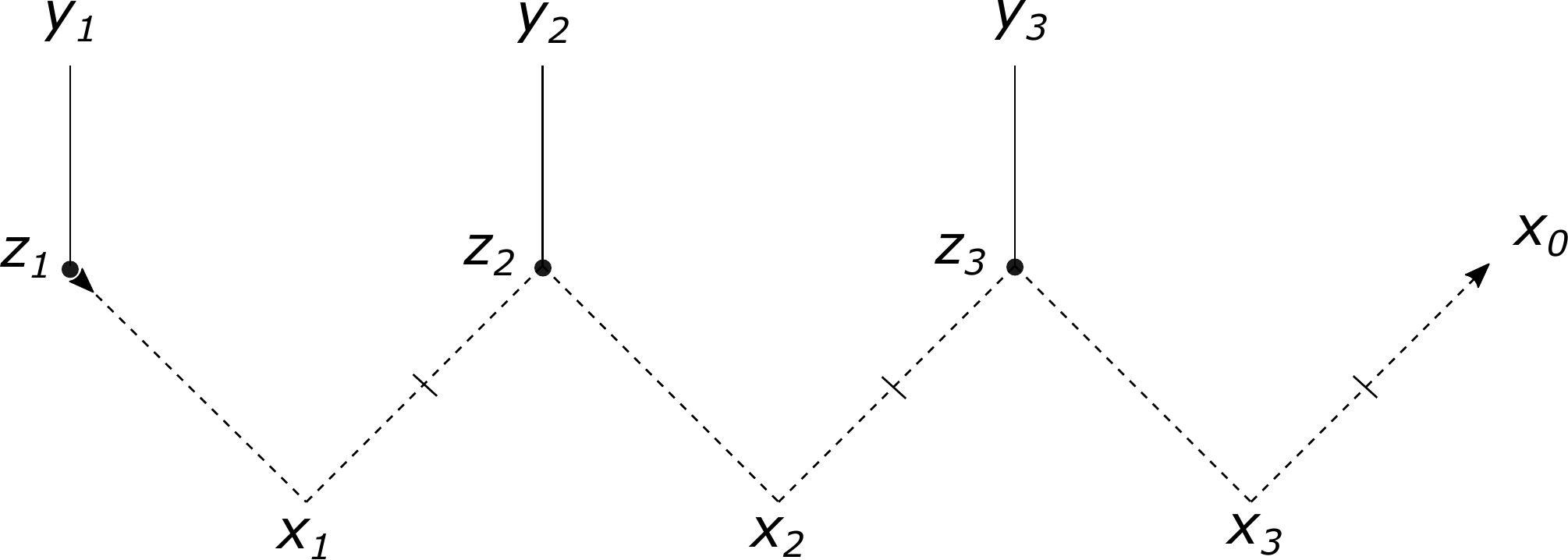}
\end{center}
\caption{On the left, the kernel $Q^+_{\nu,\ell}(x_1,x_2,x_3,x_0|y_1,y_2,y_3)$ as in formula \eqref{Q+str}. On the right, the kernel $Q^-_{\nu,\ell}(x_1,x_2,x_3,x_0|y_1,y_2,y_3)$ as in formula \eqref{Q-str}.}
\label{QkerY}
\end{figure}

We define the operators $Q^{+}_{\nu,\ell}$ by the transformation \eqref{x0'_transf} applied for a general  value of $\ell$ in $\check{Q}_{\nu,\ell}$
\begin{equation}
\label{x0'_ell}
 \check{Q}_{\nu,\ell}  \,\longrightarrow\, (x_0^2)^{1+i\frac{\lambda}{2}-i\nu}\left(\prod_{k=1}^N x_{k}^{2(2+i\lambda)}\right) Q^{+}_{\nu,\ell} \left(\prod_{k=1}^N x_{k}^{-2(2+i\lambda)}\right)[\overline{\mathbf{x_0}}]^{\ell}\,,
\end{equation}
which leads to the kernel
\begin{align}
\label{Qker_ell}
\begin{aligned}
Q^{+}_{\nu,\ell}(x_1,\dots,x_N,x_0|y_1,\dots,y_N) = \frac{[\mathbf{(\overline{x_1-y_1})(y_1-x_2)}\cdots\mathbf{(\overline{x_{N}-y_{N}})(y_{N}-x_0)}]^{\ell}}{(x_1-y_1)^{2\left(1-i\frac{\lambda}{2}+i\nu\right)}(y_1-x_2)^{2\left(1-i\frac{\lambda}{2}-i\nu\right)}\cdots(y_{N}-x_0)^{2\left(1-i\frac{\lambda}{2}-i\nu\right)}}\times\\\times 
\prod_{k=1}^{N} {(x_k-x_{k+1})^{2(-i\lambda)}}\,,
\end{aligned}
\end{align}
and $Q^{+}_{\nu,0}$ is simply related to $Q(u)$ according to \eqref{x0'_transf}. The spinor indices of $Q^{+}_{\nu,\ell}$ follow from \eqref{Qker_ell} as
\begin{equation}
{(Q^{+}_{\nu,\ell})}^{\dot{\mathbf{c}}}_{\dot{\mathbf{a}}}\,=\,{(Q^{+}_{\nu,\ell})}^{(\dot c_1,\dots,\dot c_{\ell})}_{(\dot a_1,\dots,\dot a_{\ell})}\,.
\end{equation}
An equivalent representation for the kernel $Q^+_{\nu,\ell}$ is obtained by opening the triangles $(x_k,y_k,x_k+1)$ for $k=1,\dots,N$ into star integrals (see fig.\ref{QkerY})
\begin{align}
\label{Q+str}
\begin{aligned}
&Q^{+}_{\nu,\ell}(x_1,\dots,x_N,x_0|y_1,\dots,y_N) = \left(\pi^2 \frac{\Gamma\left(1-i\frac{\lambda}{2}+i\nu +\frac{\ell}{2}\right)\Gamma\left(1-i\frac{\lambda}{2}-i\nu+\frac{\ell}{2} \right)\Gamma\left(i\frac{\lambda}{2}+\frac{\ell}{2}\right)}{\Gamma\left(1+i\frac{\lambda}{2}+i\nu +\frac{\ell}{2}\right)\Gamma\left(1+i\frac{\lambda}{2}-i\nu +\frac{\ell}{2}\right)\Gamma\left(2-i{\lambda}\right)}\right)^{-N} \times\\&\times \int\, d^4z_1\cdots d^4 z_N \frac{[\mathbf{(\overline{x_1-z_1})(z_1-x_2)}\cdots\mathbf{(\overline{x_{N}-z_{N}})(z_{N}-x_0)}]^{\ell}}{(x_1-z_1)^{2\left(1+i\frac{\lambda}{2}+i\nu\right)}(z_1-x_2)^{2\left(1+i\frac{\lambda}{2}-i\nu\right)}\cdots(z_{N}-x_0)^{2\left(1+i\frac{\lambda}{2}-i\nu\right)}}
\prod_{k=1}^{N} {(z_k-y_{k})^{2(i\lambda-2)}}\,.
\end{aligned}
\end{align}
The same transformation \eqref{x0'_ell} can be applied to $\hat{Q}_{\nu,\ell}$
\begin{equation}
\label{x0'_ell1}
 \hat{Q}_{\nu,\ell}  \,\longrightarrow\, (x_0^2)^{1+i\frac{\lambda}{2}+i\nu}\left(\prod_{k=1}^N x_{k}^{2(2+i\lambda)}\right) Q^{-}_{\nu,\ell} \left(\prod_{k=1}^N x_{k}^{-2(2+i\lambda)}\right)[{\mathbf{x_0}}]^{\ell}\,,
\end{equation}so to define the kernel
\begin{align}
\begin{aligned}
\label{Q-ker_ell}
Q^{-}_{\nu,\ell}(x_1,\dots,x_N,x_0|y_1,\dots,y_N) = \frac{[\mathbf{(y_1-x_1)(\overline{x_1-y_2})}\cdots\mathbf{({y_{N}-x_{N}})(\overline{x_{N}-x_0})}]^{\ell}}{(y_1-x_1)^{2\left(1-i\frac{\lambda}{2}-i\nu\right)}(x_1-y_2)^{2\left(1-i\frac{\lambda}{2}+i\nu\right)}\cdots(x_{N}-x_0)^{2\left(1-i\frac{\lambda}{2}+i\nu\right)}}\times \\\times \prod_{k=1}^{N-1}(x_k-x_{k+1})^{2(-i\lambda)}\,,
\end{aligned}
\end{align}
whose spinor structure is
\begin{equation}
{(Q^{-}_{\nu,\ell})}^{\mathbf{c}}_{{\mathbf{a}}}\,=\,{(Q^{-}_{\nu,\ell})}^{(c_1,\dots,c_{\ell})}_{( a_1,\dots, a_{\ell})}\,.
\end{equation}
The kernel  $Q^{-}_{\nu,\ell}$ can be rewritten after star-triangle transformation as (see fig.\ref{QkerY})
\begin{align}
\label{Q-str}
\begin{aligned}
Q^{-}_{\nu,\ell}(x_1,\dots,x_N,x_0|y_1,\dots,y_N)=  \left(\pi^2 \frac{\Gamma\left(1-i\frac{\lambda}{2}+i\nu +\frac{\ell}{2}\right)\Gamma\left(1-i\frac{\lambda}{2}-i\nu+\frac{\ell}{2} \right)\Gamma\left(i\frac{\lambda}{2}+\frac{\ell}{2}\right)}{\Gamma\left(1+i\frac{\lambda}{2}+i\nu +\frac{\ell}{2}\right)\Gamma\left(1+i\frac{\lambda}{2}-i\nu +\frac{\ell}{2}\right)\Gamma\left(2-i{\lambda}\right)}\right)^{-N} \times\\\times \int\, d^4z_1\cdots d^4 z_N \frac{[\mathbf{(z_1-x_1)(\overline{x_1-z_2})}\cdots\mathbf{({z_{N}-x_{N}})(\overline{x_{N}-x_0})}]^{\ell}}{(z_1-x_1)^{2\left(1+i\frac{\lambda}{2}-i\nu\right)}(x_1-z_2)^{2\left(1+i\frac{\lambda}{2}+i\nu\right)}\cdots(x_{N}-x_0)^{2\left(1+i\frac{\lambda}{2}+i\nu\right)}} \prod_{k=1}^{N}(z_k-y_k)^{2(i\lambda-2)}\,.
\end{aligned}
\end{align}
Now we can compute the action of ${Q}_{\nu,\ell}^{+}$ or ${Q}_{\nu,\ell}^{-}$ on the eigenfunctions of $Q_{\nu}=Q^+_{\nu,0}$, so to define their spectral representation according to \eqref{Ytrans}. As usual the computation can be done in a few graphical passages based on the identities of section \ref{sec:startri}. We are going to show that 
\begin{equation}
\label{Qhateigenv}
{Q}^{+\, (N)}_{\nu,\ell} {\Lambda}^{(N)}_{Y'}= \tau_{+}(\nu,\ell,\nu',\ell') \,{\mathbf{R}}_{\ell,\ell'}(i\nu-i\nu')\,{\Lambda}^{(N)}_{Y'} {Q}^{+\,(N-1)}_{\nu,\ell}\,,
\end{equation}
that is, in explicit spinor notation:
\begin{equation}
{\left({Q}^{+\, (N)}_{\nu,\ell}\right)}_{\mathbf{\dot{a}}}^{\mathbf{\dot c}} {\left({\Lambda}^{(N)}_{Y'}\right)}_{\mathbf{\dot{s}}}^{\mathbf{r}} = \tau_{+}(\nu,\ell,\nu',\ell') \,[{\mathbf{R}}_{\ell,\ell'}(i\nu-i\nu')]_{\mathbf{\dot{a}}\,\mathbf{\dot{s}}}^{\mathbf{\dot b}\,\mathbf{\dot d}}\,\left({\Lambda}^{(N)}_{Y'} \right)_{\mathbf{\dot{d}}}^{\mathbf{r}}\left({Q}^{+\,(N-1)}_{\nu,\ell}\right)_{\mathbf{\dot{b}}}^{\mathbf{\dot c}}\,,
\end{equation}
where
\begin{align}
\label{general_eigenv}
\begin{aligned}
\tau_+(\nu,\ell,\nu',\ell')&=\pi^2\frac{\Gamma\left( 1-i\frac{\lambda}{2}+i\nu' +\frac{\ell'}{2}\right)\Gamma\left( 1+i\frac{\lambda}{2}-i\nu +\frac{\ell}{2}\right)\Gamma\left(i(\nu-\nu') +\frac{\ell'-\ell}{2}\right)}{\Gamma\left( 1 +i\frac{\lambda}{2}-i\nu' +\frac{\ell'}{2}\right)\Gamma\left( 1 -i\frac{\lambda}{2}+i\nu +\frac{\ell}{2}\right)\Gamma\left(1 +i(\nu'-\nu) +\frac{\ell'-\ell}{2}\right)}\times\\&\times \frac{ (-1)^{\ell}}{\left(1 +i(\nu'-\nu) +\frac{\ell'+\ell}{2}\right)}.
\end{aligned}
\end{align}
In relation to the reductions of $\check{Q}_{\nu,\ell}$ defined in \eqref{squarelat} and in \eqref{Yukcheck} we point out that \eqref{general_eigenv} specializes to
\begin{equation}
\label{eigen_square}
\tau_+(-i/2,0,\nu,\ell)=\frac{4 \pi^2 }{4\nu^2+(\ell+1)^2}\,,
\end{equation}
and, respectively, to
\begin{equation}
\label{eigen_Yuk1}
\tau_+(0,1,\nu,\ell)=\frac{8\pi^2}{\left(1-2i\nu +{\ell}\right)\left(1+2i\nu-{\ell}\right)\left(3 +2i\nu +\ell\right)} \,.
\end{equation}
Let's prove formula \eqref{Qhateigenv}. Starting from its l.h.s., depicted on the left picture at $N=3$, we open the triangles in the layer ${\Lambda}$ into star integrals (right picture).
\begin{center}
\includegraphics[scale=0.35]{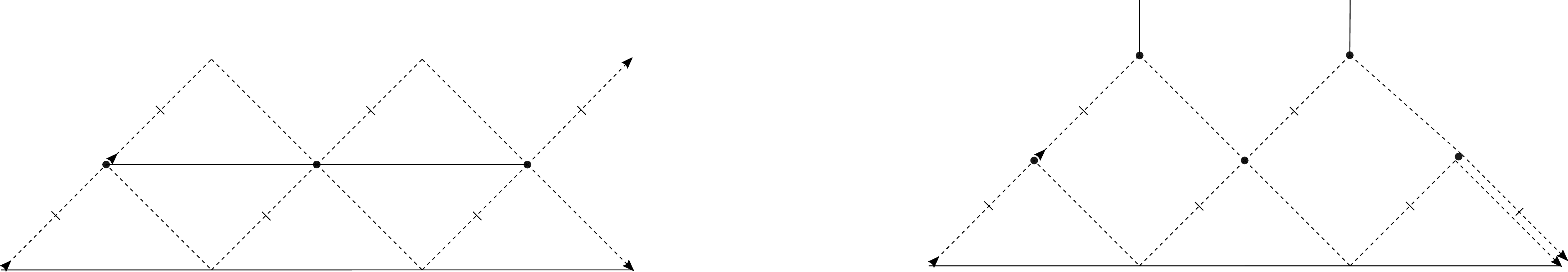}
\end{center}
We performed the star-triangle identity \eqref{STRsame} in the rightmost integration point, obtaining the left picture, containing two vertical dashed lines. The couple of vertical lines can be moved leftwards by means of the exchange relation \eqref{exch_same}, ending up with the right picture.
\begin{center}
\includegraphics[scale=0.35]{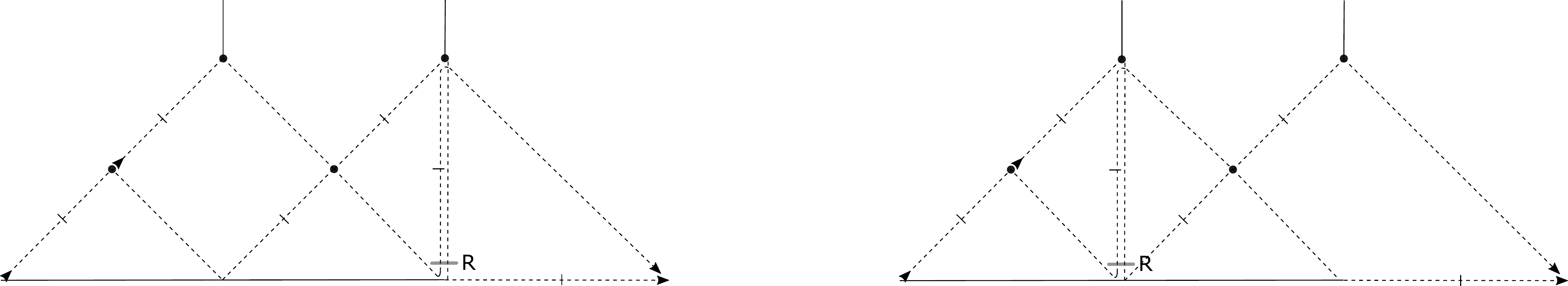}
\end{center}
Finally, we opened the triangle with basis the couple of vertical lines into a star integral (left picture). The leftover integrations can be performed by means of \eqref{STR_0} and \eqref{STRsame}, and we obtain the  the r.h.s. of \eqref{Qhateigenv} (right picture).
\begin{center}
\includegraphics[scale=0.35]{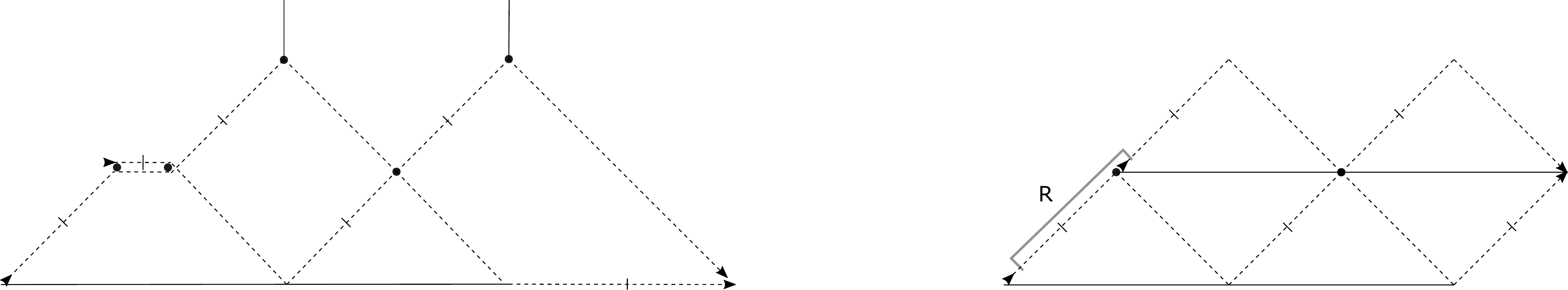}
\end{center}
It follows from \eqref{Qhateigenv} that
\begin{equation}
\label{Q+eq}
Q^+_{\nu,\ell} \,\Psi_{Y'}(\mathbf{x}) =\left(\overleftarrow{ \prod_{k=1}^N}  \tau_+(\nu,\ell,\nu'_k,\ell'_k) \,\overline{\mathbf{R}}_{k}^{\,-1}\right)\Psi_{Y'}(\mathbf{x})\,,
\end{equation}
where $\mathbf{R}_{k}=\mathbf{R}_{\ell,\ell'_k}(i\nu'_k-i\nu)$.
In a similar way one can prove that the application of ${Q}^{-}_{\nu,\ell}$ to the layer ${\Lambda}_{Y'}$ delivers the result
\begin{equation}
\label{Qcheckeigenv}
{Q}^{-\, (N)}_{\nu,\ell} {\Lambda}^{(N)}_{Y'}= \tau_-(\nu,\ell,\nu',\ell') \,{\Lambda}^{(N)}_{Y'} {Q}^{-\,(N-1)}_{\nu,\ell}\,\mathbf{R}_{\ell,\ell'}(i\nu'-i\nu) \,,
\end{equation}
or in explicit spinor notation
\begin{equation}
\left({Q}^{-\, (N)}_{\nu,\ell}\right)^{\mathbf{c}}_{\mathbf{a}} \left({\Lambda}^{(N)}_{Y'}\right)_{\mathbf{\dot s}}^{\mathbf r}= \tau_-(\nu,\ell,\nu',\ell') \,\left({\Lambda}^{(N)}_{Y'}\right)_{\mathbf{\dot{s}}}^{\mathbf{d}}\, \left({Q}^{-\,(N-1)}_{\nu,\ell}\right)_{\mathbf{{a}}}^{\mathbf{b}}\,[\mathbf{R}_{\ell,\ell'}(i\nu'-i\nu)]_{\mathbf{b}\,\mathbf{d}}^{\mathbf{c}\,\mathbf{r}} \,,
\end{equation}
where 
$\tau_-(\nu,\ell,\nu',\ell')=\tau_+(-\nu,\ell,-\nu',\ell')$. In particular for the reductions \eqref{squarelat} and \eqref{Yukhat} of $\hat{Q}_{\nu,\ell}$ the value of $\tau_-$ is given by
\begin{equation}
\tau_-(i/2,0,\nu,\ell)=\tau_+(-i/2,1,\nu,\ell)\,,\,\,\,\,\text{and}\,\,\,\,\,
\tau_-(0,1,\nu,\ell)=\tau_+(0,1,\nu,\ell)^*\,.
\end{equation}
As a consequence of \eqref{Qcheckeigenv}, the action of \eqref{Q-ker_ell} on the functions \eqref{eigenf_Y} reads
\begin{equation}
\label{Q-eq}
Q^-_{\nu,\ell} \,\Psi_{Y'}(\mathbf{x}) = \Psi_{Y'}(\mathbf{x})\,\left(\overrightarrow{ \prod_{k=1}^N}  \tau(\nu,\ell,\nu'_k,\ell'_k) \,\mathbf{R}_{k}\right)\,.
\end{equation}
The formulas \eqref{Q+eq} and \eqref{Q-eq} show that the functions \eqref{eigenf_Y} are in general not eigenfunctions of $Q^+$ and $Q^-$, due to the mixing of spinor indices by the matrices $\mathbf{R}_k$. Nevertheless, the result still has a completely factorized structure, for which the length-$N$ case coincides with the product of $N$ length-$1$ systems. In other words, the eigenfunctions $\eqref{eigenf_Y}$, via the transform \eqref{Ytrans}, realize a separation of variables for the operators $Q^+_{\nu,\ell}$ and $Q^-_{\nu,\ell}$ and therefore - after a conformal transformation - for $\hat Q_{\nu,\ell}$ and $\check Q_{\nu,\ell}$. 

\section{Conformal Fishnet Integrals}
\label{sec:final}
In this section we are going apply the spectral transform \eqref{Ytrans} to some planar Feynmann diagrams with a regular bulk topology, consisting in portions of square lattice and hexagonal Yukawa lattice. In particular the class of diagrams under study turns out to provide the sole connected Feynman integrals which contribute to specific correlators in the four-dimensional chiral conformal field theory ($\chi$CFT$_4$) arising as the double-scaling limit of $\gamma$-deformed $\mathcal{N}=4$ SYM theory~\cite{Fokken:2013aea,Caetano:2016ydc}. We recall that the $\chi$CFT$_4$ theory involves three complex massless scalar fields $\phi_k$ and three left-handed fermions $\psi_{k}$, all of which are actually $N_c\times N_c$ matrix fields transforming in the adjoint representation of \(SU(N_c)\). The Lagrangian of the theory reads
\begin{equation}\label{chiFT4}
  {\cal L}_{\chi}=N_c\Tr\left[-\frac{1}{2}\p^\mu\phi^\dagger_j\p_\mu\phi_j
  +i\bar\psi^{\dot\alpha}_{ j}
  (\tilde\sigma^{\mu})^\alpha_{\dot\alpha}\p_\mu \psi_{j\,\alpha }\right]
+{\cal L}_{\rm int}\,,
\end{equation}
where
the sum  is taken with respect to all doubly repeated indices, including \(j=1,2,3\), and the interaction part is 
\begin{equation}
\begin{aligned}
     \mathcal{L}_{\rm int} ={}N_c
\,\Tr\Bigl[\xi_1^2\,\phi_2^\dagger \phi_3^\dagger &
\phi_2\phi_3\!+\!\xi_2^2\,\phi_3^\dagger \phi_1^\dagger 
\phi_3\phi_1\!+\!\xi_3^2\,\phi_1^\dagger \phi_2^\dagger \phi_1\phi_2\!+\!i\sqrt{\xi_2\xi_3}(\psi_3 \phi_1 \psi_{ 2}+ \bar\psi_{ 3} \phi^\dagger_1 \bar\psi_2 )\\
& +i\sqrt{\xi_1\xi_3}(\psi_1 \phi_2 \psi_{ 3}+ \bar\psi_{ 1} \phi^\dagger_2 \bar\psi_3 )
 +i\sqrt{\xi_1\xi_2}(\psi_2 \phi_3 \psi_{ 1}+ \bar\psi_{ 2} \phi^\dagger_3 \bar\psi_1 )\,\Bigr].  
\label{fullL}\end{aligned}
\end{equation}
The lagrangian \eqref{chiFT4},\eqref{fullL} is not UV complete  and it should be supplied with double-trace vertices.
Remarkably, in the planar limit $N_c\to \infty$ \cite{tHooft:1973alw}, the couplings $\xi_1,\xi_2,\xi_3$ do not receive quantum corrections and the double-trace couplings have fixed points of the Callan-Symanzyk equation $\beta_{\alpha}(\alpha_{crit})=0$, which makes $\chi$CFT$_4$ a conformal invariant theory at the quantum level (see the review \cite{Kazakov2019} and references therein) \footnote{In this paper we will never need to take double-trace vertices into account, as they are sub-leading contributions in the planar limit of the correlators under study.}.
It has been showed in \cite{Kazakov2019} that this conformal theory it preserves several features of integrability and exact-solvability of the bi-scalar fishnet theory obtained setting $\xi_1=\xi_2=0$ in \eqref{fullL}
  \begin{equation}
  \label{L_fish}
       \mathcal{L}_{bi-scalar} ={}N_c
\,\Tr\Bigl[-\frac{1}{2}\partial_{\mu} \phi_1\partial^{\mu} \phi_1^{\dagger}-\frac{1}{2}\partial_{\mu} \phi_2\partial^{\mu} \phi_2^{\dagger}+ \xi^2\,\phi_1^\dagger \phi_2^\dagger \phi_1\phi_2 \Bigr]\,.
  \end{equation}
In our previous work \cite{Derkachov_Oliv} we used the methods explained in this paper for the computation of the four-point functions of the Fishnet Basso-Dixon type at finite coupling
  \begin{equation}
  \label{BDcorr}
  \bigg \langle \text{Tr}\left[(\phi_1(x_1))^N(\phi_2(x'_0))^L(\phi_1^{\dagger}(x'_1))^N(\phi_2^{\dagger}(x_0))^L\right]\bigg\rangle\,,
  \end{equation}
  for any number of fields $N$ and $L$. We recall that such a correlator in the bi-scalar CFT, there is only one connected Feynman diagram entering the perturbative expansion in the coupling $\xi^2=\xi_3^2$. It is represented for $N=3$ and generic $L$ in figure \ref{basso-dixon_intro} and coincides with a square lattice of size $L\times N$ where the external legs are pinched to the four points of \eqref{BDcorr}. First of all we observe that the correlator \eqref{BDcorr} in the planar limit of $\chi$CFT$_4$ theory still receives contributions only from the fishnet diagrams of figure \ref{chiBD_fig}, and the same is true for the more general correlator
  \begin{equation}
  \label{BDchicft}
  \bigg \langle \text{Tr}\left[(\phi_1(x_1))^N\mathcal{O}_{L_1,L_2}(x'_0) (\phi_1(x'_1))^N \mathcal{O}_{L_1,L_2}(x_0)\right]\bigg\rangle\,,
  \end{equation}
  where
  \begin{equation}
  \label{Ogeneral}
  \mathcal{O}_{L_1,L_2}(x) = (\phi_2(x))^{L_1}(\phi_3^{\dagger}(x))^{L_2}\,,
  \end{equation}
  or any other permutation of fields $\phi_2$ and $\phi_3^{\dagger}$. Secondly, we can formulate a generalization of \eqref{BDcorr} and \eqref{BDchicft} including fermionic fields in the correlator
  \begin{equation}
  \label{Full_correlator}
  \bigg \langle \text{Tr}\left[(\phi_1(x_1))^N  \,\mathcal{O}_{L_1,L_2,M_1,M_2}(x'_0) \,(\phi_1^{\dagger}(x'_1))^N \,\mathcal{O}^{\dagger}_{L_1,L_2,M_1,M_2}(x_0)\,\right]\bigg \rangle\,,
  \end{equation}
  where 
  \begin{equation}
  \label{O_operators}
  \mathcal{O}_{L_1,L_2,M_1,M_2}(x) = (\phi_2(x))^{L_1}(\phi_3^{\dagger}(x))^{L_2}({\psi}_2(x))^{M_1}(\bar{\psi}_3(x))^{M_2}\,,
  \end{equation}
  and all the considerations we are going to make are actually valid for any permutation of the fields in the r.h.s. of \eqref{O_operators}.
  
  The perturbative expansion of correlators \eqref{Full_correlator} in the couplings $\xi_i^2$ in the planar limit of \eqref{BDchicft} contains only one connected Feynman diagram, appearing at the order $\xi_1^{N(M_1+M_2)} \xi_2^{N(M_1+2L_1)} \xi_3^{N(M_2+2L_2)}$ which has the simple topology (here depicted for $N=3$)
  \begin{center}
  \includegraphics[scale=0.30]{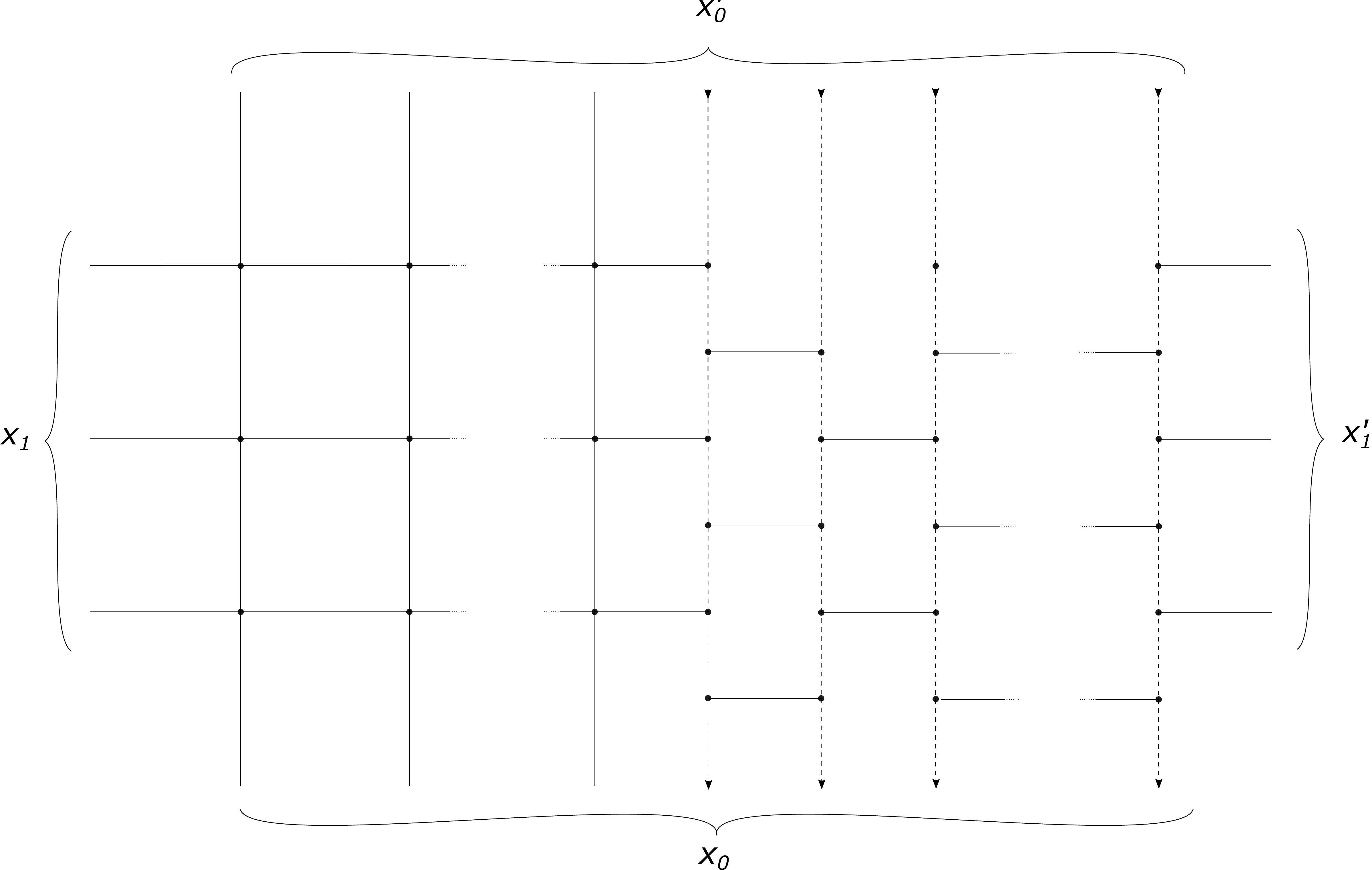}
  \end{center}
that is consists of a portion of square lattice, coinciding with the the region where at $x_2$ and $x_4$ there are fields $\phi_2$ and $\phi_3$, and portions of Yukawa hexagonal lattice corresponding to external $\psi_2$ and $\psi_3$. As explained in sect \ref{sect:Qspin}, such portions of diagrams can be realized by means of the operators \eqref{Qhat_ker} or\eqref{Qcheck_ker} for the specific choice of $\lambda=-i$ and $\ell=0$ or $\ell=1$. More precisely, the diagrams corresponding to the correlator \eqref{Full_correlator} is obtained by $L_1+L_2+1$ copies of the operator $\hat{Q}_{i/2,0}$ at length $L$, responsible for a square lattice of size $L\times (L_1+L_2)$, by $M_1$ copies of $\hat{Q}_{0,1}$ and by $M_2$ copies of $\check{Q}_{0,1}$, responsible for the hexagonal Yukawa lattice. Moreover, the in-coming external legs in $x_1$ can be written as\begin{equation*}
\lim_{x_k\to x_1} \left(\prod_{k=0}^{N}(x_k-x_{k+1})^{2} \right) \hat{Q}_{i/2,0}\,,
  \end{equation*}
 adding another copy of the scalar operator $\hat{Q}_{i/2,0}$, while the external out-coming legs should be pinched together in $x_1'$,  making the last scalar operator act on a bunch of Dirac deltas
 \begin{equation}
 \hat{Q}_{i/2,0}\left(\prod_{k=1}^{N}\delta^{(4)}(x_k-x_1') \right)\,.
 \end{equation}
 Finally, the operatorial expression for the diagram, a part from normalization constants, reads
  \begin{equation}
  \label{diagram_operatorial}
  G(x_0,x_1,x_0',x_1')=\lim_{x_k\to x_1}\left(\prod_{k=0}^{N}(x_k-x_{k+1})^{2} \right) \hat{Q}_{i/2,0}^{L_1+L_2+1} \hat{Q}_{0,1}^{M_1}  \check{Q}_{0,1}^{M_2}\left(\prod_{k=1}^{N}\delta^{(4)}(x_k-x_1') \right) \,,
  \end{equation}
  where in general $G$ has a spinorial structure inherited from the fermionic operators $(\hat{Q}_{0,1})_{\dot{a}}^a$ and $(\check{Q}_{0,1})^{\dot{a}}_a$
  \begin{equation}
  \label{spinind_corr}
   G_{a_1,\dots,a_{M_1},\dot a_1,\dots,\dot a_{M_2}}^{\dot b_1,\dots,\dot b_{M_1}, b_1,\dots, b_{M_2}}(x_0,x_1,x_0',x_1')\,,
  \end{equation}
  which encodes the spinor indices of fermionic fields in \eqref{Full_correlator}.
  The order in which the graph-building operators appear in \eqref{diagram_operatorial} is irrelevant - in agreement with the invariance of the correlator under any permutation of fields in the definition \eqref{Ogeneral}. Indeed, it follows from \eqref{l0comm} that any portion of Yukawa lattice commutes with any portion of square lattice, and also as a consequence of the relation \eqref{hatcheckcomm}
  
  , fermionic lines emitted by $\psi_2$ and $\bar{\psi}_3$ can be exchanged without producing any effect. For convenience we fix the order as in \eqref{diagram_operatorial}.
  
In order to use the formula \eqref{Ytrans}, we should first apply to \eqref{diagram_operatorial} the transformation \eqref{x0'_ell}\eqref{x0'_ell1}, so to amputate the external lines to $x_0'$. Thus, we perform the computation of the transformed graph,
  \begin{equation}
  \label{diag_transf}
   F(x_0,x_1,x_1')=\lim_{x_k\to x_1}\left(\prod_{k=0}^{N}(x_k-x_{k+1})^{2} \right) ({Q}^{-}_{i/2,0})^{L_1+L_2+1} ({Q}^{-}_{0,1})^{M_1} ({Q}^{+}_{0,1})^{M_2}\prod_{k=1}^{N}\delta^{(4)}(x_{k1'}) \,,
\end{equation}
whose graphical representation is
 \begin{center}
  \includegraphics[scale=0.30]{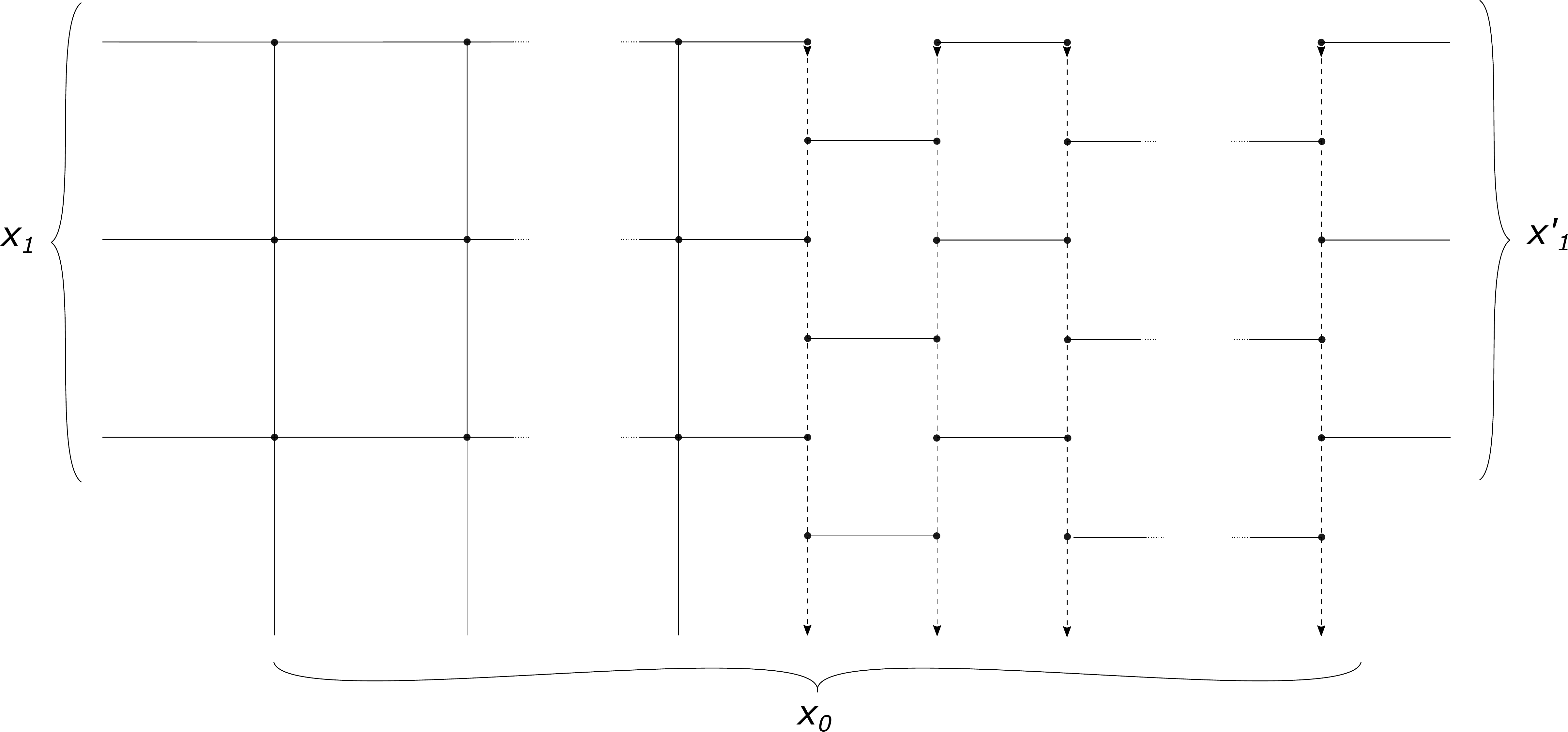}
  \end{center}
  and the spinor structure are inherited from the operators $({Q}^{-})_{a}^{c}$ and $({Q}^{+})_{\dot a}^{\dot c}$ and reads
  \begin{equation}
  F(x_0,x_1,x_1')^{c_1,\dots,c_{M_1},\dot c_1,\dots,\dot c_{M_2}}_{a_1,\dots,a_{M_1},\dot a_1,\dots,\dot a_{M_2}}\,.
  \end{equation}
 It is possible to expand the diagram \eqref{diag_transf} over the spectrum separated variables by inserting a full basis $\langle \boldsymbol{\beta}| {\Psi} _{\mathbf{Y}}(\mathbf{x})|\boldsymbol{\alpha}\rangle=\langle \beta_1,\dots,\beta_N| {\Psi}_{\mathbf{Y}} (x_1,\dots,x_N)|\alpha_1,\dots,\alpha_N\rangle$ before the first ${Q}^-_{i/2,0}$ and after the last ${Q}^+_{0,1}$. As a result \eqref{diag_transf} is cut into three pieces:\\  $\bullet$ the incoming legs \begin{equation}
  \label{ampred}
  \lim_{x_k\to x_1}\left(\prod_{k=0}^{N}(x_k-x_{k+1})^{2} \right) \langle \boldsymbol{\beta'} |\Psi_{\mathbf{Y}'}(\mathbf{x})| \boldsymbol{\alpha'}\rangle\,
  \end{equation}
$\bullet$  the regular bulk of the diagram
  \begin{align}
  \label{central_diag}
  \begin{aligned}
 \int d^4x_1\cdots d^4x_{N}\, \langle\boldsymbol{\beta'}|\Psi_{\mathbf{Y}'}(\mathbf{x})|\boldsymbol{\alpha'}\rangle^*\,({Q}_{i/2,0}^{-})^{L_1+L_2+1} ({Q}_{0,1}^{-})^{M_1}({Q}_{0,1}^+)^{M_2}\,\langle \boldsymbol{\beta}| \Psi_{\mathbf{Y}}(\mathbf{x})|\boldsymbol{\alpha}\rangle \,.
 \end{aligned}
  \end{align}
$\bullet$ the outgoing legs
  \begin{equation}
  \label{purered}
   \int d^4x_1\cdots d^4x_{N} \langle\boldsymbol{\beta}|\Psi_{\mathbf{Y}}(\mathbf{x})|\boldsymbol{\alpha}\rangle^*  \left(\prod_{k=1}^{L}\delta^{(4)}(x_k-x_1') \right) \,=\, \lim_{x_k\to x_1'}\langle\boldsymbol{\beta}|\Psi_{\mathbf{Y}}(\mathbf{x})|\boldsymbol{\alpha}\rangle^* \,.
  \end{equation}
  In this section we will make use of the shortened notation
  \begin{equation}
  \tau_{\pm}(0,Y)\equiv \tau_+(-i/2,0,\nu,\ell)\,,\,\,\,\,\,\,\,\, \tau_{\pm}(1,Y)\equiv \tau_{\pm}(0,1,\nu,\ell)\,,
  \end{equation}
  where in $ \tau_{\pm}(n,Y)$ the letter $Y=(\nu,\ell)$ is the separated variable, and $n/2$ is the spin of the  field propagating in the graph, $n=1$ for scalar fields (square lattice) and $n=1$ for fermionic fields (Yukawa lattice).
  \subsection{SoV representation of the bulk}
Let us start from the separated variables representation for the bulk. First we introduce the shortcut notations for R matrices
\begin{equation}
\mathbf{R}^+_{h,k} ={\mathbf{R}}
_{1 \ell_k}(-i\nu_k)\,,\,\,\,\,\,\,\,\,\,\mathbf{R}^-_{n,k} =\mathbf{R}_{1\ell_k}(i\nu_k)\,,
\end{equation}
where the $k=1,\dots,N$, $h = 1,\dots,M_2$ and $n = 1,\dots,M_1$ and the spinor indices of the R matrices are
\begin{equation}
{\left(\mathbf{R}^+_{h,k}\right)}^{\dot c_{h} \,(\dot r_1\dots \dot r_{\ell_k})}_{\dot a_{h}\, (\dot s_1\dots \dot  s_{\ell_k})}\,,\,\,\,\,\,\,{\left(\mathbf{R}^-_{n,k}\right)}^{c_{n} \, (r_1\dots  r_{\ell_k})}_{ a_{n}\,(s_1\dots  s_{\ell_k})}\,,
\end{equation} 
where $\dot a_h,\dot c_h$ and $c_n,a_n$ are the spinor indices of fermionic fields as in \eqref{spinind_corr}, that is the index $h$ runs over the operators $Q^+$ and the index $n$ runs over $Q^-$ in \eqref{central_diag}.
It follows from \eqref{Q+eq} and \eqref{Q-eq}, that the operators $Q^{\pm}_{\nu,\ell}$ in \eqref{central_diag} can be substituted by the corresponding eigenvalues (see \eqref{general_eigenv})
  \begin{align}
  \label{Bulk_eigen_gen}
  \begin{aligned}
 & \left(\prod_{k=1}^N\tau_+(0,Y_k)\right)^{L_1+L_2+1}\left(\prod_{k=1}^L\tau_+(1,Y_k)\right)^{M_2} \left(\prod_{k=1}^L\tau_-(1,Y_k)\right)^{M_1} \,,
  \end{aligned}
  \end{align}
  leaving us with the scalar product
   \begin{align}
    \label{scal_bulk}
  \begin{aligned}
 \int d^4x_1\cdots d^4x_{N}\, \langle\boldsymbol{\beta'}|\Psi_{\mathbf{Y}'}(\mathbf{x})|\boldsymbol{\alpha'}\rangle^*\,\langle \boldsymbol{\beta} | \mathcal{R}^{+}\, \Psi_{\mathbf{Y}}(\mathbf{x})\,\mathcal{R}^{-} |\boldsymbol{\alpha}\rangle \,,
 \end{aligned}
  \end{align}
  where we introduced the shortcut notation
  \begin{equation}
  \mathcal{R}^{+}=\overleftarrow{\prod_{h=1}^{M_2}}\overleftarrow{\prod_{k=1}^{N}}\mathbf{R}^{+}_{h,k}\,,\,\,\,\,\,\,\mathcal{R}^{-}=\overrightarrow{\prod_{n=1}^{M_1}}\overrightarrow{\prod_{k=1}^{N}}\mathbf{R}^-_{n,k}\,.
  \end{equation}
   The scalar product \eqref{scal_bulk} can be straightforwardly computed by means of \eqref{ScalProd}, and the result reads
  \begin{align}
  \begin{aligned}
  \label{Bulk_result}
  \frac{1}{N! \,\mu(\boldsymbol{Y})} \prod_{s\in \mathbb{S}_N}\delta(\mathbf{Y}-s \mathbf{Y}')  \langle \boldsymbol{\beta} | \mathcal{R}^{+} \, \mathbf{S}(\mathbf{Y})^{\dagger}|\boldsymbol{\beta}'\rangle\, \langle \boldsymbol{\alpha}' |\mathbf{S}(\mathbf{Y}) \,\mathcal{R}^{-}|\boldsymbol{\alpha}\rangle\,.
  \end{aligned}
  \end{align}
   \subsection{Incoming legs: amputation and reduction}
The computation of \eqref{ampred} can be done graphically step-by-step. Let us start from the function $\left(\prod_{k=0}^{N}(x_k-x_{k+1})^{2} \right) {Q}^-_{i/2,0} \,\langle \boldsymbol{\beta'} |\Psi_{\mathbf{Y}'}(\mathbf{x})| \boldsymbol{\alpha'}\rangle$ (left picture). We take the limit $\lim_{x_k\to x_1}$ which makes the matrices $\sig$ and $\bsig$ of the first row to simplify, as a consequence of $\mathbf{(\overline{x_1-y_k})(y_k-x_1)}=-1$ (right picture).
 \begin{center}
\includegraphics[scale=0.35]{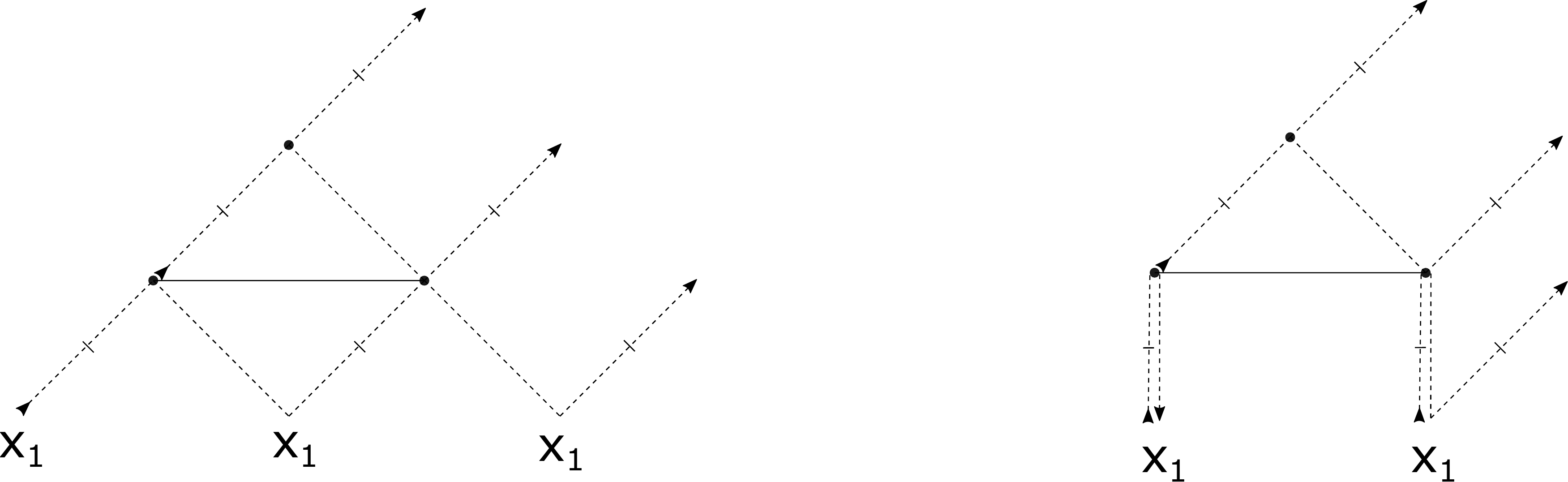}
\end{center}
Then we open the triangles in the eigenfunction by means of the identity \eqref{STR_0} (left picture) and we perform the integrations in all the points denoted by empty bullets, ending up with the right picture. 
 \begin{center}
\includegraphics[scale=0.35]{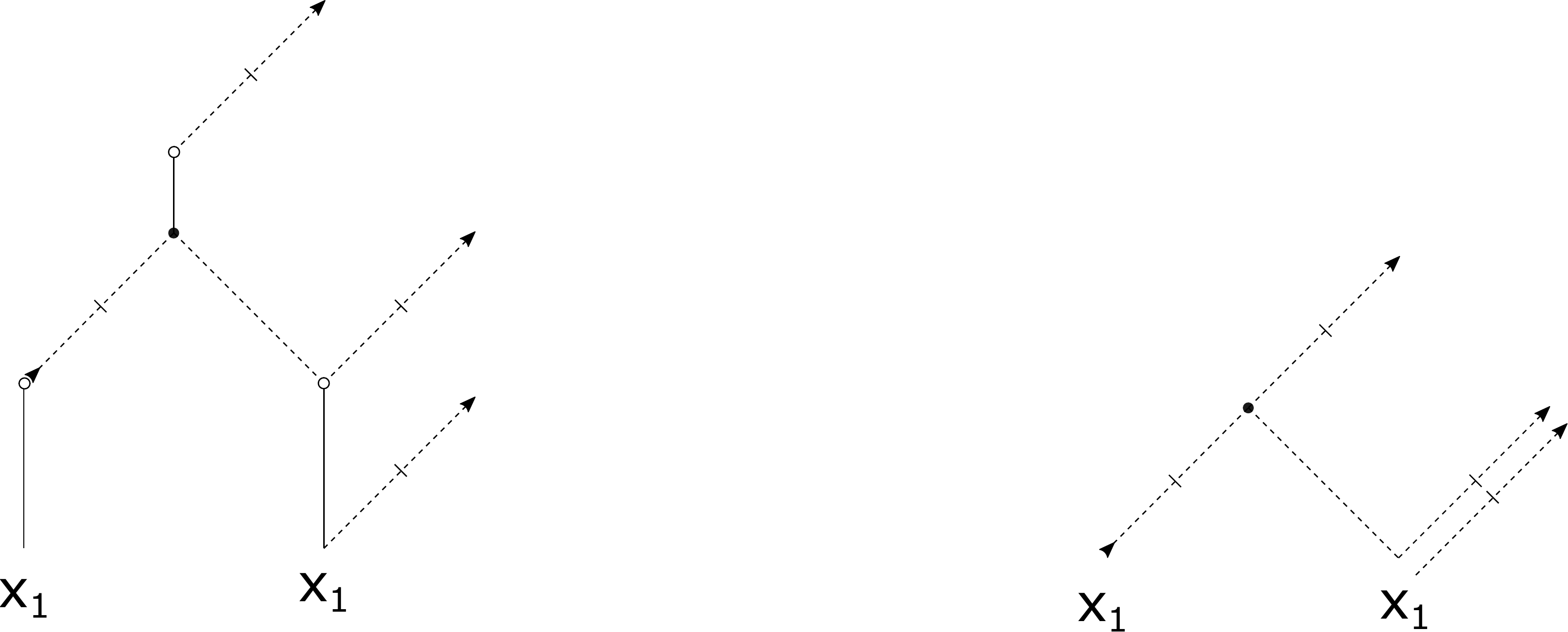}
\end{center}
The coefficient produced by these steps (depicted for $N=3$) according to \eqref{STR_0} is
\begin{equation}
\prod_{k=1}^{N-1} \pi^2 \frac{\Gamma(i\lambda)\,\Gamma\left(1-i\frac{\lambda}{2}-i\nu'_k+\frac{\ell'_k}{2} \right)\Gamma\left(1-i\frac{\lambda}{2}+i\nu'_k+\frac{\ell'_k}{2} \right)}{\Gamma(2-i\lambda)\,\Gamma\left(1+i\frac{\lambda}{2}-i\nu'_k+\frac{\ell'_k}{2} \right)\Gamma\left(1+i\frac{\lambda}{2}+i\nu'_k+\frac{\ell'_k}{2} \right)}\,,
\end{equation}
and we recognize that the last picture is the same as the initial stage, but with one row less ($N=2$). Repeating the steps until the length $N$ of the eigenfunction is reduced to one, we are left with
\begin{equation}
\label{incoming_eig}
\prod_{k=1}^N \frac{\langle \beta_k|\mathbf{(\overline{x_1-x_0})}|\alpha_k\rangle^{\ell'_k}}{(x_1-x_0)^{2(1+i\frac{\lambda}{2}+i\nu'_k)}} = \prod_{k=1}^N \langle \beta_k|\Psi_{Y_k}(x_1)|\alpha_k\rangle\,,
\end{equation}
together with the coefficient produced by the integrations and by the normalization \eqref{lambda_kernel}
\begin{equation}
\prod_{k=1}^{N}r(Y_k)^{k-1} \left(\pi^2 \frac{\Gamma(i\lambda)\Gamma\left(1-i\frac{\lambda}{2}-i\nu_k+\frac{\ell_k}{2} \right)\Gamma\left(1-i\frac{\lambda}{2}+i\nu_k+\frac{\ell_k}{2} \right)}{\Gamma(2-i\lambda)\Gamma\left(1+i\frac{\lambda}{2}-i\nu_k+\frac{\ell_k}{2} \right)\Gamma\left(1+i\frac{\lambda}{2}+i\nu_k+\frac{\ell_k}{2} \right)}\right)^{N-k}\,.
\end{equation}
Finally, specializing to the case under study $\lambda=-i$, we recognize that the last expression becomes
\begin{equation}
\label{in_coming_coeff}
\prod_{k=1}^{N}r(Y_k)^{k-1} \tau_+(0,Y_k)^{N-k}\,.
\end{equation}
 \subsection{Outgoing legs: reduction}
Let us consider the out-coming legs. In order to compute the term \eqref{purered} depicted on left with dirac deltas as dotted lines, we open the first row of triangles into star integrals according to \eqref{STR_0} and integrate out the deltas (right picture)\begin{center}
\includegraphics[scale=0.4]{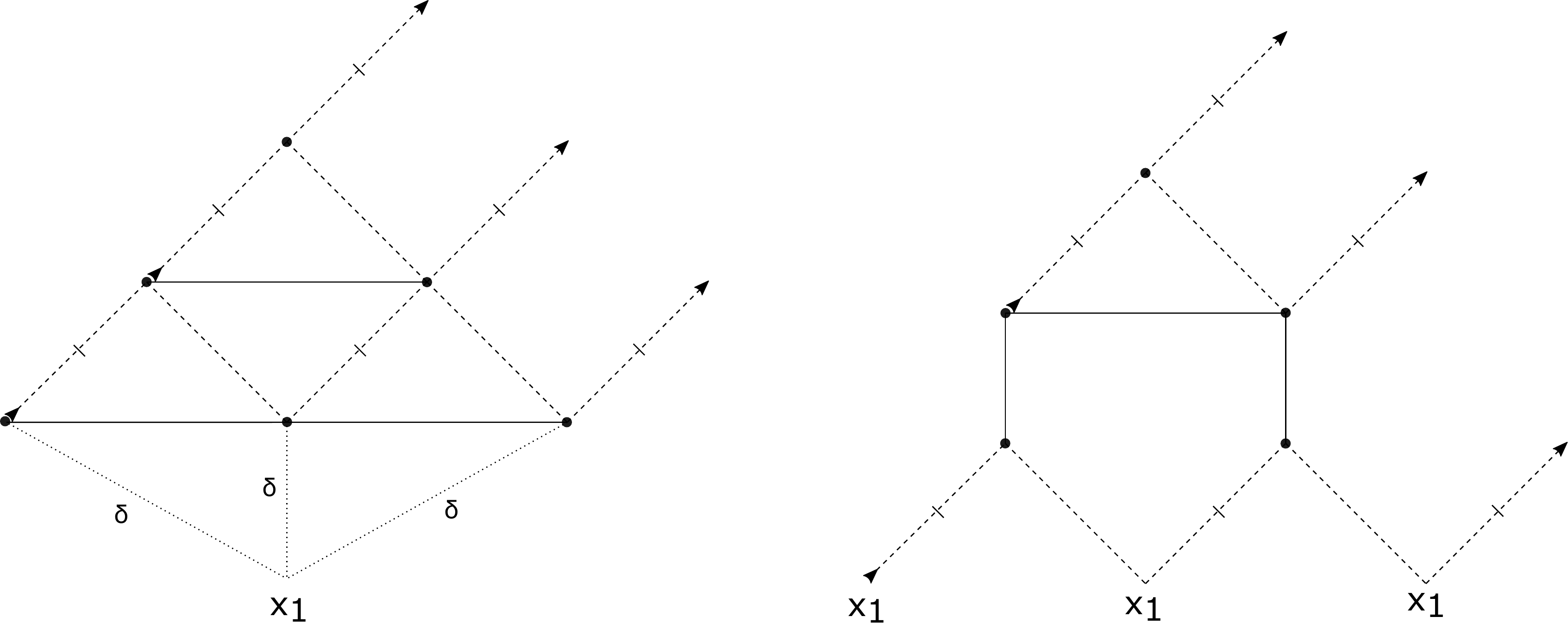}
\end{center}
As for the incoming legs, the $\sig$ and $\bsig$ matrices in the first row get paired and simplify as $\mathbf{(\overline{x_1-y_k})(y_k-x_1)}=-1$ (left picture). In the middle picture we notice that we need to apply a chain rule with powers $2+i\lambda$ and $2-i\lambda$, and recalling formula \eqref{delta} we recognize that this generates delta functions (right picture).
\begin{center}
\includegraphics[scale=0.35]{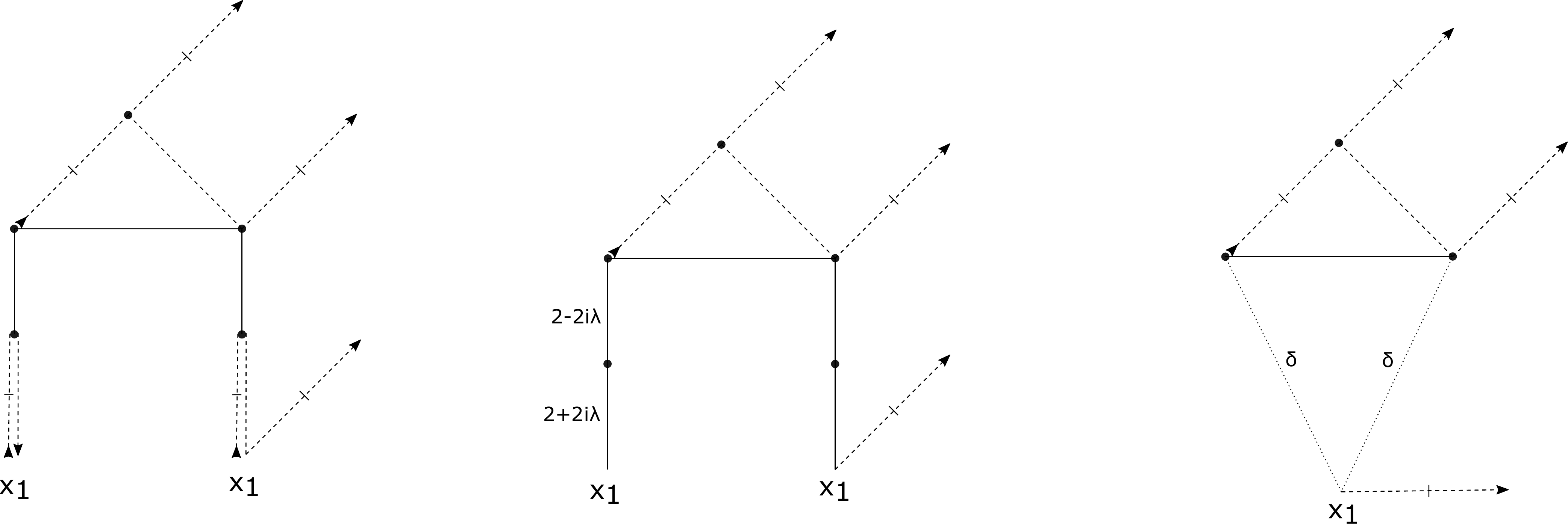}
\end{center}
The coefficient produced by the steps depicted so far (for $N=3$) is \begin{equation}
\left(\pi^2 \frac{\Gamma(i\lambda)\Gamma\left(1-i\frac{\lambda}{2}-i\nu_N+\frac{\ell_N}{2} \right)\Gamma\left(1-i\frac{\lambda}{2}+i\nu_N+\frac{\ell_N}{2} \right)}{\Gamma(2-i\lambda)\Gamma\left(1+i\frac{\lambda}{2}-i\nu_N+\frac{\ell_N}{2} \right)\Gamma\left(1+i\frac{\lambda}{2}+i\nu_N+\frac{\ell_N}{2} \right)}\right)^{N-1}\,.
\end{equation}
Finally, we notice that the last picture is the same as the initial one but with the length of the eigenfunction reduced by one (from $N=3$ to $N=2$). Therefore, the procedure can iterated until we are left with 
\begin{equation}
\label{outcoming_eig}
\prod_{k=1}^N \frac{\langle \alpha_k|\mathbf{({x'_1-x_0)}}|\beta_k\rangle^{\ell_k}}{(x'_1-x_0)^{2(1-i\frac{\lambda}{2}-i\nu_k)}} = \prod_{k=1}^N \langle\beta_k|\Psi_{Y_k}(x'_1)|\alpha_k\rangle\,,
\end{equation}
together with the coefficient produced by the integrations and the normalization \eqref{lambda_kernel}
\begin{equation}
\prod_{k=1}^{N}(r(Y_k)^*)^{k-1} \left(\pi^2 \frac{\Gamma(i\lambda)\Gamma\left(1-i\frac{\lambda}{2}-i\nu_k+\frac{\ell_k}{2} \right)\Gamma\left(1-i\frac{\lambda}{2}+i\nu_k+\frac{\ell_k}{2} \right)}{\Gamma(2-i\lambda)\Gamma\left(1+i\frac{\lambda}{2}-i\nu_k+\frac{\ell_k}{2} \right)\Gamma\left(1+i\frac{\lambda}{2}+i\nu_k+\frac{\ell_k}{2} \right)}\right)^{k-1}\,.
\end{equation}
Finally, specializing to the case under study $\lambda=-i$, we recognize that the last expression becomes
\begin{equation}
\label{outcoming_coeff}
\prod_{k=1}^{N}(r(Y_k)^*)^{k-1}\tau_+(0,Y_k)^{k-1}\,.
\end{equation}
\subsection{Final results}
At this stage we can glue together the three pieces \eqref{Bulk_result},\eqref{incoming_eig} and \eqref{outcoming_eig} by a sum over the separated variables. First we sum over primed variables $(\boldsymbol{\nu}',\boldsymbol{\ell}')$ is easily performed following leads to the simple replacement of $\mathbf{Y}'$ with $\mathbf{Y}$ according to delta functions in \eqref{Bulk_result}. Let us consider the  integration over the auxiliary spinors \begin{equation}
\int D\boldsymbol{\alpha}D\boldsymbol{\alpha}'D\boldsymbol{\beta}D\boldsymbol{\beta}'\,=\,\int D\alpha_{1}\cdots D\alpha_{N}\,  D\beta_{1}\cdots D\beta_{N}\, \int D\alpha'_{1}\cdots D\alpha'_{N}\,  D\beta'_{1}\cdots D\beta'_{N}\,,
\end{equation}
which involves the integrand 
\begin{align*}
\begin{aligned}
&\langle \boldsymbol{\alpha}|\prod_{k=1}^N \Psi^{\dagger}_{{Y_k}}(x_1)|\boldsymbol{\beta}\rangle \langle \boldsymbol{\beta}'|\prod_{k=1}^N \Psi_{{Y_{s(k)}}}(x'_1)|\boldsymbol{\alpha}'\rangle \,\sum_{s\in \mathbb{S}_N } \langle \boldsymbol{\beta} |\mathcal{R}^+\, \mathbf{S}(\mathbf{Y})^{\dagger}|\boldsymbol{\beta}'\rangle\, \langle \boldsymbol{\alpha}' |\mathbf{S}(\mathbf{Y})\, \mathcal{R}^-|\boldsymbol{\alpha}\rangle\,,
\end{aligned}
\end{align*}
and results in
\begin{equation}
\label{coord_dep}
\text{Tr}_{\ell_1,\dots,\ell_N}\Bigg [ \left(\overleftarrow{\prod_{k=1}^{N}}\Psi^{\dagger}_{{Y_k}}(x_1)\overleftarrow{\prod_{h=1}^{M_2}}\mathbf{R}^{+}_{h,k}\right)   \left(\overrightarrow{\prod_{k=1}^{N}}\Psi_{{Y_{k}}}(x'_1) \overrightarrow{\prod_{n=1}^{M_1}}\mathbf{R}^-_{n,k} \right)\Bigg ]\,,
\end{equation}
as a consequence of the completeness \cite{Perelomov:1980tt,Faddeev:1980be,Isaev:2018xcg}
\begin{equation}
\int D\alpha\,|\alpha\rangle \,\langle \alpha| = \mathbbm{1}\,,
\end{equation} 
and of the identity
\begin{equation}
\prod_{k=1}^N \Psi_{{Y_{s(k)}}}\,=\,\prod_{k=1}^N \Psi_{{Y_{k}}}\,\Longrightarrow\, \mathbf{S}(\mathbf{Y})^{-1}\, \prod_{k=1}^N \Psi_{{Y_{s(k)}}}\, \mathbf{S}(\mathbf{Y}) \, = \,\prod_{k=1}^N \Psi_{{Y_{k}}} \,.
\end{equation}
The dependence over the coordinates of the diagram is described by \eqref{coord_dep}.
Now that dependence over coordinate is set, let us collect the coefficients produced along the computation; putting together \eqref{in_coming_coeff}, \eqref{outcoming_coeff} and \eqref{Bulk_eigen_gen}, we get
\begin{align}
\label{final_eigenv}
\begin{aligned}
&\prod_{k=1}^{N} 
\tau_+(0,Y_k)^{N+L_1+L_2}\,\tau_+(1,Y_k)^{M_1} \,\tau_-(1,Y_k)^{M_2}\,.
\end{aligned}
\end{align}
The resulting expression for \eqref{diag_transf} is then given by the sum over separated variables $(\boldsymbol{\nu},\boldsymbol{\ell})$ of \eqref{coord_dep} and \eqref{final_eigenv}
\begin{align}
\label{F_graph}
\begin{aligned}
 F(x_0,x_1,&x_1')=\frac{1}{(x_1-x_0)^{2N}(x_1'-x_0)^{2N}}\times\\  &\times\sum_{\boldsymbol{\ell}}\int d\boldsymbol{\nu} \,\mu(\boldsymbol{\nu},\boldsymbol{\ell})\,\text{Tr}_{\ell_1,\dots,\ell_N} \left[\left({\prod_{k=1}^{N}}[(\mathbf{{x_1-x_0}})]^{\ell_k} \right) \mathcal{R}^+  \left({\prod_{k=1}^{N}}[(\mathbf{\overline{x'_1-x_0}})]^{\ell_k}\right)\mathcal{R}^- \right]\times \\ &\qquad\qquad\qquad\qquad\times \prod_{k=1}^{N} 
\tau_+(0,Y_k)^{N+L_1+L_2}\,\tau_+(1,Y_k)^{M_1} \,\tau_-(1,Y_k)^{M_2}\frac{(x_1'-x_0)^{2i\nu_k}}{(x_1-x_0)^{2i\nu_k}}\,  \,.
\end{aligned}
\end{align}
In addition we can express the result via the conformal invariant cross ratios $u=\frac{x_{1'0}^2 x_{10'}^2}{x_{10}^2 x_{1'0'}^2}$ and $v=\frac{x_{11'}^2 x_{00'}^2}{x_{10}^2 x_{1'0'}^2}$, which are amputated after sending $x_0'\to \infty$ according to the transormation \eqref{x0'_ell}\eqref{x0'_ell1} which as
\begin{equation}
u=\frac{x_{1'0}^2 }{x_{10}^2}\,,\,\,\,\,\,v=\frac{x_{11'}^2 }{x_{10}^2 }\,,
\end{equation}
through the complex parameters $(z,\bar z)$ solving the equations
\begin{equation}
 u=z\bar z\,,\,\,\,\,\,v=(1-z)(1-\overline{z})\,.
\end{equation}
Therefore we recognize that
\begin{equation}
\frac{(x_1'-x_0)^{2i\nu_k}}{(x_1-x_0)^{2i\nu_k}} = (z\bar z)^{i\nu_k}\,,
\end{equation}
moreover the spinor indices of the graph are carried by a scale-invariant function $W_{\boldsymbol \ell}$ of the unit vectors $(x'_1-x_0)^{\mu_k}/|x'_1-x_0|$  and $(x_1-x_0)^{\nu_k}/|x_1-x_0|$ 
\begin{align}
\label{W_spin}
\begin{aligned}
W_{\boldsymbol \ell}&\left(\mathbf{x_{10}},\overline{\mathbf{x_{1'0}}}\right)^{c_1,\dots,c_{M_1},\dot c_1,\dots,\dot c_{M_2}}_{a_1,\dots,a_{M_1},\dot a_1,\dots,\dot a_{M_2}}=\\&=\text{Tr}_{\ell_1,\dots,\ell_N} \left[\left({\prod_{k=1}^{N}}[(\mathbf{{x_1-x_0}})]^{\ell_k} \right) \mathcal{R}^+  \left({\prod_{k=1}^{N}}[(\mathbf{\overline{x'_1-x_0}})]^{\ell_k}\right)\mathcal{R}^- \right]^{c_1,\dots,c_{M_1},\dot c_1,\dots,\dot c_{M_2}}_{a_1,\dots,a_{M_1},\dot a_1,\dots,\dot a_{M_2}}=\\&=\text{Tr}_{\ell_1,\dots,\ell_N}\Bigg [\left(\overleftarrow{\prod_{k=1}^{N}}[(\mathbf{{x_1-x_0}})]^{\ell_k}\overleftarrow{\prod_{h=1}^{M_2}}\mathbf{R}^{+}_{h,k}\right)   \left(\overrightarrow{\prod_{k=1}^{N}}[(\mathbf{\overline{x'_1-x_0}})]^{\ell_k} \overrightarrow{\prod_{n=1}^{M_1}}\mathbf{R}^-_{n,k} \right)\Bigg]^{c_1,\dots,c_{M_1},\dot c_1,\dots,\dot c_{M_2}}_{a_1,\dots,a_{M_1},\dot a_1,\dots,\dot a_{M_2}}\,.
\end{aligned}
\end{align}
The r.h.s. of \eqref{W_spin}, after factoring out the coordinate dependence reads
\begin{align}
\begin{aligned}
\text{Tr}_{\ell_1,\dots,\ell_N}\Bigg [\left(\overleftarrow{\prod_{k=1}^{N}}[\sig^{\nu_1}\otimes \cdots\otimes \sig^{\nu_{\ell_k}}]\overleftarrow{\prod_{h=1}^{M_2}}\mathbf{R}^{+}_{h,k}\right)   \left(\overrightarrow{\prod_{k=1}^{N}}[\bsig^{\mu_1}\otimes \cdots\otimes \bsig^{\mu_{\ell_k}}] \overrightarrow{\prod_{n=1}^{M_1}}\mathbf{R}^-_{n,k} \right)\Bigg]^{c_1,\dots,c_{M_1},\dot c_1,\dots,\dot c_{M_2}}_{a_1,\dots,a_{M_1},\dot a_1,\dots,\dot a_{M_2}}\,,
\end{aligned}
\end{align}
and can be computed explicitly for any $N,M_1,M_2$ and $\ell_1,\dots,\ell_N$ given the explicit form of $\mathbf{R}$-matrix (see appendix \ref{app:R}) and the  simple properties of matrices $\sig$, $\bsig$ under trace. 
We point out that our result for $M_1=M_2=0$ specializes to the scalar Basso-Dixon diagram contributing to the correlators \eqref{BDcorr} and \eqref{BDchicft}. In this case, we notice that formula \eqref{coord_dep} can be worked out more explicitly as the R matrices disappear and the function $W_{\boldsymbol \ell}$ simplifies:
\begin{align}
\label{BDscalar_Gegen}
\begin{aligned}
{\text{Tr}([\mathbf{(\overline{x_1-x_0})}]^{\ell}[\mathbf{({x'_1-x_0})}]^{\ell})}=\\=&\,\text{Tr}(\bsig_{\mu_1}\sig_{\rho_1})\cdots \text{Tr}(\bsig_{\mu_\ell}\sig_{\rho_\ell})\frac{({x_1-x_0})^{\mu_1\dots\mu_\ell}{({x'_1-x_0})}^{\rho_1\dots \rho_\ell}}{|x_1'-x_0|^{\ell}|x_1-x_0|^{\ell}} =\\=&\,\frac{2^{\ell}\,{({x_1-x_0})}^{\mu_1\dots\mu_\ell}{({x'_1-x_0})}_{\mu_1\dots \mu_\ell}}{|x_1'-x_0|^{\ell}|x_1-x_0|^{\ell}}={C^{1}_\ell\left(\cos\theta\right)} \,,
 \end{aligned}
\end{align}
where $C^{1}_\ell\left(\cos\theta\right)$ is a Gegenbauer polynomial of degree $\ell$ in the variable $\cos \theta = (x_1-x_0)^{\mu}(x_1'-x_0)_{\mu}/|x_1-x_0||x_1'-x_0|$, and we recall that
\begin{equation}
C^{1}_\ell\left(\cos\theta\right) = \frac{\sin((\ell+1)\theta)}{\sin(\theta)}\,.
\end{equation}
In the variables $(z,\bar z)$ such Gegenbauer polynomials read
\begin{equation}
C^{1}_\ell\left(\cos\theta\right) = \frac{(z/\bar{z})^{\frac{\ell+1}{2}}-(z/\bar{z})^{-\frac{\ell+1}{2}}}{(z/\bar{z})^{\frac{1}{2}}-(z/\bar{z})^{-\frac{1}{2}}}\,,
\end{equation}
so that \eqref{W_spin} turns into
\begin{equation}
\label{Wl_simple}
W_{\boldsymbol \ell}(z/\bar z)=\prod_{k=1}^N \frac{(z\bar z)^{\frac{1}{2}}}{z-\bar z}\left[ {\left(\frac{z}{\bar{z}}\right)^{\frac{\ell_k+1}{2}}-\left(\frac{z}{\bar{z}}\right)^{-\frac{\ell_k+1}{2}}}\right]\,.
\end{equation}
Finally, we can write the result for \eqref{diag_transf} as
\begin{align}
\label{Final_F_graph}
\begin{aligned}
 F(&x_0,x_1,x_1')=\frac{1}{(x_1-x_0)^{2N}(x_1'-x_0)^{2N}}\times\\  &\times\sum_{\boldsymbol{\ell}}\,W_{\boldsymbol \ell}\left(\mathbf{x_{10}},\overline{\mathbf{x_{1'0}}}\right)\int d\boldsymbol{\nu} \,\mu(\boldsymbol{\nu},\boldsymbol{\ell}) \prod_{k=1}^{N} (z\bar z)^{i\nu_k}
\tau_+(0,Y_k)^{N+L_1+L_2}\,\tau_+(1,Y_k)^{M_1} \,\tau_-(1,Y_k)^{M_2}\,  \,.
\end{aligned}
\end{align}
In the case $M_1=M_2=0$, it follows from \eqref{Wl_simple} that formula \eqref{Final_F_graph} takes the simple form
\begin{align}
\begin{aligned}
\label{BD_result}
\frac{1}{(x_1-x_0)^{2N}(x_1'-x_0)^{2N}}\left[\frac{(z\bar z)^{\frac{1}{2}}}{z-\bar z}\right]^N\sum_{\boldsymbol{\ell}}\int d\boldsymbol{\nu} 
\,\mu(\boldsymbol{\nu},\boldsymbol{\ell})\, \prod_{k=1}^N\,\frac{(z\bar z)^{i\nu_k}\left[(z/\bar{z})^{\frac{\ell_k+1}{2}}-(\bar z/{z})^{\frac{\ell_k+1}{2}}\right]}{\left(\frac{(\ell_k+1)^2}{4}+\nu_k^2\right)^{N+L_1+L_2}}\,,
\end{aligned}
\end{align}
Furthermore, as a consequence of the properties \begin{align}
\begin{aligned}
&\mu(\boldsymbol{\nu},\ell_1,\dots ,-\ell_k-2,\dots,\ell_N)=-\mu(\boldsymbol{\nu},\ell_1,\dots ,\ell_k,\dots,\ell_N)\,,\\
&\mu(\boldsymbol{\nu},\ell_1,\dots ,-1,\dots,\ell_N)=0\,,
\end{aligned}
\end{align}
it is possible to rewrite \eqref{BD_result} extending the summation of $\ell_k$ over $\mathbb{Z}$ as
\begin{align}
\begin{aligned}
\label{BD_result1}
\frac{1}{(x_1-x_0)^{2N}(x_1'-x_0)^{2N}}\left[\frac{(z\bar z)^{\frac{1}{2}}}{z-\bar z}\right]^N\sum_{\boldsymbol{\ell}\in \mathbb{Z}}\int d\boldsymbol{\nu} 
\,\mu(\boldsymbol{\nu},\boldsymbol{\ell})\, \prod_{k=1}^N\,\frac{z^{i\nu_k+\frac{\ell_k+1}{2}}{\bar z}^{i\nu_k-\frac{\ell_k+1}{2}}}{\left(\frac{(\ell_k+1)^2}{4}+\nu_k^2\right)^{N+L_1+L_2}}\,,
\end{aligned}
\end{align}
which explains our previous findings \cite{Derkachov_Oliv} and reproduces the conjecture of Basso and Dixon \cite{Basso}.

In order to deliver the formula for \eqref{Full_correlator}, we need to perform on \eqref{Final_F_graph} the inverse of transformations \eqref{x0'_ell}\eqref{x0'_ell1}, from $F(x_0,x_1,x'_1)$ to $G(x_0,x_1,x'_0,x'_1)$
\begin{align}
\label{Final_G_graph}
\begin{aligned}
 G(&x_0,x_1,x'_0,x_1')=\frac{1}{(x_1-x_0)^{2N}(x_1'-x_0)^{2N}(x_0-x'_0)^{2(N+L_1+L_2-3/2(M_1+M_2))}}\times\\  &\times\sum_{\boldsymbol{\ell}}\,\tilde W_{\boldsymbol \ell}\left(x_0,x_1,x'_0,x_1'\right)\int d\boldsymbol{\nu} \,\mu(\boldsymbol{\nu},\boldsymbol{\ell}) \prod_{k=1}^{N} (z\bar z)^{i\nu_k}
\tau_+(0,Y_k)^{N+L_1+L_2}\,\tau_+(1,Y_k)^{M_1} \,\tau_-(1,Y_k)^{M_2}\, ,
\end{aligned}
\end{align}
where the polynomial ${W}_{\boldsymbol \ell}$ gets transformed into $\tilde{W}_{\boldsymbol \ell}$ according to
\begin{align}
\begin{aligned}
 &(\tilde{W}_{\boldsymbol \ell})_{a_1,\dots,a_{M_1},\dot a_1,\dots,\dot a_{M_2}}^{\dot b_1,\dots,\dot b_{M_1}, b_1,\dots, b_{M_2}}\left(x_0,x_1,x'_0,x'_1\right)=\\&=W_{\boldsymbol \ell}\left(\mathbf{\overline{x_{0'1}}x_{10}\overline{x_{00'}}}\,,\,\mathbf{x_{0'1'}\overline{x_{1'0}}x_{00'}}\right)^{c_1,\dots,c_{M_1},\dot c_1,\dots,\dot c_{M_2}}_{a_1,\dots,a_{M_1},\dot a_1,\dots,\dot a_{M_2}}\,[
\mathbf{{x_{00'}}}]_{c_1,}^{\dot b_1}\cdots [\mathbf{{x_{00'}}}]_{c_{M_1}}^{\dot b_{M_1}}[\mathbf{\overline{x_{00'}}}]_{\dot c_1}^{b_1}\cdots [\mathbf{\overline{x_{00'}}}]_{\dot c_{M_2}}^{b_{M_2}} \,=\\&=\text{Tr}_{\ell_1,\dots,\ell_N}\Bigg [\left(\overleftarrow{\prod_{k=1}^{N}}[(\mathbf{{x_{0'1}}\overline{x_{10}}})]^{\ell_k}\overleftarrow{\prod_{h=1}^{M_2}}([\mathbf{\overline{x_{00'}}}]\mathbf{R}^{+}_{h,k})_{\dot a_h}^{b_h}\right)   \left(\mathbf{\overrightarrow{\prod_{k=1}^{N}}[(x_{1'0}\overline{x_{10'}}})]^{\ell_k} \overrightarrow{\prod_{n=1}^{M_1}}(\mathbf{R}^-_{n,k} [\mathbf{{x_{00'}}}])_{a_n}^{\dot b_n}\right)\Bigg]\,.
\end{aligned}
\end{align}

\newpage
\section{Conclusions}
In this paper we studied a class of exactly-solvable planar four point correlators in the doubly-scaled $\gamma$-deformation of $\mathcal{N}=4$ SYM (also dubbed $\chi$ CFT$_4$) \cite{Gurdogan:2015csr}. The structure of $SU(N_c)$ indices of such correlators is that of a single trace which contains $N$ complex scalar fields $\phi_1$ propagating from the point $x_1$ to $x'_1$ which cross - along their propagation - a number of fields $\phi_2,\phi_3,\psi_2,\psi_3$ propagating from $x_0$ to $x'_0$. The planar limit of such correlators is dominated by only one Feynmann integral which can be regarded as the generalization of Basso-Dixon fishnet integrals \cite{Basso,Derkachov2018,Derkachov_Oliv} by the introduction of fermionic fields $\psi_2,\psi_3$ at the points $x_0$ and $x'_0$ of the diagram. Here we point out that for the choice of only fermionic fields $\psi_2,\psi_3$ at the points $x_0$ and $x'_0$, the Feynmann integrals we computed are the dominant contribution in the planar limit of the same correlators of the ``fermionic" fishnet theory studied in \cite{Pittelli2019}. We have showed that the computation of such Feynman integrals can be mapped to the diagonalization of an integrable spin magnet with $SO(1,5)$ conformal spins with open boundary conditions and a number of sites $N$ equal to the number of fields $\phi_1$ in the correlator. 

The map between Feynmann integrals and spin magnet follows from the observation \cite{Zamolodchikov:1980mb,Gurdogan:2015csr} that a square-lattice fishnet can be regarded as the result of iterative applications of an integral kernel (graph-building operator) which happens to be a commuting Hamiltonian of the integrable spin magnet. In the presence of fermionic fiels the same logic can be repeated for the Yukawa hexagonal lattice built by an integral kernel which - modulo mixing of spinor indices - is ``diagonalized" by the same eigenfunctions of the square-lattice kernel.
In this paper we give a detailed explanation of how to construct the eigenfunctions of the spin chain and prove their orthogonality, extending the logic of \cite{Derkachov2001,Derkachov2014} from $2D$ to $4D$ euclidean spacetime. Our results - derived for a spin chain in the principal series representation of $SO(1,5)$ can be analytically continued from the representation of the principal series to real scaling dimensions, recovering the graph-building operator - introduced in $2D$ by the authors and V.~Kazakov \cite{Derkachov2018} - for the Feynmann diagrams of Fishnet CFT \cite{Gurdogan:2015csr,Caetano:2016ydc} - as well as its fermionic counterpart which builds portions of Yukawa hexagonal lattice.

The construction of the spin chain's eigenfunctions, first appeared in our letter \cite{Derkachov_Oliv}, is explained in this paper via the detailed introduction of a light and handy graphical computational technique, ultimately based on the star-triangle relation of section \ref{sec:startri}. A nice feature in this respect is that the star-triangle identities for propagators with spins $\ell/2$ and $\ell'/2$ differs from the previously known scalar identity \cite{DEramo:1971hnd} by the appearance of a solution of the Yang-Baxter equation $\mathbf{R}_{\ell,\ell'}$ obtained by fusion in two channels of the $SU(2)$ Yang R matrix, responsible for mixing the spinor indices. The technique we developed should provide an extension of integration-by-parts and Gegenbauer polynomial techniques \cite{Chetyrkin:1980pr,Chetyrkin:1981qh,Kazakov:1983pk} to the computation of multi-loop massless Feynmann integrals which contain also fermionic propagators. Moreover it would be interesting to develop an analogue formalism in any space-time dimension $d>4$ and generalize our results to the computation of correlators in the $d$-dimensional fishnet theory proposed in \cite{Kazakov}. 

In our letter we conjectured \cite{Derkachov_Oliv} that the eigenstates of the spin chain magnet should be used for a first-principle derivation in the bi-scalar Fishnet CFT (or in the more general $\chi$CFT$_4$) of the cutting-and-gluing procedures based on decomposing correlators of $\mathcal{N}=4$ SYM into exactly-solvable polygonal building blocks labeled by a set of mirror excitations of the dual string theory \cite{Basso:2013vsa,Basso:2015zoa,Eden:2016xvg,Fleury:2016ykk,Fleury:2017eph,Basso2018}. In our formalism the mirror excitations labels are replaced by the quantum number of the spectrum of the $SO(1,5)$ spin chain,  which at size $N$ is factorized into $N$ equal contribution - one for each separated variable. In this respect it would be interesting to see if, in the bi-scalar fishnet reduction, the separated variables label the mirror excitation of the string-bits model proposed in \cite{Gromov2019c,Gromov2019b,Gromov2019a} as the holographic dual of the Fishnet CFT. More generally, it would be interesting to use the basis of eigenfunctions for the $SO(1,5)$ spin magnet to derive the $\chi$CFT$_4$ version of several $\mathcal{N}=4$ SYM quantities computed recently by the aforementioned decomposition procedures, with the task of unveiling details on the double-scaling limit behaviour of $\mathcal{N}=4$ SYM theory, in relation to its integrability features. In this respect we point out that, despite the correlators under study are not well defined in $\mathcal{N}=4$ SYM theory due to the lack of gauge invariance, collecting the same elementary fields in four traces (each for each point), one obtain a gauge invariant function whose perturbative expansion is essentially made by Basso-Dixon integrals\footnote{This observation was suggested by P.~Vieira.}. 

Finally, inspired by the $2D$ results of \cite{Derkachov2001}, we believe that the eigenfunctions discussed in this paper are closely related to the transformation to quantum separated variables of the periodic conformal $SO(1,5)$ spin chain which underlies the integrability of the Fishnet CFT. The separation of variables for non-compact spin magnets is a topic which recently attracted great attention~\cite{Maillet2018,Maillet2019a,Gromov2017,Cavaglia2019,Ryan2019,Gromov2019,Ryan:2020rfk}, and SoV features appear in remarkable results of AdS/CFT integrability, for instance~\cite{Kostov:2019stn,Jiang2019a}. It has not escaped our notice that the properties of the proposed eigenfunctions immediately suggest their role in the SoV of the periodic $SO(1,5)$ spin chain~\cite{Chicherin2013a} and that, in full analogy with~\cite{Derkachov2001}, the operators $\hat Q_{\nu,\ell}(u)$ or $\check Q_{\nu,\ell}(u)$ introduced in \ref{sect:Qspin} must be close relatives of the Baxter $Q$-operators.  

\begin{acknowledgments}
\label{sec:acknowledgments}

We want to thank P.~Vieira and A.~Manashov for numerous useful discussions. We are grateful to A.~Isaev, A.~Manashov and M.~Preti who read and commented our manuscript. The work of S.D. is supported by the Russian Science Foundation project No 19-11-00131. The work of E. O. was funded by the German Science Foundation (DFG) under the Research Training Group 1670 and under Germany's Excellence Strategy -- EXC 2121 ``Quantum Universe" -- 390833306. E.O. would like to thank FAPESP grant 2016/01343-7 for funding the visit to ICTP-SAIFR from February to April 2020 where part of this work was done.

\end{acknowledgments}
\section*{Appendix}
\appendix

\renewcommand{\theequation}{\Alph{section}.\arabic{equation}}
\renewcommand{\thetable}{\Alph{table}}
\setcounter{section}{0}
\setcounter{table}{0}

\section{Transformation from tensors to spinors}
\label{app:spinors}
The standard Pauli matrices $\sigma_k$, where $k=1,2,3$ have the form
$$
\sigma_1 = \left
(\begin{array}{cc}
0 & 1 \\
1 & 0 \end{array} \right )\ ;\ \sigma_2 = \left (\begin{array}{cc}
0 & -i \\
i & 0 \end{array} \right )\ ;\ \sigma_3 = \left (\begin{array}{cc}
1 & 0 \\
0 & -1 \end{array} \right )\,
$$
To convert the tensors to the spinors and back
we shall need 2-dimensional matrices
$\boldsymbol{\sigma}_{\mu}$ and $\overline{ \boldsymbol{\sigma} }_{\mu}$,
where $\mu=0,1,2,3$ and $\boldsymbol{\sigma}_{0} = \overline{\boldsymbol{\sigma}}_{0} = \II\ \,,
\boldsymbol{\sigma}_{k} = -\overline{\boldsymbol{\sigma}}_{k} = i\sigma_k$
so that $\boldsymbol{\sigma}^{\dagger}_{\mu} = \overline{ \boldsymbol{\sigma} }_{\mu}$
\begin{align}\label{E}
\boldsymbol{\sigma}_{\mu} = ( \II , i \sigma_1, i \sigma_2, i \sigma_3 ) \; ,
\qquad \overline{ \boldsymbol{\sigma} }_{\mu} = \boldsymbol{\sigma}^{\dagger}_{\mu} =
 ( \II , - i \sigma_1, - i \sigma_2, - i \sigma_3 ) \; .
\end{align}
The important relations are
\begin{align}
\tr \overline{\boldsymbol{\sigma}}_{\mu}\boldsymbol{\sigma}_{\nu} =
\tr \boldsymbol{\sigma}_{\mu}\overline{\boldsymbol{\sigma}}_{\nu} = 2\,\delta_{\mu\nu}\\
\overline{\boldsymbol{\sigma}}_{\mu}\boldsymbol{\sigma}_{\nu}+
\overline{\boldsymbol{\sigma}}_{\nu}\boldsymbol{\sigma}_{\mu} = 2\,\delta_{\mu\nu}\II
\ \,, \ \boldsymbol{\sigma}_{\mu}\overline{\boldsymbol{\sigma}}_{\nu}+
\boldsymbol{\sigma}_{\mu}\overline{\boldsymbol{\sigma}}_{\nu} = 2\,\delta_{\mu\nu}\II \\
\overline{\boldsymbol{\sigma}}_{\mu}\otimes\boldsymbol{\sigma}_{\mu} = 2\,\boldsymbol{P}\ \,, \
\boldsymbol{\sigma}_{\mu}\otimes\boldsymbol{\sigma}_{\mu} = 2(\boldsymbol{P}-\II)
\end{align}
where we suppose summation over repeated index and do not distinguish upper and lower $\mu\,,\nu$-indices because metric $\delta_{\mu\nu}$ is Euclidean.
The relations in the last line are Fierz identities: $\boldsymbol{P}$ and $\II$
are the permutation and identity operators in the tensor product of two-dimensional spaces $\mathbb{C}^2\otimes \mathbb{C}^2$.
In full analogy with formulae from the spinor representations
of the usual Lorentz group \cite{Isaev:2018xcg} it is possible to write everything in
explicit index form with dotted and un-dotted spinor indices
$\boldsymbol{\sigma}_{\mu}= ||(\boldsymbol{\sigma}_{\mu})_{a\dot{a}}||$
and $\overline{ \boldsymbol{\sigma} }_{\mu} = ||(\overline{ \boldsymbol{\sigma} }_{\mu})^{\dot{a}a}||$
\begin{align*}
\left(\overline{\boldsymbol{\sigma}}_{\mu}\right)^{\dot{a}c}
\left(\boldsymbol{\sigma}_{\nu}\right)_{c\dot{b}}+
\left(\overline{\boldsymbol{\sigma}}_{\nu}\right)^{\dot{a}c}
\left(\boldsymbol{\sigma}_{\mu}\right)_{c\dot{b}} =
2\,\delta_{\mu\nu}\,\delta^{\dot{a}}_{\dot{b}}
\ \,, \
\left(\boldsymbol{\sigma}_{\mu}\right)_{a\dot{c}}
\left(\overline{\boldsymbol{\sigma}}_{\nu}\right)^{\dot{c}b}+
\left(\boldsymbol{\sigma}_{\nu}\right)_{a\dot{c}}
\left(\overline{\boldsymbol{\sigma}}_{\mu}\right)^{\dot{c}b} =
2\,\delta_{\mu\nu}\,\delta_a^{b}\\
\left(\boldsymbol{\sigma}_{\mu}\right)_{a\dot{a}}
\left(\overline{\boldsymbol{\sigma}}_{\mu}\right)^{\dot{b}b} =
2\,\delta_a^{b}\,\delta^{\dot{b}}_{\dot{a}}\ \ ;\ \
\left(\overline{\boldsymbol{\sigma}}_{\mu}\right)^{\dot{a}a}
\left(\boldsymbol{\sigma}_{\mu}\right)_{b\dot{b}} =
2\,\delta_b^{a}\,\delta^{\dot{a}}_{\dot{b}}\\
\left(\overline{\boldsymbol{\sigma}}_{\mu}\right)^{\dot{a}a}
\left(\overline{\boldsymbol{\sigma}}_{\mu}\right)^{\dot{b}b} =
2\,\varepsilon^{ab}\,\varepsilon^{\dot{a}\dot{b}}\ \ ;\ \
\left(\boldsymbol{\sigma}_{\mu}\right)_{a\dot{a}}
\left(\boldsymbol{\sigma}_{\mu}\right)_{b\dot{b}} =
2\,\varepsilon_{ab}\,\varepsilon_{\dot{a}\dot{b}}
\end{align*}
Arbitrary tensor $t_{\mu_1\ldots\mu_n}$ can be converted to the spinor
$\psi_{a_1\ldots a_n\dot{b}_1\ldots \dot{b}_n}$
\begin{align}\label{ts}
\psi_{a_1\ldots a_n\dot{b}_1\ldots \dot{b}_n} =
\left(\boldsymbol{\sigma}_{\mu_1}\right)_{a_1 \dot{b}_1}\ldots
\left(\boldsymbol{\sigma}_{\mu_n}\right)_{a_n \dot{b}_n}\, t_{\mu_1\ldots\mu_n}
\end{align}
and there is formula for the inverse transformation
\begin{align}\label{st}
t_{\mu_1\ldots\mu_n} = 2^{-n}\,
\left(\overline{\boldsymbol{\sigma}}_{\mu_1}\right)^{\dot{b}_1 a_1}\ldots
\left(\overline{\boldsymbol{\sigma}}_{\mu_n}\right)^{\dot{b}_n a_n}\,
\psi_{a_1\ldots a_n\dot{b}_1\ldots \dot{b}_n}
\end{align}
Let us consider the spinor which is symmetric with
respect to any permutations of indices in the group $a_1\ldots a_n$ and $\dot{b}_1\ldots \dot{b}_n$ independently
$\psi_{(a_1\ldots a_n)(\dot{b}_1\ldots \dot{b}_n)}$ and convert it to the tensor using~(\ref{st})
\begin{align}
t_{\mu_1\ldots\mu_n} = 2^{-n}\,
\left(\overline{\boldsymbol{\sigma}}_{\mu_1}\right)^{\dot{b}_1 a_1}\ldots
\left(\overline{\boldsymbol{\sigma}}_{\mu_n}\right)^{\dot{b}_n a_n}\,
\psi_{(a_1\ldots a_n)(\dot{b}_1\ldots \dot{b}_n)}
\end{align}
Then from the symmetry $\psi_{(a_1\ldots a_n)(\dot{b}_1\ldots \dot{b}_n)}$ follows that $t_{\mu_1\ldots\mu_n}$ is symmetric and traceless:
the relation $\delta_{\mu_i\mu_k}\,t_{\mu_1\ldots\mu_n} = 0$
is the consequence of the Fierz identity   $\left(\overline{\boldsymbol{\sigma}}_{\mu}\right)^{\dot{a}a}
\left(\overline{\boldsymbol{\sigma}}_{\mu}\right)^{\dot{b}b} =
2\varepsilon^{ab}\,\varepsilon^{\dot{a}\dot{b}}$.

We should note that meaning of dotted and un-dotted indices and upper and lower indices in Euclidean space are different from the Minkovski space.

In Minkovski space-time the symmetry group is the Lorentz group $L^{\uparrow}_+$ and there is isomorphism $L^{\uparrow}_+ \simeq SL(2,\mathbb C)/\mathbb Z_2$.
The matrices $\boldsymbol{\sigma}_{\mu}$ and $\overline{\boldsymbol{\sigma}}_{\mu}$ in Minkovski space-time are different from their Euclidean version~(\ref{E})
$$
\boldsymbol{\sigma}_{\mu} = ( \II , \sigma_1, \sigma_2, \sigma_3 ) \; ,
\qquad \overline{ \boldsymbol{\sigma} }_{\mu} =
 ( \II , - \sigma_1, - \sigma_2, - \sigma_3 ) \;
$$
and they connected as follows
$$
\overline{ \boldsymbol{\sigma} }_{\mu} = -\varepsilon\,
\boldsymbol{\sigma}_{\mu}^{*} \varepsilon \ \ ;\ \
\varepsilon = \left
(\begin{array}{cc}
0 & 1 \\
-1 & 0 \end{array} \right )
$$
where $*$ is complex conjugation.
The same formula in index notations is
\begin{align*}
\left(\overline{\boldsymbol{\sigma}}_{\mu}\right)^{\dot{a}a} =
\varepsilon^{\dot{a}\dot{b}} \varepsilon^{ab} \left(\left(\boldsymbol{\sigma}_{\mu}\right)_{b\dot{b}}\right)^{*}
\end{align*}
so that the index raising and lowering operations are carried out using
two dimensional Levi-Chivita tensors $\varepsilon^{\dot{a}\dot{b}}$ and $\varepsilon^{a b}$ and complex conjugation interchanges dotted and un-dotted indexes.
Each matrix $U\in SL(2,\mathbb C)$ defines matrix
$\Lambda \in L^{\uparrow}_+$ by the formula
\begin{align}
U \boldsymbol{\sigma}_\mu U^{\dagger} = \boldsymbol{\sigma}_{\nu}
\Lambda^{\nu}_{\ \mu} \longrightarrow U_{a}^{\ b}
\left(U^{*}\right)_{\dot{a}}^{\ \dot{b}}
\left(\boldsymbol{\sigma}_\mu\right)_{b\dot{b}} =
\left(\boldsymbol{\sigma}_\nu\right)_{a\dot{a}}
\Lambda^{\nu}_{\ \mu}
\end{align}
so that the tensor $t^{\mu_1\cdots\mu_n}$ with standard
law of transformation
$$
t^{\mu_1\cdots\mu_n} \to
\Lambda^{\mu_1}_{\ \nu_1}\cdots \Lambda^{\mu_n}_{\ \nu_n}\,
t^{\nu_1\cdots\nu_n}
$$
is transformed by~(\ref{ts}) to the spinor with
law of transformation
\begin{align}\label{nm}
\psi_{a_1\ldots a_n\dot{b}_1\ldots \dot{b}_n } \to
U_{a_1}^{\ c_1}\cdots U_{a_n}^{\ c_n}\,
\left(U^{*}\right)_{\dot{b}_1}^{\ \dot{d}_1}\cdots \left(U^{*}\right)_{\dot{b}_n}^{\ \dot{d}_n}\,
\psi_{c_1\ldots c_n\dot{d}_1\ldots \dot{d}_n}
\end{align}
and dotted and un-dotted indices correspond to
spinors that are transformed according to
nonequivalent complex conjugated representations of $SL(2,\mathbb C)$.

In Euclidean space-time symmetry group is the rotation group
$SO(4,\mathbb R)$ and there is isomorphism $SO(4,\mathbb R) \simeq \left(SU(2)\times SU(2)\right)/Z_2$.
The matrices $\boldsymbol{\sigma}_{\mu}$ and $\overline{\boldsymbol{\sigma}}_{\mu}$ in Euclidean space-time are
$$
\boldsymbol{\sigma}_{\mu} = ( \II , i\sigma_1, i\sigma_2, i\sigma_3 ) \; ,
\qquad \overline{ \boldsymbol{\sigma} }_{\mu} =
 ( \II , - i\sigma_1, - i\sigma_2, - i\sigma_3 ) \;
$$
and they connected by hermitian conjugation
$
\overline{ \boldsymbol{\sigma} }_{\mu} = \boldsymbol{\sigma}_{\mu}^{\dagger}
$
where $\dagger$ is complex conjugation and transposition.
The same formula in index notations is
\begin{align*}
\left(\overline{\boldsymbol{\sigma}}_{\mu}\right)^{\dot{a}a} =
\left(\left(\boldsymbol{\sigma}_{\mu}\right)^t_{a\dot{a}}\right)^{*} =
\left(\left(\boldsymbol{\sigma}_{\mu}\right)_{\dot{a}a}\right)^{*}
\end{align*}
so that now the index raising and lowering operations are carried out using
complex conjugation.
Each pair of matrices $U\,,V \in SU(2)$ define matrix
$\Lambda \in SO(4,\mathbb R)$ by the formula
\begin{align}
U \boldsymbol{\sigma}_\mu V^{\dagger} = \boldsymbol{\sigma}_{\nu}
\Lambda^{\nu}_{\ \mu} \longrightarrow U_{a}^{\ b}
\left(V^{*}\right)_{\dot{a}}^{\ \dot{b}}
\left(\boldsymbol{\sigma}_\mu\right)_{b\dot{b}} =
\left(\boldsymbol{\sigma}_\nu\right)_{a\dot{a}}
\Lambda^{\nu}_{\ \mu}
\end{align}
so that the tensor $t^{\mu_1\cdots\mu_n}$ with standard
law of transformation
$$
t^{\mu_1\cdots\mu_n} \to
\Lambda^{\mu_1}_{\ \nu_1}\cdots \Lambda^{\mu_n}_{\ \nu_n}\,
t^{\nu_1\cdots\nu_n}
$$
is transformed by~(\ref{ts}) to the spinor with
law of transformation
\begin{align*}
\psi_{a_1\ldots a_n\dot{b}_1\ldots \dot{b}_n } \to
U_{a_1}^{\ c_1}\cdots U_{a_n}^{\ c_n}\,
\left(V^{*}\right)_{\dot{b}_1}^{\ \dot{d}_1}\cdots \left(V^{*}\right)_{\dot{b}_n}^{\ \dot{d}_n}\,
\psi_{c_1\ldots c_n\dot{d}_1\ldots \dot{d}_n}
\end{align*}
so that dotted and un-dotted indices correspond to
spinors that are transformed according to representations
of two different copies of $SU(2)$.

{Important note. Everywhere in the paper we work with Euclidean metric 
and it is not useful for our purposes to distinguish
carefully the upper and lower tensor indices like $\mu\,,\nu$. 
For the sake of simplicity for the cumbersome formulae we adopt the 
following notation in the paper for spinor indices
$\boldsymbol{\sigma}_{\mu}= ||(\boldsymbol{\sigma}_{\mu})_{a}^{\dot{a}}||$
and $\overline{ \boldsymbol{\sigma} }_{\mu} = ||(\overline{ \boldsymbol{\sigma} }_{\mu})_{\dot{a}}^{a}||$. The dotted and un-dotted indices distinguish 
spinors that are transformed according to representations
of two different copies of $SU(2)$ and the rules of conversion of tensors to spinors an back have the following form 
\begin{align}\label{ts-st}
\psi_{a_1\ldots a_n}^{\dot{a}_1\ldots \dot{a}_n} =
\left(\boldsymbol{\sigma}_{\mu_1}\right)_{a_1}^{\dot{a}_1}\ldots
\left(\boldsymbol{\sigma}_{\mu_n}\right)_{a_n}^{\dot{a}_n}\, t_{\mu_1\ldots\mu_n}\ \ ;\ \ 
t_{\mu_1\ldots\mu_n} = 2^{-n}\,
\left(\overline{\boldsymbol{\sigma}}_{\mu_1}\right)_{\dot{a}_1}^{a_1}\ldots
\left(\overline{\boldsymbol{\sigma}}_{\mu_n}\right)_{\dot{a}_n}^{a_n}\,
\psi_{a_1\ldots a_n}^{\dot{a}_1\ldots \dot{a}_n}
\end{align}
}
The auxiliary spinors in paper have lower indices $\alpha_{a}$ and $\beta_{\dot{a}}$ 
and the index raising operation is defined as complex conjugation $\alpha^{a} =(\alpha_{a})^*=\bar{\alpha}_{a}$ and 
$\beta^{\dot{a}} =(\beta_{\dot{a}})^*=\bar{\beta}_{\dot{a}}$ so that the rules of hermitian conjugation are (Appendix \ref{app:spinors})
\begin{align}
\boldsymbol{\sigma}^{\dagger}_{\mu} = \overline{\boldsymbol{\sigma}}_{\mu}
\ \ ;\ \ \alpha^a = \bar{\alpha}_{a} \ \ ;\ \
\beta^{\dot{a}} = \bar{\beta}_{\dot{a}} \ \ ;\ \
\langle\alpha|\boldsymbol{\sigma}_{\mu}|\beta\rangle^{\dagger} = \langle\beta|\overline{\boldsymbol{\sigma}}_{\mu}|\alpha\rangle
\end{align}
and the pairing between spinors is the standard scalar product in $\mathbb{C}^2$
\begin{align}
\langle\alpha|\alpha'\rangle =
\alpha^{a} \alpha'_{a} = \bar{\alpha}_{a} \alpha'_{a}\, ,\, \,\,\,
\langle\beta'|\beta\rangle = {\beta'}^{\dot{a}}\beta_{\dot{a}} = 
\bar{\beta}'_{\dot{a}}\beta_{\dot{a}}\,.
\end{align}
Note that vector $c_{\mu} = \langle\alpha|\boldsymbol{\sigma}_{\mu}|\beta\rangle$
is automatically a null-vector $\left(c\,, c\right)=0$ due to the Fierz identity
$\boldsymbol{\sigma}_{\mu}\otimes\boldsymbol{\sigma}^{\mu} = 2(\II-\mathbb{P})$.

\newpage

\section{Relation with representation theory of $SU(2)$}
\label{app:SU(2)}
Here we would like to demonstrate that orthogonality
and completeness relations for the eigenfunctions
in the case $N=1$ are closely related with the orthogonality of the matrix elements of the irreducible representations of
the group $SU(2)$ and the Peter-Weyl theorem \cite{Isaev:2018xcg}.

\subsubsection{Orthogonality}

Let us start from the orthogonality relation~\eqref{ort1sp} and separate from
the very beginning the radial part in our
integral $x = \rho\, \textbf{x} \,, \textbf{x}^2 = 1$
\begin{align*}
&\int \,d^4 x \,
\frac{\langle\alpha|x|\beta\rangle^n}
{x^{2(\frac{3}{2}+i\nu+\frac{n}{2})}}
\frac{\langle\beta^{\prime}|\overline{x}|\alpha^{\prime}\rangle^{n^{\prime}}}
{x^{2(\frac{1}{2}-i\nu^{\prime}+\frac{n^{\prime}}{2})}} =
\int_0^{\infty} \frac{d\rho}{\rho} \rho^{2i(\nu^{\prime}-\nu)}\,
\int \,d^4 \textbf{x} \,\delta(\textbf{x}^2-1)\,
\langle\alpha|\textbf{x}|\beta\rangle^n
\langle\beta^{\prime}|\overline{\textbf{x}}|\alpha^{\prime}
\rangle^{n^{\prime}} = \\
&\frac{2\pi^3}{n+1}\,\langle\alpha|\alpha^{\prime}\rangle^n\,
\langle\beta^{\prime}|\beta\rangle^n\,
\delta_{n n^{\prime}}\,\delta(\nu-\nu^{\prime})
\end{align*}
so that this formula is equivalent to the formula for integral over sphere
$\textbf{x}^2 = 1$
\begin{align}\label{g}
\int \,d^4 \textbf{x} \,\delta(\textbf{x}^2-1)\,
\langle\alpha|\textbf{x}|\beta\rangle^n
\langle\beta^{\prime}|\overline{\textbf{x}}|\alpha^{\prime}\rangle^{n^{\prime}}
= \frac{2\pi^2}{n+1}\,\langle\alpha|\alpha^{\prime}\rangle^n\,
\langle\beta^{\prime}|\beta\rangle^n\,
\delta_{n n^{\prime}}
\end{align}
We shall show that this relation is nothing else as the
orthogonality relation for the matrix elements of irreducible
representations of $SU(2)$.
The standard parametrization of the matrix from $SU(2)$
\begin{align*}
g = \left(
\begin{array}{cc}
a & b\\
-\bar{b} & \bar{a}\\
\end{array}
\right)\ \,,\ a\bar{a}+b\bar{b} = 1\,
\end{align*}
in terms of two complex numbers $a$ and $b$ is equivalent to the parametrization in terms of unit four-dimensional vector
$\textbf{x} =(x_0,\vec{x}) \,, \textbf{x}^2 = x_0^2+x_1^2+x_2^2+x_3^2 = 1 $:
\begin{align*}
g = \left(
\begin{array}{cc}
x_0+i\,x_3 & x_2+i\,x_1\\
-x_2+i\,x_1 & x_0 -i\,x_3\\
\end{array}
\right)\  = x_0\,\II + x_1\,i\sigma_1+x_2\,i\sigma_2+x_3\,i\sigma_3 =
x_{\mu}\sigma_{\mu} \,.
\end{align*}
In this parametrization the normalized invariant integral
over the group $SU(2)$ is exactly the integral over the sphere
\begin{align}
\int d g = \frac{1}{2\pi^2} \int \,d^4 \textbf{x} \,\delta(\textbf{x}^2-1)\,
\end{align}
The orthogonality theorem for the matrix elements of operators acting in irreducible representations of the compact group $G$ states:
let $\mathbb{T}^{(\alpha)}$ irreducible unitary representations
of the group $G$ and $\dim \mathbb{T}^{(\alpha)} = d_{\alpha}$ then
we have for the matrix elements
\begin{equation}\label{Tg}
\int_{G} d g\,
\overline{\mathrm{T}}_{ik}^{(\alpha)}(g)\,\mathrm{T}_{jl}^{(\beta)}(g) =
\frac{1}{d_{\alpha}}\,\delta_{\alpha\beta}\,\delta_{ij}\,\delta_{kl}\,.
\end{equation}
In our case $G$ is $SU(2)$ and the irreducible representation $\mathbb{T}^{(n)}$ with dimension $d_n = n+1$ is realized by restriction of the n-th tensor
power of the defining two-dimensional representation
$g^{\otimes n}$ to the space of symmetric spinors
of the rank $n$. The quantity $\langle\alpha|g|\beta\rangle^n$ is the generating function for the matrix elements $\mathrm{T}_{ik}^{(n)}(g)$ so
that the relation~(\ref{g}) contains all relations~(\ref{Tg}).

Now we are going to the explicit formulae and first of all
introduce the generating function for the basis $|e_{i}\rangle$ in the space of the symmetric spinors of the rank $n$
\begin{align*}
|\beta\rangle\otimes \cdots
\otimes |\beta\rangle = \sum_{i=0}^{n} \binom n {i}^{\frac{1}{2}}\,
\beta_2^{i}\,\beta_1^{n-i}\,|e_{i}\rangle =
\sum_{i=0}^{n} \psi_{n,i}(\beta)\,|e_{i}\rangle\ \ ;\ \
|\beta \rangle = \left(\begin{array}{c}
\beta_1 \\
\beta_2
\end{array}\right) =
\beta_1\,|\uparrow \rangle+
\beta_2\,|\downarrow \rangle
\end{align*}
where $\psi_{n,i}(\beta) = \binom n {i}^{\frac{1}{2}}\,
\beta_2^{i}\,\beta_1^{n-i}$.
By construction $\langle e_k|e_i\rangle = \delta_{ik}$, the matrix
element of operator of representation is defined in a usual way
$T^{(n)}_{ik}(g) = \langle e_i|g^{\otimes n}|e_k\rangle$
so that we have
\begin{align*}
\langle\alpha|g|\beta\rangle^n =
\sum_{i,k=0}^{n}\,\psi_{n,i}(\bar{\alpha})\psi_{n,k}(\beta)
T^{(n)}_{ik}(g) \ \ ;\ \
\langle\beta^{\prime}|g^{\dagger}|\alpha^{\prime}\rangle^{n^{\prime}} =
\sum_{j,l=0}^{n^{\prime}}\,
\psi_{n^{\prime},l}(\bar{\beta}^{\prime})
\psi_{n^{\prime},j}(\alpha^{\prime})\,
\overline{T}^{(n^{\prime})}_{jl}(g)
\end{align*}
Now it is easy to calculate the integral~(\ref{g}) using the orthogonality of the matrix elements~(\ref{Tg})
\begin{align*}
&\int d g\, \langle\alpha|g|\beta\rangle^n \langle\beta^{\prime}|g^{\dagger}|\alpha^{\prime}\rangle^{n^{\prime}}  =
\sum_{i,k=0}^{n}\,\,\psi_{n,i}(\bar{\alpha})\psi_{n,k}(\beta)
\sum_{j,l=0}^{n^{\prime}}\,\psi_{n^{\prime},l}(\bar{\beta}^{\prime})
\psi_{n^{\prime},j}(\alpha^{\prime})\,
\int d g\,
T_{ik}^{(n)}(g)\,\overline{T}_{jl}^{(n^{\prime})}(g) =\\
&\sum_{i,k=0}^{n}\,\,\psi_{n,i}(\bar{\alpha})\psi_{n,k}(\beta)
\sum_{j,l=0}^{n^{\prime}}\,\psi_{n^{\prime},l}(\bar{\beta}^{\prime})
\psi_{n^{\prime},j}(\alpha^{\prime})\,
\frac{1}{n+1}\,\delta_{n n^{\prime}}\delta_{ij}\,\delta_{kl}=\\
&\frac{1}{n+1}\,\delta_{n n^{\prime}}\,
\sum_{i=0}^{n}\,
\binom n {i}\,
\left(\bar{\alpha}_2\alpha_2^{\prime}\right)^{i}\,
\left(\bar{\alpha}_1\alpha_1^{\prime}\right)^{n-i}\,
\sum_{k=0}^{n}\,
\binom {n} {k}\,
\left(\bar{\beta}_2^{\prime}\beta_2\right)^{k}\,
\left(\bar{\beta}_1^{\prime}\beta_1\right)^{n-k} = \\
&\frac{1}{n+1}\,\delta_{n n^{\prime}}\,
\left(\bar{\alpha}_1\alpha_1^{\prime}+
\bar{\alpha}_2\alpha_2^{\prime}\right)^{n}\,
\left(\bar{\beta}_1^{\prime}\beta_1+
\bar{\beta}_2^{\prime}\beta_2\right)^{n} =
\frac{1}{n+1}\,\delta_{n n^{\prime}}\,
\langle\alpha|\alpha^{\prime}\rangle^n\,
\langle\beta^{\prime}|\beta\rangle^n\,
\end{align*}

\subsubsection{Completeness and Peter-Weyl theorem}
\label{app:PeterWeyl}
Now we are going to the completeness relation~\eqref{compspin1}
and again separate from the very beginning the radial part
$x = \rho_1\, \textbf{x} \,, y = \rho_2\, \textbf{y} \,, \textbf{x}^2 = \textbf{y}^2 = 1$
\begin{align*}
\int\limits_{-\infty}^{+\infty} d\nu\,
\frac{1}
{\rho_1^{2(1+\frac{i\lambda}{2}+i\nu)}}
\frac{1}
{\rho_2^{2(1-\frac{i\lambda}{2}-i\nu)}}
\,\sum_{n\geq 0}\, \frac{n+1}{2\pi^3\,(n!)^2}\,
\int D\alpha D\beta\,
\langle\alpha|\boldsymbol{x}|\beta\rangle^n
\langle\beta|\overline{\boldsymbol{y}}|\alpha\rangle^{n} = \\
\frac{\pi}{\rho_1^3}\,\delta(\rho_1-\rho_2)\sum_{n\geq 0}\, \frac{n+1}{2\pi^3\,(n!)^2}\,
\int D\alpha D\beta\,
\langle\alpha|\boldsymbol{x}|\beta\rangle^n
\langle\beta|\overline{\boldsymbol{y}}|\alpha\rangle^{n} =
\delta^{4}\left(x - y\right)\,.
\end{align*}
After extraction of the radial part from the four-dimensional $\delta$-function
\begin{align}
\delta^{4}\left(x - y\right) =
\frac{1}{\rho_1^3}\,\delta(\rho_1-\rho_2)\,\delta(\textbf{x} - \textbf{y})
\end{align}
we reduce the completeness relation to the form
\begin{align}\label{compspin1}
\sum_{n\geq 0}\, \frac{n+1}{2\pi^2\,(n!)^2}\,
\int D\alpha D\beta\,
\langle\alpha|\boldsymbol{x}|\beta\rangle^n
\langle\beta|\overline{\boldsymbol{y}}|\alpha\rangle^{n} =
\delta(\textbf{x} - \textbf{y})\,.
\end{align}
which has the natural $SU(2)$ counterpart.
The orthogonality relation
\begin{align*}
\int_{G} d g\,
\overline{\mathrm{T}}_{ik}^{(\alpha)}(g)\,\mathrm{T}_{jl}^{(\beta)}(g) =
\frac{1}{d_{\alpha}}\,\delta_{\alpha\beta}\,\delta_{ij}\,\delta_{kl}
\end{align*}
shows that matrix elements $\mathrm{T}_{ik}^{(\alpha)}(g)$ of all irreducible representations form an orthogonal set of functions on the compact group $G$.
Peter-Weyl theorem states that this basic is complete for any compact group
and this statement can be reformulated using Fourier transformation.
Let us expand $f(g)$ -- function on the group $G$ over considered basis
(we assume summation over $i$ and $k$)
\begin{align*}
f(g) = \sum_{\alpha} d_{\alpha}\, F^{(\alpha)}_{ik}\,  \overline{\mathrm{T}}_{ik}^{(\alpha)}(g)
\end{align*}
where coefficients $F^{(\alpha)}_{ik}$ can be obtained
from the orthogonality relation
\begin{align*}
F^{(\alpha)}_{ik} = \int_{G} d g\,f(g)\,\mathrm{T}_{ik}^{(\alpha)}(g)
\end{align*}
The set of coefficients $F^{(\alpha)}_{ik}$ contains all information about the function $f(g)$ and the Fourier transformation is defined as a map
\begin{align}
f(g) \to F^{(\alpha)}_{ik} =
\int_{G} d g\,f(g)\,\mathrm{T}_{ik}^{(\alpha)}(g)
\end{align}
so that Fourier-transformation maps $f(g)$
to operator $F^{(\alpha)}$
\begin{align}
f(g) \to F^{(\alpha)} = \int_{G} d g\,f(g)\,\mathrm{T}^{(\alpha)}(g)
\end{align}
and inverse transformation is given by
\begin{align}\label{F1}
f(g) = \sum_{\alpha} d_{\alpha}\, F^{(\alpha)}_{ik}\,  \overline{\mathrm{T}}_{ik}^{(\alpha)}(g) =
\sum_{\alpha} d_{\alpha}\, \tr\left(F^{(\alpha)}
\mathrm{T}^{(\alpha)}\left(g^{-1}\right)\right)
\end{align}
where we used the unitarity of the representation $\overline{\mathrm{T}}_{ik}^{(\alpha)}(g) =
\left(\mathrm{T}^{(\alpha)}\right)^{\dagger}_{ki}\left(g\right) =
\mathrm{T}^{(\alpha)}_{ki}\left(g^{-1}\right)$.

In explicit notations we have
\begin{align}
f(g) = \int_{G} d h\,f(h)\,
\sum_{\alpha} d_{\alpha}\, \tr\left(\mathrm{T}^{(\alpha)}(h)
\mathrm{T}^{(\alpha)}\left(g^{-1}\right)\right)
\end{align}
so that we obtain the completeness relation
\begin{align}
\delta(h,g) = \sum_{\alpha} d_{\alpha}\, \tr\left(\mathrm{T}^{(\alpha)}(h)
\mathrm{T}^{(\alpha)}\left(g^{-1}\right)\right) = \sum_{\alpha} d_{\alpha}\, \tr\left(\mathrm{T}^{(\alpha)}\left(h g^{-1}\right)\right)
\end{align}
where $\delta$-function $\delta(h,g)$ is defined by the standard relation
$f(g) = \int_{G} d h\,f(h)\,\delta(h,g)$.

The last relation in the case of $SU(2)$
is exactly the relation~(\ref{compspin1})
\begin{align*}
\sum_{n\geq 0}\, \frac{n+1}{2\pi^2\,(n!)^2}\,
\int D\alpha D\beta\,
\langle\alpha|\boldsymbol{x}|\beta\rangle^n
\langle\beta|\overline{\boldsymbol{y}}|\alpha\rangle^{n} =
\delta(\textbf{x} - \textbf{y})\,.
\end{align*}
where $h =\boldsymbol{x} = \textbf{x}_{\mu}\boldsymbol{\sigma}_{\mu}
\ \,, g^{\dagger} = g^{-1} = \overline{\boldsymbol{y}} = \textbf{y}_{\mu}\overline{\boldsymbol{\sigma}}_{\mu}$ and
$\delta(h,g) = \delta(\textbf{x} - \textbf{y})$.
To show the coincidence of two relations one has to calculate Gaussain integrals using relation
\begin{align}
\int D\alpha\,\psi_{n,i}(\bar{\alpha})\,\psi_{n,k}(\alpha) = n!\,\delta_{ik}
\end{align}
We have
\begin{align*}
&\int D\alpha D\beta\, \langle\alpha|h|\beta\rangle^n \langle\beta|g^{-1}|\alpha\rangle^{n}  =
\int D\alpha D\beta\,
\sum_{i,k=0}^{n}\,\,\psi_{n,i}(\bar{\alpha})\psi_{n,k}(\beta)
\sum_{j,l=0}^{n}\,\psi_{n,l}(\bar{\beta})
\psi_{n,j}(\alpha)\,
T_{ik}^{(n)}(h)\,T_{lj}^{(n)}(g^{-1}) =\\
&(n!)^2\,\sum_{i,k=0}^{n}\,
T_{ik}^{(n)}(h)\,T_{ki}^{(n)}(g^{-1}) =
(n!)^2\,\tr\left(\mathrm{T}^{(n)}(h)
\mathrm{T}^{(n)}\left(g^{-1}\right)\right)
\end{align*}

\newpage

\section{Appendix: R-matrix and fusion procedure}
\label{app:R}

The Yang-Baxter equation has the following form
\begin{equation}
R_{12}(u) R_{13}(u+v)
R_{23}(v) = R_{23}(v) R_{13}(u+v) R_{12}(u)
\end{equation}
All operators act in a tensor product of three spaces
$V_1\otimes V_2\otimes V_3$.
Indices show that $R_{12}(u)$ acts nontrivially
in the space $V_1\otimes V_2$ and is identity operator
in the space $V_3$. The meaning of another indices is the same.
The space $V_k$  is the space of the symmetric
spinors $\Psi_{(a_1\ldots a_{n_k})}$ and it is $(n_k+1)$-dimensional representation of the group $\mathrm{SU}(2)$
with spin $\frac{n_k}{2}$.
The general operator $R_{12}(u)$ acts in a tensor product of two representations with spins $\frac{n_1}{2}$ and $\frac{n_2}{2}$. It is the space of the spinors $\Psi_{(a_1\ldots a_{n_1})(b_1\ldots b_{n_2})}$ which are symmetric with respect to any permutations inside two groups of indices separately and in the matrix notations we have
\begin{align}
\left[R_{12}(u)\Psi\right]_{(a_1\ldots a_{n_1})(b_1\ldots b_{n_2})} =
R^{(c_1\ldots c_{n_1})(d_1\ldots d_{n_2})}_{(a_1\ldots a_{n_1})(b_1\ldots b_{n_2})}(u)\,
\Psi_{(c_1\ldots c_{n_1})(d_1\ldots d_{n_2})}\,,
\end{align}
where the summation over repeated indices is assumed.
For simplicity we skip indices $12$ in the matrix of operator $R_{12}(u)$.
The simplest solution of Yang-Baxter equation is Yang R-matrix
which acts in the tensor product of two-dimensional representations
of the spin $\frac{1}{2}$
\begin{equation}
R^{c\, d}_{a\, b}(u) = \frac{1}{u+1}\,\left(u\, \delta_{a}^{c}\,\delta_{b}^{d}+
\delta_{a}^{d}\,\delta_{b}^{c}\right)
\end{equation}
The standard procedure for constructing finite-dimensional higher-spin $\mathrm{R}$-operators out of
the Yang R-matrix is the {\it fusion procedure}.
Following the recipe from \cite{Kulish1981,Kulish82} we of the product of the Yang R-matrices
\begin{equation}
\mathbf{R}_{(a_1\ldots a_{n_1})\,b}^{(c_1 \ldots c_{n_1})\,d}
\textstyle\left(u+\frac{n_1-1}{2}\right) =
\mathrm{Sym}\,
\mathbf R^{c_1\, d_1}_{a_1\, b}(u)\,\mathbf R^{c_2\, d_2}_{a_2\, d_1}(u+1)
\,\cdots\,\mathbf R^{c_{n_1}\, d}_{a_{n_1}\, d_n}(u+n_1-1),
\end{equation}
where $\mathrm{Sym}$ implies symmetrization with respect to groups of indices $a_1\ldots a_{n_1}$ and $c_1 \ldots c_{n_1}$.
In such a way one obtains an operator acting on the space of symmetric rank $n_1$
spinors, i.e. on the space of spin $\frac{n_1}{2}$ representation, and on the
two-dimensional space of spin $\frac{1}{2}$ representation.
Next the R-operator~(\ref{n_1}) is used as building block and
repetition of the the same procedure increases spin of representation in the second space from $\frac{1}{2}$ to $\frac{n_2}{2}$.

In realization of this program we prefer not to deal with a multitude of spinor indices but rewrite everything using holomorphic representation~\cite{Perelomov:1980tt,Faddeev:1980be,Isaev:2018xcg}.
The usual matrix-like action of operators in a space of symmetric spinors $\Psi_{(a_1 \ldots a_n)}$  has the form
\begin{equation} \label{TPsi}
\left[A\,\Psi\right]_{(a_1\ldots a_n)}
= A_{(a_1\ldots a_n)}^{(b_1 \ldots b_n)}\, \Psi_{(b_1 \ldots b_n)}\, ,
\end{equation}
where the summation over repeated indices is assumed.
We introduce auxiliary spinors $\alpha=(\alpha_1,\alpha_2),\,\beta=(\beta_1,\beta_2)$
and contract them with the tensors
\begin{equation} \label{Tlam}
\Psi(\alpha) = \Psi_{(a_1 \ldots a_n)}\,
\alpha^{a_1} \cdots \alpha^{a_n} ,\;\;
A(\alpha\,,\beta) = \bar{\alpha}^{a_1} \cdots \bar\alpha^{a_n}\,
A_{(a_1\ldots a_n)}^{(b_1 \ldots b_n)}
\,\beta_{b_1} \cdots \beta_{b_n}\,,
\end{equation}
where in fact the contraction of $a$-indices is performed with complex 
conjugate spinor due to our 
convention about index raising operation as complex conjugation 
$\alpha^{a} =(\alpha_{a})^*=\bar{\alpha}_{a}$.

The symmetization over spinor indices is taken into account automatically.
Henceforth, in place of the tensors we work with the corresponding generating functions which are homogeneous polynomials of degree $n$ of two variables
\begin{equation} \label{PsiHom}
\Psi(\alpha) = \Psi(\bar{\alpha}_1,\bar{\alpha}_2) \ \ ,\ \
\Psi(\lambda\bar{\alpha}_1,\lambda\bar{\alpha}_2) = \lambda^n\,
\Psi(\bar{\alpha}_1,\bar{\alpha}_2)\,.
\end{equation}
In this way formula~(\ref{TPsi}) acquires a rather compact form
\begin{equation} \label{TPsispinor}
\left[A\,\Psi\right](\alpha) =
\frac{1}{n!}\,\int D\beta\, A(\alpha\,|\,\beta)
\,\Psi(\beta)
\end{equation}
due to Gaussian integral~(\ref{Gauss})
\begin{align}
\int D\beta\,
\beta_{b_1}\cdots\beta_{b_n}\,
\beta^{a_1}\cdots\beta^{a_n} =
n!\,\hat{S}\,
\delta^{a_1}_{b_1}\,\delta^{a_2}_{b_2}\cdots
\delta^{a_n}_{b_n}
\end{align}
so that operator $A$ is described as an integral operator with the kernel $A(\alpha\,,\beta)$.
We should note that the left argument in the kernel is spinor with upper index $\alpha^a$ so that in fact it is complex conjugate spinor $\alpha^a = \bar{\alpha}_a$.
The kernel of the product of operators $A$ and $B$ is
given by the convolution of the kernels
\begin{equation} \label{convol}
\left[A\,B\right](\alpha\,,\beta) =
\frac{1}{n!}\,\int D\gamma\, A(\alpha\,|\,\gamma)
\,B(\gamma\,|\,\beta)
\end{equation}
The kernel of identity operator is
$\langle\alpha|\beta\rangle^n$, where $\langle\alpha|\beta\rangle = \bar{\alpha}_{a}\beta_a$
\begin{equation} \label{Id}
\Psi(\alpha) =
\frac{1}{n!}\,\int D\beta\, \langle\alpha|\beta\rangle^n \,\Psi(\beta)
\end{equation}
There are two equivalent formulae for the kernel of 
the general R-operator acting in a tensor product of two representations with spins $\frac{n_1}{2}$ and $\frac{n_2}{2}$
\begin{align}\label{R1}
\langle \alpha_1,\alpha_2|\mathbf R_{n_1,n_2}(u)|\beta_1,\beta_2\rangle\, =\,
\rho(u)\,\langle\alpha_2|\beta_1\rangle^{-u+\frac{n_1+n_2}{2}}
\partial_{s_1}^{n_1}
\partial_{s_2}^{n_2}
\langle \alpha_2+s_1\alpha_1|
\beta_1+s_2\beta_2
\rangle^{u+\frac{n_1+n_2}{2}} \\
\label{R2}
\,=\, \rho(u)\,
\langle\alpha_1|\beta_2\rangle^{-u+\frac{n_1+n_2}{2}}\,
\partial_{s_1}^{n_1}
\partial_{s_2}^{n_2}
\langle \alpha_1+s_2\alpha_2|\beta_2+s_1\beta_1
\rangle^{u+\frac{n_1+n_2}{2}}
\end{align}
The normalization is fixed by requirement $\mathbf{R}(u)\,\mathbf{R}(-u)=\II$
\begin{align}
\rho(u) = \frac{1}{\left(u+\frac{n_1-n_2}{2}+1\right)_{n_2}
\left(u+\frac{n_2-n_1}{2}+1\right)_{n_1}}
\end{align}
where $(a)_n$ is Pochhammer symbol: $(a)_n = a(a+1)\cdots(a+n-1)$.
Above and in the following we agree that in all formulas of such type
one should to put $s_1=s_2=0$ after all differentiations.
The formula~(\ref{R2}) is obtained from~(\ref{R1}) by rescaling
$s_k \to s_k\,\langle\alpha_2|\beta_1\rangle\,
\langle\alpha_1|\beta_2\rangle^{-1}$ and
there are useful representations for R-operator
\begin{align*}
\langle \alpha_1,\alpha_2|\mathbf R_{n_1,n_2}(u)|\beta_1,\beta_2\rangle =
\frac{1}{\left(u+\frac{n_2-n_1}{2}+1\right)_{n_1}}\,
\langle\alpha_2|\beta_1\rangle^{-u+\frac{n_1+n_2}{2}}
\partial_{s}^{n_1}
\langle\alpha_{21}(s)|
\beta_1\rangle^{u+\frac{n_1-n_2}{2}}\,
\langle\alpha_{21}(s)|\beta_2
\rangle^{n_2} = \\
\frac{1}{\left(u+\frac{n_1-n_2}{2}+1\right)_{n_2}}\,
\langle\alpha_2|\beta_1\rangle^{-u+\frac{n_1+n_2}{2}}
\partial_{s}^{n_2}
\langle\alpha_1|
\beta_{12}(s)
\rangle^{n_1}
\langle\alpha_2|\beta_{12}(s)
\rangle^{u+\frac{n_2-n_1}{2}} = \\
\frac{1}{\left(u+\frac{n_2-n_1}{2}+1\right)_{n_1}}\,
\langle\alpha_1|\beta_2\rangle^{-u+\frac{n_1+n_2}{2}}\,
\partial_{s}^{n_1}
\langle \alpha_1|\beta_{21}(s)
\rangle^{u+\frac{n_1-n_2}{2}}\,
\langle\alpha_2|\beta_{21}(s)
\rangle^{n_2}\\
\frac{1}{\left(u+\frac{n_1-n_2}{2}+1\right)_{n_2}}\,
\langle\alpha_1|\beta_2\rangle^{-u+\frac{n_1+n_2}{2}}\,
\partial_{s}^{n_2}
\langle\alpha_{12}(s)|\beta_1
\rangle^{n_1}\,
\langle\alpha_{12}(s)|\beta_2\rangle^{u+\frac{n_2-n_1}{2}}
\end{align*}
where we use compact notations
$\alpha_{ik}(s) = \alpha_i+s\alpha_k\,,\beta_{ik}(t) = \beta_i+t\beta_k$ for simplicity.
It is easy to check that for $n_1=n_2=1$ this expression coincides
with the kernel of Yang R-matrix~(\ref{RYang})
\begin{align*}
& \langle \alpha_1,\alpha_2|\mathbf R_{1,1}(u)|\beta_1,\beta_2\rangle =
\frac{1}{u+1}\,\left(u\,\langle\alpha_1|
\beta_1\rangle\,\langle\alpha_2|\beta_2\rangle
+\langle\alpha_1|\beta_2\rangle\,\langle\alpha_2|\beta_1\rangle\right)\,, \end{align*}
In our knowledge this compact expression for the kernel of R-operator is new; the most similar representation is obtained by E.K. Sklyanin in \cite{Sklyanin1988ClassicalLO}.
Let us perform the first step of the fusion procedure according to~(\ref{n_1})
\begin{align*}
&\int D \beta\, \langle \alpha_1,\alpha_2|\mathbf R_{1,1}(u)|\beta_1,\beta\rangle
\,\langle \alpha_1,\beta|\mathbf R_{1,1}(u+1)|\beta_1,\beta_2\rangle \to \\
&\langle\alpha_2|\beta_1\rangle^{-u+1}\,
\partial_{s}
\langle\alpha_{21}(s)|
\beta_1\rangle^{u}\,
\int D \beta\,\langle\alpha_{21}(s)|\beta\rangle\,
\,\langle \alpha_1,\beta|\mathbf R_{1,1}(u+1)|\beta_1,\beta_2\rangle = \\
&\langle\alpha_2|\beta_1\rangle^{-u+1}\,
\partial_{s}\langle\alpha_{21}(s)|
\beta_1\rangle^{u}
 \langle \alpha_1,\alpha_{21}(s)|\mathbf R_{1,1}(u+1)|\beta_1,\beta_2\rangle \to
\\
&\langle\alpha_2|\beta_1\rangle^{-u+1}\,\partial_{s}
\underline{\langle\alpha_{21}(s)|
\beta_1\rangle^{u}\,
\langle\alpha_{21}(s)|\beta_1\rangle^{-u}}\,
\partial_{t_1}
\partial_{t_2}
\langle \alpha_2+(s+t_1)\alpha_1|
\beta_1+t_2\beta_2\rangle^{u+2} = \\
&\langle\alpha_2|\beta_1\rangle^{-u+1}\,
\partial^2_{t_1}
\partial_{t_2}
\langle \alpha_2+t_1\alpha_1|
\beta_1+t_2\beta_2\rangle^{u+2} \to \langle \alpha_1,\alpha_2|\mathbf R_{2,1}\left(u+\frac{1}{2}\right)|\beta_1,\beta_2\rangle\,,
\end{align*}
where we ignore all overall constants for simplicity.
All calculation is based on the explicit formula \eqref{Id} for 
the kernel of identity operator and crucial simplificaion of the marked product.
The general procedure is clear and in this way we construct
\begin{align*}
\langle \alpha_1,\alpha_2|\mathbf R_{n_1,1}(u)|\beta_1,\beta_2\rangle \to
\langle\alpha_2|\beta_1\rangle^{-u+\frac{n_1+1}{2}}\,
\partial_{s_1}^{n_1}
\partial_{s_2}
\langle \alpha_2+s_1\alpha_1|
\beta_1+s_2\beta_2
\rangle^{u+\frac{n_1+1}{2}}
\end{align*}
For the fusion procedure increasing spin in the second space we shall
use the equivalent formula
\begin{align*}
\langle \alpha_1,\alpha_2|\mathbf R_{n_1,1}(u)|\beta_1,\beta_2
\to
\langle\alpha_1|\beta_2\rangle^{-u+\frac{n_1+1}{2}}\,
\partial_{s}
\langle \alpha_1+s\alpha_2|
\beta_2\rangle^{u+\frac{1-n_1}{2}}
\langle \alpha_1+s\alpha_2|\beta_1
\rangle^{n_1}
\end{align*}
and again for example we perform the first step of the fusion procedure
\begin{align*}
&\int D \beta\, \langle \alpha_1,\alpha_2|\mathbf R_{n_1,1}(u)|\beta,\beta_2\rangle
\,\langle \beta,\alpha_2|\mathbf R_{n_1,1}(u+1)|\beta_1,\beta_2 \rangle \to  \\
&\langle\alpha_1|\beta_2\rangle^{-u+\frac{n_1+1}{2}}\,
\partial_{s}
\langle\alpha_{12}(s)|
\beta_2\rangle^{u+\frac{1-n_1}{2}}
\int D \beta\,
\langle\alpha_{12}(s)|\beta
\rangle^{n_1}\,
\langle \beta,\alpha_2|\mathbf R_{n_1,1}(u+1)|\beta_1,\beta_2 \rangle = \\
&\langle\alpha_1|\beta_2\rangle^{-u+\frac{n_1+1}{2}}\,
\partial_{s}
\langle\alpha_{12}(s)|
\beta_2\rangle^{u+\frac{1-n_1}{2}}
\langle \alpha_{12}(s),\alpha_2|\mathbf R_{n_1,1}(u+1)|\beta_1,\beta_2 \rangle\to
\\
&\langle\alpha_1|\beta_2\rangle^{-u+\frac{n_1+1}{2}}\,
\partial_{s}
\underline{\langle\alpha_{12}(s)|
\beta_2\rangle^{u+\frac{1-n_1}{2}}
\langle\alpha_{12}(s)|\beta_2\rangle^{-u+\frac{n_1-1}{2}}}\,
\partial_{t_1}^{n_1}
\partial_{t_2}
\langle \alpha_1+(s+t_2)\alpha_2|
\beta_2+t_1\beta_1
\rangle^{u+\frac{n_1+3}{2}}\\
&
\langle\alpha_1|\beta_2\rangle^{-u+\frac{n_1+1}{2}}\,
\,
\partial_{t_1}^{n_1}
\partial_{t_2}^2
\langle \alpha_1+t_2\alpha_2|
\beta_2+t_1\beta_1
\rangle^{u+\frac{n_1+3}{2}}
\to \langle \alpha_1,\alpha_2|\mathbf R_{n_1,2}\left(u+\frac{1}{2}\right)|\beta_1,\beta_2\rangle
\end{align*}
We provide the proof of the identity $\mathbf R(u)\,\mathbf R(-u) = \II$, where for simplicity reasons
we ignore all coefficients:
\begin{align*}
&\int D \gamma_1\,D \gamma_2\,\langle \alpha_1,\alpha_2|\mathbf R_{n_1,n_2}\left(u\right)|\gamma_1,\gamma_2\rangle
\,\langle \gamma_1,\gamma_2|\mathbf R_{n_1,n_2}\left(-u\right)|\beta_1,\beta_2\rangle\to \\
&
\int D \gamma_2\,\langle\alpha_1|\gamma_2\rangle^{-u+\frac{n_1+n_2}{2}}\,
\partial_{s}^{n_2}
\langle\alpha_{12}(s)|
\gamma_2\rangle^{u+\frac{n_2-n_1}{2}}
\int D \gamma_1\,
\langle\alpha_{12}(s)|\gamma_1
\rangle^{n_1}\,
\langle \gamma_1,\gamma_2|\mathbf R_{n_1,n_2}\left(-u\right)|\beta_1,\beta_2\rangle = \\
&
\int D \gamma_2\,\langle\alpha_1|\gamma_2\rangle^{-u+\frac{n_1+n_2}{2}}\,
\partial_{s}^{n_2}
\langle\alpha_{12}(s)|
\gamma_2\rangle^{u+\frac{n_2-n_1}{2}}
\langle \alpha_{12}(s),\gamma_2|\mathbf R_{n_1,n_2}\left(-u\right)|\beta_1,\beta_2\rangle\to
\\
&
\int D \gamma_2\,\langle\alpha_1|\gamma_2\rangle^{-u+\frac{n_1+n_2}{2}}\,
\partial_{s}^{n_2}
\langle\alpha_{12}(s)|
\gamma_2\rangle^{u+\frac{n_2-n_1}{2}}\,
\langle\alpha_{12}(s)|\beta_2\rangle^{u+\frac{n_1+n_2}{2}}\,\partial_{t}^{n_1}
\langle\alpha_{12}(s)|
\beta_{21}(t)
\rangle^{-u+\frac{n_1-n_2}{2}}
\langle \gamma_2|\beta_{21}(t)
\rangle^{n_2}\to \\
&
\partial_{s}^{n_2}\partial_{t}^{n_1}
\langle\alpha_1|\beta_{21}(t)\rangle^{-u+\frac{n_1+n_2}{2}}\,
\underline{\langle\alpha_{12}(s)|
\beta_{21}(t)\rangle^{u+\frac{n_2-n_1}{2}}}\,
\langle\alpha_{12}(s)|\beta_2\rangle^{u+\frac{n_1+n_2}{2}}\,
\underline{\langle\alpha_{12}(s)|\beta_{21}(t)
\rangle^{-u+\frac{n_1-n_2}{2}}} \to \\
&\partial_{s}^{n_2}\partial_{t}^{n_1}
\langle\alpha_1|\beta_{21}(t)\rangle^{-u+\frac{n_1+n_2}{2}}\,
\langle\alpha_{12}(s)|\beta_2\rangle^{u+\frac{n_1+n_2}{2}}\to
\langle\alpha_1|\beta_1\rangle^{n_1}\,
\langle\alpha_2|\beta_2\rangle^{n_2}\,.
\end{align*}
All the ignored coefficients can be restored as they are presented in the various explicit representations for the kernel of R-operator 
listed in the beginning of this section.
Note that in this section we present the formulae for the un-dotted spinors 
and of course there exists corresponding  analogues for the dotted spinors.
We skip all such evident formulae for simplicity. 
Moreover during the whole paper we use $\alpha$-spinors as 
undotted and $\beta$-spinors as dotted but in this sections 
all $\alpha\,,\beta\,,\gamma\,,\ldots$ spinors are un-dotted. 
We hope that will not leads to any misunderstanding.

\section{Derivation of the integral identities}
\subsection{Derivation of the chain-rule \eqref{chain_rule} and star-triangle identities \eqref{STRsame} \eqref{STRopp}}
\label{app:chainstar}
In this section we provide a detailed derivation of the formulae \eqref{chain_rule}, \eqref{STRsame} and \eqref{STRopp}, which lye at the basis of most of our results. In order to do so, we write the propagators of formula \eqref{propags} by an explicit projection of $\sig^{\otimes n_1},\,\bsig^{\otimes n_2}$ matrices over the space of symmetric spinors, and fix $a=\Delta_1+\frac{n_1}{2}$, $b=\Delta_2+\frac{n_2}{2}$:
\begin{equation}
\frac{\langle\alpha_1|\mathbf{x-z}|\beta_1\rangle^{n_1} }{(x-z)^{2(a-n_1/2)}}\,,\,\,\,\,\, \frac{\langle\beta_2|\overline{\mathbf{z-y}}|\alpha_2\rangle^{n_2}}{ (y-z)^{2(b-n_2/2)}}\,,
\end{equation}
where $|\alpha_k\rangle$ and $|\beta_k \rangle$ are the left-handed and right-handed auxiliary spinors introduced in appendix \ref{app:spinors} 
\begin{equation}
(\alpha_k)_{a} \,,\,\,\,\,(\beta_k)_{\dot{a}}\,,\,\,\,\,\,\,\,\, k=1,2\,,\,a,\dot{a}=1,2.
\end{equation}
First of all we represent such propagators in a suitable form
\begin{align}
&\frac{\langle \alpha_1 | \mathbf{x-z}| \beta_1\rangle^{n_1} }{(x-z)^{2(a-n_1/2)}} = \frac{\Gamma(a-n_1)}{2^{n_1} \Gamma(a)} \partial_{t_1}^{n_1} \frac{1}{(x-z-t_1\langle \alpha_1|\sig|\beta_1\rangle)^{2(a-n_1)}}\,,\,\,\, t_1=0\,,\\&\frac{\langle \beta_2 | \overline{\mathbf{z-y}}| \alpha_2\rangle^{n_2}}{(y-z)^{2(b-n_2/2)}} =\frac{\Gamma(b-n_2)}{2^{n_2} \Gamma(b)} \partial_{t_2}^{n_2} \frac{1}{(y-z+t_2\langle \beta_2|\bsig|\alpha_2\rangle)^{2(b-n_2)}}\,,\,\,\,t_2=0\,.
\end{align}
Secondly, we compute the convolution of propagators in such representation by the standard chain rule for scalar propagators
\begin{equation}
\int d^4z \frac{1}{(x-z)^{2a} (y-z)^{2b}} = \pi^2 \frac{\Gamma(2-a)\Gamma(2-b)\Gamma(a+b-2)}{\Gamma(a)\Gamma(b)\Gamma(4-a-b)} \frac{1}{(x-y)^{2(a+b-2)}}\,.
\end{equation}
The computation reads as follows:
\begin{align*}
&\int d^4z \frac{\langle \alpha_1 | \mathbf{x-z}| \beta_1\rangle^{n_1} \langle \beta_2 | \overline{\mathbf{z-y}}| \alpha_2\rangle^{n_2}}{(x-z)^{2(a-n_1/2)} (y-z)^{2(b-n_2/2)}} = \frac{\Gamma(a-n_1)\Gamma(b-n_2)}{2^{n_1+n_2} \Gamma(a)\Gamma(b)} \times\\ &\times\partial_{t_1}^{n_1} \partial_{t_2}^{n_2}  \int  \frac{d^4 z}{(x-z-t_1\langle \alpha_1|\sig|\beta_1\rangle)^{2(a-n_1)}(y-z+t_2\langle \beta_2|\bsig|\alpha_2\rangle)^{2(b-n_2)}}=\\&=\,\pi^2  \frac{\Gamma(2-a+n_1)\Gamma(2-b+n_2)\Gamma(a+b-2-n_1-n_2)}{2^{n_1+n_2} \Gamma(a)\Gamma(b)\Gamma(4-a-b+n_1+n_2)}\times\\&\times\partial_{t_1}^{n_1} \partial_{t_2}^{n_2} (x-y-t_1 \langle\alpha_1|\sig|\beta_1\rangle-t_2  \langle \beta_2|\bsig|\alpha_2\rangle )^{2(2-a-b+n_1+n_2)}.
\end{align*}
Furthermore we can compute explicitly the following square
\begin{align}
\notag &(x-y-t_1 \langle\alpha_1|\sig|\beta_1\rangle-t_2 \langle \beta_2|\bsig|\alpha_2\rangle )^2= \\&=(x-y)^2 -2t_1 |x-y|  \langle\alpha_1|\mathbf{x-y}|\beta_1\rangle-2 t_2 |x-y|  \langle \beta_2|\overline{\mathbf{x-y}}|\alpha_2\rangle + 4t_1 t_2 \langle\alpha_1|\alpha_2\rangle \langle\beta_2|\beta_1\rangle \,.
\end{align}
After the change of variables
\begin{equation}
t_i \mapsto t_i \frac{2}{ |x-y|}\,,\,\,\,i=1,2\,,
\end{equation}
we can rewrite the result of the integration as
\begin{align}
\notag &\pi^2 \frac{\Gamma(2-a+n_1)\Gamma(2-b+n_2)\Gamma(a+b-2-n_1-n_2)}{\Gamma(a)\Gamma(b)\Gamma(4-a-b+n_1+n_2)}\frac{1}{(x-y)^{2\left(a+b-2-\frac{n_1+n_2}{2}\right)}}\times\\&\times\partial_{t_1}^{n_1} \partial_{t_2}^{n_2} (1-t_1 \langle\alpha_1|\mathbf{x-y}|\beta_1\rangle-t_2 \langle \beta_2|\overline{\mathbf{{x}-{y}}}|\alpha_2\rangle +t_1 t_2 \langle \beta_2|\beta_1\rangle \langle \alpha_1 |\alpha_2\rangle)^{(2-a-b+n_1+n_2)}.
\end{align}
We can conveniently redefine the spinors $\alpha$ in order to get a more compact expression
\begin{equation}
|\alpha \rangle \to ({\mathbf{y-x}})|\beta' \rangle \,,
\end{equation}
So that
\begin{align}
\notag &\int d^4z\, \frac{\langle \beta'_1 |(\overline{\mathbf{y-x}}) (\mathbf{x-z})| \beta_1\rangle^{n_1} \langle \beta_2 | (\overline{\mathbf{z-y}})(\mathbf{y-x})| \beta'_2\rangle^{n_2}}{(x-z)^{2(a-n_1/2)} (y-z)^{2(b-n_2/2)}}=\\= &\pi^2 \frac{\Gamma(2-a+n_1)\Gamma(2-b+n_2)\Gamma(a+b-2-n_1-n_2)}{\Gamma(a)\Gamma(b)\Gamma(4-a-b+n_1+n_2)}\frac{1}{(x-y)^{2\left(a+b-2-\frac{n_1+n_2}{2}\right)}}\times\\&\times\partial_{t_1}^{n_1} \partial_{t_2}^{n_2} (1+t_1 \langle\beta'_1|\beta_1\rangle+t_2 \langle \beta_2|\beta'_2\rangle +t_1 t_2 \langle \beta_2|\beta_1\rangle \langle \beta'_1 |\beta'_2\rangle)^{(2-a-b+n_1+n_2)}\,.\notag
\end{align}
The last formulation of the star-triangle relation can be recast in an even more compact form by use of the fused $R$-matrix of $Y(\mathfrak{su}(2))$. The result is the fundamental relation that we use through the paper
\begin{align}
\notag &\int d^4z\, \frac{\langle \beta'_1 |(\overline{\mathbf{y-x}}) (\mathbf{x-z})| \beta_1\rangle^{n_1} \langle \beta_2 | (\overline{\mathbf{z-y}})(\mathbf{y-x})| \beta'_2\rangle^{n_2}}{(x-z)^{2(a-n_1/2)} (y-z)^{2(b-n_2/2)}}=\\= &\pi^2 \frac{\Gamma(2-a+n_1)\Gamma(2-b+n_2)\Gamma(a+b-2-n_1-n_2)}{\Gamma(a)\Gamma(b)\Gamma(4-a-b+n_1+n_2)}\frac{1}{(x-y)^{2\left(a+b-2-\frac{n_1+n_2}{2}\right)}}\times\\& \frac{\Gamma\left(3-a-b+n_1+n_2\right)\Gamma\left(3-a-b+n_1+n_2\right)}{\Gamma\left(3-a-b+n_1\right)\Gamma\left(3-a-b+n_2\right)}\langle \beta_1',\beta_2 |\mathbf{R}_{n_1,n_2}\left(2-a-b+\frac{n_1+n_2}{2}\right)|\beta_1,\beta_2'\rangle\,.\notag
\end{align}
We can rewrite it shifting $a\to a+n_1/2$, $b\to b+n_2/2$ as
\begin{align}
\notag &\int d^4z\, \frac{\langle \beta'_1 |(\overline{\mathbf{y-x}}) (\mathbf{x-z})| \beta_1\rangle^{n_1} \langle \beta_2 | (\overline{\mathbf{z-y}})(\mathbf{y-x})| \beta'_2\rangle^{n_2}}{(x-z)^{2a} (y-z)^{2b}}=\\= &\pi^2 \frac{\Gamma\left(2-a+\frac{n_1}{2}\right)\Gamma\left(2-b+\frac{n_2}{2}\right)\Gamma\left(a+b-2-\frac{n_1+n_2}{2}\right)}{\Gamma\left(a+\frac{n_1}{2}\right)\Gamma\left(b+\frac{n_2}{2}\right)\Gamma\left(4-a-b+\frac{n_1+n_2}{2}\right)}\frac{1}{(x-y)^{2(a+b-2)}}\times\\& \times\frac{\Gamma\left(3-a-b+\frac{n_1+n_2}{2}\right)\Gamma\left(3-a-b+\frac{n_1+n_2}{2}\right)}{\Gamma\left(3-a-b+\frac{n_1-n_2}{2}\right)\Gamma\left(3-a-b+\frac{n_2-n_1}{2}\right)}\langle \beta_1',\beta_2 |\mathbf{R}_{n_1,n_2}\left(2-a-b\right)|\beta_1,\beta_2'\rangle\,.\notag
\end{align}
It  is possible to simplify the $\Gamma$-functions containing $a+b$ which appear in the r.h.s. of the chain-rule identity (in such a way the symmetry $(a,b,n_1,n_2) \to (b,a,n_2,n_1)$ is not anymore manifest):
\begin{align*}
\frac{(-1)^{n_2}}{\left(3-a-b+\frac{n_1+n_2}{2}\right)}\frac{\Gamma\left(a+b-2+\frac{n_2-n_1}{2}\right)}{\Gamma\left(3-a-b+\frac{n_2-n_1}{2}\right)}\,.
\end{align*}
So, finally
\begin{align}
\label{chain_proof}
\begin{aligned}
&\int d^4z\, \frac{\langle \beta'_1 |(\overline{\mathbf{y-x}}) (\mathbf{x-z})| \beta_1\rangle^{n_1} \langle \beta_2 | (\overline{\mathbf{z-y}})(\mathbf{y-x})| \beta'_2\rangle^{n_2}}{(x-z)^{2a} (y-z)^{2b}}=\\&=\pi^2 \frac{\Gamma\left(2-a+\frac{n_1}{2}\right)\Gamma\left(2-b+\frac{n_2}{2}\right)\Gamma\left(a+b-2+\frac{n_2-n_1}{2}\right)}{\Gamma\left(a+\frac{n_1}{2}\right)\Gamma\left(b+\frac{n_2}{2}\right)\Gamma\left(3-a-b+\frac{n_2-n_1}{2}\right)}\frac{(-1)^{n_2}}{(x-y)^{2(a+b-2)}}\times\\& \times\frac{\langle \beta_1',\beta_2 |\mathbf{R}_{n_1,n_2}\left(2-a-b\right)|\beta_1,\beta_2'\rangle}{\left(3-a-b+\frac{n_1+n_2}{2}\right)}\,.
\end{aligned}
\end{align}
We introduce some notation:
\begin{equation}
A_{n_1,n_2}(a,b)=\frac{\Gamma\left(2-a+\frac{n_1}{2}\right)\Gamma\left(2-b+\frac{n_2}{2}\right)\Gamma\left(a+b-1+\frac{n_2-n_1}{2}\right)}{\Gamma\left(a+\frac{n_1}{2}\right)\Gamma\left(b+\frac{n_2}{2}\right)\Gamma\left(3-a-b+\frac{n_2-n_1}{2}\right)}\,.
\end{equation}
It is useful for our scopes to recast the uniqueness identity in the star-triangle form following a conformal inversion respect to the origin $x_0^{\mu}=0$ followed by a translation of vector $-t^{\mu}$. First we notice that
\begin{equation}
I(x^{\mu}) = \frac{x^{\mu}}{x^2} \,\Rightarrow\, I(\mathbf{x-y}) = \mathbf{-x(\overline{x -y})y} =\mathbf{y(\overline{y-x})x}\,, \text{and we recall that} \,\,\, \mathbf{(x-y)(\overline{x-y})}=\mathbbm{1}\,.
\end{equation}
We can redefine the spinors and write
\begin{align}
\label{STR_inter}
\begin{aligned}
 &\int d^4z\, \frac{\langle \beta'_1 | \mathbf{(\overline{x-z})}\mathbf{(z-t)}| \beta_1\rangle^{n_1} \langle \beta_2 | (\mathbf{\overline{ t-z}})(\mathbf{z-y})| \beta'_2\rangle^{n_2}}{(x-z)^{2a} (y-z)^{2b}(z-t)^{2(4-a-b)}}=\\= &\pi^2 \frac{\Gamma\left(2-a+\frac{n_1}{2}\right)\Gamma\left(2-b+\frac{n_2}{2}\right)\Gamma\left(a+b-2+\frac{n_2-n_1}{2}\right)}{\Gamma\left(a+\frac{n_1}{2}\right)\Gamma\left(b+\frac{n_2}{2}\right)\Gamma\left(3-a-b+\frac{n_2-n_1}{2}\right)}\frac{(-1)^{n_1}}{(x-y)^{2(a+b-2)}(x-t)^{2(2-b)}(y-t)^{2(2-a)}}\times\\& \times\frac{\langle \beta_1'(\overline{\mathbf{ x- y}})(\mathbf{y-t}),\beta_2 |\mathbf{R}_{n_1,n_2}\left(2-a-b\right)|\beta_1,(\overline{\mathbf{t-x}})(\mathbf{ x-y})\beta_2'\rangle}{\left(3-a-b+\frac{n_1+n_2}{2}\right)}\,.
\end{aligned}
\end{align}
It follows that
\begin{align}
\notag &\int d^4z\, \frac{\langle \beta'_1 |(\overline{\mathbf{t-y}})(\mathbf{ y-x}) (\overline{\mathbf{x-z}})\mathbf{(z-t)}| \beta_1\rangle^{n_1} \langle \beta_2 | (\overline{\mathbf{t-z}})(\mathbf{z-y})(\overline{\mathbf{y-x}})(\mathbf{ x-t})| \beta'_2\rangle^{n_2}}{(x-z)^{2a} (y-z)^{2b}(z-t)^{2(4-a-b)}}=\\= &\pi^2 \frac{\Gamma\left(2-a+\frac{n_1}{2}\right)\Gamma\left(2-b+\frac{n_2}{2}\right)\Gamma\left(a+b-2+\frac{n_2-n_1}{2}\right)}{\Gamma\left(a+\frac{n_1}{2}\right)\Gamma\left(b+\frac{n_2}{2}\right)\Gamma\left(3-a-b+\frac{n_2-n_1}{2}\right)}\frac{(-1)^{n_1}}{(x-y)^{2(a+b-2)}(x-t)^{2(2-b)}(y-t)^{2(2-a)}}\times\\& \times\frac{\langle \beta_1',\beta_2 |\mathbf{R}_{n_1,n_2}\left(2-a-b\right)|\beta_1,\beta_2'\rangle}{\left(3-a-b+\frac{n_1+n_2}{2}\right)}\,.\notag
\end{align}
We can redefine the spinors and write
\begin{align}
\notag &\int d^4z\, \frac{\langle \beta'_1 | (\overline{\mathbf{x-z}})\mathbf{(z-t)}| \beta_1\rangle^{n_1} \langle \beta_2 | (\overline{\mathbf{t-z}})(\mathbf{z-y})| \beta'_2\rangle^{n_2}}{(x-z)^{2a} (y-z)^{2b}(z-t)^{2(4-a-b)}}=\\= &\pi^2 \frac{\Gamma\left(2-a+\frac{n_1}{2}\right)\Gamma\left(2-b+\frac{n_2}{2}\right)\Gamma\left(a+b-2+\frac{n_2-n_1}{2}\right)}{\Gamma\left(a+\frac{n_1}{2}\right)\Gamma\left(b+\frac{n_2}{2}\right)\Gamma\left(3-a-b+\frac{n_2-n_1}{2}\right)}\frac{(-1)^{n_1}}{(x-y)^{2(a+b-2)}(x-t)^{2(2-b)}(y-t)^{2(2-a)}}\times\\& \times\frac{\langle \beta_1'(\overline{\mathbf{x-y}})(\mathbf{y-t}),\beta_2 |\mathbf{R}_{n_1,n_2}\left(2-a-b\right)|\beta_1,(\overline{\mathbf{t-x}})(\mathbf{ x-y})\beta_2'\rangle}{\left(3-a-b+\frac{n_1+n_2}{2}\right)}\,.\notag
\end{align}
We notice that in the numerator of the integrand the flow of $\sigma$ matrices is opposite in the two tensors: the one of degree $n_1$ starts from $x$ to $t$, while the other (of degree $n_2$) flows from $t$ to $y$. We can actually derive an identity for flows in the same direction, after a redefinition of spinors
\begin{equation}
|\beta_2 \rangle \to (\overline{\mathbf{x-y}})|\alpha_2\rangle \,,\,\,\,|\beta'_2 \rangle \to (\overline{\mathbf{x-y}})|\alpha'_2\rangle \,,
\end{equation}
in the chain rule identity \eqref{chain_proof}, so to obtain 
\begin{align}
\notag &\int d^4z\, \frac{\langle \beta'_1 |(\overline{\mathbf{y-x}})(\mathbf{x-z})| \beta_1\rangle^{n_1} \langle \alpha_2'|(\mathbf{x-y}) (\overline{\mathbf{y-z}})| \alpha_2\rangle^{n_2}}{(x-z)^{2a} (y-z)^{2b}}=\\= &\pi^2 \frac{\Gamma\left(2-a+\frac{n_1}{2}\right)\Gamma\left(2-b+\frac{n_2}{2}\right)\Gamma\left(a+b-2+\frac{n_2-n_1}{2}\right)}{\Gamma\left(a+\frac{n_1}{2}\right)\Gamma\left(b+\frac{n_2}{2}\right)\Gamma\left(3-a-b+\frac{n_2-n_1}{2}\right)}\frac{1}{(x-y)^{2(a+b-2)}}\times\\& \times\frac{\langle \beta'_1  ,\alpha_2' (\mathbf{ x-y})|\mathbf{R}_{n_1,n_2}\left(2-a-b\right)|\beta_1,(\overline{\mathbf{y-x}})\alpha_2\rangle}{\left(3-a-b+\frac{n_1+n_2}{2}\right)}\,.\notag
\end{align}
and the star-triangle follows by the usual inversion and translation 
\begin{align}
\notag &\int d^4z\, \frac{\langle \beta'_1 |(\overline{\mathbf{t-y}})(\mathbf{ y- x})(\overline{\mathbf{x-z}})(\mathbf{z-t})| \beta_1\rangle^{n_1} \langle \alpha_2'| (\mathbf{t-x})(\overline{\mathbf{x-y}})(\mathbf{y-z}) (\overline{\mathbf{z-t}})| \alpha_2\rangle^{n_2}}{(x-z)^{2a} (y-z)^{2b}(z-t)^{2(4-a-b)}}=\\ &=\pi^2 \frac{\Gamma\left(2-a+\frac{n_1}{2}\right)\Gamma\left(2-b+\frac{n_2}{2}\right)\Gamma\left(a+b-2+\frac{n_2-n_1}{2}\right)}{\Gamma\left(a+\frac{n_1}{2}\right)\Gamma\left(b+\frac{n_2}{2}\right)\Gamma\left(3-a-b+\frac{n_2-n_1}{2}\right)}\frac{(-1)^{n_1+n_2}}{(x-y)^{2(a+b-2)}(x-t)^{2(2-b)}(y-t)^{2(2-a)}}\times\\& \times\frac{\langle \beta'_1  ,\alpha_2' (\mathbf{t-x})(\overline{\mathbf{x-y}})(\mathbf{y-t})|\mathbf{R}_{n_1,n_2}\left(2-a-b\right)|\beta_1,(\overline{\mathbf{t-y}})(\mathbf{y-x})(\overline{\mathbf{x-t}})\alpha_2\rangle}{\left(3-a-b+\frac{n_1+n_2}{2}\right)}\,.\notag
\end{align}
By redefinition of spinors $|\beta_1'\rangle\to (\mathbf{\overline{t-y}})|\alpha_1\rangle$ and $|\alpha_2'\rangle \to \mathbf{({t-x})(\overline{x-y})}|\alpha_2'\rangle$ we can re-write the last identity as
\begin{align}
\notag &\int d^4z\, \frac{\langle \alpha_1 |(\mathbf{ y- x})(\overline{\mathbf{x-z}})(\mathbf{z-t})| \beta_1\rangle^{n_1} \langle \alpha_2'| (\mathbf{y-z}) (\overline{\mathbf{z-t}})| \beta_2\rangle^{n_2}}{(x-z)^{2a} (y-z)^{2b}(z-t)^{2(4-a-b)}}=\\= &\pi^2 \frac{\Gamma\left(2-a+\frac{n_1}{2}\right)\Gamma\left(2-b+\frac{n_2}{2}\right)\Gamma\left(a+b-2+\frac{n_2-n_1}{2}\right)}{\Gamma\left(a+\frac{n_1}{2}\right)\Gamma\left(b+\frac{n_2}{2}\right)\Gamma\left(3-a-b+\frac{n_2-n_1}{2}\right)}\frac{1}{(x-y)^{2(a+b-2)}(x-t)^{2(2-b)}(y-t)^{2(2-a)}}\times\\& \times\frac{\langle \beta_1'  ,\alpha_2' |\mathbf{R}_{n_1,n_2}\left(2-a-b\right)|(\mathbf{y-t})\beta_1,(\mathbf{y-x})(\overline{\mathbf{x-t}})\beta_2\rangle}{\left(3-a-b+\frac{n_1+n_2}{2}\right)}\,.\notag
\end{align}
\begin{figure}
\includegraphics[scale=0.47]{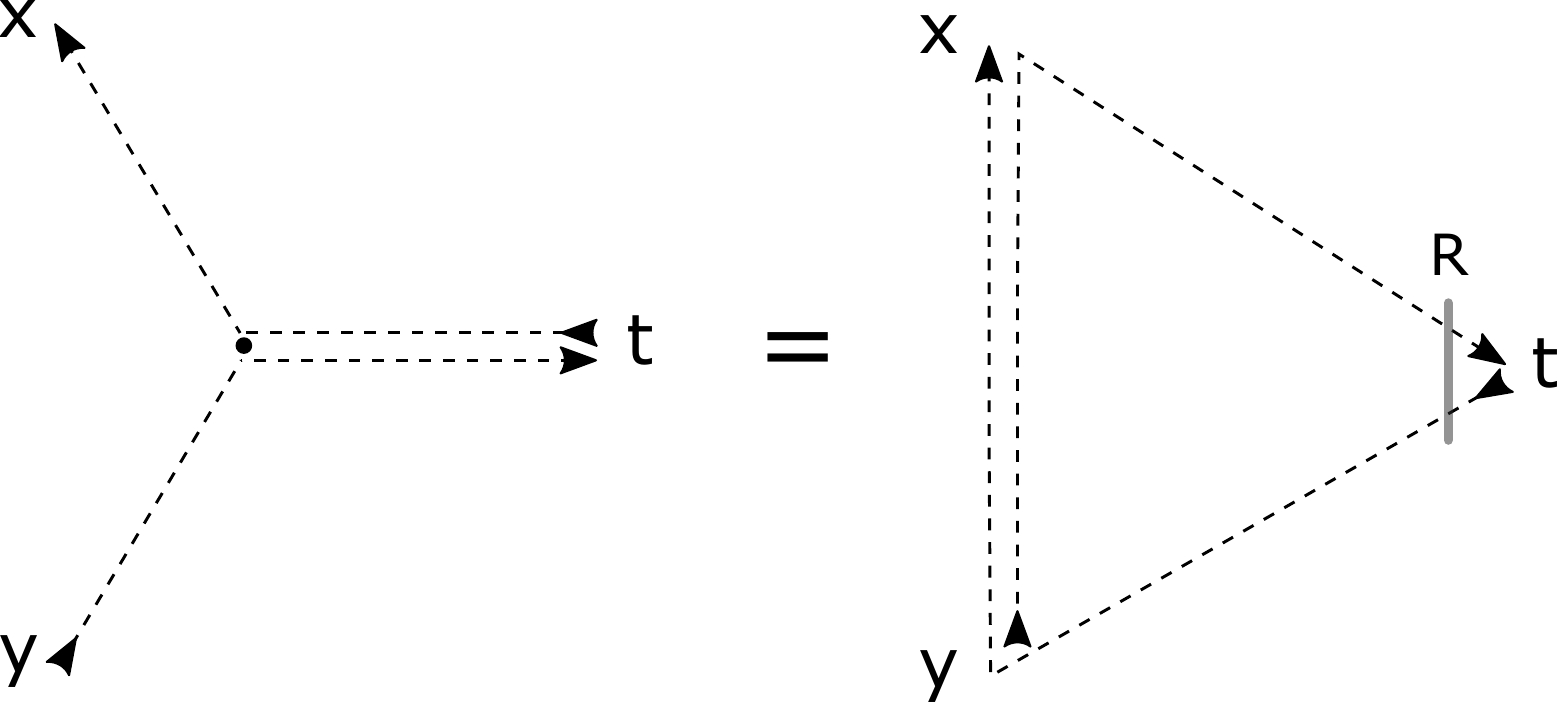}\hspace*{5mm}
\includegraphics[scale=0.47]{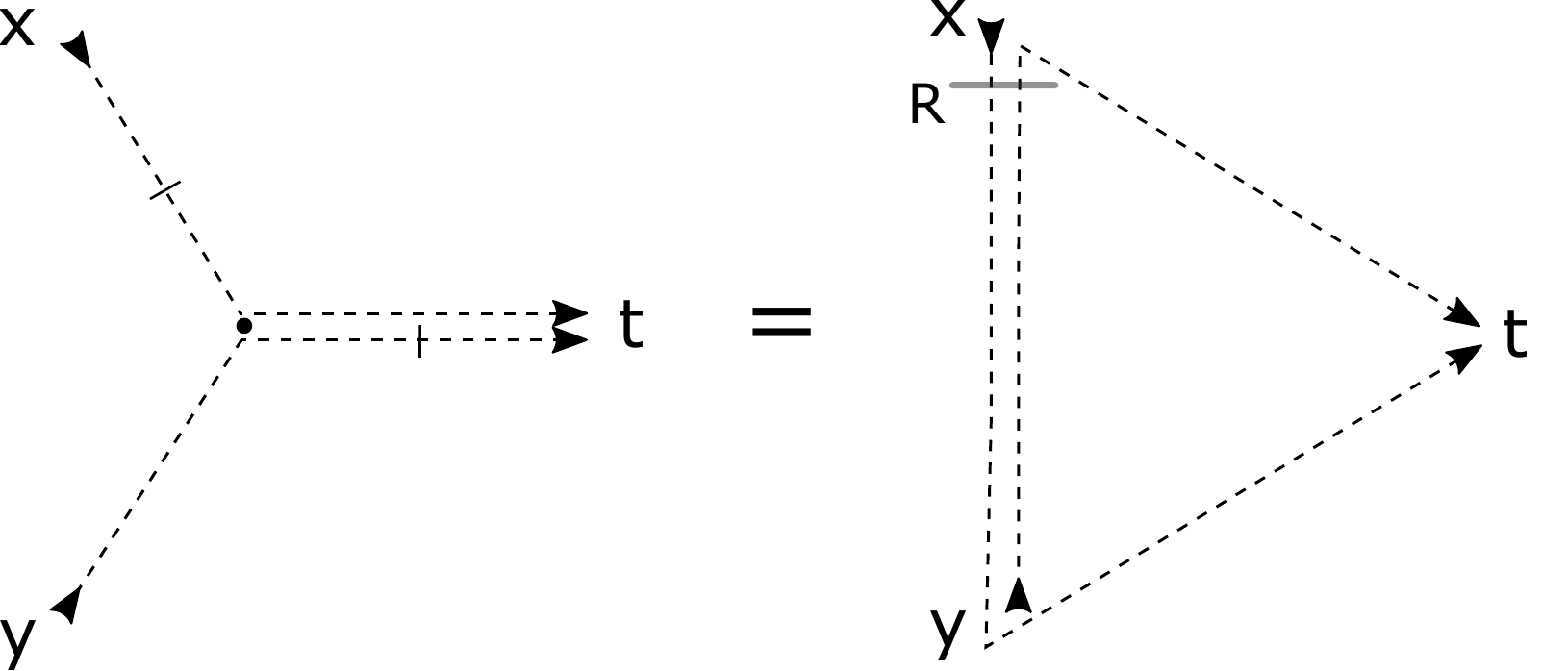}
\caption{Star-triangle relations for "opposite $\sig$ flow" and "same $\sigma$ flow", where the order of matrices $\sig$, $\bsig$ in the product is given by the arrows. The action of the $\mathbf R$-matrix is represented along the spinorial structure by a grey line.}
\end{figure}
Finally we consider some particular case for which the uniqueness and star-triangle relations simplify. First we set $n_2=0$ (or $n_1=0$) in \eqref{STR_inter}, obtaining the relation for one single polynomial in the numerator
\begin{align}
\notag &\int d^4z\, \frac{\langle \beta'_1 | (\overline{\mathbf{x-z}})\mathbf{(z-t)}| \beta_1\rangle^{n_1}}{(x-z)^{2a} (y-z)^{2b}(z-t)^{2(4-a-b)}}=\\= &\pi^2 \frac{\Gamma\left(2-a+\frac{n_1}{2}\right)\Gamma\left(2-b\right)\Gamma\left(a+b-2+\frac{n_1}{2}\right)}{\Gamma\left(a+\frac{n_1}{2}\right)\Gamma\left(b\right)\Gamma\left(4-a-b+\frac{n_1}{2}\right)}\frac{\langle \beta_1'|(\overline{\mathbf{x-y}})(\mathbf{ y-t}) |\beta_1\rangle^{n_1}}{(x-y)^{2(a+b-2)}(x-t)^{2(2-b)}(y-t)^{2(2-a)}}\,,
\end{align}
and setting further $n_1=0$ one recovers the well-known scalar star-triangle relation
\begin{align}
\notag &\int  \frac{d^4z}{(x-z)^{2a} (y-z)^{2b}(z-t)^{2(4-a-b)}}=\\&=\frac{\Gamma\left(2-a\right)\Gamma\left(2-b\right)\Gamma\left(a+b-2\right)}{\Gamma\left(a\right)\Gamma\left(b\right)\Gamma\left(4-a-b\right)}\frac{\pi^2}{(x-y)^{2(a+b-2)}(x-t)^{2(2-b)}(y-t)^{2(2-a)}}\,.
\end{align}

\subsection{Derivation of the exchange relation \eqref{exch_same}}
\label{app:exchange_I}
\begin{align}
\label{exch_proof_app} &\int d^4z\, \frac{[(\mathbf{x_0- z})(\mathbf{\overline{z-x}})(\mathbf{x-x_0'})]^{\ell} [(\mathbf{x_0-y}) (\mathbf{\overline{y-z}})(\mathbf{z-x_0'})]^{\ell'}}{(x_0-y)^{2(2-b')}(y-z)^{2(2-a')}(x_0-z)^{2a}(x-z)^{2b}(z-x_0')^{2(2-b')}(x-x_0')^{2(2-b)}}=\\&= \notag\frac{\Gamma\left(2-b+\frac{\ell}{2}\right)\Gamma\left(2-a+\frac{\ell}{2}\right)\Gamma\left(b'+\frac{\ell'}{2}\right)\Gamma\left(a'+\frac{\ell'}{2}\right)}{\Gamma\left(a+\frac{\ell}{2}\right)\Gamma\left(b+\frac{\ell}{2}\right)\Gamma\left(2-a'+\frac{\ell'}{2}\right)\Gamma\left(2-b'+\frac{\ell'}{2}\right)}\times\\&\times \int d^4z\, \frac{\mathbf{R}\,[\mathbf{(x_0-z)}(\overline{\mathbf{z-x}})(\mathbf{ x- x_0'})]^{\ell'}[(\mathbf{x_0-y})(\overline{\mathbf{y-z}})(\mathbf{z-x_0'})]^{\ell} \,\mathbf{R}^{-1}}{(x_0-y)^{2(2-b)}(y-z)^{2(2-a)}(x_0-z)^{2a'}(x-z)^{2b'}(z-x_0')^{2(2-b)}(x-x_0')^{2(2-b')}}\,,
\end{align}
where $\mathbf{R}=\mathbf{R}_{\ell,\ell'}(b-b')$.
The proof goes as follows: first we can focus on the contribution given by the triangle $(z,x,x_0')$, which according to the uniqueness identity can be rewritten as a star integral:
\begin{align}
&\frac{[(\overline{\mathbf{z-x}})(\mathbf{ x-x_0'})]^{\ell}[(\mathbf{ z-x_0'})]^{\ell'}}{(z-x_0')^{2(2-b')}(z-x)^{2b}(x-x_0')^{2(2-b)}}=\frac{[(\overline{\mathbf{ z- x}})(\mathbf{ x-x_0'})]^{\ell}[(\mathbf{z-x_0'})]^{\ell'}\mathbf{R} \,\mathbf{R}^{-1}}{(z-x_0')^{2(2-b')}(z-x)^{2b}(x-x_0')^{2(b'-b+2-b')}} =\\&= \frac{C_1}{(x-x_0')^{2(2-b')}} \int d^4 z' \frac{[(\overline{\mathbf{z-z'}})(\mathbf{z'-x_0'})]^{\ell}[(\mathbf{ z- z'})(\overline{\mathbf{z'- x}})(\mathbf{ x- x_0'})]^{\ell'}\,\mathbf{R}^{-1}}{(z-z')^{2(2+b-b')}(z'-x)^{2b'}(z'-x_0')^{2(2-b)}}\,,
\end{align}
at this point we can consider the star integral in $z$:
\begin{align}
&\int d^4 z\,\frac{[(\mathbf{x_0-z})(\overline{\mathbf{z-z'}})]^{\ell}[(\mathbf{ x_0-y})\mathbf{(\overline{ y - z})}\mathbf{(z-z')}]^{\ell'}}{(z-x_0)^{2a}(z-y)^{2(2-a')}(z-z')^{2(2+b-b')}(x_0-y)^{2(2-b'+b-b)}}=\notag \\&=C_2\,\frac{\mathbf{R}\, [(\mathbf{x_0-y})(\overline{\mathbf{y-z'}})]^{\ell}[\mathbf{(x_0-z')}]^{\ell'}}{(y-x_0)^{2(2-b)}(z'-y)^{2(2-a)}(x_0-z')^{2a'}}\,.
\end{align}
The coefficients $C_1$ and $C_2$ are given by the star-triangle as
\begin{equation}
C_1^{-1}=\frac{\pi^2(-1)^{\ell}\,A_{\ell,\ell'}(b,2-b')}{\left(1+b-b'+\frac{\ell+\ell'}{2}\right)\left(b'-b+\frac{\ell'-\ell}{2}\right)} \,,\,\text{and}\,\,C_2=\frac{\pi^2\,(-1)^{\ell} A_{\ell,\ell'}(a,2-a')}{\left(1-a+a'+\frac{\ell+\ell'}{2}\right)\left(a-a'+\frac{\ell'-\ell}{2}\right)} \,,
\end{equation}
so the resulting coefficient is
\begin{equation}
C_1 C_2 = \frac{\Gamma\left(2-b+\frac{\ell}{2}\right)\Gamma\left(2-a+\frac{\ell}{2}\right)\Gamma\left(b'+\frac{\ell'}{2}\right)\Gamma\left(a'+\frac{\ell'}{2}\right)}{\Gamma\left(a+\frac{\ell}{2}\right)\Gamma\left(b+\frac{\ell}{2}\right)\Gamma\left(2-a'+\frac{\ell'}{2}\right)\Gamma\left(2-b'+\frac{\ell'}{2}\right)}\,.
\end{equation}
\begin{center}
\begin{figure}
\includegraphics[scale=0.45]{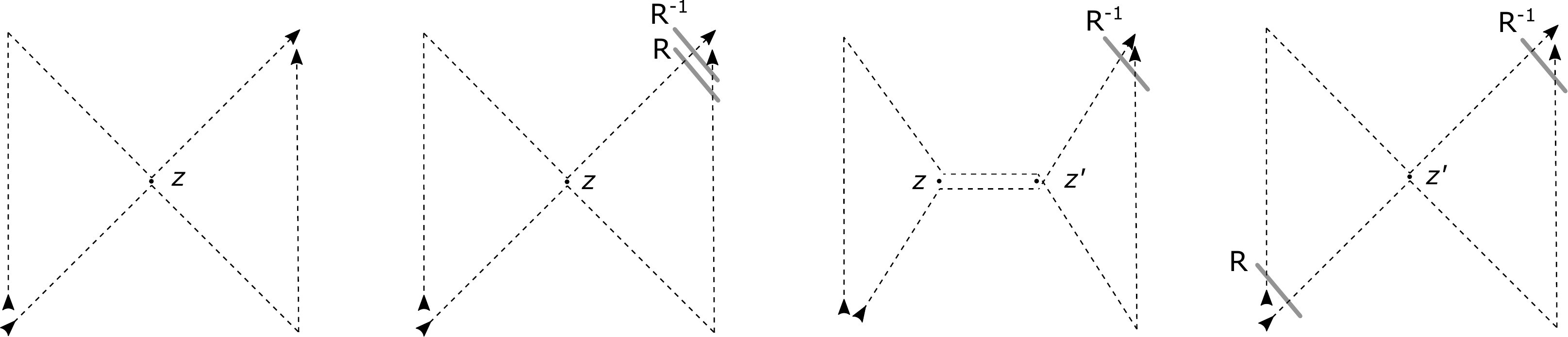}
\caption{The four steps of the proof of the exchange relation \eqref{exch_proof_app}, from the left to the right. Initial expression (1); insertion of identity in the space of spinors as $\mathbf{R}_{\ell,\ell'}(b-b') {\mathbf{R}_{\ell,\ell'}(b'-b)}$ (2); opening of the triangle on the right side into a star with integration in $z'$ (3); computation of the star integral in $z$ (4). }
\end{figure}
\end{center}
Finally, putting all pieces together, we are left with the r.h.s. of \eqref{exch_proof_app}
\begin{align}
\frac{C_1C_2}{(x-x_0')^{2(2-b')}}&\int d^4 z' \frac{\mathbf{R}\,[(\mathbf{x_0-y})(\overline{\mathbf{y- z'}})(\mathbf{z'-x_0'})]^{\ell}[\mathbf{(x_0-z')}(\overline{\mathbf{z'- x}})(\mathbf{ x- x_0'})]^{\ell'}\mathbf{R}^{-1}}{(y-x_0)^{2(2-b)}(z'-y)^{2(2-a)}(x_0-z')^{2a'}(z-z')^{2(2+b-b')}(z'-x)^{2b'}(z'-x_0')^{2(2-b)}}\,.
\end{align}

\subsection{Derivation of the exchange relation \eqref{exch_opp_1}}
\label{app:exchange_II}
\subsubsection{A new type of star-triangle identity }
In this section we provide the proof of another, not equivalent to \eqref{STRsame} or \eqref{STRopp}, star triangle relation which plays a crucial role in the proof of the exchange relation \eqref{exch_opp_1}. The integral identity we are going to prove holds under the uniqueness constraint ($a+b+c=4$), and reads
\begin{align}
\begin{aligned}
\label{STRRRR}
&\int d^4 v  \, \frac{\langle \lambda_1 |( x - v)
\overline{v}|\mu_1\rangle^{n}\,
\langle \mu_2 |(v-z)|\lambda_2\rangle^{m}}
{(x-v)^{2\left(a+\frac{n+m}{2}\right)}
v^{2\left(b+\frac{n}{2}\right)}
(v-z)^{2\left(c+\frac{m}{2}\right)}}\, =
\frac{\pi^2}{2^{n+m}}
\frac{\Gamma(2-a+\frac{n-m}{2})}{\Gamma(a+\frac{n+m}{2})}
\frac{\Gamma(2-b-\frac{n}{2})}{\Gamma(b+\frac{n}{2})}
\frac{\Gamma(2-c+\frac{m}{2})}{\Gamma(c+\frac{m}{2})}\\
&\frac{1}{z^{2\left( 2-a+\frac{n}{2}\right)}
x^{2\left(2-c+\frac{m}{2}\right)}}
\,\partial_s^n\partial_t^{m}\,
\frac{\beta^{a-2+\frac{m-n}{2}}}
{\left(x -ABz\right)^{2\left(2-b-\frac{n}{2}\right)}}
\end{aligned}
\end{align}
where
\begin{align*}
A_{\mu\nu} = \delta_{\mu\nu} +
s \langle \lambda_1|\sigma_{\mu}\overline{\sigma}_{\nu}|\mu_1\rangle
\ \ \ ; \ \ \
B_{\mu\nu} = \delta_{\mu\nu} +
t\langle \mu_2|\mathbf{z} \overline{\sigma}_{\nu}\sigma_{\mu}|\lambda_2\rangle
\end{align*}
and these matrices are orthogonal up to normalization
\begin{align*}
A\,A^{t} = \alpha\,\II \ \ ;\ \ A^{-1} = \alpha^{-1}\,A^{t}
\ \ ;\ \ \det A = \alpha^2
\ \ ;\ \ \alpha = 1+2s\langle\lambda_1|\mu_1\rangle\\
B\,B^{t} = \beta\,\II \ \ ;\ \ B^{-1} = \beta^{-1}\,B^{t}
\ \ ;\ \ \det B = \beta^2
\ \ ;\ \ \beta = 1+2t\langle\lambda_2|\mathbf{z}|\mu_2\rangle
\end{align*}
As usual we can represent the numerators through derivatives, so that the l.h.s. of \eqref{STRRRR} can be rewritten as
\begin{align*}
&\int d^4 v  \, \frac{\langle \lambda_1 |( x - v)
\overline{v}|\mu_1\rangle^{n}
\langle \mu_2 |(v-z)|\lambda_2\rangle^{m}}
{(x-v)^{2\left(a+\frac{n+m}{2}\right)}
v^{2\left(b+\frac{n}{2}\right)}
(v-z)^{2\left(c+\frac{m}{2}\right)}}\, = \frac{\Gamma(c-\frac{m}{2})}{2^m\Gamma(c+\frac{m}{2})}
\frac{\Gamma(a+\frac{m-n}{2})}{2^n\Gamma(a+\frac{m+n}{2})}\\
&
\partial_s^n\partial_t^{m}\,
\int d^4 v  \, \frac{1}
{\left(x-v-s a\right)^{2\left(a+\frac{m-n}{2}\right)}
v^{2\left(b+\frac{n}{2}\right)}
\left(v-z- t b \right)^{2\left(c-\frac{m}{2}\right)}}\,,
\end{align*}
where
\begin{align*}
a_{\mu} = \langle \lambda_1 |\sigma_{\mu}
\overline{v}|\mu_1\rangle \ \ ;\ \ b_{\mu} = \langle \mu_2 |\sigma_{\mu}|\lambda_2\rangle\,.
\end{align*}
Then we notice that
\begin{align*}
(x-v)_{\mu}-s\langle \lambda_1 |\sigma_{\mu}
\overline{v}|\mu_1\rangle =
x_{\mu}-A_{\mu\nu}v_{\nu}\ \ \ ;\ \ \
\left(x-A v\right)^{2} =
\alpha\left(A^{-1}x - v\right)^{2}\,,
\end{align*}
and now it is possible to use the standard scalar star-triangle relation
\begin{align*}
&\int d^4 v  \, \frac{1}
{\left(A^{-1}x - v\right)^{2\left(a+\frac{m-n}{2}\right)}
v^{2\left(b+\frac{n}{2}\right)}
\left(v-z -t b\right)^{2\left(c-\frac{m}{2}\right)}} =
\pi^2\frac{\Gamma(2-a+\frac{n-m}{2})}{\Gamma(a+\frac{m-n}{2})}
\frac{\Gamma(2-b-\frac{n}{2})}{\Gamma(b+\frac{n}{2})}
\frac{\Gamma(2-c+\frac{m}{2})}{\Gamma(c-\frac{m}{2})}
\times \\&\,\,\,\,\,\,\,\,\,\times \frac{1}{\left(A^{-1}x\right)^{2(2-c+\frac{m}{2})}
\left(z+t b \right)^{2(2-a+\frac{n-m}{2})}
\left(z +t b - A^{-1}x\right)^{2(2-b-\frac{n}{2})}}\,,
\end{align*}
so that
\begin{align*}
& \frac{\Gamma(c-\frac{m}{2})}{2^m\Gamma(c+\frac{m}{2})}
\frac{\Gamma(a+\frac{m-n}{2})}{2^n\Gamma(a+\frac{m+n}{2})} \partial_s^n\partial_t^{m}\,\frac{1}{\alpha^{a+\frac{m-n}{2}}}
\int d^4 v  \, \frac{1}
{\left(A^{-1}x - v\right)^{2\left(a+\frac{m-n}{2}\right)}
v^{2\left(b+\frac{n}{2}\right)}
\left(v-z -t b\right)^{2\left(c-\frac{m}{2}\right)}} 
= \\
&
\frac{\pi^2}{2^{n+m}}
\frac{\Gamma(2-a+\frac{n-m}{2})}{\Gamma(a+\frac{n+m}{2})}
\frac{\Gamma(2-b-\frac{n}{2})}{\Gamma(b+\frac{n}{2})}
\frac{\Gamma(2-c+\frac{m}{2})}{\Gamma(c+\frac{m}{2})}\,\times &\\
&\qquad\qquad\qquad\qquad\qquad\qquad\qquad\qquad \times \partial_s^n\partial_t^{m}\,
\frac{\alpha^{2-a-c+\frac{n}{2}}}{x^{2(2-c+\frac{m}{2})}
\left(z+t b\right)^{2(2-a+\frac{n-m}{2})}
\left(z-A^{-1}x\right)^{2(2-b-\frac{n}{2})}}\,.
\end{align*}
Finally, we can re-write 
\begin{equation*}
\left(z+t b- A^{-1}x\right)_{\mu} =
A^{-1}_{\mu\nu}\,\left(x -A\left(z+t b\right)\right)_{\nu} =
A^{-1}_{\mu\nu}\,\left(x_{\nu}-A_{\nu\beta}
\left(\delta_{\beta\gamma}+t|z|^{-1}
\langle\mu_2|\mathbf{z}\overline{\sigma}_{\gamma}
\sigma_{\beta}|\lambda_2\rangle\right)z_{\gamma}\right) \,,
\end{equation*}
and
\begin{equation*}
\left(z+t b\right)^{2} = \left(z_{\mu}+t\langle\mu_2|
\sigma_{\mu}|\lambda_2\rangle\right)^{2} =
z^2\left(1+2t|z|^{-1}\langle\mu_2|\mathbf{z}|\lambda_2\rangle\right)\,,
\end{equation*}
so that after the re-scaling $t \to t |z|$ we are left with the r.h.s. of \eqref{exch_opp_1}.
\subsubsection{Proof of the exchange relation}
In order to give a proof of the exchange relation \eqref{exch_opp_1} we need to start from the star-triangle \eqref{STRRRR} proved in the previous section
\begin{align*}
&\int d^4 v  \, \frac{\langle \lambda_1 |( x - v)
\overline{v}|\mu_1\rangle^{n}\,
\langle \mu_2 |(v-z)|\lambda_2\rangle^{m}}
{(x-v)^{2\left(a+\frac{n+m}{2}\right)}
v^{2\left(b+\frac{n}{2}\right)}
(v-z)^{2\left(c+\frac{m}{2}\right)}}\, =
\frac{\pi^2}{2^{n+m}}
\frac{\Gamma(2-a+\frac{n-m}{2})}{\Gamma(a+\frac{n+m}{2})}
\frac{\Gamma(2-b-\frac{n}{2})}{\Gamma(b+\frac{n}{2})}
\frac{\Gamma(2-c+\frac{m}{2})}{\Gamma(c+\frac{m}{2})}\\
&\frac{1}{z^{2\left( 2-a+\frac{n}{2}\right)}
x^{2\left(2-c+\frac{m}{2}\right)}}
\,\partial_s^n\partial_t^{m}\,
\frac{\beta^{a-2+\frac{m-n}{2}}}
{\left(x -ABz\right)^{2\left(2-b-\frac{n}{2}\right)}}\,,
\end{align*}
where
\begin{equation*}
A_{\mu\nu} = \delta_{\mu\nu} +
s \langle \lambda_1|\sigma_{\mu}\overline{\sigma}_{\nu}|\mu_1\rangle
\ \ \ ; \ \ \
B_{\mu\nu} = \delta_{\mu\nu} +
t\langle \mu_2|\mathbf{z} \overline{\sigma}_{\nu}\sigma_{\mu}|\lambda_2\rangle\,.
\end{equation*}
We can rewrite the last integral identity in a different form, using the following integral
\begin{align}
\label{replac}
\int d^4 v \frac{\delta^{(4)}\left(x - B v\right)}
{v^{2\left(2-c+\frac{m}{2}\right)}(v-Az)^{2\left(2-b-\frac{n}{2}\right)}} =
\frac{\beta^{a-2+\frac{m-n}{2}}}
{x^{2\left(2-c+\frac{m}{2}\right)}
\left(x- AB z\right)^{2\left(2-b-\frac{n}{2}\right)}}\,,
\end{align}
which is based on the general formula for the $\delta$-function
$$
\delta^{(4)}\left(x - B v\right) =
\frac{1}{\det B}\,\delta^{(4)}\left(B x - v\right)\,,
$$
where we recall that $\det B = \beta^2$, and $B$ is an orthogonal matrix $B^{T}=B^{-1}$ for which $A B= BA$:
\begin{align*}
\int d^4 v \frac{\delta^{(4)}\left(Bv-x\right)}
{v^{2\left(2-c+\frac{m}{2}\right)}(v-Az)^{2\left(2-b-\frac{n}{2}\right)}} =
\int d^4 v \frac{\beta^{-2}\,\delta^{(4)}\left(v-B^{-1} x\right)}
{v^{2\left(2-c+\frac{m}{2}\right)}(v-Az)^{2\left(2-b-\frac{n}{2}\right)}} = \\
\frac{\beta^{-2}}
{\left(B^{-1}x\right)^{2\left(2-c+\frac{m}{2}\right)}
\left(B^{-1}x-Az\right)^{2\left(2-b-\frac{n}{2}\right)}} =
\frac{\beta^{2-b-c+\frac{m-n}{2}}}
{x^{2\left(2-c+\frac{m}{2}\right)}
\left(x-B Az\right)^{2\left(2-b-\frac{n}{2}\right)}}\,.
\end{align*}
Thus, the star-triangle rewritten with the replacement \eqref{replac} reads
\begin{align*}
&\int d^4 v  \, \frac{\langle \lambda_1 |( x - v)
\overline{v}|\mu_1\rangle^{n}\,
\langle \mu_2 |(v-z)|\lambda_2\rangle^{m}}
{(x-v)^{2\left(a+\frac{n+m}{2}\right)}
v^{2\left(b+\frac{n}{2}\right)}
(v-z)^{2\left(c+\frac{m}{2}\right)}}\, =
\frac{\pi^2}{2^{n+m}}
\frac{\Gamma(2-a+\frac{n-m}{2})}{\Gamma(a+\frac{n+m}{2})}
\frac{\Gamma(2-b-\frac{n}{2})}{\Gamma(b+\frac{n}{2})}
\frac{\Gamma(2-c+\frac{m}{2})}{\Gamma(c+\frac{m}{2})}\\
&
\frac{1}{z^{2\left( 2-a+\frac{n}{2}\right)}}
\,\partial_s^n\partial_t^{m}\,
\int d^4 v \frac{\delta^{(4)}\left(x - B v\right)}
{v^{2\left(2-c+\frac{m}{2}\right)}(v-Az)^{2\left(2-b-\frac{n}{2}\right)}}\,.
\end{align*}
The next step is the convolution with an arbitrary propagator: we multiply both sides of obtained equality by $(y-x)^{-2\gamma}$ and integrate over $x$.
In the right hand side the integral is easily calculated due to delta-function and in the left hand side one should use the chain rule \eqref{chain_rule}, which in this case reads
\begin{align*}
\int d^4 x \frac{\langle \lambda_1 |( x - v)
\overline{v}|\mu_1\rangle^{n}}
{(y-x)^{2\gamma}(x-v)^{2\left(a+\frac{n+m}{2}\right)}} = \\
\pi^2
\frac{\Gamma(2-\gamma)}{\Gamma(\gamma)}
\frac{\Gamma(2-a+\frac{n-m}{2})}{\Gamma\left(a+\frac{n+m}{2}\right)}
\frac{\Gamma\left(a+\gamma+\frac{n+m}{2} -2\right)}
{\Gamma\left(4-a-\gamma+\frac{n-m}{2}\right)}
\frac{\langle \lambda_1 |( y - v)
\overline{v}|\mu_1\rangle^{n}}
{(y-v)^{2\left(a+\gamma+\frac{n+m}{2} -2\right)}}\,.
\end{align*}
After this convolution we have
\begin{align*}
&
\frac{\Gamma(2-\gamma)}{\Gamma(\gamma)}
\frac{\Gamma\left(a+\gamma+\frac{n+m}{2} -2\right)}
{\Gamma\left(4-a-\gamma+\frac{n-m}{2}\right)}
\int d^4 v  \, \frac{\langle \lambda_1 |( y - v)
\overline{v}|\mu_1\rangle^{n}\,
\langle \mu_2 |(v-z)|\lambda_2\rangle^{m}}
{(y-v)^{2\left(a+\gamma+\frac{n+m}{2}-2\right)}
v^{2\left(b+\frac{n}{2}\right)}
(v-z)^{2\left(c+\frac{m}{2}\right)}}\, =\\
&\frac{1}{2^{n+m}}
\frac{\Gamma(2-b-\frac{n}{2})}{\Gamma(b+\frac{n}{2})}
\frac{\Gamma(2-c+\frac{m}{2})}{\Gamma(c+\frac{m}{2})}
\frac{1}{z^{2\left( 2-a+\frac{n}{2}\right)}}
\,\partial_s^n\partial_t^{m}\,
\int d^4 v \frac{1}
{\left(y-Bv\right)^{2\gamma}v^{2\left(2-c+\frac{m}{2}\right)}
(v-Az)^{2\left(2-b-\frac{n}{2}\right)}}\,.
\end{align*}
Now we calculate derivatives using the rule
\begin{align}
\left.\partial_t^n \frac{1}{(x-y - t a)^{2A}}\right|_{t=0} = \frac{2^n\Gamma(A+n)}{\Gamma(A)}\,\frac{(x-y,a)^n}{(x-y)^{2(A+n)}}\,,
\end{align}
so that
\begin{align*}
&\left(y-Bv\right)_{\mu} = (y-v)_{\mu} -
t\langle \mu_2|\mathbf{z} \overline{v}\sigma_{\mu}|\lambda_2\rangle
\rightarrow \left.\partial_t^m
\frac{1}{\left(y-Bv\right)^{2\gamma}}\right|_{t=0} = \frac{2^m\Gamma(\gamma+m)}{\Gamma(\gamma)}\,
\frac{\langle \mu_2|\mathbf{z} \overline{v}(y-v)|\lambda_2\rangle^m}{(y-v)^{2(\gamma+m)}}\,,\\
&\left(v-Az\right)_{\mu} = (v-z)_{\mu} - s \langle \lambda_1|\sigma_{\mu}\overline{z}|\mu_1\rangle
\rightarrow
\left.\partial_t^n \frac{1}{\left(v-Az\right)^{2\left(2-b-\frac{n}{2}\right)}}\right|_{t=0} = \frac{2^n\Gamma\left(2-b+\frac{n}{2}\right)}{\Gamma\left(2-b-\frac{n}{2}\right)}
\,\frac{\langle \lambda_1|(v-z)\overline{z}|\mu_1\rangle^n}{(x-y)^{2\left(2-b+\frac{n}{2}\right)}}\,,
\end{align*}
and we obtain
\begin{align*}
&
\frac{\Gamma(2-\gamma)}{\Gamma(\gamma+m)}
\frac{\Gamma\left(a+\gamma+\frac{n+m}{2} -2\right)}
{\Gamma\left(4-a-\gamma+\frac{n-m}{2}\right)}
\int d^4 v  \, \frac{\langle \lambda_1 |( y - v)
\overline{v}|\mu_1\rangle^{n}\,
\langle \mu_2 |(v-z)|\lambda_2\rangle^{m}}
{(y-v)^{2\left(a+\gamma+\frac{n+m}{2}-2\right)}
v^{2\left(b+\frac{n}{2}\right)}
(v-z)^{2\left(c+\frac{m}{2}\right)}}\, =\\
&
\frac{\Gamma(2-b+\frac{n}{2})}{\Gamma(b+\frac{n}{2})}
\frac{\Gamma(2-c+\frac{m}{2})}{\Gamma(c+\frac{m}{2})}
\frac{1}{z^{2\left( 2-a+\frac{n}{2}\right)}}
\int d^4 v \frac{\langle \lambda_1|(v-z)\overline{z}|\mu_1\rangle^n\,
\langle \mu_2|\mathbf{z} \overline{v}(y-v)|\lambda_2\rangle^m}
{\left(y-v\right)^{2(\gamma+m)}v^{2\left(2-c+\frac{m}{2}\right)}
(v-z)^{2\left(2-b+\frac{n}{2}\right)}}\,.
\end{align*}
We can rewrite the last formula normalizing the numerators as
\begin{align*}
&
\int d^4 v  \, \frac{\langle \lambda_1 |\mathbf{( y - v)
\overline{v}}|\mu_1\rangle^{n}\,
\langle \mu_2 |\mathbf{(v-z)}|\lambda_2\rangle^{m}}
{(y-v)^{2\left(a+\gamma+\frac{m}{2}-2\right)}
v^{2\left(b\right)}
(v-z)^{2\left(c\right)}}\, =\,
\frac{C}{z^{2\left( 2-a\right)}}
\int d^4 v \frac{\langle \lambda_1|\mathbf{(v-z)
\overline{z}}|\mu_1\rangle^n\,
\langle \mu_2|\mathbf{z} \mathbf{\overline{v}(y-v)}|\lambda_2\rangle^m}
{\left(y-v\right)^{2(\gamma+m/2)}v^{2\left(2-c\right)}
(v-z)^{2\left(2-b\right)}}\,,
\end{align*}
where
\begin{equation}
C=\frac{\Gamma(2-b+\frac{n}{2})}{\Gamma(b+\frac{n}{2})}
\frac{\Gamma(2-c+\frac{m}{2})}{\Gamma(c+\frac{m}{2})}\frac{\Gamma(\gamma+m)}{\Gamma(2-\gamma)}
\frac{\Gamma\left(4-a-\gamma+\frac{n-m}{2}\right)}{\Gamma\left(a+\gamma+\frac{n+m}{2} -2\right)}
\,.
\end{equation}
As a last step, we have to perform some conformal transformations in order to bring the exchange relation to its final form.
Firstly we translate the points $(y,z,v)$ by $x$, that is
\begin{align*}
&
\int d^4 v  \, \frac{\langle \lambda_1 |\mathbf{( y - v)
(\overline{v-x})}|\mu_1\rangle^{n}\,
\langle \mu_2 |\mathbf{(v-z)}|\lambda_2\rangle^{m}}
{(y-v)^{2\left(a+\gamma+\frac{m}{2}-2\right)}
(v-x)^{2\left(b\right)}
(v-z)^{2\left(c\right)}}\, =\\
&
\frac{C}{(z-x)^{2\left( 2-a\right)}}
\int d^4 v \frac{\langle \lambda_1|\mathbf{(v-z)(\overline{z-x})}|\mu_1\rangle^n\,
\langle \mu_2|\mathbf{(z-x)}\mathbf{(\overline{v-x})(y-v)}|\lambda_2\rangle^m}
{\left(y-v\right)^{2(\gamma+m/2)}(v-x)^{2\left(2-c\right)}
(v-z)^{2\left(2-b\right)}}\,,
\end{align*}
and secondly we perform the inversion of $y,z,v,x$ around the origin
\begin{align*}
&
\int d^4 v  \, \frac{\langle \lambda_1 |\mathbf{y(\overline{y-v})
(v-x)\overline{x}}|\mu_1\rangle^{n}\,
\langle \mu_2 |\mathbf{z(\overline{v-z})v}|\lambda_2\rangle^{m}}
{(y-v)^{2\left(a+\gamma+\frac{m}{2}-2\right)}
(v-x)^{2\left(b\right)}
(v-z)^{2\left(c\right)}v^{2\left(2-\gamma-m/2\right)}}\, = y^{2(2-a)}\times\\
&
\frac{C}{(z-x)^{2\left( 2-a\right)}}
\int d^4 v \frac{\langle \lambda_1|\mathbf{v(\overline{v-z})(z-x)\overline{x}}|\mu_1\rangle^n\,
\langle \mu_2|\mathbf{z(\overline{\mathbf{z-x}}) 
(v-x)(\overline{y-v})y}|\lambda_2\rangle^m}
{\left(y-v\right)^{2(\gamma+m/2)}(v-x)^{2\left(2-c\right)}
(v-z)^{2\left(2-b\right)}v^{2\left(c+b-\gamma-\frac{m}{2}\right)}}\,.
\end{align*}
We can rewrite the last relation without explicit spinors, and perform a last translation of all points by the vector $w$, so to obtain
\begin{align*}
&
\frac{1}{(y-w)^{2(2-a)}}\int d^4 v  \, \frac{[\mathbf{(w-y)(\overline{y - v})
(v-x)}]^{n}\,
[\mathbf{(\overline{z-v})(v-w)(\overline{w-y})}]^{m}}
{(y-v)^{2\left(a+\gamma+\frac{m}{2}-2\right)}
(v-x)^{2\left(b\right)}
(v-z)^{2\left(c\right)}(v-w)^{2\left(2-\gamma-m/2\right)}}\, = \\
&
\frac{C}{(z-x)^{2\left( 2-a\right)}}
\int d^4 v \frac{[\mathbf{(w-v)(\overline{v-z})(z-x)}]^n\,
[\mathbf{(\overline{\mathbf{z-x})} 
(x-v)(\overline{v-y})}]^m}
{\left(y-v\right)^{2(\gamma+m/2)}(v-x)^{2\left(2-c\right)}
(v-z)^{2\left(2-b\right)}(v-w)^{2\left(c+b-\gamma-\frac{m}{2}\right)}}\,.
\end{align*}
\section{Orthogonality at $N=3$}
\label{app:N3scalar}
In the case $N=3$ the iterative application of the relation \eqref{exchorto1} and then use of the formula \eqref{N1scalar} gives
\begin{align*}
&\left(\bar{\Lambda}^{(1)}_{Y'_1}\right)_{\mathbf a_{1}}^{\dot{\mathbf a}_{1}}
\left(\bar{\Lambda}^{(2)}_{Y'_2}\right)_{\mathbf a_{2}}^{\dot{\mathbf a}_{2}}
\left(\bar{\Lambda}^{(3)}_{Y'_3} \right)_{\mathbf a_{3}}^{\dot{\mathbf a}_{3}}\,
\left(\Lambda^{(3)}_{Y_3}\right)^{\mathbf b_3}_{\dot{\mathbf b}_3} =
\lambda\left(Y_3,Y'_3\right)\,\lambda\left(Y_3,Y'_2\right)\,\\
&\mathbf{R}^{\dot{\mathbf{a}}_3\dot{\mathbf{s}}}_{\dot{\mathbf{c}_3}
\dot{\mathbf{b}}_3} (Y'_3,Y_3)\,
\mathbf{R}^{\dot{\mathbf{a}}_2\dot{\mathbf{s}}_1}_{\dot{\mathbf{c}_2}
\dot{\mathbf{s}}} (Y'_2,Y_3)\,
\mathbf{R}^{\mathbf{c}_2\mathbf{s}}_{\mathbf{a}_2\mathbf{s}_1} (Y_3,Y'_2)\,
\mathbf{R}^{\mathbf{c}_3\mathbf{b}_3}_{\mathbf{a}_3\mathbf{s}} (Y_3,Y'_3)\,
\underline{\left(\bar{\Lambda}^{(1)}_{Y'_1}\right)_{\mathbf a_{1}}^{\dot{\mathbf a}_{1}}
\left(\Lambda^{(1)}_{Y_3}\right)^{{\mathbf s}_1}_{\dot{\mathbf s}_1}}
\left(\bar{\Lambda}^{(1)}_{Y'_2}\right)_{\mathbf c_{2}}^{\dot{\mathbf c}_{2}}
\left(\bar{\Lambda}^{(2)}_{Y'_3}\right)_{\mathbf c_{3}}^{\dot{\mathbf c}_{3}} \\
& = \lambda\left(Y_3,Y'_3\right)\,\lambda\left(Y_3,Y'_2\right)\,
\frac{2 \pi^3}{\ell_3+1}\,\delta_{\ell_3,\ell^{\prime}_1}\,\delta(\nu_3-\nu'_1)\,\\
&\mathbf{R}^{\dot{\mathbf{a}}_3\dot{\mathbf{s}}}_{\dot{\mathbf{c}_3}
\dot{\mathbf{b}}_3} (Y'_{3},Y'_{1})\,
\mathbf{R}^{\dot{\mathbf{a}}_2\dot{\mathbf{a}}_1}_{\dot{\mathbf{c}_2}
\dot{\mathbf{s}}} (Y'_{2},Y'_{1})\,
\mathbf{R}^{\mathbf{c}_2\mathbf{s}}_{\mathbf{a}_2\mathbf{a}_1} (Y'_{1},Y'_{2})\,
\mathbf{R}^{\mathbf{c}_3\mathbf{b}_3}_{\mathbf{a}_3\mathbf{s}} (Y'_{1},Y'_{3})\,\left(\bar{\Lambda}^{(1)}_{Y'_2}\right)_{\mathbf c_{2}}^{\dot{\mathbf c}_{2}}
\left(\bar{\Lambda}^{(2)}_{Y'_3}\right)_{\mathbf c_{3}}^{\dot{\mathbf c}_{3}}
\end{align*}
We see that everything is reduced to the case $N=2$ so that
it is possible to use the $N=2$ orthogonality to obtain
\begin{align*}
&\left(\bar{\Lambda}^{(1)}_{Y'_1}\right)_{\mathbf a_{1}}^{\dot{\mathbf a}_{1}}
\left(\bar{\Lambda}^{(2)}_{Y'_2}\right)_{\mathbf a_{2}}^{\dot{\mathbf a}_{2}}
\left(\bar{\Lambda}^{(3)}_{Y'_3} \right)_{\mathbf a_{3}}^{\dot{\mathbf a}_{3}}\,
\left(\Lambda^{(3)}_{Y_3}\right)^{\mathbf b_3}_{\dot{\mathbf b}_3}
\left(\Lambda^{(2)}_{Y_2}\right)^{\mathbf b_2}_{\dot{\mathbf b}_2}
\left(\Lambda^{(1)}_{Y_1}\right)^{\mathbf b_1}_{\dot{\mathbf b}_1} =
\lambda\left(Y_3,Y'_3\right)\lambda\left(Y_3,Y'_2\right)\,
\frac{2 \pi^3}{\ell_3+1}\,\delta_{\ell_3,\ell^{\prime}_1}\,\delta(\nu_3-\nu'_1)\,\\
&\mathbf{R}^{\dot{\mathbf{a}}_3\dot{\mathbf{s}}}_{\dot{\mathbf{c}_3}
\dot{\mathbf{b}}_3} (Y'_{3},Y'_{1})\,
\mathbf{R}^{\dot{\mathbf{a}}_2\dot{\mathbf{a}}_1}_{\dot{\mathbf{c}_2}
\dot{\mathbf{s}}} (Y'_{2},Y'_{1})\,
\mathbf{R}^{\mathbf{c}_2\mathbf{s}}_{\mathbf{a}_2\mathbf{a}_1} (Y'_{1},Y'_{2})\,
\mathbf{R}^{\mathbf{c}_3\mathbf{b}_3}_{\mathbf{a}_3\mathbf{s}} (Y'_{1},Y'_{3})\,\\
&(2 \pi^3)^2\,\frac{\delta_{\ell_2,\ell'_2}\delta_{\ell_1,\ell'_3}}
{(\ell_1+1)(\ell_2+1)}\,
\delta(\nu_2-\nu'_2)\delta(\nu_1-\nu'_3)\,
\lambda(Y_2,Y_1)\,
\mathbf{R}^{\dot{\mathbf{c}}_3\dot{\mathbf{c}}_2}_{\dot{\mathbf{b}_1}
\dot{\mathbf{b}}_2} (Y'_{3},Y'_{2})\,\mathbf{R}^{\mathbf{b}_1\mathbf{b}_2}_{\mathbf{c}_3\mathbf{c}_2} (Y'_{2},Y'_{3}) = \\
&(2 \pi^3)^3\,\frac{\delta_{\ell_3,\ell'_1}\delta_{\ell_2,\ell'_2}\delta_{\ell_1,\ell'_3}}
{(\ell_1+1)(\ell_2+1)(\ell_3+1)}\,
\delta(\nu_3-\nu'_1)\delta(\nu_2-\nu'_2)\delta(\nu_1-\nu'_3)\,
\lambda\left(Y_3,Y_1\right)\,\lambda\left(Y_3,Y_2\right)\,\lambda(Y_2,Y_1)\\
&\mathbf{R}^{\dot{\mathbf{a}}_3\dot{\mathbf{s}}}_{\dot{\mathbf{c}_3}
\dot{\mathbf{b}}_3} (Y'_{3},Y'_{1})\,
\mathbf{R}^{\dot{\mathbf{a}}_2\dot{\mathbf{a}}_1}_{\dot{\mathbf{c}_2}
\dot{\mathbf{s}}} (Y'_{2},Y'_{1})\,
\mathbf{R}^{\dot{\mathbf{c}}_3\dot{\mathbf{c}}_2}_{\dot{\mathbf{b}_1}
\dot{\mathbf{b}}_2} (Y'_{3},Y'_{2})\,
\mathbf{R}^{\mathbf{c}_2\mathbf{s}}_{\mathbf{a}_2\mathbf{a}_1} (Y'_{1},Y'_{2})\,
\mathbf{R}^{\mathbf{c}_3\mathbf{b}_3}_{\mathbf{a}_3\mathbf{s}} (Y'_{1},Y'_{3})\,\mathbf{R}^{\mathbf{b}_1\mathbf{b}_2}_{\mathbf{c}_3\mathbf{c}_2} (Y'_{2},Y'_{3})
\end{align*}
These examples demonstrate the general structure
of the final expression and suggest the clear iterative construction of eigenfunction for any size $N$ of the system.
\bibliographystyle{nb}
\bibliography{SoV_biblio}

\end{document}